\documentclass[5p,times]{elsarticle}

\usepackage{tabularx}
\usepackage{graphicx}
\usepackage{latexsym}
\usepackage{amsmath,bm}
\usepackage{txfonts}
\usepackage{helvet}
\usepackage{braket}
\usepackage{amscd}
\usepackage{longtable}
\usepackage{epsfig}
\usepackage{lipsum}
\usepackage[dvipsnames]{xcolor}
\usepackage{textgreek}
\usepackage{mathtools}
\usepackage{cuted}
\usepackage{caption}
\usepackage{siunitx}
  
\usepackage{tikz}
\usetikzlibrary{calc,arrows}

\newcommand{\tikzmark}[1]{%
\tikz[overlay,remember picture] \node (#1) {};}

\usepackage[hidelinks]{hyperref}
\hypersetup{
  colorlinks   = true, %Colours links instead of ugly boxes
  urlcolor     = blue, %Colour for external hyperlinks
  linkcolor    = black, %Colour of internal links
  citecolor   = red %Colour of citations
}

\usepackage[english]{babel}
\addto\extrasenglish{%
\def\sectionautorefname~#1\null{Sec.\,#1\null}
\def\subsectionautorefname~#1\null{Sec.\,#1\null}
\def\subsubsectionautorefname~#1\null{Sec.\,#1\null}
\def\equationautorefname~#1\null{Eq.\,(#1)\null}
\def\appendixautorefname~#1\null{#1\null}
\def\figureautorefname~#1\null{Fig.\,#1\null}
} 

\newcommand{\abs}[1]{|#1|}

\renewcommand{\vec}[1]{\mathbf{#1}}

\newcommand{\oper}[1]{\hat{\bm{\mathrm{#1}}}}
\newcommand{\sixJ}[6]{%
    \begin{Bmatrix}
        #1 & #2 & #3 \\
        #4 & #5 & #6
    \end{Bmatrix}
}
\newcommand{\FA}{F_{\rm XA}}
\newcommand{\FB}{F_{\rm B}}
\newcommand{\KA}{K_{\rm A}}
\newcommand{\KB}{K_{\rm B}}
\newcommand{\KAB}{K_{\rm AB}}

\usepackage[intoc]{nomencl}
\makenomenclature

\journal{Progress in Nuclear Magnetic Resonance}
\begin{document}

\begin{frontmatter}

\title{Zero- to ultralow-field nuclear magnetic resonance}
%\title{A review article that will never be published}

\author[1,2]{Danila~A.~Barskiy}
		\address[1]{Helmholtz-Institut Mainz, 55099 Mainz, Germany; GSI Helmholtzzentrum f{\"u}r Schwerionenforschung GmbH, 64291 Darmstadt, Germany}
		\address[2]{Johannes Gutenberg-Universit{\"a}t  Mainz, 55099 Mainz, Germany}
		
\author[3]{John~W.~Blanchard}%\corref{cor1}
		
		\address[3]{Quantum Technology Center, University of Maryland, College Park, MD 20742, USA}

\author[1,2,4]{Dmitry~Budker}
		%\address[1]{Helmholtz-Institut Mainz, GSI Helmholtzzentrum f{\"u}r Schwerionenforschung, 55128 Mainz, Germany}
		%\address[2]{Johannes Gutenberg-Universit{\"a}t  Mainz, 55099 Mainz, Germany}
		\address[4]{Department of Physics, University of California, Berkeley, CA 94720-7300 USA}

\author[5]{James~Eills}%\9orref{cor1}
		
		\address[5]{Institute for Bioengineering of Catalonia, 08028 Barcelona, Spain}

\author[6]{Szymon~Pustelny}
        \address[6]{Jagiellonian University in Krak\'ow, 30-348 Krak\'ow, Poland}

\author[7]{Kirill~F.~Sheberstov}
		\address[7]{Laboratoire des Biomolécules, LBM, Département de Chimie, École Normale Supérieure, PSL University, Sorbonne Université, CNRS, 75005 Paris, France}
		
\author[8]{Michael~C.~D.~Tayler} %\email{michael.tayler@icfo.eu}
        \address[8]{ICFO-Institut de Ci\`encies Fot\`oniques, The Barcelona Institute of Science and Technology, 08860 Castelldefels (Barcelona), Spain}

\author[9]{Andreas~H.~Trabesinger}
    \address[9]{Brown Boveri Platz 4, 5400 Baden, Switzerland}

\date{\today}

\onecolumn
Dedication: Dedicated to Prof. Alexander Pines (1945--2024), pioneer of zero-field NMR and extraordinary mentor who continues to inspire our work.
\twocolumn

\begin{abstract}

%Zero- and ultralow-field nuclear magnetic resonance (ZULF NMR) is a form of NMR spectroscopy conducted in magnetic fields at which the spin--spin interactions within molecules and materials dominate the Zeeman shifts. 
%\AT{Zero and ultralow-field nuclear magnetic resonance (ZULF NMR) is a form of NMR spectroscopy \DB{This leaves out imaging. I propose to use Alex Pines' word ``modality'': ... is an NMR modality where...} where experiments are performed in fields at which spin--spin interactions within molecules and materials are stronger than Zeeman interactions.}
Zero and ultralow-field nuclear magnetic resonance (ZULF NMR) is an NMR modality where experiments are performed in fields at which spin--spin interactions within molecules and materials are stronger than Zeeman interactions.
This typically occurs at external fields of microtesla strength or below, considerably smaller than Earth's field.
In ZULF NMR, the measurement of spin--spin couplings and spin relaxation rates provides a nondestructive means for identifying chemicals and chemical fragments, and for conducting sample or process analyses. 
The absence of the symmetry imposed by a strong external magnetic field enables experiments that exploit terms in the nuclear spin Hamiltonian that are suppressed in high-field NMR, which in turn opens up new capabilities in a broad range of fields, from the search for dark matter to the preparation of hyperpolarized contrast agents for clinical imaging. 
Furthermore, as in ZULF NMR the Larmor frequencies are typically in the audio band, the nuclear spins can be manipulated with d.c. magnetic field pulses, and 
highly sensitive magnetometers are used for detection.
In contrast to high-field NMR, the low-frequency signals readily pass through conductive materials such as metals, and heterogeneous samples do not lead to resonance line broadening, meaning that high-resolution spectroscopy is possible.
Notable practical advantages of ZULF NMR spectroscopy are the low cost and relative simplicity and portability of the spectrometer system.
%, and the portability that has allowed for proof-of-principle ZULF NMR measurements for oil-well exploration and in chemical labs for in-line sample quality control.
In recent years ZULF NMR has become more accessible, thanks to improvements in magnetometer sensitivity and commercial availability, and the development of hyperpolarization methods that provide a simple means to boost signal strengths by several orders of magnitude. 
%\AT{Here we review these topics and present a perspective on potential future avenues of ZULF-NMR research.} \DB{A ``passive'' variant: These topics and reviewed and a perspective on potential future avenues of ZULF-NMR research are presented. (No ``Here'' or ``we''.)}
These topics are reviewed and a perspective on potential future avenues of ZULF-NMR research is presented.

\end{abstract}


\end{frontmatter}

\tableofcontents

% \newpage

% \begin{table}[h]
% \centering
% \begin{tabular}{|@{\hspace{1cm}}c@{\hspace{1cm}}|@{\hspace{1cm}}c@{\hspace{1cm}}|}
% \hline
% Date & Time \\ [0.5ex] 
% \hline \hline
%  & \\
% \hline
% \end{tabular}
% \caption{A reminder of our upcoming meetings, German time.}
% \label{table:0}
% \end{table}

% \begin{table}[h]
% \centering
% \begin{tabular}{|@{\hspace{1cm}}l@{\hspace{1cm}}|@{\hspace{1cm}}l@{\hspace{1cm}}|}
% \hline
% Section & Responsible \\ [0.5ex] 
% \hline\hline
% %\autoref{Subsec:AMR} & DB\\
% %\autoref{Subsec:Radioactive} & DAB/DB?\\
% % \autoref{Subsec:Relative} & JE/JB/all\\
% % \autoref{Subsec:Geometry} & JB\\
% %\autoref{Subsec:2D} & AT\\
% \autoref{Subsubsec:DNPvivo} & DAB\\
% % \autoref{subsec:LLSS} &KS\\
% % \autoref{Subsec:Partial} &KS\\
% \autoref{Subsec:Batteries} & AT/DB\\
% %\autoref{Subsubsec:quantumsim} & AT\\
% %\autoref{Sec:Conclusions} & DB/all\\
% \hline
% \end{tabular}
% \caption{Sections that need more text or substantial rewriting.}
% \end{table}

%\begin{figure}
%    \centering
%    \includegraphics{schedule.png}
%    \addtocounter{figure}{-1}
%    \caption{A reminder of our upcoming meetings, German time.}
%    \label{fig:my_label}
%\end{figure}

\section{Introduction}
\label{Sec:Intro}

\subsection{What is ZULF NMR and what it is good for}
\label{Subsec:Why}

In the context of nuclear magnetic resonance (NMR) spectroscopy, we define the zero- to ultralow-field (ZULF) regime as the magnetic field range in which `internal' spin interactions dominate `external' ones. That is, in ZULF NMR couplings to magnetic fields originating from the sample itself are stronger than couplings to magnetic fields generated by the experimental apparatus \cite{Blanchard2016emagres,Blanchard2021_LtL,JIANG2021ZULF}. 
As a direct consequence, and in stark contrast to `conventional' NMR experiments, in the ZULF regime nuclear spin dynamics are not dominated by the imposed symmetry %\SP{symmetry or geometry?} \AT{I think that it should be symmetry. (Dima agrees)} 
 of a large magnetic field. 

Operation under the ZULF conditions means in particular that all information encoded in the interaction tensors is preserved.
% Hamiltonians are not truncated by the imposed symmetry of a large magnetic field. This means in particular that all information encoded in the interaction tensors is preserved.
The access to terms of the interaction tensors that are suppressed---or truncated---in high-field spectra opens up unique ways to exploring spin physics that remains largely inaccessible in high-field NMR (see Sec.\,\ref{Subsec:Untruncated}). We consider this feature to be the main motivation for research in this traditionally neglected corner of the NMR landscape, and the driver of the renewed interest in it since the turn of the millennium. 

Progress in ZULF NMR has been fuelled by partially independent developments that have substantially changed how ZULF NMR experiments are performed. Advances have been made in two areas in particular. First, versatile external sources of nuclear spin polarization have become available (see Sec.\,\ref{Sec:Spin_pol}), removing the dependence on a strong magnetic field to polarize the spin ensembles. Second, there have been major developments in magnetometry, providing detection modalities that are not limited by a direct dependence of signal strength on precession frequency, and therefore external field strength, as traditional inductive detection does (see Sec.\,\ref{Sec:Detection}).

Beyond access to untruncated interaction Hamiltonians, further motivation for ZULF NMR arises from advantages traditionally associated with NMR at `low field'. 
%\SP{I think we should use stronger arguments here, as the technical arguments are not tempting enough (for me). We need to strengthen this part with new capabilities (can we mentiong here some of the arguments promised by Anderas in abstract?) MAybe we should even reverse the order of this and next apragraph} \AT{My suggestion is to keep the focus on the ZULF regime, where the unique selling point, if you will, is access to the full interaction tensors. The arguments regarding decreased technological complexity, increased skin depth and so apply equally for almost all low-field experiments, I would think. However, I agree that the narrative could be strengthened by mentioning that access to all terms in the interaction tensor enables searches for beyond-SM physics and the preparation of hyperpolarized agents.} 
These include decreased complexity and increased portability of spectrometers that do not require a strong magnet at their core. In addition, many applications benefit from the absence of line broadening induced by magnetic susceptibility, which is especially advantageous for studying inhomogeneous samples. Moreover, in the typical frequency range of ZULF NMR experiments---hundreds of hertz and below---the skin effect is negligible for most materials, 
%including electrically conductive ones, 
so that for example probing through metal enclosures becomes possible.

The elephant-in-the-room question is, however, what spectroscopic insight can be gained in a regime where chemical-shift effects are negligible, by definition (see Sec.\,\ref{Subsec:Regimes}). Chemical shifts dominate the spectroscopic content in high-field liquid-state NMR, but have no role in ZULF NMR. However, the indirect spin--spin (or, $J$-coupling) interaction is independent of external field, and therefore is present even at zero field. This interaction is exquisitely sensitive to changes in local molecular geometry, conformation, and electronic structure \cite{Blanchard2016emagres,Blanchard2021_LtL,JIANG2021ZULF}. Measuring $J$-coupling values with high spectral resolution provides therefore valuable physicochemical information in the ZULF regime, and
%\AT{
%The information content of ZULF-NMR spectra is distinct from that of its high-field counterparts. Information about the local electronic environment of the nuclei, which often reflect chemical function, is lost in ZULF-NMR experiments. \DB{I am not sure, it is fully lost. It is still there in the strength of J-coupling, no?}  \MCDT{I agree with Dima.  There is probably a better way of rephrasing this, however, I feel it is a redundant sentence that adds nothing to what has already been said earlier in the paragraph.  I suggest deleting it.}
%\MCDT{keep this sentence though: }
the spectra will depend most sensitively on the topology of the $J$-coupling network (see also \autoref{Subsec:Notations}) and changes thereof, for example due to chemical reactions (see \autoref{Sec:Chemistry}).
%}

In the present review, when we discuss spectroscopic capabilities, we mainly refer to \textit{J-spectroscopy} \cite{Lin2018} in the liquid state, whereas we largely ignore solid-state NMR and nuclear quadrupole resonance (NQR). The latter is sometimes referred to as the original zero-field NMR and is a separate well-developed field that we do not cover here.

\subsection{Magnetic field regimes}
\label{Subsec:Regimes}

\begin{figure}[t]
\centering
	\includegraphics[width=0.9\columnwidth]{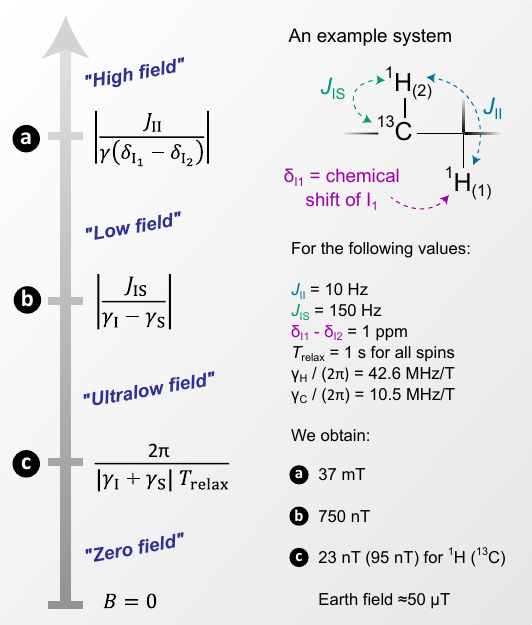}
	\vspace{-6pt}
	\caption{Practical definitions of the different field regimes. The boundaries between regimes vary significantly depending on the system under study.}%\DAB{NEEDS UPDATE} \DAB{Change (a) and (b) to $2*\pi*J$ to match the text. (c) no change.}
%\SP{Is there a particular reason why none of the comments (now commented) were taken into account?} \AT{Yes, I started working with a graphic designer on redrawing the figure, but that work stalled. I shall restart it.}
 %\AT{Figure to be redrawn.}
% \MCDT{AT is ``on to this''. Figure is being redrawn by a design professional.}
%\AT{Szymek suggested changing the colour along the vertical axis. Personally, I prefer the colour gradient.}
%\SP{Potential editorial suggestions. 1) Why we don't change color of 'High field', 'low field', etc. according to the corresponding range? It would look fancy. 2) I would reverse the order a, b, and c in the box so that top in the plot is top in the field.} %\MCDT{Picky remark: there is no label for the vertical axis, e.g.\ ``Magnetic field strength, $B$''.  Then you could dispose of the ``$|B|=$'' at each of the points a,b,c. }.
	%\MCDT{Again, call me picky but there is a mixture of serif and sans-serif fonts in this figure, as well as bold, italic.  It makes me uncomfortable!  Needs aesthetic improvement.}
	
	\label{fig:RegimeDefinition}
\end{figure}

The hierarchy of interactions provides the basis for a practical definition of typically used denominations for regimes of the magnetic field strength, $\abs{B}$, in NMR: `high field', `low field', `ultralow field' and `zero field'. Let us consider here a solution-state experiment with a molecule containing two different nuclear spin species, $I$ and $S$, characterized by following parameters (see Fig.\,\ref{fig:RegimeDefinition}): homo- and heteronuclear $J$-couplings ($J_{\rm II}$ and $J_{\rm IS}$), differences in homonuclear chemical shifts ($\delta_{\rm I_1} - \delta_{\rm I_2}$) and heteronuclear gyromagnetic ratios ($\gamma_{I} - \gamma_{S}$), and characteristic spin relaxation times ($T^{\{I,S\}}_{\rm relax}$). Here, the high-field regime is where the Zeeman interaction dominates, $\abs{\gamma_{I}(\delta_{\rm I_1} - \delta_{\rm I_2})B} \gg 2\pi \abs{J_{\rm II}}$ and the boundary to the low-field regime is where the two terms are comparable. Once the difference in Larmor frequency between spins of different species becomes comparable to their mutual $J$-coupling $J_{IS}$, we are entering the ultralow-field regime, with the boundary at $\abs{(\gamma_{I} - \gamma_{S})B}\approx \abs{2\pi J_{IS}}$. On further reduction of the magnetic field strength, we reach the regime where the Larmor precession period $1/\abs{\gamma_{I,S} B}$ becomes comparable for both nuclear species to their respective relaxation time $T^{\{I,S\}}_{relax}$. We define the zero-field regime as the magnetic field range in which spin polarization relaxes to thermal equilibrium faster than it precesses about any residual magnetic field. In Fig.\,\ref{fig:RegimeDefinition}, we give a typical example with numerical values for the different regimes.

%%%%%%%%%%%%%%%%%%%%%%%%%%%%%%%%%%%%%%%
%%%%%%%%%%%%%%%%%%%%%%%%%%%%%%%%%%%%%%%

For a sample without measurable $J$-couplings (e.g., a bulk sample of water), the only regimes are high- and zero-field; it would be meaningless to talk about low or ultralow fields in this case. Note that in other areas of research, different naming conventions and definitions are used, and our definitions would also need modification if dipolar or quadrupolar couplings were considered. 

Our definition of field regimes is guided exclusively by the hierarchy of spin interactions and ignores technological considerations. For example, field ranges can be categorized also according to the most suitable detection modality at the corresponding Larmor frequency \cite{MYERS2007182}. Historically, however, whether a magnetic field is considered `high' or `low' typically depended on how that field was generated, that is, by a superconducting, resistive or permanent magnet---or by the Earth's core. By contrast, operation in the ZULF regime typically requires shielding of the Earth's and laboratory fields.
  
\subsection{A short, incomplete history}
\label{Subsec:History}

The history of NMR is, in large part, a history of pushing towards ever higher field strengths, and for good reasons. In a traditional NMR experiment the sample is polarized by the magnetic field in which it is immersed, spectral dispersion is dominated by chemical shift differences, and the signal is detected by electromagnetic induction. All these interactions typically take place in one and the same magnet, and they all scale favorably with increasing field strength. 

In the shadow of the technical development and scientific exploitation of spectrometers built around increasingly powerful magnets, a small but intriguing body of work emerged over the decades dedicated to experiments in the Earth-field range ($\sim$50\,$\mu$T). Many of these experiments were motivated by sensing the magnetic field of the Earth. In 1954, Martin Packard and Russell Varian first demonstrated free nuclear induction as a tool for measuring the geomagnetic field \cite{Packard1954}. In their experiment, they polarized a 500 cm$^{3}$ water sample in a field of 10\,mT, which was applied perpendicular to the Earth's field. Subsequent nonadibatic reduction of the polarizing field to zero set off precession of the nuclear magnetization, which they detected inductively. The approach was soon adopted for geophysical surveying \cite{Waters1955, Waters1958}. As a notable spin-off, Erwin Hahn proposed in 1960 that sea‐water motion could be detected by recording, with his eponymous echo sequences, phase shifts due to movement of spins in a gradient field \cite{Hahn1960}. This method later inspired high-field experiments for quantifying blood flow and elastic deformations.

The instrumental simplicity of Earth-field NMR means that it can be employed for the study of unusually large samples. Spectacular applications include studies of diffusion in Antarctic ice \cite{Callaghan2007} and experiments with a helicopter-borne NMR coil system---six meters in diameter and weighing some 1000\,kg---with the aim of searching for oil spills under ice \cite{altobelli2019helicopter}. Moreover, the homogeneity of the Earth's magnetic field inspired a series of magnetic resonance imaging (MRI) experiments \cite{Stepisnik1990,Halse2006}, some of which involved the combination with the measurement of dispersion due to slow molecular motions \cite{Bene1980,Mohoric2009}.

As argued in Sec.\,\ref{Subsec:Regimes}, for spins without measurable couplings to each other, the only field regimes are high- and zero-field. The majority of Earth's field experiments, including those described above, therefore belong in the former category. However, there are several works in which heteronuclear $J$-coupling interactions between ${}^{1}$H-${}^{29}$Si \cite{Appelt2006}, ${}^{1}$H-${}^{19}$F \cite{Qiu2008,Qiu2009} and ${}^{1}$H-${}^{31}$P \cite{Liao2010} species were resolved. These fall into the low-field regime.

Independently of the developments in Earth-field NMR, a few groups tackled the challenge of performing NMR at field strengths in which intramolecular interactions dominate. With NQR and electron paramagnetic resonance (EPR), such experiments are readily possible even at zero magnetic field, as nuclear quadrupolar interactions and zero-field splittings arising from dipole--dipole interactions between unpaired electrons are often sufficiently large to make inductive detection feasible. However, we are aware of only one case where magnetic resonance signals of non-quadrupolar nuclei have been detected in the ZULF regime using `conventional NMR': in experiments with solid hydrogen, as first performed by Frederick Reif and Edward Purcell in 1953 \cite{Reif1953earlyzero} using the resonance absorption technique \cite{Pound1953}. The unusually large line splittings due to dipolar interactions, combined with high polarization at cryogenic temperatures, made it possible to obtain resolved spectra at zero magnetic field.

It would take 30 years until the conceptual ideas and experimental tools were in place to enable zero-field NMR with less extreme samples. Seeking methods for removing orientational broadening from so-called powder spectra while retaining resolved dipole--dipole and quadrupole couplings, Daniel Weitekamp and co-workers in the laboratory of Alexander Pines at Berkeley started to develop time-domain zero-field NMR \cite{Pines2007recollections}. The motivation was as follows: The broad spectra of powder samples are a consequence of the truncation of local interactions due to the preferred axis imposed by a strong magnetic field. Magic-angle spinning and multiple-pulse techniques commonly employed in solid-state NMR can be used to remove such anisotropy-induced broadening. However, eliminating the anisotropic contributions of the local spin interactions necessarily also removes potentially valuable information. An elegant solution is to remove the external field, so that there is no privileged direction and all crystallites of a polycrystalline sample or all sites of an amorphous solid become equivalent. In this way resolution can be improved, comparable to the case of an oriented single crystal in high-field NMR, without sacrificing the information encoded in local spin couplings.

\begin{figure}[t]
\centering
	\includegraphics[width=\columnwidth]{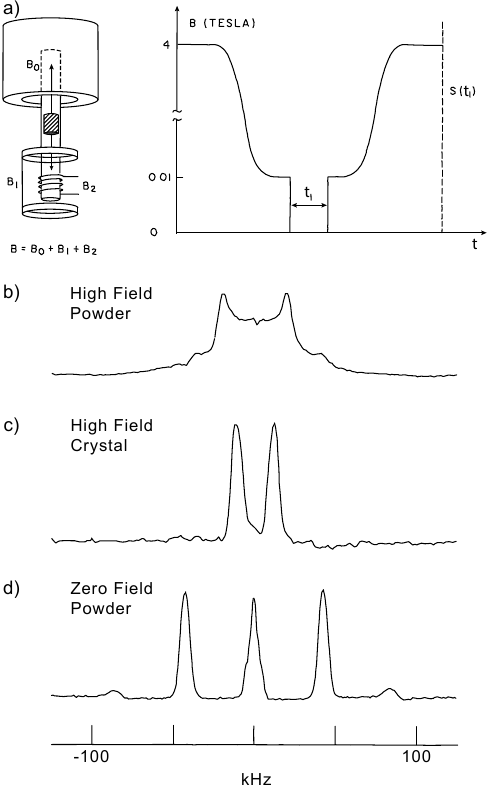}
	\vspace{-6pt}
	\caption{
	(a) Schematic diagram of the field-cycling apparatus and the time-dependent field profile for early zero-field NMR experiments.
	(b-d) ${}^{1}$H NMR spectra of barium chlorate hydrate [Ba(ClO$_3$)$_2$ $\cdot$ H$_2$O].
	Figure reprinted with permission from Ref.\ \cite{Weitekamp1983} by the American Physical Society.
}
	\label{fig:History}
\end{figure}

The challenges related to polarizing the samples and detecting a signal were addressed by adopting field-cycling methods (see Fig.\,\ref{fig:History}a). In this approach the initial spin state is prepared in a high-field spectrometer, where also the final spin state is read out \cite{Weitekamp1983,Zax85}. In between, the sample is shuttled out of the superconducting magnet into an intermediate field. Spin evolution is then initiated by a sudden transition to zero magnetic field, mirroring the strategy employed by Packard and Varian in their Earth's field experiments \cite{Packard1954}. After a time $t_1$, the magnetic field is turned back on, preserving the components of the nuclear spin magnetization that are projected along the magnetic field. The zero-field spectrum is then obtained by performing the experiment at several values of $t_1$ and Fourier transforming the resulting time-domain data. As an example, panels b and d of Fig.\,\ref{fig:History} show the high-field and zero-field spectra of powdered barium chlorate hydrate [Ba(ClO$_3$)$_2$ $\cdot$ H$_2$O], respectively. As expected, the zero-field linewidth is comparable to that of the high-field single-crystal spectrum (Fig.\,\ref{fig:History}c).

Exploring this approach further, Pines and co-workers introduced a series of isotropic averaging pulse sequences that selectively average nuclear spin Hamiltonians while preserving invariance with respect to sample orientation \cite{Lee1987,Llor1991,Llor1995a,Llor1995b} (see also \autoref{Subsubsec:Decoupling}). In parallel, the group started to develop alternative ways for detecting NMR at low fields. Pines teamed up with the fellow Berkeley group of John Clarke, a pioneer in the field of superconducting quantum interference devices (SQUIDs), and in particular in their use as magnetometers (see \autoref{Subsec:SQUID}). The availability of a magnetometer sensitive enough to detect nuclear magnetization offered the intriguing possibility of recording NMR signals with a sensitivity that does not depend on precession frequency, and therefore on the strengths of the field present during detection. The potential of using SQUIDs (or, more generally, Josephson junctions) for the detection of NMR signals had been demonstrated already in seminal experiments starting from the late 1960s \cite{Silver1967,Day1972,Greenberg1998}. But the Pines--Clarke collaboration pushed the approach considerably further, for example by combining SQUID detection with laser-polarized noble gases---another area in which the Pines group made substantial advances starting from the 1990s \cite{Navon1996,Room1997}---and generating images of $^{3}$He and $^{129}$Xe samples in millitesla fields \cite{Augustine1998}. In these fields, xenon chemical shifts in different physico-chemical environments were resolved as well \cite{WongFoy2002}, benefiting from the exceptionally large chemical shift range of xenon due to the highly polarizable electron cloud of the atom.

By the early 2000s, steady advances made it possible to detect with SQUIDs NMR signals in microtesla fields from milliliter amounts of liquid samples \cite{McDermott2002, Trabesinger2004}. In these experiments the samples were thermally prepolarized in fields of a few millitesla, so that a wide range of samples and nuclear species could be studied. Importantly, in the microtesla field range the \textit{absolute} homogeneity of even the simplest solenoids proved to be good enough to obtain NMR linewidths on the order of one hertz (despite the \textit{relative} homogeneity being in the percent range). This, in turn, enabled the extension of SQUID-detected NMR to the detection of heteronuclear $J$-couplings in neat liquids \cite{McDermott2002} and $^{13}$C-enriched solutes \cite{Trabesinger2004}, including some venturing towards the ultralow-field regime as defined in Sec.\,\ref{Subsec:Regimes} \cite{Trabesinger2004}. The `microtesla platform' was also exploited for MRI \cite{McDermott2004}, demonstrating several distinct advantages of operating at such low fields, including a large dispersion of longitudinal relaxation times \cite{Lee2005} and the imaging of samples next to or inside metallic objects \cite{SQUID-Pepper-Can}.

The Clarke group, however, was not the only one developing sensitive magnetometers in Berkeley. The group led by one of us (D. Budker) was working to push the sensitivity of atomic magnetometers \cite{Budker2007} (see \autoref{Subsec:OPM}). Such optically pumped magnetometers (OPMs) had been used already in the late 1960s by Claude Cohen-Tannoudji and co-workers for the detection of nuclear magnetization, in their case, from a gas of nuclear polarized $^3$He atoms \cite{Cohen-Tannoudji1969}. Here too, the ambition of the newly formed Pines--Budker collaboration was to push the approach much further and to develop the basis for using atomic magnetometers as versatile detectors for NMR in the ZULF regime. Starting from the mid-2000s, the team did just that, from the initial demonstration of detecting hyperpolarized gaseous xenon with an optical atomic magnetometer \cite{yashchuk2004hyperpolarized} to the detection of intramolecular \textsuperscript{13}C--\textsuperscript{1}H $J$-couplings in the ZULF regime \cite{ledbetter2009optical}.

Nowadays, when we talk about `ZULF-NMR experiments', we typically refer to direct detection of NMR using magnetometers, rather than indirect detection using field-cycling. In the following subsection we describe a generic `modern' ZULF-NMR experiment.

\subsection{A generic ZULF-NMR experiment}
\label{Subsec:GenericExperiment}

A typical NMR experiment is composed of three stages: spin polarization, encoding, and detection. In conventional NMR, each of these stages depends on the presence of a magnetic field on the order of hundreds of millitesla to tens of tesla. In ZULF NMR, by contrast, the need for strong persistent magnets is eliminated at every stage. This is possible owing to, first, the development of hyperpolarization techniques that do not require strong magnetic fields (Sec.\,\ref{Sec:Spin_pol}); second, the realization that scalar spin--spin or $J$-coupling provides for spectroscopic encoding at ZULF, without having to rely on chemical shifts (Sec.\,\ref{Sec:SpinEvolution}); and, third, the availability of sensitive detection modalities that enable non-inductive sensing (Sec.\,\ref{Sec:Detection}).

\begin{figure}[t]
\centering
	\includegraphics[width=\columnwidth]{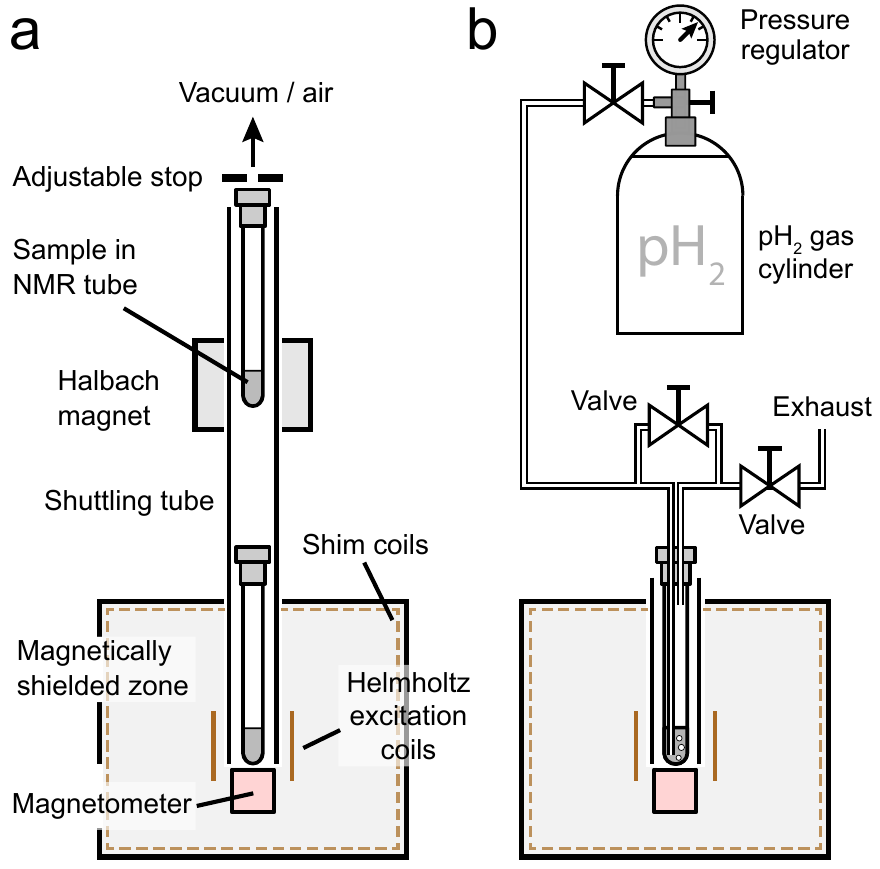}
	\vspace{-6pt}
	\caption{
Typical examples of setups for liquid-state ZULF NMR. Some of the key components are shown: a magnetic shield and shim coils to create a near-zero-field region; a magnetometer for signal detection; a sample to be measured; a source of polarization to create detectable magnetization in the sample. (a) An apparatus in which the sample is thermally polarized in a permanent magnet, and is pneumatically shuttled to the zero-field region for detection. (b) An apparatus in which the sample is polarized by chemical reaction with parahydrogen (\textit{p}H$_2$) gas that is bubbled through the sample.
Reused from Ref.\ \cite{Tayler2017}, with the permission of AIP Publishing.
%\SP{We should have additional coils in b.} \MCDT{James has the figure and agreed to add the coils in (b)}
	}
	\label{fig:ZULF_NMR_Appa}
\end{figure}

The experimental setup for ZULF NMR can be relatively simple, see Fig.\,\ref{fig:ZULF_NMR_Appa}. Layers of mu-metal and/or ferrite shielding can be used to attenuate the ambient magnetic field by several orders of magnitude to create the ZULF region. The residual field can be further reduced by `shimming', that is, by applying compensating magnetic fields through coils inside the magnetic shield. 
%\SP{We did not comment on polarization and there even are two versions of the polarization scheme shown in Fig. 3 that are not commented in the main text, but only in the caption.} 

The sample is polarized using one of the various methods available (see Sec.\,\ref{Sec:Spin_pol}), for example by thermal polarization in a magnet outside the ZULF region (Fig.\,\ref{fig:ZULF_NMR_Appa}a and Sec.\,\ref{Subsec:Thermal_pol}) or a hyperpolarization technique such as parahydrogen-induced polarization (PHIP; Fig.\,\ref{fig:ZULF_NMR_Appa}b and Sec.\,\ref{Subsec:PHIP}).
A signal can be initiated by applying brief magnetic field pulses to the sample in the ZULF region, for instance with a Helmholtz pair.
%\MCDT{A set of Helmholtz coils, or simply a Helmholtz coil? The word ``pair'' seems to imply 2 sets of coils}
%\AT{Changed to "a Helmholtz pair".}
The signal is then measured, typically with an atomic magnetometer (see Sec.\,\ref{Subsec:OPM}). Whereas traditionally home-built atomic magnetometer have been used, recently commercial magnetometers have become available that are suitable for this task, both in terms of sensitivity and functionality \cite{Blanchard2020}, thereby removing one of the main hurdles for new groups entering the field of ZULF NMR.

%\DB{DB should add about eigenstates and quantum beats and about 100\% ``chemical shift.}
%%%%%%%%%%%%%%%%%%%%%%%%%%%%%%%%%%%%%%%
%%%%%%%%%%%%%%%%%%%%%%%%%%%%%%%%%%%%%%%

\begin{figure}[t]
\centering
%	\includegraphics[width=\columnwidth]{Progress_in_NMR_ZULF/SD_zqc_evolution_variant4.png}
	%\vspace{6pt} 
 \includegraphics[width=\columnwidth]{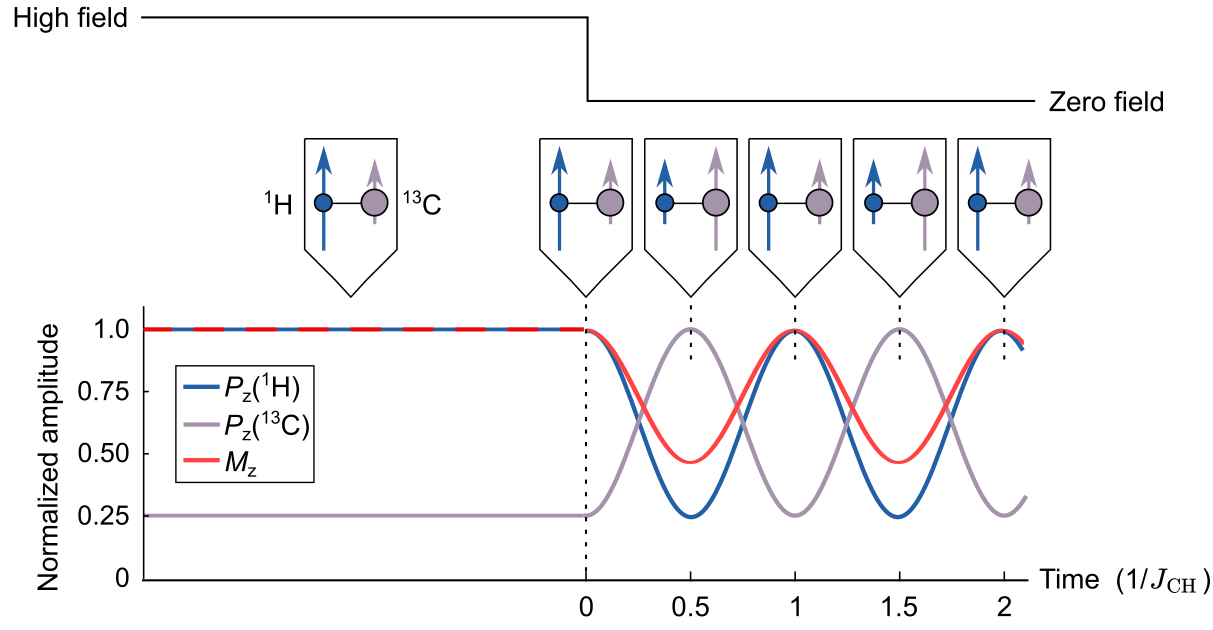}
	\caption{
A minimal illustration of ZULF NMR $J$-spectroscopy. A $^1$H--$^{13}$C spin pair with spin--spin scalar coupling $J_{\rm CH}$ is thermally polarized in a high-field regime, leading to $^1$H and $^{13}$C spin ordering along the field axis in the ratio 1:0.25. The field is then suddenly switched off, and the polarization difference proceeds to oscillate about the mean at the frequency $J_{\rm CH}$.  While the sum of polarizations remains constant in the ZULF regime (ignoring relaxation effects) the total magnetization does not, due to the different gyromagnetic ratios of the spin species --- see \autoref{Sec:Theory}.  The along-axis magnetization oscillation is typically sampled directly using a magnetometer -- see \autoref{Sec:Detection}.}
%\DB{We need the red line on the left as well! The quantities in the legend box need to be italic.} \MCDT{Done!}
	\label{fig:ZULF_explanation}
\end{figure}

In order to illustrate the basic principles underlying ZULF $J$-spectroscopy, let us consider the evolution of the thermally polarized state of a coupled $^1$H-$^{13}$C spin pair. This is illustrated in Fig.\,\ref{fig:ZULF_explanation}. The thermal equilibrium state at high field corresponds to spin polarization along the external field, where the polarization of $^1$H spins is approximately four times higher than that of $^{13}$C spins, due to the difference in gyromagnetic ratio. This state does not evolve with time.
%, see Fig.\,\ref{fig:ZULF_explanation}c
%\SP{There is no figure c and d. We should call it differently.} \AT{Thank you. I have gently rephrased the text to take this comment into account.} 
However, if the field is suddenly removed, the state starts to evolve (see Fig.\,\ref{fig:ZULF_explanation}). The original $^1$H polarization coherently transfers to the $^{13}$C spins, while the original $^{13}$C polarization transfers to the $^1$H spins. The oscillation occurs at the $J$-coupling frequency (for instance, a typical value for the one-bond $^1$H-$^{13}$C $J$-coupling in a formyl group, \textsuperscript{1}H\textsuperscript{13}COO--, is 210\,Hz \cite{Theis2013,Emondts2014,Garcon2019}).
%\MCDT{[cite formic acid, methyl formate papers?]} \AT{Yes, I think that would be helpful. Would you have the references at hand?} 
The result is an observable oscillation of the sample magnetization, from which the $J$-spectrum is obtained.

The origin of the oscillation can be understood as follows. While the field is on, the Zeeman interaction dominates the Hamiltonian and the spin system is in an eigenstate; it does not evolve. When the field is suddenly switched off, the state is initially unchanged, but it is no longer an eigenstate of the Hamiltonian, which is now dominated by the $J$-coupling interaction. In fact, the initial state corresponds to a superposition of the two eigenstates of the `new' Hamiltonian, whose energies in our example are separated by $J$ (see \autoref{Subsec:VectorModel}). The temporal evolution of the superposition of eigenstates corresponds to quantum beats, and these are exactly the oscillations sketched in Fig.\,\ref{fig:ZULF_explanation}.  

\subsection{Notations for spin systems in ZULF NMR}
\label{Subsec:Notations}

%\AT{\textbf{This subsection might also be shifted to the beginning of \autoref{Sec:Theory}.}}\\

In high-field NMR experiments, spin systems are typically classified according to the Pople notation \cite{pople1957notation}. Therein, each nucleus is assigned a capital letter of the Roman alphabet. Pairs of weakly coupled nuclei of the same species (i.e., \mbox{$\abs{\gamma_{I}(\delta_{I_1} - \delta_{I_2})B}$} $\gg 2\pi \abs{J_{II}}$) are assigned letters with a large distance between them in the alphabet, such as A and X. If the coupling between the spins is on the order of or larger than the absolute value of the chemical-shift difference, then adjacent letters of the alphabet are used, for example A and B. Nuclei that are chemically equivalent but magnetically nonequivalent are given the same letter but with a prime added to one of them (e.g., AA'XX' or AA'BB').

In ZULF NMR, chemical shift differences between like spins vanish, and for heteronuclear pairs $\abs{(\gamma_{I} - \gamma_{S})B}\lesssim \abs{2\pi J_{IS}}$, by definition (see \autoref{Subsec:Regimes}). Therefore, typically a modified Pople notation is used, with letters that are well separated in the alphabet denoting exclusively heteronuclear pairs. Adjacent letters stand for chemically nonequivalent nuclei of the same species and parentheses group together the spins with the largest coupling. For instance, (XA$_n$)B$_m$ denotes a spin system with two spin species, where X represents for example a ${}^{13}$C nucleus, A$_n$ a set of $n$ equivalent ${}^{1}$H nuclei coupled to X, and B$_m$ a chemically distinct set of $m$ equivalent ${}^{1}$H nuclei that are less strongly coupled to X and A$_n$.

An alternative way to classify spin systems is according to the topology of the $J$-coupling network \cite{levitt1985topology,radloff1989topology,levitt2013spin}.
%An example of such a representation is given in \autoref{fig:topo}. 
Whereas to our knowledge this approach has not been explored systematically in the context of ZULF NMR, we expect that this classification of spin systems might be useful across a broad range of ZULF-NMR scenarios.

\section{Theory}
\label{Sec:Theory}

Whereas ZULF NMR naturally shares many commonalities with its conventional counterpart \cite{Ernst1987NMR1D2D}, the fact that spin--spin interactions originating from the sample itself---rather than the Zeeman interaction---dominate the dynamics of the nuclear spin system leads to a number of unique properties and behaviors. A general description of these features, basic and more advanced, is explained in detail in the following section, together with experimental examples.

\subsection{Nuclear spin Hamiltonians}
\label{Subsec:Nucl_Spin_Ham}

Free evolution of nuclear spins is governed by a Hamiltonian $\hat{\mathcal{H}}$ that, in the case of NMR of isotropic liquids, can be split into two terms. The first term,  $\hat{\mathcal{H}}_Z$, describes the Zeeman interaction with the external magnetic field $\bm{\mathrm{B}}_0$, and the second term,  $\hat{\mathcal{H}}_J$, the indirect spin--spin interaction ($J$-coupling):
\begin{eqnarray}\label{eq:HamIso}
    \nonumber\hat{\mathcal{H}}_{iso}&=&\hat{\mathcal{H}}_z+\hat{\mathcal{H}}_J\notag\,,\\
    \hat{\mathcal{H}}_Z&=&-\hbar\sum_{i}\gamma_i(1-\delta_i)\bm{\mathrm{B}}_0 \cdot\hat{\bm{\mathrm{I}}}_i\,, \\
    \nonumber \hat{\mathcal{H}}_J&=&2 \pi \hbar  \sum_{i>j} J_{ij} \hat{\bm{\mathrm{I}}}_i \cdot \hat{\bm{\mathrm{I}}}_j\,.
\end{eqnarray}
%\MCDT{It seems there is a duplication in the last equation.  Cannot we remove the second = sign and term thereafter? ($\hat{\mathcal{H}}_J$)}
%\SP{I do not like $2\pi$ in the last equation. I would pull it into $J$ to keep it symmetric.}
%\KS{In NMR, $J$-couplings are  given in Hz and gyromagnetic ratio is given in rad/s, so it is a standard practice.}\DB{I am with Kirill on this one!}
Here the summation is taken over the entire system of coupled spins; $\hbar=h/(2\pi)$ denotes the reduced Planck constant; $\gamma_i$ is the gyromagnetic ratio of the $i^\textrm{th}$ spin; $\delta_i$ is the isotropic chemical shift for the $i^\textrm{th}$ spin; $\hat{\bm{\mathrm{I}}}_i$  denotes a spin operator of the $i^\textrm{th}$ spin; and $J_{ij}$ is the isotropic spin--spin coupling constant between spins $i$ and $j$. We note that while $J$-coupling is written as a scalar interaction in Eq.\,\eqref{eq:HamIso}, it is generally a tensor interaction \cite{Bryce2000,Vaara2002} (see also Sec.\,\ref{Subsec:Untruncated}).  

If the sample under study is anisotropic, then other internal interactions may manifest themselves in ZULF NMR. For example, direct spin--spin interactions have been observed \cite{Weitekamp1983,Blanchard2015}. These dipole--dipole interactions are described with the Hamiltonian
\begin{eqnarray}\label{eq:HamDip}
    \nonumber\hat{\mathcal{H}}_{DD}^{(i,j)}&=& -\hbar \hat{\bm{\mathrm{I}}}_i\cdot\bm{\mathrm{D}}^{(i,j)}\cdot\hat{\bm{\mathrm{I}}}_j = \hbar \sum_{p,q} \hat{I}_{p,i} \hat{I}_{q,j} D_{pq}^{(i,j)}\,,\\
    D_{pq}^{(i,j)}&=& -\frac{\mu_0}{4\pi}\frac{\gamma_i \gamma_j \hbar}{r_{ij}^3}\frac{1}{2}
    \left(\frac{3 r_{ij}^{(p)} r_{ij}^{(q)}}{r_{ij}^2}-\delta_{pq}
    \right)\label{eq:HamDD}\,,\\
    \nonumber p,q &\in& \{ x,y,z \}\,.
\end{eqnarray}
Here, $\bm{\mathrm{D}}$ represents the second-rank tensor of the dipole--dipole interaction between spins $i$ and $j$ which depends on the orientation of $\bm{\mathbf{r}}_{i j}$---the vector joining the centers of the two nuclei with the distance $r_{ij} \equiv |\bm{\mathbf{r}}_{i j}|$ between them---with respect to the quantization axis $z$; $\mu_0$ denotes the magnetic permeability of vacuum;  the quantity $\delta_{pq}$ is the Kronecker delta.

Recently the spin-1 nuclei \textsuperscript{14}N and \textsuperscript{2}H were observed in ZULF NMR \cite{Bevilacqua2017jpcl,picazo2024zero,Alcicek2021,Bodenstedt2022jpcl}. In these cases, the quadrupolar interaction between the nuclear spin and the gradient of the molecular electric field at the position of the nucleus is relevant. It is given by the Hamiltonian
%\begin{eqnarray}
%    \nonumber\mathcal{H}_{Q}&=& %\frac{eQ}{2I\left(2I - 1 %\right)}\hat{\bm{\mathrm{I}}}_i\cdot\mathbf{V%}\cdot\hat{\bm{\mathrm{I}}}_i = %\frac{eQ}{2I\left(2I - 1 \right)} %\sum_{p,q} \hat{I}_{p,i}\hat{I}_{q,i} %V_{pq}\,,\\
%    p,q &\in& \{ x,y,z \}\,. %\label{eq:HamQ}
%\end{eqnarray}
\begin{eqnarray}
\nonumber\hat{\mathcal{H}}_{Q}^{(i)}&=& \frac{eQ^{(i)}}{2I_i\left(2I_i - 1 \right)}\hat{\bm{\mathrm{I}}}_i\cdot\bm{\mathrm{V}}^{(i)}\cdot\hat{\bm{\mathrm{I}}}_i =\\ &=&\frac{eQ^{(i)}}{2I_i\left(2I_i - 1 \right)} \sum_{p,q} \hat{I}_{p,i}\hat{I}_{q,i} V_{pq}^{(i)}\,,\\
\nonumber p,q &\in& \{ x,y,z \}\,.
\end{eqnarray}
Here $eQ^{(i)}$ denotes the nuclear quadrupole moment of the $i^\textrm{th}$ spin and the second-rank tensor $\bm{\mathrm{V}}^{(i)}$ corresponds to the gradient tensor of the electric field evaluated at the nucleus $i$.

The relative strengths of the interactions described by $\hat{\mathcal{H}}_Z$ and $\hat{\mathcal{H}}_J$, $\hat{\mathcal{H}}_{DD}$ and $\hat{\mathcal{H}}_Q$ determine the dynamics of a spin system.  This is because the operators (for instance, $\hat{\mathcal{H}}_J$ and $\hat{\mathcal{H}}_Z$) generally do not commute with one another, unless all of the spins are magnetically equivalent.  This situation can be dealt with by using a perturbation approximation when one Hamiltonian is much larger than the other.  As an example, in conventional NMR %($\abs{B_0} >  0.1$\,T) 
the differences in Larmor frequencies are much larger than the $J$-couplings, so that the Zeeman Hamiltonian is the leading one ($|\gamma B_0 (\delta_i-\delta_j)| \gg 2\pi\abs{J_{ij}}$ for homonuclear spins or $|(\gamma_i - \gamma_j)| B_0 \gg 2\pi \abs{J_{ij}}$ for heteronuclear spins.  This defines the high-field regime (see \autoref{Subsec:Regimes}).  Eigenstates of the spin system are to a good approximation those of $\hat{\mathcal{H}}_Z$ plus the commuting part of $\hat{\mathcal{H}}_J$.  The noncommuting part of $\hat{\mathcal{H}}_J$ can be considered as a first-order perturbation; in some NMR textbooks, the dropping of the noncommuting term entirely is called the secular or weak-coupling approximation.    

The ZULF NMR scenario is opposite to the regime described above, in the sense that it 
%typically 
involves $J$-couplings that are larger than Zeeman frequencies.  Here the ZULF spin eigenstates are those of $\hat{\mathcal{H}}_J$ plus---if it is nonzero---the commuting part of $\hat{\mathcal{H}}_Z$.   If the noncommuting part of $\hat{\mathcal{H}}_Z$ is relatively small, this can now be applied as a first-order perturbation \cite{Ledbetter2011near}.

Compact analytical expressions for the commuting and noncommuting parts of $\hat{\mathcal{H}}_Z$ and $\hat{\mathcal{H}}_J$ in ZULF NMR are found for the case where the spin system contains two spin-{\textonehalf} nuclei.  Ignoring chemical shifts, these are the permutation-symmetric and permutation-antisymmetric operators
\begin{subequations}
\begin{align}
 \hat{\mathcal{H}}_{Z,{\rm commuting}} &= - \hbar \frac{\gamma_1+\gamma_2}{2}\bm{\mathrm{B}}_0\cdot(\hat{\bm{\mathrm{I}}}_1+\hat{\bm{\mathrm{I}}}_2)\,, \label{eq:2SScommuting} \\
 \hat{\mathcal{H}}_{Z,{\rm noncommuting}} &= - \hbar \frac{\gamma_1-\gamma_2}{2}\bm{\mathrm{B}}_0\cdot(\hat{\bm{\mathrm{I}}}_1-\hat{\bm{\mathrm{I}}}_2)\,, \label{eq:2SSnoncommuting}
\end{align}
\end{subequations}
%\DB{Unfortunately, it seems that we continue to have a mess with units. What are the units of the Hamiltonians? Equations (1) and (4) differ in units by $\hbar$} 
with the commutation property defined by
\begin{subequations}
\begin{align}
%  \hat{\mathcal{H}}_{J} &= 2\pi J_{12} \hat{\bm{\mathrm{I}}}_1\cdot \hat{\bm{\mathrm{I}}}_2\,,\\
  \left [\hat{\mathcal{H}}_J, \hat{\mathcal{H}}_{Z,{\rm commuting}} \right ] &= 0\,,\\
  \left [\hat{\mathcal{H}}_J, \hat{\mathcal{H}}_{Z,{\rm noncommuting}} \right ] &\neq 0\,.
\end{align}
\end{subequations}
%\DB{Michael: is the notation consistent with Eq. (1)? }\MCDT{Yes, now it is.}
Operator expressions similar to those in \autoref{eq:2SScommuting} and \autoref{eq:2SSnoncommuting} can be generated and used to determine the near-zero-field energy states of simple molecules, such as those of the XA$_n$ type (\autoref{Subsec:XAn}) \cite{Ledbetter2011near}. Perturbation theory can also be applied to more complex spin systems with one dominant $J$-coupling and several smaller couplings, such as those between distant spins, as discussed in \autoref{XAnBm} and references \cite{Butler2013, Theis2013}.
%\SP{are we putting stop after refs, as I think we have both versions in the manuscript.} \AT{\textit{Nature} style is without full stop, but as we use US English, we probably should include it.}

As a final remark, the noncommuting term or permutation-antisymmetric term [\autoref{eq:2SSnoncommuting}] can have a major role in redistributing spin order among ZULF spin eigenstates.  The spin-{\textonehalf} pair provides insights into this process as the dynamics can be solved exactly, without the use of perturbation approximations, as discussed in Sec.\,\ref{Subsec:VectorModel}.  More general behavior under pulsed fields is the topic of \autoref{Sec:SpinEvolution}.

\subsection{Density-operator formalism}

\subsubsection{Spin density operator}
The state of an NMR ensemble is described by a spin density operator $\hat{\rho}$, defined as
\begin{equation}
    \hat{\rho} = \overline{| \psi \rangle\langle \psi|}\,,
\end{equation}
where $|\psi\rangle$ is the wavefunction  of a single member of the ensemble and the overbar denotes ensemble averaging.  

In the following discussion, we use the Liouvillian formalism, where coherent evolution of the spin system in time is given, for the particular case of a time-independent Hamiltonian, by the equation
\begin{equation}	
\hat{\rho}\left(t\right) = \rm{exp} \left(-\frac{i \hat{\mathcal{H}} t}{\hbar}\right) \hat{\rho}\left(0\right) \rm{exp} \left(\frac{i \hat{\mathcal{H}} t}{\hbar}\right)\,.
    \label{eq:RhoPropagation}
\end{equation}
%The above equation is commonly written in shorthand form as
%\begin{equation}
% \hat{\rho}\left(0\right) \xrightarrow{\hat{\mathcal{H}} t} \hat{\rho}\left(t\right) \,.
%\end{equation}

%\MCDT{We also introduce the operator trace between $\hat{\rho}$ and another spin operator $\hat{Q}$
\noindent The trace of the product of a spin operator $\hat{Q}$ and $\hat{\rho}$
\begin{equation}
    {\rm Tr} \left[\hat{Q}\hat{\rho}\right] = \overline{\langle \psi | \hat{Q} | \psi \rangle}\,,
    \label{eq:psirho}
\end{equation}
has several uses.\footnote{In matrix algebra, the trace for a product of two (or more) operators is invariant with respect to a cyclic permutation of the operators.}  First, the trace allows the definition of a matrix representation of the density operator, with elements
\begin{equation}
\rho_{uv}=\left\langle v | \hat{\rho} | u \right\rangle = {\rm Tr}\left[|u\rangle\langle v|\hat{\rho}\right]\,,\label{eq:rhoij}
\end{equation}
where $|u\rangle$ and $|v\rangle$ are orthonormal basis states of $|\psi\rangle$.  Second, one can use the trace to determine the expectation value of $\hat{Q}$ at a given time $t$,
\begin{equation}
    \langle \hat{Q}(t) \rangle = \textrm{Tr}\left[\hat{Q}\hat{\rho}(t)\right]\,,
    \label{eq:expectationQ}
\end{equation}
%\DB{I still do not get why we switch from $ {\rm Tr} \left[\hat{Q}\hat{\rho}\right]$ to $\textrm{Tr}\left[\hat{\rho}(t)\hat{Q}\right]$} \MCDT{OK, now I see there was a duplication using different sybmol order.  It does not matter, so I changed \autoref{eq:expectationQ}}
and also the projection $p(\hat{Q},\,\hat{\rho})$ of $\hat{\rho}$ onto $\hat{Q}$,
\begin{equation}
    p(\hat{Q},\, \hat{\rho}) = \frac{\textrm{Tr}\left[\hat{Q}^\dagger\hat{\rho}\right]}{\textrm{Tr}\left[\hat{Q}^\dagger\hat{Q}\right]}\,,
    \label{eq:projectionQrho}
\end{equation}
where $\hat{Q}^\dagger$ is the adjoint of $\hat{Q}$.

These textbook formulae \cite{Ernst1987NMR1D2D} are used in the following sections.% \DB{Michael, a request for you from the consensus: kindly add a reference to a textbook where we can read the formulas with this dagger convention...} \MCDT{Done.}  %\DAB{I would add that cyclic permutation of matrices under the trace, conserves the trace. This is why it does not matter in which order to write rho and O.}\DB{Yes, I agree, but I want to make sure we are not missing some point here that Michael was trying to make...}\MCDT{ I also agree.  So far I think we don't actually use this property but it would be good to mention it, maybe: ``Like in matrix algebra, the trace for a product of two (or more) operators is invariant to a (cyclic) permutation of the operators''.  This should avoid any confusion between cyclic permutation and cyclic commutation.}

\subsubsection{Observable quantity}
\label{Subsubsec:Observable}

%In order to calculate NMR signals, an observable operator 
%\AT{To me, 'observation operator' sounds a bit strange. Ernst, Bodenhausen and Wokaun use 'observable operator'.} 
%$\hat{O}$ needs to be introduced. The observed quantity (observable) is given by the expectation value: \SP{We repeat something that was said before. But I think this last part in 2.2.1 was something that was recently added.}
%\begin{equation}
%    \langle \hat{O}(t) \rangle = \textrm{Tr}\left[\hat{\rho}(t)\hat{O}\right]\,.
%    \label{eq:ObservationOperator}
%\end{equation}
 
Similar to conventional NMR, ZULF NMR relies on detection of magnetic dipole fields originating from nuclear magnetization.  The measured quantity is the magnetic field at the location of the detector. 

In general, the field at the detector depends on the sample magnetization and the relative position of the detector with respect to the sample \cite{Xu2022}.
However, 
provided the detector is displaced from the sample along the sensitive axis of the sensor, the signal is determined purely by the magnetization component in that direction.
%if the detector is displaced from the sample in the direction to which it is sensitive, then it only detects the signal proportional to magnetization in the same direction. 
In the following, for simplicity, we assume this situation.

The field is proportional to the total magnetization of the sample, with the magnetization operator $\hat{\bm{\mathrm{M}}} = (\hat{M}_x,\hat{M}_y,\hat{M}_z)$ given by
\begin{equation}
	\hat{\bm{\mathrm{M}}} \propto  \sum_{i=1}^N\gamma_i\hat{\bm{\mathrm{I}}}_i\,,
	\label{eq:MagnetizationMz1}
\end{equation}
where the summation runs over all spins in the sample.

Equations\,\eqref{eq:RhoPropagation}--\eqref{eq:MagnetizationMz1} can be used to calculate time-dependent NMR signals. For example, the signal observed along an axis $p$ ($p \in \{x, y, z\}$) is
\begin{equation}
	s_p\left(t\right) \propto {\rm Tr}[\hat{M}_p \hat{\rho}(t)] = \sum_{u,\,v} \bra{u}\hat{M}_{p}\ket{v} \rho_{uv}(0) \textnormal{exp} \left(-{\rm i} \omega_{uv} t\right)\,.
	\label{eq:TimeDomaineSignal}
\end{equation}
Here the double summation is made over all pairs of eigenstates of the Hamiltonian, $\{ |u\rangle$, \ldots $|v\rangle \}$, including self pairs, $\omega_{uv} \equiv (\langle u|\hat{\mathcal{H}} | u \rangle - \langle v|\hat{\mathcal{H}} | v \rangle)/\hbar$ denotes corresponding transition frequencies in angular units.
%, and in analogy to \autoref{eq:rhoij}, $M_{p,vu} \equiv \bra{u} \hat{M}_{p}\ket{v}$. 
%Expressions for the signals measured along other spatial directions can be introduced by replacing $x$ with the corresponding axis label. 

%\DB{Problematic part here:} The detected signal depends on the sensitive axis of the magnetometer, its location with respect to the sample, and on initial state. Indeed, even for the sample polarized in the same manner, the magnetization measured along the $x$-axis is typically different from the one measured along the $y$- or $z$-axes\MCDT{citation needed} \JB{Maybe something from Thomas or Billion?} \DAB{\cite{Xu2022}}. 

%One of the applications of this property is the direct
%determination of the sign of the gyromagnetic ratio by ZULF NMR. This can be done, for instance, by simultaneous detection of two orthogonal components of spin precession \MCDT{citation needed}\cite{eills2023enzymatic}. Many atomic magnetometers have this capability \cite{Seltzer204Unshielded}. By contrast, determination of absolute sense of rotation with high-field NMR is a nontrivial problem due to cross-talk between detection coils. 

The presented approach is general and provides means for calculating ZULF NMR spectra of various compounds. For that one needs to determine the initial density matrix and the form of $\bra{u} \hat{M}_p\ket{v}$. In some cases, this can be done exactly and analytically, in others perturbative approaches have to be implemented. Determining the spectra of complex molecules might require numerical calculation.

In \autoref{Subsec:XAn} we turn our attention to systems with XA$_n$ spin topology, which can be solved exactly within the framework introduced above. Before doing so, we discuss in \autoref{Subsec:VectorModel} a vector-model description of XA spin systems, with a goal of providing an intuitive geometrical approach to understanding key features of ZULF NMR.

\subsection{Vector-model description of XA spin systems} 
\label{Subsec:VectorModel}
%\MCDT{The theory is adapted from our 2017 RSI article \cite{Tayler2017}.  I think it is a shame that the math was written in line with the text rather than as numbered equations, as it probably went unnoticed by many people.}
%\AT{This has been \autoref{Subsec:XAn} before.}

Many of the basic features of ZULF NMR that might seem unusual to the uninitiated can be understood by analyzing the dynamics of the density operator for a spin-{\textonehalf} pair.  
Of the $2^2~\times~2^2=16$ possible orthogonal basis operators, the following five %(nonnormalized) 
% \DAB{(maybe we just normalize them properly?)} \MCDT{They should not be normalized! That's not how this works.  The important thing is that they obey the commutation relations.  If you start changing the amplitudes, those break down.} 
operators are of interest:
\begin{subequations}
\begin{align}
    \hat{A}_1 &= \hat{I}_{1x}\hat{I}_{2x} + \hat{I}_{1y}\hat{I}_{2y}\,, \\
    \hat{A}_2 &= \frac{\hat{I}_{1z} - \hat{I}_{2z}}{2}\,, \\
    \hat{A}_3 &= \hat{I}_{1x}\hat{I}_{2y} - \hat{I}_{1y}\hat{I}_{2x}\,, \\
    \hat{A}_4 &= \hat{I}_{1z}\hat{I}_{2z}\,, \\
    \hat{A}_5 &= \frac{\hat{I}_{1z} + \hat{I}_{2z}}{2}\,.
\end{align}\label{eq:Aoperators}
\end{subequations}

These are five of the six zero-quantum (ZQ) operators for the spin pair (the last one is the identity operator).  A ZQ operator is invariant under rotation of the whole system about the $z$-axis \hbox{($\hat{A}_j = \exp[-{\rm i} \theta (\hat{I}_{1z} + \hat{I}_{2z}) ] \hat{A}_j \exp[{\rm i} \theta (\hat{I}_{1z} + \hat{I}_{2z})  ]$} for arbitrary real $\theta$) and has zero overall coherence order. 
A useful property of the basis in \autoref{eq:Aoperators} is the cyclic commutation of the three operators $\hat{A}_1$, $\hat{A}_2$, and $\hat{A}_3$, which obey the relationship \cite{levitt2013spin}
\begin{subequations}
\begin{align}
\left[\hat{A}_1,\hat{A}_2\right] &= \mathrm{i} \hat{A}_3\,, \\
\left[\hat{A}_2,\hat{A}_3\right] &= \mathrm{i} \hat{A}_1\,, \\ 
\left[\hat{A}_3,\hat{A}_1\right] &= \mathrm{i} \hat{A}_2\,. 
\end{align} \label{eq:cycliccommutationdefinition}
\end{subequations}

In addition, all operators in \autoref{eq:Aoperators} commute with $\hat{A}_4$ and $\hat{A}_5$. Therefore, the only noncommuting operators of the ZQ basis are $\hat{A}_1$, $\hat{A}_2$, and $\hat{A}_3$. 

\begin{figure}
    \centering
    \includegraphics[width=\columnwidth]{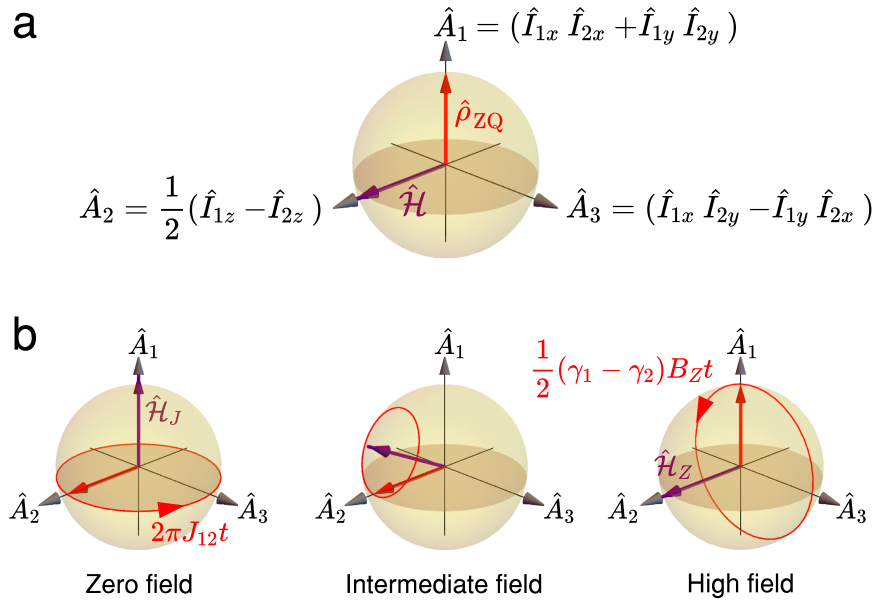}
    \caption{Bloch sphere representation of the fictitious-spin-{\textonehalf} subspace of zero-quantum operators for a spin-{\textonehalf} pair: (a) gray arrows correspond to basis vectors representing the cyclically commuting spin operators $\hat{A}_1$, $\hat{A}_2$ and $\hat{A}_3$, defined by \autoref{eq:Aoperators}; (b) precession dynamics in the zero-quantum density operator subspace, illustrated for three distinct magnetic-field regimes (see also \autoref{fig:RegimeDefinition}). The purple arrow indicates the total field vector, of magnitude $\sqrt{(2\pi J_{12})^2+(\gamma_1-\gamma_2)^2B_z^2/4}$ given by the vector sum of the components in red.}
    \label{fig:ZQvectormodel}
\end{figure}

A consequence of the commutation properties in \autoref{eq:cycliccommutationdefinition} is that the ZQ operator basis of the spin pair behaves overall like the operator basis of a single fictitious spin-{\textonehalf},
%\MCDT{Do we call these operators a \href{https://en.wikipedia.org/wiki/Homomorphism}{homomorphism} of spin-1/2 basis operators, or avoid such jargon that readers will probably not understand?}\DB{I would say, skip this terminology, but we can discuss on Wednesday} 
where operators $\hat{A}_1$, $\hat{A}_2$, and $\hat{A}_3$ evolve in a way that is analogous to single-spin Cartesian operators $\hat{I}_z$, $\hat{I}_x$, and $\hat{I}_y$, respectively. % \SP{I wonder if there is a particular reason for order of indices here or this is just a chance that the appear in this order? It seems the latter, so maybe we should say 'respectively'?}.  %N.B. The operators evolve commonly at a single nonzero eigenfrequency, while all other operators are stationary.  
Therefore, as is the case for a single spin-{\textonehalf}, the ZULF spin dynamics of a heteronuclear pair can be represented conveniently and graphically using a Bloch sphere (\autoref{fig:ZQvectormodel}).  

A point $\mathbf{r_\rho}$ in the sphere corresponds to a given projection of $\hat{\rho}$ onto each fictitious-spin-{\textonehalf} basis operator:
\begin{equation}
    \mathbf{r_\rho} = \left( \frac{{\rm Tr}[\hat{A}_1^\dagger\hat{\rho}]}{{\rm Tr}[\hat{A}_1^\dagger\hat{A}_1]}, 
    \frac{{\rm Tr}[\hat{A}_2^\dagger\hat{\rho}]}{{\rm Tr}[\hat{A}_2^\dagger\hat{A}_2]},
    \frac{{\rm Tr}[\hat{A}_3^\dagger\hat{\rho}]}{{\rm Tr}[\hat{A}_3^\dagger\hat{A}_3]}
    \right)\,.
\end{equation}
%\DB{Here you used daggers. No daggers are used in Eq.\,(11,12). This needs an explanation...}\MCDT{Operator projection is now defined in the previous subsection.  Not the same thing as expectation value!}
Similarly, a spin Hamiltonian 
\begin{equation}
    \hat{\mathcal{H}} = \sum_j{\omega_j}\hat{A}_j\,
\end{equation}
can be represented as the 3-vector
\begin{equation}
    \bm{\mathrm{r}_\mathcal{H}} = \left( \frac{{\rm Tr}[\hat{A}_1^\dagger\hat{\mathcal{H}}]}{{\rm Tr}[\hat{A}_1^\dagger\hat{A}_1]}, 
    \frac{{\rm Tr}[\hat{A}_2^\dagger\hat{\mathcal{H}}]}{{\rm Tr}[\hat{A}_2^\dagger\hat{A}_2]},
    \frac{{\rm Tr}[\hat{A}_3^\dagger\hat{\mathcal{H}}]}{{\rm Tr}[\hat{A}_3^\dagger\hat{A}_3]}
    \right) \equiv (\omega_1, \omega_2, \omega_3) \,. \label{eq:vectoromega}
\end{equation}
Under these conditions, the time evolution of $\hat{\rho}$ is equivalent to precession of $\bm{\mathrm{r}_\rho}$ about $\bm{\mathrm{r}_\mathcal{H}}$, in analogy to the classical vector model of NMR.  The motion can be summarized by the equation
\begin{equation}
    \frac{{\rm d}\bm{\mathrm{r}}_\rho}{{\rm d} t}  =  \bm{\mathrm{r}}_\rho \times \bm{\mathrm{r}}_{\mathcal{H}} \,, \label{eq:ZULFvectormodeltwospinsystem}
\end{equation}
for which two extreme cases are illustrated in \autoref{fig:ZQvectormodel}b.  One of these (right side of \autoref{fig:ZQvectormodel}b) corresponds to the familiar high-field NMR picture, where a magnetic field is applied along the $z$-axis.  The Zeeman Hamiltonian can be identified with
\begin{equation}
    \hat{\mathcal{H}}_{Z} = \omega_2 \hat{A}_2 + \omega_5 \hat{A}_5\,,
\end{equation}
for $\omega_2 = (\gamma_1 - \gamma_2) B_z$ and $\omega_5 = (\gamma_1 + \gamma_2) B_z$.  The time-integrated form of the precession equation can be written in this case as
\begin{equation}
        \hat{A}_1 \xrightarrow{\hat{\mathcal{H}}_Z t} \hat{A}_1\cos(\omega_2 t) - \hat{A}_3\sin(\omega_2 t)\,, \label{eq:cyclic2}
\end{equation}
giving the familiar high-field result that the vector $\bm{\mathrm{r}_\rho}$ undergoes precession at a frequency proportional to the difference in gyromagnetic ratio for the two spins.  The opposite extreme
(left side of \autoref{fig:ZQvectormodel}b) is evolution at zero field, where the Hamiltonian comprises only the $J$-coupling:
\begin{equation}
    \hat{\mathcal{H}}_{J} = \omega_1 \hat{A}_1 + \omega_4 \hat{A}_4\,,
\end{equation}
with $\omega_1 = \omega_4 = 2\pi J_{12}$.  Here the precession equation is
\begin{equation}
    \hat{A}_2 \xrightarrow{\hat{\mathcal{H}}_J t} \hat{A}_2\cos(\omega_1 t) + \hat{A}_3\sin(\omega_1 t)\,. \label{eq:cyclic}    
\end{equation}
We note that an initial density operator produced by thermal polarization or hyperpolarization of the spin system, as discussed in \autoref{Sec:Spin_pol}, may contain a component proportional to ${\hat{M}_z \equiv (\gamma_1 - \gamma_2) \hat{A}_2 + (\gamma_1 + \gamma_2) \hat{A}_5}$, as defined in \autoref{eq:MagnetizationMz1}.  Applying \autoref{eq:cyclic} gives
\begin{equation}
    \begin{split}
    (\gamma_1\hat{I}_{1z} + \gamma_2\hat{I}_{2z}) \xrightarrow{\hat{\mathcal{H}}_{\rm J} t}& 
    \frac{1}{2}(\gamma_1 - \gamma_2)(\hat{I}_{1z} - \hat{I}_{2z})\cos(2\pi J_{12} t)  \\ 
    & + (\gamma_1 - \gamma_2)(\hat{I}_{1x}\hat{I}_{2y} - \hat{I}_{1y}\hat{I}_{2x})\sin(2\pi J_{12} t) \\
    & + \frac{1}{2}(\gamma_1 + \gamma_2)(\hat{I}_{1z} + \hat{I}_{2z})\,,\label{eq:cyclic1a}
    \end{split}
\end{equation} 
where the $\hat{A}_2$ component evolves back and forth into $\hat{A}_3$ at the frequency of the inter-pair coupling, $J_{12}$ \cite{Ledbetter2008,Emondts2014}.
In a heteronuclear spin pair, $\gamma_1\neq \gamma_2$, the operator $\hat{A}_2$ corresponds to a net magnetic moment along the $z$-axis; therefore, part of the initial magnetization oscillates at a single, nonzero frequency, as illustrated in \autoref{fig:ZULF_explanation}. The rationale for using heteronuclear compounds for zero-field NMR can be understood based on the fact that the magnitudes of both time-dependent terms in \autoref{eq:cyclic1a} are zero for a homonuclear spin system, i.e., for $\gamma_1=\gamma_2$.

Returning to \autoref{eq:cyclic2}, the vector model can be used to determine the pair evolution under pulsed $z$ magnetic fields.  Evolution of scalar spin order, defined for the $I_1$--$I_2$ pair as $\hat{\bm{\mathrm{I}}}_1\cdot\hat{\bm{\mathrm{I}}}_2 = \hat{I}_{1x}\hat{I}_{2x} + \hat{I}_{1y}\hat{I}_{2y} + \hat{I}_{1z}\hat{I}_{2z} \equiv (\hat{A}_1+\hat{A}_4)$, is one important example, as a form of spin order that can be long-lived compared to the longitudinal relaxation time $T_1$ \cite{Emondts2014} at low fields, and one that is often a product of spin-hyperpolarization procedures involving parahydrogen \cite{Theis2011} (see \autoref{Subsec:PHIP}).  The scalar order is a stationary form of spin order at zero field, but according to \autoref{eq:cyclic2} it evolves under $\hat{\mathcal{H}_Z}$ as
\begin{equation}
    \begin{split}
        \hat{\bm{\mathrm{I}}}_1 \cdot \hat{\bm{\mathrm{I}}}_2 \xrightarrow{\hat{\mathcal{H}}_{Z} t}&
        (\hat{I}_{1x}\hat{I}_{2x}+\hat{I}_{1y}\hat{I}_{2y})\cos[(\gamma_1 - \gamma_2) B_z t] \\
        &- (\hat{I}_{1x}\hat{I}_{2y}-\hat{I}_{1y}\hat{I}_{2x})\sin[(\gamma_1 - \gamma_2) B_z t] \\
        & + \hat{I}_{1z}\hat{I}_{2z}\,.\label{eq:cyclic2a}
    \end{split}
\end{equation}
Although none of the terms on the right side of \autoref{eq:cyclic2a} correspond to a net magnetic moment of the sample, the one on the second line (proportional to $\hat{A}_3$) may evolve into the observable moment $\hat{A}_2$ under $\hat{\mathcal{H}}_J$, as described above, upon switching the total field back to zero.  Cyclic permutation of $x$, $y$, and $z$ indices in \autoref{eq:cyclic2a} allows similar results to be obtained for d.c. fields applied along $x$- and $y$-axes. 

In summary, \autoref{eq:ZULFvectormodeltwospinsystem} provides a framework for discussing transformations between zero-quantum forms of spin order in which conventional NMR concepts and parlance (e.g., flip angle) can be used.  In addition to providing a formalism for describing the nature of spin evolution in the extremes of high and zero magnetic fields, it can provide analytical solutions for spin evolution in the low-field spin-dynamics regime, also illustrated in \autoref{fig:ZQvectormodel}b. 
Although the formalism is only valid for a spin-{\textonehalf} pair, it can often be successfully applied as a first-order approximation of more complex spin systems, including those where there is a dominant $J$-coupling between a pair of heteronuclear spins and smaller $J$-couplings with all other nuclei (e.g.,\ [$^{13}$C$_1$]-benzene).

\subsection{XA$_n$ spin systems at zero field}
\label{Subsec:XAn}
\subsubsection{Energy-level structure \label{Sec:XAn}}

%\SP{@Andreas, can you please add some introduction. We may here or in below say something about XAn  systems and then adapt what John wrote in 2.3.1. What do you think?}

%\SP{Let us read this and check if this fits} 
%As discussed above, in conventional NMR, the spin evolution occurs under the dominant Zeeman interaction.  
In the ZULF regime, %this is no longer the case and 
the evolution of a spin is primarily governed by the $J$-coupling Hamiltonian.  The eigenbasis of a heteronuclear spin system is therefore different from that in the high-field case. We first identify the energy-level structure of such a system.
%Here, we discuss a spin system evolving under the dominant $J$-coupling and analyze how such a system is perturbed by additional, weaker couplings.  This allows us to derive some general properties of spin systems at zero and ultralow magnetic fields.

In isotropic liquids at truly zero field, a heteronuclear spin system evolves exclusively due to the indirect spin--spin interaction.  
The simplest system that supports such evolution is a system with the XA$_n$ spin topology, which consists of a set of $n$ magnetically equivalent spins A (typically \textsuperscript{1}H nuclei) with pairwise coupling $J_{\rm AA}$ among them. The A-spins are furthermore coupled to a heteronuclear spin X (e.g., ${}^{13}$C, ${}^{15}$N, ${}^{31}$P) with the coupling constant $J_{\rm XA}$.  
The corresponding spin operators are $\hat{\bm{\mathrm{S}}}$, with the quantum number $S$, for the X-spin, and $\hat{\bm{\mathrm{I}}}_{{\rm A},i}$, with quantum number $I_{A,i}$, for each of the $n$ A-spins. 
%\MCDT{These are not eigenvalues. I would say that they are associated quantum numbers of the total angular momentum, where eigenvalue of $\hat{S}^2$ is $S(S+1)$ for example}.  
The general coupling Hamiltonian is of the form
% \begin{eqnarray}
%     \hat{\mathcal{H}}_{J{\rm XA}_n} &=&  2 \pi \hbar\sum_{i} J_{\rm XA} \hat{\bm{\mathrm{S}}} \cdot \hat{\bm{\mathrm{I}}}_{{\rm A},i} \nonumber \\ &=& 2 \pi \hbar J_{\rm XA} \hat{\bm{\mathrm{S}}} \cdot \hat{\bm{\mathrm{K}}}_{\rm A}\,, 
%     \label{eq:H0XAn}
% \end{eqnarray}
% \MCDT{\begin{equation}
%     \begin{split}
%     \hat{\mathcal{H}}_{J{\rm XA}_n} =&  2 \pi \hbar \left( \sum_{i} J_{\rm XA} \hat{\bm{\mathrm{S}}} \cdot \hat{\bm{\mathrm{I}}}_{{\rm A},i} + \sum_{i, j>i} J_{\rm AA} \oper{I}_{{\rm A},i} \cdot\oper{I}_{{\rm A},j} \right) \\ 
%     =& 2 \pi \hbar \left\{ J_{\rm XA} \hat{\bm{\mathrm{S}}} \cdot \hat{\bm{\mathrm{K}}}_{\rm A} + \frac{1}{2}J_{\rm AA} \left[ \oper{K}_{\rm A}^2 - n I (I+1) \hat{1}_{\rm A}\right]  \right\} \,, 
%     \label{eq:H0XAn}
%     \end{split}
% \end{equation}}
\begin{equation}
    \begin{split}
    \hat{\mathcal{H}}_{J,{\rm XA}_n} =&\,  2 \pi \hbar \left( \sum_{i} J_{\rm XA} \hat{\bm{\mathrm{S}}} \cdot \hat{\bm{\mathrm{I}}}_{{\rm A},i} + \sum_{i, j>i} J_{\rm AA} \oper{I}_{{\rm A},i} \cdot\oper{I}_{{\rm A},j} \right) \\ 
    =\,& 2 \pi \hbar \left( J_{\rm XA} \hat{\bm{\mathrm{S}}} \cdot \hat{\bm{\mathrm{K}}}_{\rm A} + \frac{J_{\rm AA}}{2} \left[ \oper{K}_{\rm A}^2 - n \oper{I}_{{\rm A}}^2\right]  \right) \,, 
    \label{eq:H0XAn}
    \end{split}
\end{equation}
where $\hat{\bm{\mathrm{K}}}_{\rm A}=\sum_i \hat{\bm{\mathrm{I}}}_{{\rm A},i}$ is the total spin operator of the A-nuclei with the eigenvalue $K_{\rm A}$, and $\oper{I}_{{\rm A}}^2 \equiv \oper{I}_{{\rm A},1}^2$. %\AT{In the second term on  the second line of \autoref{eq:H0XAn}, should it be $\oper{I}_{{\rm A}}^2$ instead of $\oper{I}_{{\rm A},1}^2$ then?} %\SP{This second term is only present when there is heteronuclear coupling. Shall we comment that explicitly?} \MCDT{COMMENT: You need this to avoid ambiguity about whether there has been an implicit summation over index $i$ (there should not be)}.  
%\KS{I would say that $I^2$ operator above and below (and also $F^2, K_A^2$, etc) is scalar and therefore should not be put in bold. Or there is some other notation I am missing like bold denoted zero rank spherical tensor operator?} 
Note for an XA system ($n=1$) there is no $J_{\rm AA}$ and the second term in \autoref{eq:H0XAn} should be dropped.  In this case $\oper{K}_{\rm A} =\oper{I}_{\rm A}$, it is zero anyway.

In order to evaluate the scalar product in \autoref{eq:H0XAn}, one can introduce the total angular momentum operator of the heteronuclear pair: $\hat{\bm{\mathrm{F}}}_{{\rm XA}}=\hat{\bm{\mathrm{K}}}_{{\rm A}}+\hat{\bm{\mathrm{S}}}$. %\AT{The ``A'' in $\hat{\bm{\mathrm{F}}}_{{\rm A}}$ caused some confusion in discussion we had during our meetings, as $\hat{\bm{\mathrm{F}}}_{{\rm A}}$ includes also a X-spin term. Any chance that we can rename it? $\hat{\bm{\mathrm{F}}}_{{\rm XA}}$?} 
Because of the expansion
\begin{equation}\label{Eq.FA2}
    \oper{F}_{\rm XA}^2=\left( \hat{\bm{\mathrm{K}}}_{{\rm A}}+\hat{\bm{\mathrm{S}}} \right)^2 \equiv \oper{K}_{\rm A}^2+\oper{S}^2+2\hat{\bm{\mathrm{K}}}_{{\rm A}}\cdot\hat{\bm{\mathrm{S}}}\,,
\end{equation}
we can write
\begin{equation}\label{Eq.KAS}
\oper{S}\cdot \oper{K}_{\rm A} =\frac{1}{2}\left( \oper{F}_{\rm XA}^2 - \oper{K}_{\rm A}^2 - \oper{S}^2 \right)\,.
\end{equation}

The eigenstates of the Hamiltonian are the eigenstates of $\oper{F}_{\rm XA}^2$ and $\hat{F}_{{\rm XA},z}$ and they can be written as $| F_{\rm XA}(K_{\rm A}, S), m_{F_{\rm XA}}\rangle$, %\SP{We should probably comment why we kept $S$ and drop $I_A$. \MCDT{It is because S and KA respectively describe the total angular momentum of the X and A spin species.  IA is the individual angular momentum. It's simply that for XAn we are taking the simple case (X1An) of a larger class of systems: XmAn. So it happens that the total S angular momentum is the same as the individual spin angular momentum for m=1. }} 
where $K_{\rm A}$, $S$, $F_{\rm XA}$, and $m_{F_{\rm XA}}$ are the quantum numbers corresponding to the eigenvalues of $\hat{\bm{\mathrm{K}}}_{{\rm A}}$, $\hat{\bm{\mathrm{S}}}$, $\hat{\bm{\mathrm{F}}}_{{\rm XA}}$, and $\hat{F}_{{\rm XA},z}$, respectively. 
%(for brevity, the quantum number $K_{\rm A}$ and $S$ are hereon omitted in the notation of the states).

\begin{figure*}
\begin{center} 
    \includegraphics[width=1.6\columnwidth]{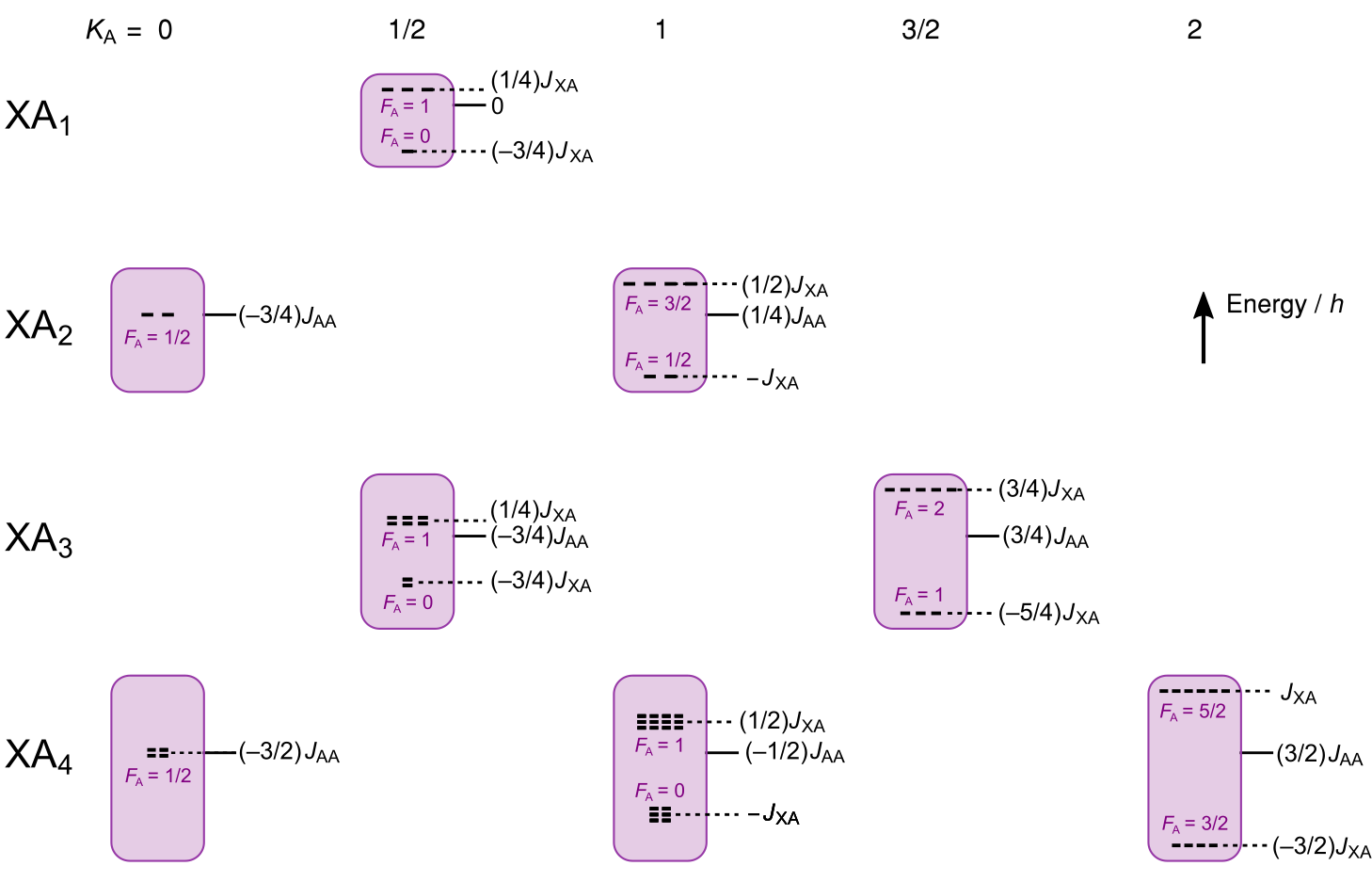}
    \end{center}
    \caption{
    % \SP{@Michael: We are using slightly different marking of $J$s. Across the text, we are using $J_\textrm{XA}$, while in the figure we have $J_\textrm{AX}$. It would be worth chaining in the figure, for consistency of the manuscript.} \MCDT{Done, figure and caption now read $J_\textrm{XA}$. These comments can be removed by DB.}    
    Energy-level diagrams for small XA$_{\rm n}$ spin-1/2 systems at zero magnetic field (${\rm n}\leq4$, $I=1/2$, $S=1/2$).  States are grouped by total $I$-spin angular momentum quantum number $K_{\rm A}$ as shown by the shaded regions. Outside each region, a solid horizontal line indicates the reference energy for each $K_{\rm A}$ manifold due to the homonuclear spin coupling $J_{\rm AA}$. Dotted lines indicate the energy-level shifts within each manifold due to the heteronuclear coupling $J_{\rm XA}$. %\MCDT{I like this figure much better because it mirrors the structure of \autoref{tab:multiplicityAN}.}
    }
    \label{Fig:XAn2}
\end{figure*}

The energy of the state $|F_{\rm XA}(K_{\rm A},S),m_{F_{\rm XA}}\rangle$ is given by
\begin{strip}
    \begin{equation}
        \begin{split}
            E^{(0)}_{F_{\rm XA}(K_{\rm A},S) m_{F_{\rm XA}}}
            =\,& \left\langle F_{\rm XA}(K_{\rm A}, S), m_{F_{\rm XA}}\right| \hat{\mathcal{H}}^{(0)}_{{\rm XA}_n}\left| F_{\rm XA}(K_{\rm A}, S), m_{F_{\rm XA}}\right\rangle  \\
            =\,& \pi\hbar \left\langle F_{\rm XA}(K_{\rm A}, S), m_{F_{\rm XA}}\right| \Big [ J_{\rm XA} \left( \oper{F}_{\rm XA}^2 - \oper{K}_{\rm A}^2 - \oper{S}^2 \right) + J_{\rm AA} \left( \oper{K}_{\rm A}^2 - n \oper{I}_{\rm A}^2 \right) \Big]\left| F_{\rm XA}(K_{\rm A}, S), m_{F_{\rm XA}}\right\rangle \\
            =\,& \pi\hbar J_{\rm XA}\Big[F_{\rm XA}(F_{\rm XA}+1)-K_{\rm A}(K_{\rm A}+1)-S(S+1)\Big]+\pi\hbar J_{\rm AA}\Big[K_{\rm A}(K_{\rm A}+1)- n I_{\rm A}(I_{\rm A} + 1)\Big]\,, \label{Eq.E0}
        \end{split}
    \end{equation}
\end{strip}
where $m_{F_{\rm XA}}$ is the magnetic quantum number in the XA basis and superscript $(0)$ indicates that the system evolution is governed only by the internal interactions (unperturbed system). The equation allows one to calculate energy-level diagrams of the XA$_n$ spin system.  Diagrams for XA, XA$_2$, XA$_3$ and XA$_4$ spin systems are shown in \autoref{Fig:XAn2}. In addition, a diagram for XA$_6$ is shown in reference \cite{Alcicek2021Ogranophosphorus}.  

It should be noted that, as in high-field NMR, the coupling $J_{\rm AA}$ between the magnetically equivalent A-spins can almost always be ignored, even if it has a nonzero value. This is because observable NMR transitions can only occur between states of a common $K_{\rm A}$ (see \autoref{XAnSelectionRules}) and while the energy differences between $K_{\rm A}$ manifolds can depend on $J_{\rm AA}$, the transition frequencies within them do not.

At zero field, each $F_{\rm XA}$ state is $(2F_{\rm XA}+1)$ times degenerate due to the magnetic sublevel structure.  The permutation symmetry (i.e., magnetic equivalence) of the A-spins can also cause an additional degeneracy, as multiple states can have the same $K_{\rm A}$.  We call this second `degeneracy' the multiplicity, $g_{K_{\rm A}}$.  As an example, there are two $K_A=1/2$ manifolds for a methyl group (CH$_3$), and $g_{1/2} = 2$, while $g_{3/2}=1$.  As a consequence, the zero-field eigenstates $|F_{\rm XA}(K_{\rm A},S),m_{F_{\rm XA}}\rangle = |1(1/2,1/2),m_{F_{\rm XA}}\rangle$ and $|0(1/2,1/2),0\rangle$ of an XA\textsubscript{3} spin system ($S=I_i= 1/2$) are doubled (see \autoref{Fig:XAn2}).  Values of $g_{K_{\rm A}}$ in other (A$_n$) spin systems are tabulated in \autoref{tab:multiplicityAN}.

% \begin{figure}
% \begin{center} 
%    %\includegraphics[width=\columnwidth]{XAn.pdf}
%     \includegraphics[width=\columnwidth]{XAnShift.pdf}
%     \end{center}
%     \caption{
%     Energy-level diagrams of all-spin-1/2 XA, XA$_2$, and XA$_3$ systems at zero magnetic field. Transitions corresponding to changes in total $z$-axis magnetization are indicated with arrows. %\SP{This is along the $z$-axis}.
%     The doubled-line levels in the XA$_3$ system indicate the manifolds of dual multiplicity. \SP{Note that for better visibility, the energy-level shift due to homonuclear spin coupling $J_{\rm AA}$ was exaggerated.}}
%     \label{Fig:XAn}
% \end{figure}

\begin{table*}[]
    \centering
    \begin{tabular}{{l}*{12}{p{0.9cm}}}
    \hline
   & \multicolumn{10}{c}{A-species total angular momentum quantum number, $K_{\rm A}$}\\
    & 0 & 1/2 & 1 & 3/2 & 2 & 5/2 & 3 & 7/2 & 4 & 9/2 & \ldots \\ \hline \hline\\Spin system& \multicolumn{10}{c}{Multiplicity, $g_{K_A}$}\\\\
A &  & $\tikzmark{s1} 1 $ \\
A\textsubscript{2} & $\tikzmark{s2} 1$ && $1$  \\
A\textsubscript{3} && $2$ && $1$  \\
A\textsubscript{4} & $\tikzmark{s3}2$ && $3$ && $1$  \\
A\textsubscript{5} && $5$ && $4$ && $1$  \\
A\textsubscript{6} & $\tikzmark{s4}5$ && $9$ && $5$ && $1$& \\
A\textsubscript{7} && $14$ && $14$ && $6$ && $1$ \\
A\textsubscript{8} & $\tikzmark{s5}14$ && $28$ && $20$ && $7$&& $1$ \\
A\textsubscript{9} && $42$ && $48$ && $27$ && $8$&& $1$  \\
&&&$\tikzmark{e5}$&& $\tikzmark{e4}$ && $\tikzmark{e3}$ && $\tikzmark{e2}$ && $\tikzmark{e1}$ \\
Sequence rule&\multicolumn{3}{c}{$n(n-1)(n-2)(n-7)/24$}&\multicolumn{3}{c}{$n(n-1)(n-5)/2$} &\multicolumn{2}{l}{$n(n-3)/2$} & $(n-1)$ && 1\\
\hline
\begin{tikzpicture}[overlay,remember picture]
\foreach \i in {1,2,3,4,5}
\draw[->] ($(s\i.north west)+(0.01,0.1)$) -- ($(e\i.south east)+(-0.2,0.3)$);
\end{tikzpicture} % see https://www.quora.com/LyX-How-to-draw-diagonal-arrow-in-matrix
    \end{tabular}
    \caption{Multiplicity of the $K_{\rm A}$ energy-level manifolds in a spin system comprising $n$ magnetically equivalent spin-{\textonehalf} particles.  These map directly to the multiplicity of the zero-field eigenstates for XA$_n$ spin system, where $I_{\rm A}=1/2$.
    %, shown in diagrams \autoref{Fig:XAn2}--\autoref{fig:OrganoPhosphorous}
    The value of $g_{K_{\rm A}}$ in any row (A$_n$) of the table may be derived in a manner reminiscent of Pascal's triangle, by addition of the multiplicities $g_{K_{\rm A}+1/2}$ and $g_{K_{\rm A}-1/2}$ from the previous row (A$_{n-1}$). Alternatively, the sequence formulae at the bottom can be used to calculate $g_{K_{\rm A}}$ along the diagonal arrows.  Note that in addition each $K_{\rm A}$ angular momentum manifold is $(2K_{\rm A} + 1)$-degenerate due to the projection quantum number.  The total number of levels is given by summation of $g_{K_A}$ multiplied by the degeneracy, and equals an expected result: $\sum_{K_{\rm A}}{g_{K_{\rm A}}(2K_{\rm A}+1)} = 2^n$.}
    \label{tab:multiplicityAN}
\end{table*}

\subsubsection{Selection rules}\label{XAnSelectionRules}

% \MCDT{The subsection has been entirely rewritten and should be reviewed by DB/all authors.  It is not colored so as to see outstanding comments/unfinished bits more easily.}
As discussed in \autoref{Subsubsec:Observable}, NMR experiments detect magnetic fields associated with the magnetization of the nuclear ensemble. 
For a given state of the ensemble, we can find the magnetization vector by taking the trace of the product of the magnetization operator and the density operator, see \autoref{eq:TimeDomaineSignal}. We may ask what elements in the density matrix lead to nonzero magnetization. The answer is that non-oscillating magnetization is associated with the populations of eigenstates, while oscillating and rotating magnetization 
is associated with off-diagonal matrix elements (coherences) between certain nondegenerate eigenstates. 
Selection rules establish the relation between the coherences and magnetization, as well as which coherences are excited by application of an external magnetic field.
% Selection rules tell us which coherences are associated with a given magnetization direction \SP{I am not sure about this statement.}. %\DB{This is all fine. I wonder, however, if we should mention rotating rather than oscillating fields that are more closely associated with single-quantum coherences? Let us quickly discuss.} 
% % \MCDT{Let us not get into a deep, extended discussion about it.  If we have an XA system, say formic acid, the overall magnetization at zero field will oscillate (periodically change magnitude but not axis) hence it does not rotate.  Since most of the recent literature on zero-field NMR refers to Z as the axis of magnetization (hence the vertical lines representing allowed transitions in \autoref{Fig:XAn} and others), the coherences are actually zero quantum.  That is my reasoning for the choice of words.  If it's not wrong, let's keep it and move on?}
% These are the same for both excitation of the spin system by an external magnetic field and for generation of the detected field. 
More formally, this can be described as follows.

In order to determine the amplitude of the ZULF-NMR signal given in Eq.\,\eqref{eq:TimeDomaineSignal}, one has to evaluate the specific matrix elements of the initial density operator $\rho_{uv}(0)$ and the matrix element of the magnetization operator between the corresponding eigenstates.  For the XA$_n$ system,
%\begin{equation}
%    \begin{split}
%        M_{p,\, uv}=&\langle F_{\rm A}'(K_{\text A}',S'),m_{F_{\rm A}'}|\hat M_p|F_{\rm A}(K_{\text A},S),m_{F_{\rm A}}\rangle\\
%        =&\left\langle F_{\rm A}'(K_{\text A}',S'),m_{F_{\rm A}'}\left| \gamma_{\rm X}\hat S_p +\gamma_{\rm A}\hat K_{{\text A},p}\right|F_{\rm A}(K_{\text A},S),m_{F_{\rm A}}\right\rangle\\
%        =&\left\langle F_{\rm A}'(K_{\text A}',S'),m_{F_{\rm A}'}\left| \gamma_{\rm X}\hat F_{{\rm A},p}+(\gamma_{\rm X}-\gamma_{\rm A})\hat{S}_p\right|F_{\rm A}(K_{\text A},S),m_{F_{\rm A}}\right\rangle\,.
%        \label{eq:TransitionAmplitudeMz}
%    \end{split}
%\end{equation}
\begin{equation}
        \bra{u} \hat M_p \ket{v} = \bra{u} \gamma_{\rm A}\hat F_{{\rm XA},p}+(\gamma_{\rm X}-\gamma_{\rm A})\hat{S}_p \ket{v} \,,
        \label{eq:TransitionAmplitudeMz}
\end{equation}
where 
%the second equality follows from the definition of $\hat{\textbf{F}}_{\rm XA}$: 
$\hat{F}_{{\rm XA},p} \equiv \hat{K}_{{\rm A},p}+\hat{S}_{p}$. 

%\KS{Please check, I think there is a mistake in the equation above: it should be $gamma_A * K_A + gamma_X * S = gamma_A (S + K_A) + (gamma_X - gamma_A)*S$.}
%\MCDT{Yes, you are right.  Fixed now.}

\begin{figure}
    \centering
    \includegraphics{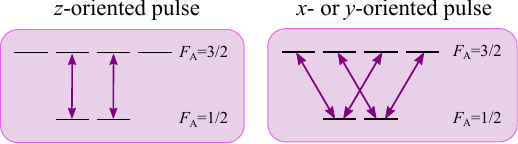}
    \caption{Transition rules for $x$-, $y$-, and $z$-oriented magnetic-field pulses in the representative XA$_2$ spin system.}
    \label{fig:TransitionRules}
\end{figure}

A convenient approach to evaluate $\braket{u|\hat{F}_{{\rm XA},p}|v}$ and $\braket{u|\hat{S}_p|v}$ is to use the algebra of tensor operators.  A spherical tensor operator $\hat{T}_{\kappa,q}^{(k)}$ is a superposition of spin operators for the subset of spins $k$ (e.g., X- or A-spins) with well-defined total angular momentum (or `tensor rank') $\kappa$ and projection quantum numbers $q$ with respect to global rotations of the system (i.e., rotation of all spins). %\SP{I still think this is unclear.} \DB{Not sure how to make it clearer. Tensors are always a bit abstract... The text reads OK to me.} %\SP{This is unclear. I also prefer $\kappa$ and $q$ instead of $L$ and $M$ (particularly the latter is dangerously close to magnetization symbol.}\MCDT{I am aware there are alternative conventions, k and q sounds fine.  Or $\lambda$ and $\mu$}.  
The quantum numbers $\kappa$ and $q$ mix the XA$_{n}$ eigenstates in a predictable and relatively simple way, using the rules of addition of angular momentum.  As an example, in the case of $\braket{u|\hat{S}_p|v}$ ($k=\rm{X}$), the relevant spherical tensor operators are those with rank $\kappa=1$:
\begin{subequations}
    \begin{align}
        \hat{S}_z &\equiv \frac{\hat{T}_{1,0}^{(\rm X)}}{\sqrt{2}} \,, \\
        \hat{S}_x &\equiv \frac{\hat{T}_{1,1}^{(\rm X)} + \hat{T}_{1,-1}^{(\rm X)}}{2} \,.
    \end{align}
\end{subequations}
The quantity $\hat{S}_p \ket{v}$ is found using a textbook approach \cite{Varshalovich1988}:
\begin{equation}
    \begin{split}
\hat{T}_{\kappa,q}^{(\rm{X})}\ket{F_{\rm XA}(K_{\rm A},S),m_F} =& (-1)^{2F_{\rm XA}+\kappa}\sqrt{(2F_{\rm XA}+1)(2\kappa+1)} \\
& \times\sum_{F_{\rm XA}'} 
\begin{Bmatrix}
    K_{\rm A} & S & F_{\rm XA}\\
    \kappa & F_{\rm XA}' & S
\end{Bmatrix} 
C_{F_{\rm XA} m_F, \kappa q}^{F_{\rm XA}' m_F'}\\ 
& \times\ket{F_{\rm XA}'(K_{\rm A}, S),m_F'}\,,
\label{eq:TLMSmatrixelement}
    \end{split}
\end{equation}
where $m_F$ is the magnetic quantum number, $C_{F_{\rm XA} m_F, \kappa q}^{F_{\rm XA}' m_F'}$ is a Clebsch--Gordan coefficient, and the term in curly brackets is a Wigner $6j$-symbol, and prime indicates the partner state in a coherence/transition.
Using \autoref{eq:TLMSmatrixelement}, the matrix elements are therefore
\begin{subequations}
    \begin{align}
        \braket{v|\hat{S}_z|u} &=\braket{F_{\rm XA}'(K_{\rm A}',S'),m_F'| \hat{S}_z | F_{\rm XA} (K_{\rm A},S) ,m_F} \nonumber \\
           &=(-1)^{2F_{\rm XA}+1}\sqrt{\frac{6F_{\rm XA}+3}{2}} 
            \begin{Bmatrix}
                K_{\rm A} & S & F_{\rm XA}\\
                1 & F_{\rm XA}' & S
            \end{Bmatrix} \nonumber \\ &\ \ \ \ \ \times C_{F_{\rm XA} m_F,10}^{F_{\rm XA}'m_F'} \delta_{S'S}\delta_{K_{\rm A}'K_{\rm A}}\,,
    \label{eq:T10Smatrixelement}
\\
\braket{v|\hat{S}_x|u} &= \braket{F_{\rm XA}'(K_{\rm A}',S'),m_F'| \hat{S}_z | F_{\rm XA} (K_{\rm A},S) ,m_F} \nonumber \\
        &= (-1)^{2F_{\rm XA}+1}\sqrt{\frac{6F_{\rm XA}+3}{2}} 
        \begin{Bmatrix}
            K_{\rm A} & S & F_{\rm XA}\\
            1 & F_{\rm XA}' & S
        \end{Bmatrix} \nonumber \\ &\ \ \ \ \ \times \left( C_{F_{\rm XA} m_F,1-1}^{F_{\rm XA}'m_F'} + C_{F_{\rm XA} m_F,11}^{F_{\rm XA}'m_F'} \right) \delta_{S'S}\delta_{K_{\rm A}'K_{\rm A}}\,.
    \label{eq:T11Smatrixelement}
    \end{align}
\end{subequations}
From these formulas, the selection rules for the changes of the quantum numbers $K_{\rm A}$, $S$ and $m_{F_{\rm XA}}$, associated with transitions induced by particularly oriented magnetic fields, can be determined. Specifically, a necessary condition for an allowed transition, arising from the $6j$-term, is for $\ket{u}$ and $\ket{v}$ to have common values of $S$ and $K_{\rm A}$ ($\Delta S=0$, $\Delta K_{\rm A}=0$), which implies that $F_{\rm XA}$ changes by 0 or 1 ($\Delta F_{\rm XA}=0,\pm 1$).  In addition, the Clebsch--Gordan coefficient ensures that $m_F$ is conserved for the transitions induced by magnetic field oriented along $z$ ($\Delta m_{F_{\rm XA}}=0$), but it may change by $\pm$1 for a field along $x$ or $y$ ($\Delta m_{F_{\rm XA}}=\pm 1$).  
These selection rules are summarized graphically in \autoref{fig:TransitionRules}.

The expression on the right-hand side of \autoref{eq:T10Smatrixelement} typically evaluates to between 0.4 and 0.5 for XA$_{n}$ spin systems with $S = 1/2$ and $I = 1/2$ (an exact result $|\braket{v|S_z|u}| = 1/2$ is in fact obtained for half-integer $K_{\rm A}$ and $m_F=\pm1/2$).  Selected values are provided in \autoref{tab:3j6jsymbol} for reference, where the mean value is also given to compare the overall strength of transitions between $F_{\rm XA}$ and $F_{\rm XA}'$.  Allowed transitions are therefore detected with approximately equal weighting.  It follows that the intensities of the zero-field spectral lines---each of which results from a superposition of coherences with the same energy difference---are approximately proportional to the multiplicity factor $g_{K_A} (2m_F+1)$.  An experimental example of this is shown in \autoref{fig:OrganoPhosphorous}, which is discussed in \autoref{subsec:axn}.  Additional dependence on $\rho_{uv}$ is discussed in \autoref{sec:Polarization1}.
%\DB{The statement in this paragraph needs more explanation. While $\rho_{uv}$ can at least be understood from Eq.\,\eqref{eq:TimeDomaineSignal} (need to mention this), the appearance of degeneracy is not obvious. I also think that all selection rules can be explained w/o any 6j and C-G coefficients. I wonder, if we shpuld do it here.} \SP{I had similar problems with this part. I am not sure if there is something specific we want to say by that. The problem also is that the discussion of density matrix and signal is in the next section.}
%\DB{Michael: please add several sentences to make the text clearer!} \MCDT{Please check whether it now makes sense.  The point is that yes, selection rules tell you whether something should be zero or nonzero.  But we can speak quantitatively as well, because we can actually calculate the transition dipole moments.  

\begin{table*}[h]
    \centering
       \begin{tabular}{ccccccccccccc}
        \hline
        \multicolumn{4}{c}{Angular momentum} & \multicolumn{8}{c}{Projection}\\
        $K_{\rm{A}}$ & $S$ & $F_{\rm{XA}}$ & $F_{\rm{XA}}'$ & \multicolumn{8}{c}{$|m_F|$} \\
        &&&&  0 & 1/2 & 1 & 3/2 & 2 & 5/2 & 3 & Mean \\
        \hline\hline
        1/2 & 1/2 &  0   &  1         & 0.5 &&&&&&& 0.5\\ 
        1   & 1/2 &  1/2 &  3/2       && 0.471  &&&&&& 0.471\\ 
        3/2 & 1/2 &  1   &  2         & 0.5 && 0.433 &&&&&0.455\\
        2   & 1/2 &  3/2 &  5/2       && 0.490 && 0.4 &&&& 0.445\\ 
        5/2 & 1/2 &  2   &  3         & 0.5 && 0.471 && 0.373 &&& 0.438\\     
        3   & 1/2 &  3/2 &  5/2       && 0.495 && 0.452 && 0.350 && 0.432\\
        7/2 & 1/2 &  3   &  4         & 0.5 && 0.484 && 0.433 && 0.331 &0.428\\ \hline
        1/2 & 1 &  1/2 &  3/2       && 0.236 &&&&&& 0.236\\ 
        1   & 1 &  0   &  1         & 0.408  &&&&&&& 0.408\\ 
        1   & 1 &  1   &  2         & 0.289 && 0.25 &&&&& 0.263\\
        3/2 & 1 &  1/2 &  3/2       && 0.373 &&&&&& 0.373\\ 
        3/2 & 1 &  3/2 &  5/2       && 0.3 && 0.245 &&&& 0.272\\     
        2   & 1 &  1   &  2         & 0.387 && 0.335 &&&&& 0.353\\
        2   & 1 &  2   &  3         & 0.316 && 0.298 && 0.236 &&& 0.277\\
        5/2 & 1 &  3/2 &  5/2       && 0.374 && 0.306 &&&& 0.334\\
        5/2 & 1 &  5/2 &  7/2       && 0.312 && 0.292 && 0.226 && 0.279\\ 
        3   & 1 &  2   &  3         & 0.378 && 0.356 && 0.282 && & 0.331\\
        3   & 1 &  3   &  4         & 0.327 && 0.317 && 0.283 && 0.216 & 0.279\\\hline
\end{tabular}
\caption{Numerical values for the right-hand side of \autoref{eq:T10Smatrixelement} using $S=1/2$ or $S=1$, which represents the relative amplitudes of nominally allowed transitions between magnetic sublevels $\ket{F_{\rm XA} (K_{\rm A}, S), m_F} $ and $ \ket{F'_{\rm A} (K_{\rm A}, S), m_F }$ under $\hat{S}_z$. The mean (i.e.,\ the average taken over all sublevels from $m_F=-F_\textrm{XA}$ to $m_F=F_\textrm{XA}$) is provided to indicate the relative probability of the transition $F_{\rm XA}\leftrightarrow F_{\rm XA}'$ at zero magnetic field.}
\label{tab:3j6jsymbol}
\end{table*}

We also note that the selection rules can be understood intuitively by realizing that magnetic dipole transitions are in fact processes where a single photon with zero orbital angular momentum is absorbed or emitted in the course of the transition.  As the photon is a spin-1 particle, this immediately leads to the selection rule that the total angular momentum and its projection can only change by 0 or $\pm$1.  The selection rules for the `intermediate' quantum numbers can also be understood `without equations' in analogy to how this is done in atomic physics, see, for example, the tutorial discussion in reference \cite{auzinsh2010optically}.

\begin{table}[h]
    \centering
       \begin{tabular}{ccccccc}
        \hline
        System &$p$ & \multicolumn{3}{c}{Total angular momentum} & Projection\\
         & & $\Delta K_{\rm A}$ & $\Delta S$ & $\Delta F_{\rm XA}$, $\Delta F$ & $\Delta m_{F_{\rm XA}}$, $\Delta m_{F}$ \\
        \hline\hline
        XA$_n$ &  $x,y$ & 0 & 0 & $0,\pm 1$ & $\pm 1$ \\ 
         & $z$ & 0 & 0 & $\pm 1$ & 0 \\ \hline
        (XA$_n$)B$_m$  &  $x,y$ &0&0& $0,\pm1$ & $\pm1$ \\
        & $z$ & 0 & 0 & $\pm 1$ & 0 \\ \hline
\end{tabular}
\caption{Allowed changes in angular momentum quantum numbers for $p$-axis-observable transitions in the zero-field NMR spectra of XA$_n$ and (XA$_n$)B$_m$ systems.}
\label{tab:example}
\end{table}

\subsubsection{Zero-field-NMR signal for a thermally polarized system\label{sec:Polarization1}}

In \autoref{eq:TimeDomaineSignal}, the values of the matrix elements $\rho_{uv}(0)$ depend on the initial polarizations of the spins (various methods to produce polarization are discussed in \autoref{Sec:Spin_pol}). As in conventional NMR, the conceptually simplest method is to polarize the spin system thermally.  This is done by placing the sample in a strong field $\bm{\mathrm{B}}_{\rm pol}$ (typically at least 1\,T), generated, for example, with a permanent magnet situated outside of the detection region.  After a time of at least 3$T_1$, the spin ensemble is close to thermal equilibrium, $\rho_{uv}(0) = \rho_{\rm th}$.  For $\bm{\mathrm{B}}_{\rm pol}$ oriented along the $z$-axis, the thermal equilibrium density operator of an XA$_n$ system is 
% \SP{We have a small problem here, as we use $I$ which was a reserved spin notation for $A$ spins.} \MCDT{There is no obvious choice for a replacement symbol here. What do you propose?}
% \begin{equation}
%     \begin{split}
% 	    \hat{\rho}_{th}=&\frac{1}{Z}\exp\left(\frac{\hat{\mathcal{H}}}{k_B T}\right)
%         \approx\frac{1}{Z}\exp\left(-\frac{\hbar B_{\rm pol}}{k_BT}\sum_i\gamma_i\hat{I}_{z,i}\right)\\
% 	    \approx& \frac{1}{Z}\mathbb{I} - \frac{1}{Z}\frac{\hbar B_{\rm pol}}{k_BT} 
%      \left[ 
%      (\gamma_{\text X}-\gamma_{\text A}) \hat{S}_z + 
%      \gamma_{\text A} \hat{F}_{{\text XA},z}
%      \right]\,,
%     \end{split}
%     \label{eq:DensityMatrixThermal}
% \end{equation}
\begin{equation}
    \begin{split}
	    \hat{\rho}_{\rm th}=&\frac{1}{Z}\exp\left(\frac{\hat{\mathcal{H}}}{k_B T}\right)\\
        \approx&\frac{1}{Z}\exp\left[-\frac{\hbar B_{\rm pol}}{k_BT}\left(\gamma_\textrm{X}\hat{S}_z+\sum_i\gamma_A\hat{I}_{z,i}\right)\right]\\
	    \approx& \frac{1}{Z}\mathbb{I} - \frac{1}{Z}\frac{\hbar B_{\rm pol}}{k_BT} 
     \left[ 
     (\gamma_{\text X}-\gamma_{\text A}) \hat{S}_z + 
     \gamma_{\text A} \hat{F}_{{\text XA},z}
     \right]\,,
    \end{split}
    \label{eq:DensityMatrixThermal}
\end{equation}
where $B_{\rm pol} = |\bm{\mathrm{B}}_{\rm pol}|$ and $Z=\Pi_i(2I_i+1)$ is the partition function. If both spin species X and A are spin-{\textonehalf}, then $Z=2^{n+1}$. It should be noted that there are two approximations used in  \autoref{eq:DensityMatrixThermal}. The first one is neglecting the $J$-coupling contribution to the high-field Hamiltonian and the second one is the assumption of operation in the low-polarization limit, in which only the first nontrivial expansion coefficient is retained.

The sample is then rapidly ($\delta t \ll T_1$) shuttled into the zero- or ultralow-field detection region, where its magnetization is manipulated and detected.  Under such conditions, one can assume that the high-field thermal equilibrium state is maintained, so that $\rho_{uv}(0) = \rho_{\rm th}$.  If one is interested in ZULF-NMR signals of nonzero frequency, only the term proportional to $(\gamma_{\text X}-\gamma_{\text A})$ in \autoref{eq:DensityMatrixThermal} is important. This follows because the matrix elements of eigenoperators $\hat{F}_{{\text XA},z}$ are zero for $\ket{u}\neq\ket{v}$, and because the transition frequency is zero for $\ket{u}=\ket{v}$.  The observable NMR signal measured along the $z$-axis, ignoring constant offset terms and relaxation, is therefore
\begin{eqnarray}
    s_z(t) &\propto& (\gamma_{\text X}-\gamma_{\text A})^2 \sum_{u,v\neq u}  \braket{v|S_z|u}^2 \mathrm{e}^{-\mathrm{i} \omega_{uv} t} \,,
\end{eqnarray}
or in terms of the allowed angular momentum quantum numbers [\autoref{eq:T10Smatrixelement}]
\begin{eqnarray}
    s_z(t) &\propto& (\gamma_{\text X}-\gamma_{\text A})^2 \sum_{F_\textrm{XA'}\neq F_\textrm{XA}} \sum_{F_\textrm{XA}, K_A, S, m_F}  \frac{6F_\textrm{XA} + 3}{2}  g_{K_A} \nonumber \\ 
    && \times
\left(\begin{Bmatrix}
    K_A & S & F_\textrm{XA}\\
    1 & F_\textrm{XA'} & S
\end{Bmatrix}
C_{F_\textrm{XA} m_F,10}^{F_\textrm{XA'} m_F} \right)^2  \nonumber \\ 
&& \times 
\mathrm{e}^{-2\pi \mathrm{i} J_\textrm{XA} [F_\textrm{XA}(F_\textrm{XA}+1) - F_\textrm{XA'}(F_\textrm{XA'}+1)] t} \,. \label{eq:stF}
\end{eqnarray}
Similar relations for the signals measured along the $x$- or $y$-axis can be derived by combining \autoref{eq:T11Smatrixelement} and \autoref{eq:DensityMatrixThermal} with \autoref{eq:TimeDomaineSignal}.

In \autoref{eq:stF}, we see that the amplitude of the zero-field NMR signal starting from a thermally polarized sample is proportional to the square of the difference of the gyromagnetic ratios of spins X and A. Other dependencies, however, can be generated starting from the same thermally polarized initial state.  Spin-species-selective inversion of the A- or X-spins prior to nonadiabatic switching converts the initial state to
\begin{equation}
    \begin{split}
	    \hat{\rho}_{\rm th} \xrightarrow[]{(\pi)_{\rm A}}\frac{1}{Z}\mathbb{I} - \frac{1}{Z}\frac{\hbar B_{\rm pol}}{k_BT} 
     \left[ 
     (\gamma_{\text X}+\gamma_{\text A}) \hat{S}_z + 
     \gamma_{\text A} \hat{F}_{{\text XA},z}
     \right]\,,
    \end{split}
    \label{eq:DensityMatrixThermalafterXpulse}
\end{equation}
or 
\begin{equation}
    \begin{split}
	    \hat{\rho}_{\rm th} \xrightarrow[]{(\pi)_{\rm X}}\frac{1}{Z}\mathbb{I} - \frac{1}{Z}\frac{\hbar B_{\rm pol}}{k_BT} 
     \left[ 
     (-\gamma_{\text X}-\gamma_{\text A}) \hat{S}_z + 
     \gamma_{\text A} \hat{F}_{{\text XA},z}
     \right]\,,
    \end{split}
    \label{eq:DensityMatrixThermalafterApulse}
\end{equation}
respectively.  
The result is to re-scale the component of $\rho_{uv}$ corresponding to time-dependent magnetization, from $(\gamma_{\rm A}-\gamma_{\rm X})$ to $(-\gamma_{\rm A}-\gamma_{\rm X})$ or $(\gamma_{\rm A}+\gamma_{\rm X})$, respectively \cite{Tayler2016}.  This means that the signal is larger by a factor of $|\gamma_{\rm A}+\gamma_{\rm X}|/|\gamma_{\rm A}-\gamma_{\rm X}|$ when the gyromagnetic ratios for X and A are of the same sign.
The same enhancement factors can be obtained when the spin system is adiabatically transferred to zero field under a magnetic field along $z$ and a pulsed magnetic field along the $x$- or $y$-axis is applied before measurement (see reference \cite{Emondts2014} and its supplementary materials).

\subsubsection{Experimental spectra}
\label{subsec:axn}

\begin{figure}
\begin{center} 
    \includegraphics[width=\columnwidth]{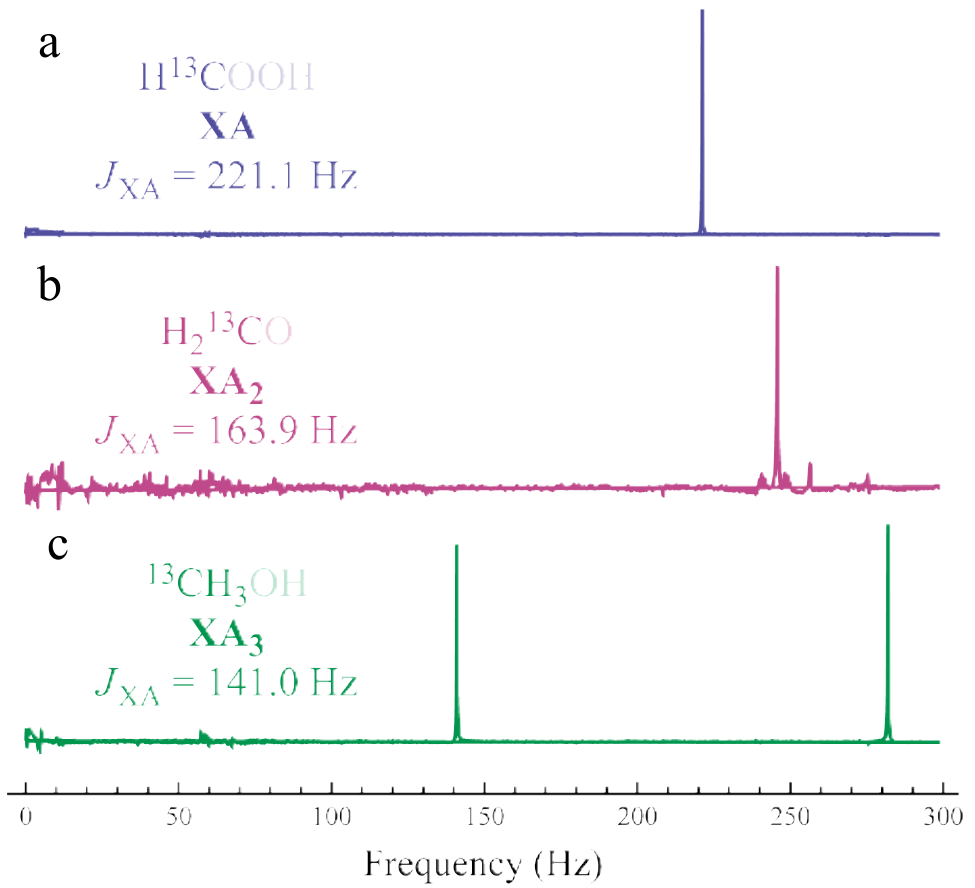} \end{center}
    \caption[Zero-field $J$-spectra of XA$_n$ systems for $n=1,2,3$]{Zero-field $J$-spectra of [$^{13}$C]-formic acid (a), [$^{13}$C]-formaldehyde (b), and [$^{13}$C]-methanol (c), as examples of XA, XA$_2$, and XA$_3$ spin systems, respectively.  Reprinted (adapted) from reference \cite{Theis2013}, Copyright (2013), with permission from Elsevier.}
    \label{Fig:XAn-Exp}
\end{figure}

\begin{figure}
    \centering
    \includegraphics[width=\columnwidth]{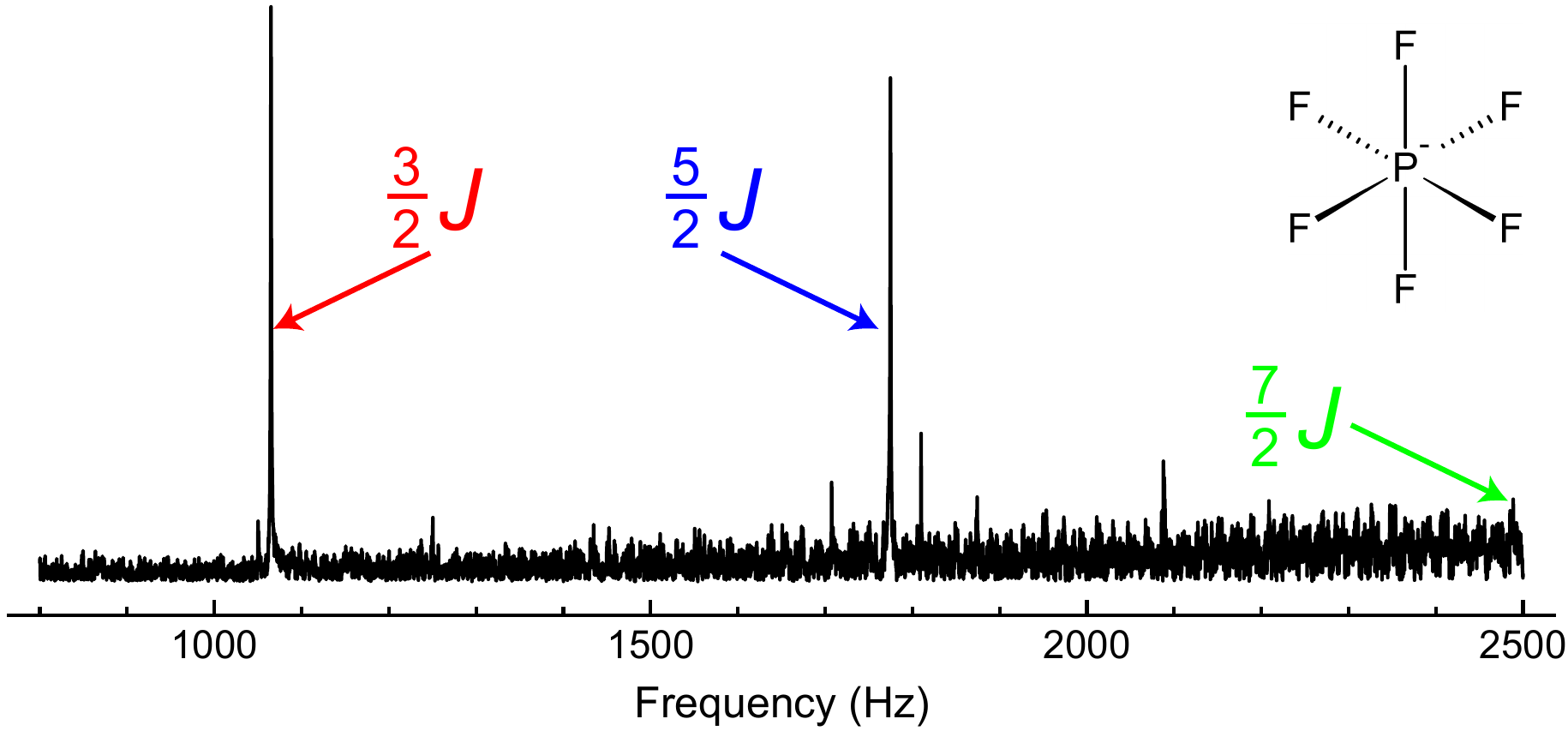}
    \caption{\label{fig:BMIMPF6} ZULF-NMR spectrum of PF$_6^-$ in the ionic liquid 1-butyl-3-methylimidazolium hexafluorophosphate [BMIm][PF$_6$]. For an XA$_6$ spin system, peaks are expected at $3\,J_{\rm{XA}}/2$, $5\,J_{\rm{XA}}/2$, and $7\,J_{\rm{XA}}/2$. The presence of multiple resonances allows the one-bond ${}^{31}$P--${}^{19}$F coupling, ${}^{1}J_{\rm{PF}} =709.974(2)$\,Hz, to be extracted with high precision. 
The increased noise floor at higher frequencies is related to the relatively small detector bandwidth ($\sim$\,200\,Hz) 
---correcting for the roughly Lorentzian frequency-dependent response leads to digitization noise being multiplied by an exponentially increasing calibration factor.
    Reprinted from reference \cite{Blanchard2021_LtL} Copyright (2021), with permission from Elsevier.
}   
\end{figure}

\begin{figure} 
    \centering
    \includegraphics[width=\columnwidth]{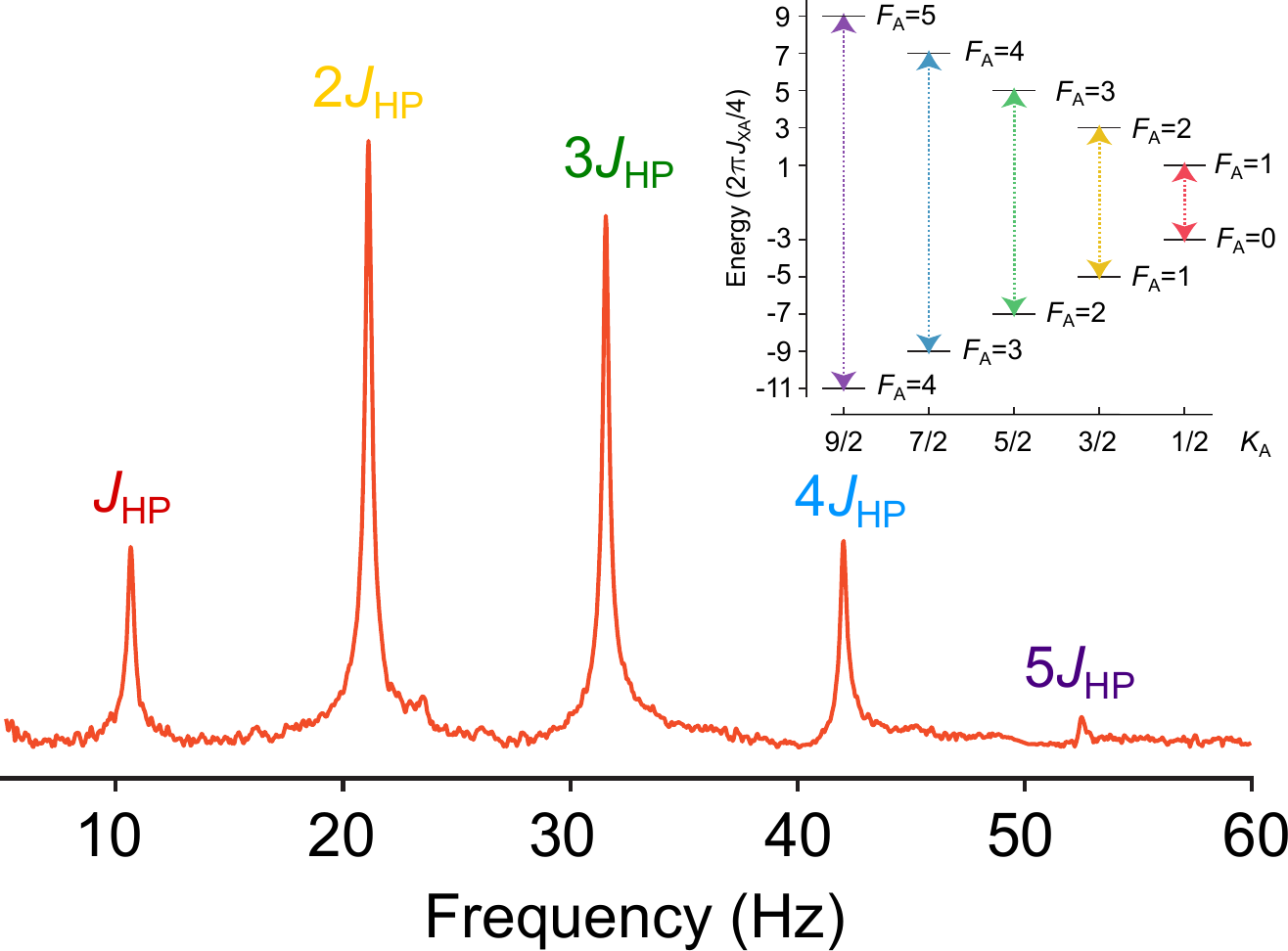}
    \caption{\label{fig:OrganoPhosphorous} ZULF NMR spectrum of trimethyl phosphite (CH$_3$O)$_3$P. For an (XA$_3$)$_3$ spin system, peaks are expected at $J_{\rm XA}$, $2J_{\rm XA}$, $3J_{\rm XA}$, $4J_{\rm XA}$, and  $5J_{\rm XA}$, where $J_{\rm XA}$ is a three-bond coupling of 10.5\,Hz between $^1$H and $^{31}$P. %For simplicity, a degeneracy of the energy levels was negligible. 
    Relative line intensities approximately follow the multiplicity pattern for the $K_\text{A}$ eigenstates of an A\textsubscript{9} system, given in \autoref{tab:multiplicityAN}.
    Adapted from reference \cite{Alcicek2021Ogranophosphorus} under terms of the Creative Commons CC-BY license 4.0.  %\MCDT{Let us adapt the figure by relabeling the x axis; in this review, we use $K_A$ instead of $I_A$.}
    }
\end{figure}

Examples  of experimental zero-field spectra of XA$_n$ spin systems are shown in \autoref{Fig:XAn-Exp}, for simple compounds available in neat liquid form, which can be regarded as `standard ZULF-NMR samples': [\textsuperscript{13}C]-formic acid (H\textsuperscript{13}COOH, XA), [\textsuperscript{13}C]-formaldehyde (\textsuperscript{13}CH\textsubscript{2}O, XA$_2$), and [\textsuperscript{13}C]-methanol (\textsuperscript{13}CH\textsubscript{3}OH, XA$_3$).  In the simplest case of an XA spin system at zero field, the only possible quantum numbers are $S=1/2$ and $K_{\rm A} = 1/2$ (as per the first row in \autoref{Fig:XAn2}), giving two sets of states, $F_{\rm XA}=0$ and $F_{\rm XA}=1$.  There is only one $\Delta F_{\rm XA}=1$ and $\Delta K_{\rm A}=0$ transition at nonzero frequency, at the value of the $J_{\rm XA}$-coupling constant.  In formic acid, this is at around 220\,Hz (Fig.\,\ref{Fig:XAn-Exp}a). For the XA$_2$ spin system, there are three sets of states with distinct $K_{\rm A}$ and $F_{\rm XA}$ (see \autoref{Fig:XAn2}).  Two of these ($F_{\rm XA} = 1/2$ and $F_{\rm XA}=3/2$) are derived from $K_{\rm A} = 1$ and $S=1/2$ and share an allowed transition at $3J_{\rm XA}/2$.  As a result, the zero-field spectrum of [\textsuperscript{13}C]-formaldehyde consists of a single line at around $3J_{\rm XA}/2$ = 250\,Hz (\autoref{Fig:XAn-Exp}b).  The XA$_3$ system is more interesting: there are now two $K_{\rm A}$ manifolds containing nondegenerate energy states.  As the third row of \autoref{Fig:XAn2} shows, the $K_{\rm A}=1/2$ manifold contains two doubly degenerate states with $F_{\rm XA}=0$ and $F_{\rm XA}=1$, while the $K_{\rm A}=3/2$ manifold comprises states with $F_{\rm XA}=1$ and $F_{\rm XA}=2$. This level structure leads to two transitions at nonzero frequency, at $J_{\rm XA}$ and $2J_{\rm XA}$ for $K_{\rm A}=1/2$ and $K_{\rm A}=3/2$ respectively, which in [\textsuperscript{13}C]-methanol occur at 140\,Hz and 280\,Hz, respectively (\autoref{Fig:XAn-Exp}c).  As a general rule, the zero-field NMR spectrum of a spin-{\textonehalf} XA$_n$ system exhibits transition frequencies at even integer multiples of $J_{\rm XA}/2$ for odd $n$, and odd integer multiples of $J_{\rm{XA}}/2$ for even $n$, up to a maximum frequency of $(n+1)J_{\rm XA}/2$.  \autoref{fig:BMIMPF6} shows the experimental zero-field spectrum of the octahedral anion PF$_6^-$ (XA$_6$), and \autoref{fig:OrganoPhosphorous} that of the XA$_9$ system (CH$_3$O)$_3$P. %\DAB{(`pseudo' refers here to the fact that we neglect the $J$-coupling interactions between the three methyl groups)}.  %\MCDT{I found this paper: \href{https://doi.org/10.1039/J19690001972}{Finer+Harris JCSA} may have something relevant for ULF NMR of (XA$_3$)$_3$.}

\begin{figure}
  \begin{center} \includegraphics[width=\columnwidth]{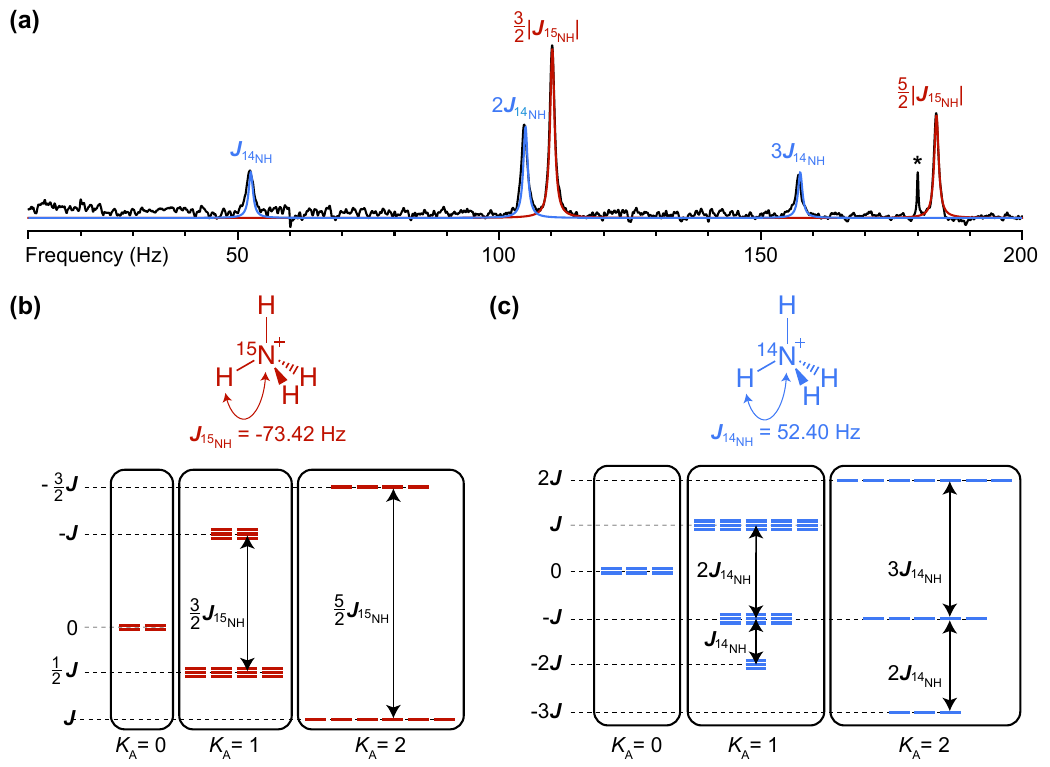} \end{center}
  %\begin{center} \includegraphics[width=6.5 in, resolution=50]{Figures/J-Spectroscopy/NH4.pdf} \end{center}
  \caption[(a) Zero-field $J$-spectrum of a mixture of $\rm {}^{14}NH_4Cl$ and $\rm {}^{15}NH_4Cl$.]{The zero-field $J$-spectrum of a 60:40 mixture of \textsuperscript{15}NH\textsubscript{4}Cl and \textsuperscript{14}NH\textsubscript{4}Cl dissolved in an acidic aqueous solution ($\sim6$\,M NH$_4$Cl) as an example of an XA\textsubscript{4} spin systems containing both spin-1 and spin-{\textonehalf} X-nuclei.  Energy-level diagram for (b) $^{15}$NH$_4^{+}$ and (c) $^{14}$NH$_4^{+}$ cations. Energy-level shifts due to the \textsuperscript{1}H--\textsuperscript{1}H coupling $J_{\rm HH}$ are ignored. Adapted from reference \cite{picazo2024zero} under terms of the Creative Commons CC-BY license 4.0.}
  \label{Fig:NH4}
\end{figure}

The cases considered so far involve spin-{\textonehalf} nuclei, but the theory also applies to nuclei with a quadrupolar moment ($S\geq1$), for instance \textsuperscript{14}N and \textsuperscript{2}H (both $S=1$). 
Rapid spin relaxation due to nuclear quadrupole interactions typically hinders observation of their signals, although this is not always the case. One example is the ammonium ion \textsuperscript{14}NH\textsubscript{4}\textsuperscript{+}, which has tetrahedral symmetry, so that the electric field  gradients at the quadrupolar nitrogen nucleus and, correspondingly, $\hat{\mathcal{H}}_Q^{({}^{14} \rm N)}$, nearly vanish, resulting in the suppression of quadrupolar relaxation.  This spin system is one of the few cases where zero-field $J$-spectra of coupled spin systems containing quadrupolar nuclei have been detected, with line widths comparable to isotopologs containing only spins-{\textonehalf}.  As a result, ammonium is a XA$_4$ spin system that can be studied for X being either spin-1 or spin-{\textonehalf} \cite{picazo2024zero}, see \autoref{Fig:NH4}. The transitions in \autoref{Fig:NH4}a are readily understood from the theory presented in \autoref{XAnSelectionRules}. Specifically, the total \textsuperscript{1}H angular momentum results in three manifolds, $K_A=0,1,2$, where the last two contain one or more nonzero-frequency transitions due to the $\Delta K_A=0$ selection rule.  The observable transitions in $\rm {}^{15}NH_4Cl$ occur at 3$|J_{^{15}\rm NH}|$/2 and 5$|J_{^{15}\rm NH}|$/2 (\autoref{Fig:NH4}b), while in $\rm {}^{14}NH_4Cl$ resonances are observed at $|J_{^{14}\rm NH}|$, 2$|J_{^{14}\rm NH}|$, and 3$|J_{^{14}\rm NH}|$ (\autoref{Fig:NH4}c). The difference in the $\rm {}^{14}N$-H and $\rm {}^{15}N$-H $J$-coupling frequencies is primarily due to the different gyromagnetic ratios of the two nuclei. From multiple measurements of the same $\rm NH_4Cl$ sample, the ratio of the $J$-couplings is $\left|J_{\rm {}^{15}NH}/J_{\rm {}^{14}NH}\right|=1.4012(4)$, as compared to $\left|\gamma_{\rm {}^{15}N-H}/\gamma_{\rm {}^{14}N-H}\right|=1.4027548(5)$. The small difference between the ratios of $\left|J_{\rm {}^{15}NH}/J_{\rm {}^{14}NH}\right|$ and $\left|\gamma_{\rm {}^{15}NH}/\gamma_{\rm {}^{14}NH}\right|$ manifested as a primary isotope effect of $-58$ mHz is likely due to the different zero point energies of $\rm {}^{14}NH_4^+$ and $\rm {}^{15}NH_4^+$, leading to a slight difference in electron densities, although further studies are necessary to verify this claim.

\subsection{(XA$_n$)B$_m$ spin systems at zero field}\label{XAnBm}
\label{Subsec:XAnBm}
The theory presented in \autoref{Sec:XAn} describes the zero-field NMR behavior for a nuclear spin system of simple topology.  With additional $J$-couplings between spins and more elaborate topologies, zero-field NMR spectra become increasingly more complex and harder to analyze rigorously.  In special cases, however, simplifications can be made.  In this context, we now discuss (XA$_n$)B$_m$ spin systems (the modified Pople notation that we use throughout this review is explained in \autoref{Subsec:Notations}).

As one-bond $J$-couplings are usually substantially larger than couplings to more distant spins, the zero-field NMR spectra of (XA$_n$)B$_m$ spin system can be evaluated in leading order by isolating the XA$_n$ subgroup from the overall spin network.  To deal with the higher-order corrections, we analyze the system using perturbation theory \cite{Butler2013}.

\subsubsection{Energy-level structure}

The interaction Hamiltonian of an (XA$_n$)B$_m$ spin system at zero field is given by
\begin{equation}
    \hat{\mathcal{H}}_{J_{{\rm (XA}_n){\rm B}_m}}=2\pi\hbar \left( 
    J_{\rm XA} \oper{K}_{\rm A} \cdot \hat{\bm{\mathrm{S}}} 
    + J_{\rm XB} \oper{K}_{\rm B} \cdot \hat{\bm{\mathrm{S}}} 
    + J_{\rm AB} \oper{K}_{\rm A} \cdot \oper{K}_{\rm B}\right)\,,
\end{equation}
where couplings between magnetically equivalent spins are ignored, and $\oper{K}_{\rm B}$ is the total angular momentum operator of the B-spin subsystem.

Now, we assume that the $J_{\rm XA}$ coupling is the dominant interaction in the system ($|J_{\rm XA}| \gg |J_{\rm AB}|, |J_{\rm XB}|$). This allows us to identify a reference Hamiltonian (i.e., that of the XA$_n$ subsystem, $\hat{\mathcal{H}}^{(0)}=\hat{\mathcal{H}}_{J,{\rm XA}_n}$) to approximate the eigenstates at lowest perturbation order.  The perturbation Hamiltonian $\hat{\mathcal{H}}_{\rm per}$ is then defined as %$\hat{\mathcal{H}}_{\rm per} = \hat{\mathcal{H}}_{J{{\rm (XA}_n){\rm B}_m}} - \hat{\mathcal{H}}_{J{{\rm (XA}_n)}}$ where
\begin{equation}\label{Eq.H1}
    \hat{\mathcal{H}}_{\rm per} =  2\pi\hbar \left( J_{\rm XB}  \hat{\bm{\mathrm{S}}}\cdot\oper{K}_{\rm B} +  J_{\rm AB} \oper{K}_{\rm A} \cdot \oper{K}_{\rm B} \right)\,.
\end{equation}

The conventional formula for the first-order energy corrections to each reference eigenstate, i.e., $\Delta E^{(1)}_u=\langle u|\hat{\mathcal{H}}_{\rm per}| u\rangle$, does not immediately provide a compact analytical result. This is so because neither scalar product term in \autoref{Eq.H1} commutes with $\hat{\mathcal{H}}_{J,{\rm XA}_n}$.  Therefore, to simplify the analysis, it is convenient to replace $\hat{\mathcal{H}}_{\rm per}$ with an approximation.  First, it can be assumed that $\oper{K}_{\rm B}$ couples directly to $\oper{F}_{\rm XA}$, instead of coupling independently to $\oper{K}_{\rm B}$ and $\oper{S}$. Hence, B effectively interacts with the projections of $\oper{K}_{\rm B}$ and $\oper{S}$ onto $\oper{F}_{\rm XA}$, and one can write the first-order perturbation Hamiltonian as
\begin{equation}\label{Eq:H1v2}
    \begin{split}
    \hat{\mathcal{H}}^{(1)}=& 2\pi\hbar(J_{\rm XB}\oper{S}^\|\cdot\oper{K}_{\rm B}+J_{\rm AB}\oper{K}_{\rm A}^\|\cdot\oper{K}_{\rm B})\\
    =&2\pi\hbar \left( J_{\rm XB}^{(1)} + J_{\rm AB}^{(1)} \right) \oper{F}_{\rm XA} \cdot \oper{K}_{\rm B}\,,
    \end{split}
\end{equation}
where $\oper{S}^\|$ and $\oper{K}_{\rm A}^\|$ are components of $\oper{S}$ and $\oper{K}_{\rm A}$ parallel to $\oper{F}_{\rm XA}$ and 
\begin{subequations}
    \begin{align}\label{Eq.Jfirst}
        J_{\rm XB}^{(1)}&=J_{\rm XB}\left[\frac{F_{\rm XA} \left(F_{\rm XA}+1\right)+ S\left(S+1\right)-K_{\rm A} \left(K_A + 1 \right)}{2F_{\rm XA}\left(F_{\rm XA}+1\right)}\right]\,,\\
        J_{\rm AB}^{(1)}&=J_{\rm AB}\left[\frac{F_{\rm XA} \left(F_{\rm XA}+1\right)+K_{\rm A} \left(K_{\rm A} + 1 \right) - S\left(S+1\right)}{2F_{\rm XA}\left(F_{\rm XA}+1\right)}\right]\,
    \end{align}
\end{subequations}
are the effective $J$-coupling constants specific to each state. \autoref{Eq:H1v2} allows simple evaluation of the first-order energy corrections by using the total-spin operator $\oper{F}$ 
\begin{equation}\label{Eq.FFAIB}
    \oper{F}=\oper{F}_{\rm XA}+\oper{K}_{\rm B}\,,
\end{equation}
% and then an approach analogous to that used to derive \autoref{Eq.KAS} allows \autoref{Eq:H1v2} to be rewritten as
in which the Hamiltonian \eqref{Eq:H1v2} can be rewritten as
\begin{equation}\label{Eq:H1v3}
    \hat{\mathcal{H}}^{(1)} =  \pi h \left( J_{\rm XB}^{(1)} + J_{\rm AB}^{(1)} \right) \left(\oper{F}^2-\oper{F}_{\rm XA}^2-\oper{K}_{\rm B}^2\right)\,.
\end{equation}
With this form of the Hamiltonian and the general formula for the first-order energy corrections, $\Delta E^{(1)}=\langle u|\hat{\mathcal{H}}^{(1)}| u\rangle$, the resulting correction for the energy of the state $|u\rangle$ is
\begin{align}\label{Eq.E1}
    &\Delta E^{(1)}_u = \\
    & \pi\hbar\left(J_{\rm BX}^{(1)}+J_{\rm AB}^{(1)}\right)\left[F(F+1)-F_{\rm XA}\left(F_{\rm XA}+1\right)-K_{\rm B}\left(K_{\rm B}+1\right)\right]\,.\nonumber
\end{align}
Here, $F$ and $K_{\rm B}$ are the quantum numbers corresponding to the $\oper{F}$ and $\oper{K}_{\rm B}$ operators, respectively.  
%Note the special case of an (XA)B\textsubscript{m} system with $K_{\rm{A}}=S=1/2$, where the effective couplings are simply $J_{\rm XB}^{(1)} = J_{\rm XB}/2$ and $J_{\rm AB}^{(1)} = J_{\rm AB}/2$ and independent of $F_{\rm A}$. \SP{No sure who add it, but I am not sure what is so special about this case. Shall we really keep this?}

The first-order perturbation approximation is usually sufficient to describe the (XA$_n$)B$_m$ system, especially in situations where $|J_{\rm XB} + J_{\rm AB}|$ is much smaller than $|J_{\rm XA}|$.  For a better estimate, however, higher-order energy corrections can also be calculated.  A second-order correction is obtained using the generic result of perturbation theory with the untruncated form of the perturbation Hamiltonian $\hat{\mathcal{H}}_{\rm per}$ [\autoref{Eq.H1}] instead of $\hat{\mathcal{H}}^{(1)}$:
\begin{equation}
    \Delta E^{(2)}_u = \sum_{v \neq u}\frac{|\bra{u} \hat{\mathcal{H}}_{\rm per} \ket{v}|^2}{ E^{(0)}_u - E^{(0)}_v}\,, \label{eq:E2XAnBm2}
\end{equation}
where the summation runs over all zero-order eigenstates $\ket{v}$ different in energy to $\ket{u}$ and of the same value of $F$.

To calculate the numerator of \autoref{eq:E2XAnBm2}, one needs to calculate the specific matrix elements
\begin{equation}
    \bra{v}\hat{\mathcal{H}}_{\rm per}\ket{u}=2\pi \hbar \left( J_{\rm XB}\bra{v}\oper{S}\cdot\oper{K}_{\rm B}\ket{u}  +  J_{\rm AB}\bra{v}\oper{K}_{\rm A}\cdot\oper{K}_{\rm B}\ket{u} \right)\,,
    \label{eq:Hper}
\end{equation}
which is not trivial because the two coupling operators $\oper{S}\cdot\oper{K}_{\rm B}$ and $\oper{K}_{\rm A}\cdot\oper{K}_{\rm B}$ do not share an eigenbasis.  To avoid this problem, however, one can evaluate the $\bra{u}\oper{S}\cdot\oper{K}_{\rm B}\ket{v}$ and $\bra{u}\oper{K}_{\rm A}\cdot\oper{K}_{\rm B}\ket{v}$ terms separately in their respective eigenbases, and add them together afterwards \cite{Butler2013}. The following results can be used: 
\begin{subequations}
\begin{equation}
    \begin{split}
        \bra{v}\oper{S}\cdot\oper{K}_{\rm B}\ket{u}\,\,=&\sum_{\FB=|S-\KB|}^{S+\KB}\left(\FB+\frac{1}{2}\right)\sqrt{(2\FA+1)(2\FA'+1)}\\
        &\times\sixJ{S}{\KB}{\FB}{F}{\KA}{\FA}\sixJ{S}{\KB}{\FB}{F}{\KA}{\FA'}\\
        &\times[\FB(\FB+1)-S(S+1)-\KB(\KB+1)]\,,
    \end{split}
\end{equation}
\begin{equation}
\label{eq:SKB}\\
    \begin{split}
        \bra{v}\oper{K}_{\rm A}\cdot\oper{K}_{\rm B}\ket{u}\,\, =&\sum_{\KAB=|\KA-\KB|}^{\KA+\KB}\left(\KAB+\frac{1}{2}\right)\sqrt{(2\FA+1)(2\FA'+1)}\\
        &\times\sixJ{\KA}{\KB}{\KAB}{F}{S}{\FA}\sixJ{\KA}{\KB}{\FB}{F}{S}{\FA'}\\
        &\times[\KAB(\KAB+1)-\KA(\KA+1)-\KB(\KB+1)]\,.
    \end{split}
%    \label{eq:KAKB}    
\end{equation}
\end{subequations}

An illustrative example of this perturbation approach is methyl-[1-\textsuperscript{13}C]-formate.  The system can be regarded as a strongly coupled \textsuperscript{13}C--\textsuperscript{1}H spin pair XA (the formate group) coupled to the methyl group, B$_3$, so the overall topology of the molecule is (XA)B$_3$.  Energy levels of the XA system are shown in \autoref{Fig:methylformate}a.  Upon coupling to the B$_3$ subsystem, these form two manifolds ($\KB=1/2$ and $\KB=3/2$), each with a rich energy-level structure, as shown in \autoref{Fig:methylformate}b.  
Analytical expressions for the first- and second-order corrected energies are provided in \autoref{Table:XAB3-PT}.  From these arise detectable shifts and/or splittings of the zero-field NMR lines as discussed further in \autoref{sec:XAnBmspectra}.

\begin{table}
    \begin{center}
    \begin{tabular}{cccccc}
    \hline
        $F$  & $F_{\rm XA}$ & $K_{\rm B}$	&	$E^{(0)} / h $	&	$\Delta E^{(1)} / h$		&	$\Delta E^{(2)} / h $ \\
        \hline\hline \\[-6pt]
        5/2 & 1 &	3/2 &	$\frac{ J_{\rm XA}}{4}$	&	$\frac{3(J_{\rm XB}+J_{\rm AB})}{4}$	&	$0$\\[4pt]
        3/2 & 1 &	3/2 &	$\frac{ J_{\rm XA}}{4}$	&	$-\frac{(J_{\rm XB}+J_{\rm AB})}{2}$	&	$\frac{15(J_{\rm XB}-J_{\rm AB})^2}{16 J_{\rm XA}}$\\[4pt]
       1/2 & 1 &	3/2 &	$\frac{ J_{\rm XA}}{4}$	&	$-\frac{5(J_{\rm XB}+J_{\rm AB})}{4}$	&	$0$\\[4pt]
        3/2 & 1 &	1/2 	&	$\frac{ J_{\rm XA}}{4}$	&   $\frac{(J_{\rm XB}+J_{\rm AB})}{4}$	&	$0$\\[4pt]
        1/2 & 1 &	1/2 	&	$\frac{ J_{\rm XA}}{4}$	&	$-\frac{(J_{\rm XB}+J_{\rm AB})}{2}$	&	$\frac{3(J_{\rm XB}-J_{\rm AB})^2}{16J_{\rm XA}}$\\[4pt]
       3/2 & 0 &	3/2 &	$-\frac{3 J_{\rm XA}}{4}$	&	$0$	&	$-\frac{15(J_{\rm XB}-J_{\rm AB})^2}{16J_{\rm XA}}$\\[4pt]
        1/2 & 0 &	1/2 &	$-\frac{3 J_{\rm XA}}{4}$	&	$0$	&	$-\frac{3(J_{\rm XB}-J_{\rm AB})^2}{16J_{\rm XB}}$\\[6pt]
        \hline
    \end{tabular}
    \end{center}
\caption{The lowest-order perturbation corrections to the zero-field energy levels $\ket{F(F_{\rm XA},\,K_{\rm B})}$ of an (XA)B$_3$ spin system of $K_A=1/2$ and $S=1/2$.}\label{Table:XAB3-PT}
\end{table}

For [\textsuperscript{13}C]-methyl formate and most other (XA$_n$)B$_m$ systems studied so far (see, for example, references \cite{Butler2013,Theis2013}), the second-order perturbation theory appears to reproduce the energy-level structure quite well.  However, higher precision may be desired, for which one may need to numerically solve the eigenvalues of the Hamiltonian to find the exact result.  This approach can be also implemented when more complex spin systems containing several inequivalent spins are considered \cite{Put2023Detection}. In this context, quantum simulation provides intriguing medium-term perspectives (see \autoref{Subsubsec:quantumsim}).

\subsubsection{Selection rules}

The rules for the change of quantum numbers associated with allowed transitions in (XA$_n$)B$_m$ systems are a generalization of the rules presented in \autoref{XAnSelectionRules}. To consider them let us first note that the total magnetization of the considered system in the $p$ direction is
\begin{equation}
        \bra{u} \hat M_p \ket{v} = \bra{u} \gamma_{\rm X}\hat{S}_p + \gamma_{\rm A} \hat{K}_{\rm{A},p} + \gamma_{\rm B} \hat{K}_{\rm{B},p} \ket{v} \,,
        \label{eq:TransitionAmplitudeMzXAnBm}
\end{equation}
where there is an additional contribution from the B-spins with respect to the XA$_n$ system. In the low-order perturbation approximation, we can keep the eigenbases of the uncoupled reference systems: $\ket{v_{XA}}\ket{v_{\rm B}} = |F_{\rm XA}'(K_{\rm A}'),m_F'\rangle |K_{\rm B}',m_B'\rangle$ and $\ket{u_{\rm XA}}\ket{u_{\rm B}}  = |F_{\rm XA}(K_{\rm A}),m_F\rangle |K_{\rm B},m_{\rm B}\rangle$, which allows us to rewrite \autoref{eq:TransitionAmplitudeMzXAnBm} as
\begin{equation}
    \begin{split}
        \bra{u}M_{p}\ket{v} =& \bra{u_{\rm XA}} 
        \gamma_{\rm A}\hat F_{{\rm XA},p}+(\gamma_{\rm X}-\gamma_{\rm A})\hat{S}_p 
        \ket{v_{\rm XA}} \braket{u_{\rm B} | v_{\rm B}} \\
        &+ \braket{u_{\rm XA} | v_{\rm XA}}\bra{u_{\rm B}} 
        \gamma_{\rm B}\hat K_{{\rm B},p} 
        \ket{v_{\rm B}}\,.
    \end{split}
\end{equation}
This formula shows that in addition to the previously introduced selection rules for the XA$_n$ spin system: $\Delta K_{\rm A}=0$, $\Delta F_{\rm XA}=0,\pm 1$ and (trivially) $\Delta S = 0$, there is an additional rule: $\Delta K_{\rm B}=0$. In addition, we have $\Delta F = 0,\pm1$ because the overall magnetization remains a vector quantity. Finally, if we choose to define the detection axis as $z$, $\Delta m_{F}=0$, and for the other two directions $\Delta m_{F}=\pm 1$. In the (XA$_n$)B$_m$ spin system, the last selection rules arise instead of the rule for the $m_{F_{\rm XA}}$ in the XA$_n$ spin system.  For a detailed derivation, we refer the reader to reference \cite{Butler2013}.

Two groups of transitions are produced in the (XA$_n$)B$_m$ case.  One of these is a set of high-frequency transitions observed near $J_{\rm XA}$ or its full- or half-integer multiples.  A second group, absent in the XA system, arises at lower frequency due to the transitions between states associated with the B-spin coupling. The frequency of the latter is determined by the $J_{\rm XB}$ and $J_{\rm AB}$ coupling constants. 

The transition frequencies of a representative (XA)B$_3$ spin system are summarized in \autoref{Table:XAB3-freq}.  As described in \autoref{sec:XAnBmspectra}, this complex energy-level structure and the selection rules described lead to information-rich zero-field NMR spectra.

\begin{table*}
    \begin{center}
    \begin{tabular}{cccccc}
    \hline
    & Lower state & Upper state & \multicolumn{3}{c}{Frequency} \\
        &	$\left| F_{\rm XA}, K_{\rm B}, F\right\rangle$	&	$\left| F_{\rm XA}, K_{\rm B}, F\right\rangle$		& Reference &	1\textsuperscript{st} order correction	&	2\textsuperscript{nd} order correction \\
        \hline\hline \\[-6pt]
        $\nu_1$	&	$\left| 1, \frac{1}{2}, \frac{1}{2} \right\rangle$	&	$\left| 1, \frac{1}{2}, \frac{3}{2} \right\rangle$	&	$0$ & $\frac{3}{4}(J_{\mathrm{BX}}+J_{\mathrm{AB}})$	&	$-\frac{3 (J_{\mathrm{BX}}-J_{\mathrm{AB}})^2 }{16J_{\mathrm{XA}}}$\\[4pt]
        $\nu_2$	&	$\left| 1, \frac{3}{2}, \frac{1}{2} \right\rangle$	&	$\left| 1, \frac{3}{2}, \frac{3}{2} \right\rangle$	&	$0$ & $\frac{3}{4}(J_{\mathrm{BX}}+J_{\mathrm{AB}})$	&	$+\frac{15 (J_{\mathrm{BX}}-J_{\mathrm{AB}})^2 }{16J_{\mathrm{XA}}}$\\[4pt]
        $\nu_3$	&	$\left| 1, \frac{3}{2}, \frac{3}{2} \right\rangle$	&	$\left| 1, \frac{3}{2}, \frac{5}{2} \right\rangle$	&	$0 $ & $\frac{5}{4}(J_{\mathrm{BX}}+J_{\mathrm{AB}})$	&	$-\frac{15 (J_{\mathrm{BX}}-J_{\mathrm{AB}})^2 }{16J_{\mathrm{XA}}}$\\[4pt]
        $\nu_4$	&	$\left| 0, \frac{3}{2}, \frac{3}{2} \right\rangle$	&	$\left| 1, \frac{3}{2}, \frac{1}{2} \right\rangle$	&	$J_{\mathrm{XA}}$ & $-\frac{5}{4}(J_{\mathrm{BX}}+J_{\mathrm{AB}})$	&	$+\frac{15 (J_{\mathrm{BX}}-J_{\mathrm{AB}})^2 }{16J_{\mathrm{XA}}}$\\[4pt]
        $\nu_5$	&	$\left| 0, \frac{1}{2}, \frac{1}{2} \right\rangle$	&	$\left| 1, \frac{1}{2}, \frac{1}{2} \right\rangle$	&	$J_{\mathrm{XA}}$&$-\frac{1}{2}(J_{\mathrm{BX}}+J_{\mathrm{AB}})$		&	$+\frac{3 (J_{\mathrm{BX}}-J_{\mathrm{AB}})^2 }{8J_{\mathrm{XA}}}$\\[4pt]
        $\nu_6$	&	$\left| 0, \frac{3}{2}, \frac{3}{2} \right\rangle$	&	$\left| 1, \frac{3}{2}, \frac{3}{2} \right\rangle$	&	$J_{\mathrm{XA}}$&$-\frac{1}{2}(J_{\mathrm{BX}}+J_{\mathrm{AB}})$	&	$+\frac{15 (J_{\mathrm{BX}}-J_{\mathrm{AB}})^2 }{8J_{\mathrm{XA}}}$\\[4pt]
        $\nu_7$	&	$\left| 0, \frac{1}{2}, \frac{1}{2} \right\rangle$	&	$\left| 1, \frac{1}{2}, \frac{3}{2} \right\rangle$	&	$J_{\mathrm{XA}}$&$\frac{1}{4}(J_{\mathrm{BX}}+J_{\mathrm{AB}})$	&	$+\frac{3 (J_{\mathrm{BX}}-J_{\mathrm{AB}})^2 }{16J_{\mathrm{XA}}}$\\[4pt]
        $\nu_8$	&	$\left| 0, \frac{3}{2}, \frac{3}{2} \right\rangle$	&	$\left| 1, \frac{3}{2}, \frac{5}{2} \right\rangle$	&	$J_{\mathrm{XA}}$&$\frac{3}{4}(J_{\mathrm{BX}}+J_{\mathrm{AB}})$	&	$+\frac{15 (J_{\mathrm{BX}}-J_{\mathrm{AB}})^2 }{16J_{\mathrm{XA}}}$\\[6pt]
        \hline
    \end{tabular}
    \end{center}
\caption{Frequencies of allowed transitions for an (XA)B$_3$ system in zero magnetic field ($K_{\rm A} = 1/2$, $S = 1/2$).} \label{Table:XAB3-freq}
\end{table*}

\subsubsection{Experimental spectra} \label{sec:XAnBmspectra}

\begin{figure}
  \begin{center} \includegraphics[width=\columnwidth]{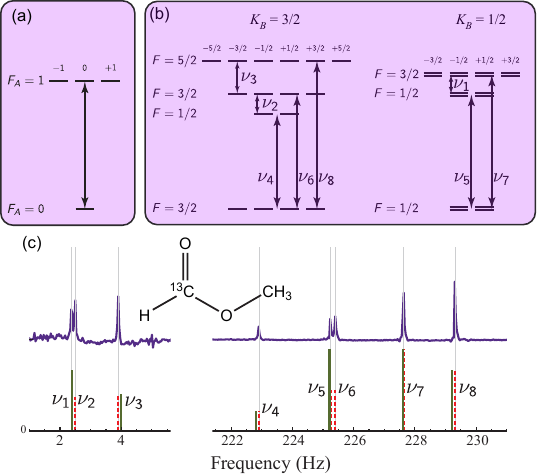} \end{center}
  \caption{Zero-field NMR spectroscopy of methyl [$^{13}$C]formate.  Spin eigenstates and allowed transitions are shown for the XA ($^{1}$H-$^{13}$C) pair in formate (a) when isolated from the methyl group and (b) when coupled to the three methyl protons (B$_3$), producing separate manifolds with quantum numbers $K_B=3/2$ and $K_B=1/2$.  (c) The experimental spectrum (top) and expected spectra based on first- and second-order perturbation corrections (bottom), represented by solid black and dashed red vertical lines, respectively. Reprinted (adapted) from reference \cite{Butler2013} with the permission of AIP publishing.}
  \label{Fig:methylformate}
\end{figure}

As a representative example of an (XA$_n$)B$_m$ spin system, we consider methyl [1-$^{13}$C]-formate (\textsuperscript{1}H-\textsuperscript{13}COOCH\textsubscript{3}), a compound with (XA)B$_3$ spin topology. The experimental zero-field NMR spectrum of the pure [1-$^{13}$C]-enriched compound in liquid form is shown in \autoref{Fig:methylformate}c. To zeroth order, the \textsuperscript{13}C-\textsuperscript{1}H spin pair forms an XA system with an energy-level structure shown in \autoref{Fig:methylformate}a, identical to that given in \autoref{Fig:XAn2} and discussed in \autoref{Subsec:XAn}.  The formyl \textsuperscript{13}C-\textsuperscript{1}H pair is chosen as the reference system because its $J$-coupling constant is by far the largest in the molecule (around 226.8\,Hz). The three magnetically equivalent \textsuperscript{1}H nuclei in the methyl group are the B-spins with $\KB=1/2$ and $\KB=3/2$ manifolds, which couple to the XA reference system through scalar couplings $^3J_{\rm HC} \equiv J_{\rm BX} = 4.0$\,Hz and $^4J_{\rm HH} \equiv J_{\rm AB} = -0.8$\,Hz \cite{Butler2013}. The energy-level structures incorporating interaction with the B nuclei are shown in \autoref{Fig:methylformate}b. The spectrum simulated using first- and second-order perturbation theory largely agree with the experimental one, although full numerical diagonalization reveals higher-order effects that cause small frequency shifts and lift degeneracies between some of the energy levels. The numerical approach is also adopted for larger spin systems where concise analytical expressions for the perturbation corrections are not generally available (see, for example, references~\cite{Blanchard2013JACS,Alcicek2021Ogranophosphorus}).

\begin{figure*}
\centering
	\includegraphics[width=0.9\textwidth]{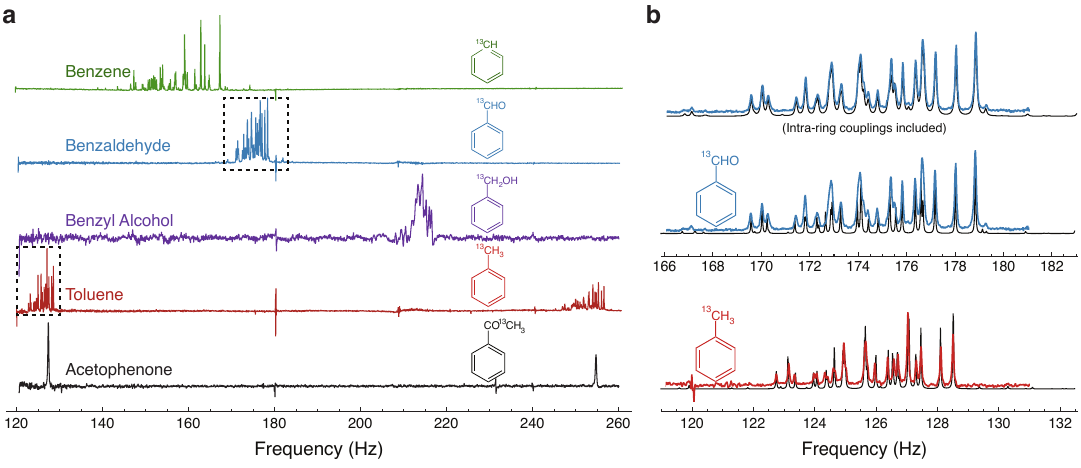}
	\vspace{-8pt}
 \caption{(a) Zero-field NMR $J$-spectra of a series of ${}^{13}$C-labeled benzene derivatives, showing the effects of different $^{13}$CH$_n$ groups and their increasing displacement from the aromatic ring.
 Multiplets appear at $J$ for $^{13}$CH groups, at $3J/2$ for $^{13}$CH$_2$ groups, and at $J$ and $2J$ for $^{13}$CH$_3$ groups. 
 The spread of the peaks within the multiplets decreases as the number of bonds from the $^{13}$C label to the aromatic ring protons increases. 
 (b) Comparison of $K_A = \sum I_A = 1/2$ peaks of benzaldehyde-$\alpha$-$^{13}$C$_1$ (red trace) and toluene-$\alpha$-$^{13}$C$_1$ (blue trace) spectra. 
 %The benzaldehyde-$\alpha$-$^{13}$C$_1$ spectrum \DAB{has been shifted to lower frequency for the purpose of illustration}.
 The structure of the spectra is similar because of the identical spin topology, with small differences arising due to geometric and substituent effects.
 The simulation is in reasonable agreement with the benzaldehyde spectrum even when differences in the intra-ring couplings are not included.
 Reprinted (adapted) with permission from reference \cite{Blanchard2013JACS}. Copyright 2013 American Chemical Society.
 }
	\label{fig:Aromatics-Combined}
		\vspace{-10pt}
\end{figure*}

The spectra of the aromatic compounds shown in \autoref{fig:Aromatics-Combined}, further illustrate important general points regarding ZULF NMR spectra of multi-spin systems.  First, different spin topologies generally result in dramatically different spectral patterns (\autoref{fig:Aromatics-Combined}a). Second, manifolds with same total angular momentum can give rise to similar partial spectral patterns, such as in \autoref{fig:Aromatics-Combined}b, where a phenyl (-C\textsubscript{6}H\textsubscript{5}) group is coupled to the $K_{\rm A}=1/2$ manifold of a \textsuperscript{13}CHO and a \textsuperscript{13}CH\textsubscript{3} group, respectively.  Third, even though certain spectral patterns (or partial patterns) for different compounds may be similar (\autoref{fig:Aromatics-Combined}b), chemical information can still be obtained from the absolute frequencies, as is done with chemical shifts in high-field NMR.

\subsection{XA$_n$ spin systems in ultralow magnetic field}

\subsubsection{Energy-level structure}

We now consider an XA$_n$ system subjected to a magnetic field of strength $B_{\rm ULF}$ along the $z$-axis. The general Hamiltonian of such a system is
\begin{equation}
        \hat{\mathcal{H}}_{{\rm ULF,XA}_n} =\,  \hat{\mathcal{H}}_{{J,\rm XA}_n} - \hbar \left( \gamma_{\rm X} \hat{S}_{z} + \gamma_{\rm A} \sum_{i=1}^n\hat{I}_{z,i} \right) B_{\rm ULF} \,.
    \label{eq:ULFgeneral2}
\end{equation}
%We assume initially that the magnetic field is weak, i.e., the Larmor frequency of each spin is much smaller than $J_{\rm XA}$, so that 
In the ultralow-field regime (see \autoref{Subsec:Regimes}), the energy-level structure is obtained using the `reference' zero-field Hamiltonian [the first term \autoref{eq:ULFgeneral2}] and the finite-field perturbation Hamiltonian $\hat{\mathcal{H}}_{{\rm per}}$ [the second term in \autoref{eq:ULFgeneral2}].
% \begin{equation}
%         \hat{\mathcal{H}}_{{\rm per}}= - \hbar \left( \gamma_{\rm X}\hat{S}_{z}+\gamma_{\rm A}\sum_{i=1}^n\hat{I}_{z,i} \right) B_{\rm ULF} \,.
% \end{equation}
In the first-order approximation, the noncommuting part of $\hat{\mathcal{H}}_{{\rm per}}$ is ignored, resulting in a simpler expression for the perturbation Hamiltonian:
\begin{equation}
    \hat{\mathcal{H}}^{(1)}=-\hbar\gamma_{{\rm XA}_n}(\FA,\KA)B_{\rm ULF}\hat{F}_{{\rm XA}, z}\,,
    \label{eq:HamiltonianBULFlinear}
\end{equation}
where we introduced the quantity $\gamma_{{\rm XA}_n}$ as an effective gyromagnetic ratio for the zero-field eigenstate manifolds of common $\FA$, $\KA$, and $S$. Values of $\gamma_{{\rm XA}_n}$ can be determined through the formula
\begin{equation}
        \gamma_{{\rm XA}_n} = \frac{1}{\FA}\sum_{m_{K_{\rm A}} = - K_A}^{K_A} \sum_{m_{S} = - S}^{S} (\gamma_A m_{K_{\rm A }} + \gamma_X m_S) \left(C_{K_{\rm A}m_{K_{\rm A}},S m_S}^{F_{\rm XA},F_{\rm XA}}\right)^2 \,,
    \label{eq:EffectiveGyromagneticRatio}
\end{equation}
and some of these are tabulated in \autoref{tab:EffectiveGyromagneticRatioMT}, where we note that the identity $\gamma_{{\rm XA}_n} \equiv \gamma_{\rm X}$ holds for the special scenario of total magnetic equivalence: $\gamma_{\rm A}=\gamma_{\rm X}$. The matrix representation of $\hat{\mathcal{H}}^{(1)}$ in the zero-field eigenbasis is therefore diagonal, and the shifted energy levels correct to first order are given by
\begin{equation}
    \begin{split}
        E^{(1)}_{\FA(\KA,S) m_{\FA}} =\,& \pi J_{\rm AX} \left[F_{\rm XA}(F_{\rm XA}+1)-K_A(K_A+1)-S(S+1)\right]\\
        &- \gamma_{{\rm XA}_n}(\FA,\KA) B_{\rm ULF} m_{\FA}\,.
     \label{eq:EnergyFirstOrderULFMT}
     \end{split}
\end{equation}

\begin{table}[t]
    \centering
\begin{tabularx}{\columnwidth}{cc c c >{\centering\arraybackslash} X}
\hline
        &$\KA$ & $S$ & $\FA$ & $\gamma_{{\rm XA}_n}$\\
        \hline\hline
        &0   & 1/2 & 1/2 & $\gamma_{\rm X}$\\ 
        &1/2 & 1/2 & 1 & $(\gamma_{\rm X}+\gamma_{\rm A})/2$\\
        &1 & 1/2 & 1/2 & $(-\gamma_{\rm X}+4\gamma_{\rm A})/3$\\
        &1 & 1/2 & 3/2 & $(\gamma_{\rm X}+2\gamma_{\rm A})/3$\\
        &3/2 & 1/2 & 1 & $(-\gamma_{\rm X}+5\gamma_{\rm A})/4$\\
        &3/2 & 1/2 & 2 & $(\gamma_{\rm X}+3\gamma_{\rm A})/4$\\
        &2 & 1/2 & 3/2 & $(-\gamma_{\rm X}+6\gamma_{\rm A})/5$\\
        &2 & 1/2 & 5/2 & $(\gamma_{\rm X}+4\gamma_{\rm A})/5$\\
        \hline
        &0   & 1 & 1 & $\gamma_{\rm X}$\\ 
        &1/2 & 1 & 1/2 & $(4\gamma_{\rm X}-\gamma_{\rm A})/3$\\
        &1/2 & 1 & 3/2 & $(2\gamma_{\rm X}+\gamma_{\rm A})/3$\\ 
        &1 & 1 & 1 & $(\gamma_{\rm X}+\gamma_{\rm A})/2$\\
        &1 & 1 & 2 & $(\gamma_{\rm X}+\gamma_{\rm A})/2$\\
        &3/2 & 1 & 1/2 & $(-2\gamma_{\rm X}+5\gamma_{\rm A})/3$\\
        &3/2 & 1 & 3/2 & $(4\gamma_{\rm X}+11\gamma_{\rm A})/15$\\
        &3/2 & 1 & 5/2 & $(2\gamma_{\rm X}+3\gamma_{\rm A})/5$\\
        &2 & 1 & 1 & $(-\gamma_{\rm X}+3\gamma_{\rm A})/2$\\
        &2 & 1 & 2 & $(\gamma_{\rm X}+5\gamma_{\rm A})/6$\\
        &2 & 1 & 3 & $(\gamma_{\rm X}+2\gamma_{\rm A})/3$\\  
 \hline
    \end{tabularx}
    \caption{Effective gyromagnetic ratios for XA$_n$ spin manifolds of total angular momentum quantum numbers $K_{\rm A}$, $S$ and $F_{\rm XA}$ in the ultralow-field regime.  
%    Note that apart from through the quantum numbers, $\gamma_{\rm XA_n}$ does not depend explicitly on $n$. \AT{The second sentence is rather convoluted. How about
Note that $\gamma_{\rm XA_n}$ does not depend explicitly on $n$, only through the quantum numbers. %\MCDT{OK.}
}
    \label{tab:EffectiveGyromagneticRatioMT}
\end{table}

Beyond the first-order regime, the energy dependence of $m_{F_\text{XA}}$ sublevels on magnetic field is generally nonlinear.
%SP{It should be noted that the linear dependence of the energies of Zeeman sublevels on the magnetic field [\autoref{eq:EnergyFirstOrderULF}] is only an approximation because the Hamiltonian $\hat{\mathcal{H}}_{\rm per}$ is only diagonal in the uncoupled basis. As a result, there are higher-order contributions that lead to nonlinear spliting of Zeeman sublevels.} 
For example, for any XA$_n$ spin system with $S$=1/2, diagonalization of the Hamiltonian $\hat{\mathcal{H}}_{{\rm ULF,XA}_n}$ gives an exact expression for the energies of the levels as a function of the magnetic field $B_{\rm ULF}$:
% \begin{equation}
%     \begin{split}
%         E_{\FA(\KA,S) m_{\FA}} =& -\frac{J_{\rm XA}}{4}-\gamma_{\rm A} m_{F_{\rm A}} B_{\rm ULF}\\
%         &+SJ_{\rm XA}\sqrt{\left(I+\frac{1}{2}\right)^2-2m_F x_{\rm ULF}+x_{\rm ULF}^2}\,,
%     \end{split}
%     \label{eq:BreitRabi}
% \end{equation}
\begin{equation}
    \begin{split}
        &E_{\FA(\KA,S) m_{\FA}} = -\hbar\gamma_{\rm A} m_{F_{\rm XA}} B_{\rm ULF}\\
        &\hspace{1cm} +2\pi\hbar J_{\rm XA}\left[-\frac{1}{4}\pm S\sqrt{\left(\KA+\frac{1}{2}\right)^2-2m_F x_{\rm ULF}+x_{\rm ULF}^2\,}\right]\,,
    \end{split}
    \label{eq:BreitRabi}
\end{equation}
where $x_{\rm ULF}=(\gamma_{\rm X}-\gamma_{\rm A})B_{\rm ULF}/(2\pi J_{\rm XA})$ and $\pm$ in the second term corresponds to parallel/antiparallel orientation of $\vec{S}$ and $\vec{K}_{\rm A}$ (i.e., to $\FA=\KA\pm S$).  Further information on the derivation of the nonlinear term under different field conditions can be found in reference \cite{Appelt2010}. %\MCDT{How general is this?  Something does not look quite right because $F_A$ must appear explicitly as well.  For example, take the singlet-triplet basis states of an spin-1/2-pair XA system.  You would expect to recover the (-3/4)$J_{\rm XA}$ (for $F_A=0$) and (+1/4)$J_{\rm XA}$ (for $F_A=1$) energy levels for the zero-field system, which \autoref{eq:BreitRabi} does not do.} {\color{blue}This is general. However, there was a problem previously with the ``+'' sign in the last term which is suppose to be $\pm$. Now everything should work.} 
For a weak field, \autoref{eq:BreitRabi} is the famous Breit--Rabi formula; it can be expanded as a power series in $x_{\rm ULF}$, which, up to second order, takes the form
% \begin{equation}
%     \begin{split}
%         E_{\FA(\KA,S) m_{\FA}} \approx& \left[S\left(I+\frac{1}{2}\right)-\frac{1}{4}\right]J_{\rm XA}\\
%         &+\left[\gamma_{\rm A}+\frac{S}{I+1/2}\left(\gamma_{\rm X}-\gamma_{\rm A}\right)\right]m_{F_{\rm A}}B_{\rm ULF}\\
%         &+S\frac{(I+1/2)^2-m_{F_{\rm A}}^2}{2(I+1/2)^3}\frac{(\gamma_{\rm X}-\gamma_{\rm A})^2B_{\rm ULF}^2}{J_{\rm ULF}}\,.
%     \end{split}
%     \label{eq:}
% \end{equation}
\begin{equation}
    \begin{split}
        E_{\FA(\KA,S) m_{\FA}} \approx& J_{\rm XA}\pi \hbar\left[-\frac{1}{2}\pm S(2\KA+1)\right]\\
        &-\left[\gamma_{\rm A} m_{F_{\rm XA}}\mp\frac{S(\gamma_{\rm A}-\gamma_{\rm X})}{2\KA+1}\right]\hbar B_{\rm ULF}\\
        &\pm S\hbar\frac{(2\KA+1)^2-4m_{F_{\rm XA}}^2}{4(2\KA+1)^3}\frac{(\gamma_{\rm A}-\gamma_{\rm X})^2}{\pi J_{\rm XA}}B_{\rm ULF}^2\,.
    \end{split}
    \label{eq:EnergyZeeman}
\end{equation}
The second line corresponds to the previously introduced linear Zeeman splitting of the energy levels [see the second term in \autoref{eq:EnergyFirstOrderULFMT}] and the third line accounts for the quadratic Zeeman splitting.

\begin{figure*}
    \centering
    \includegraphics[width=\textwidth]{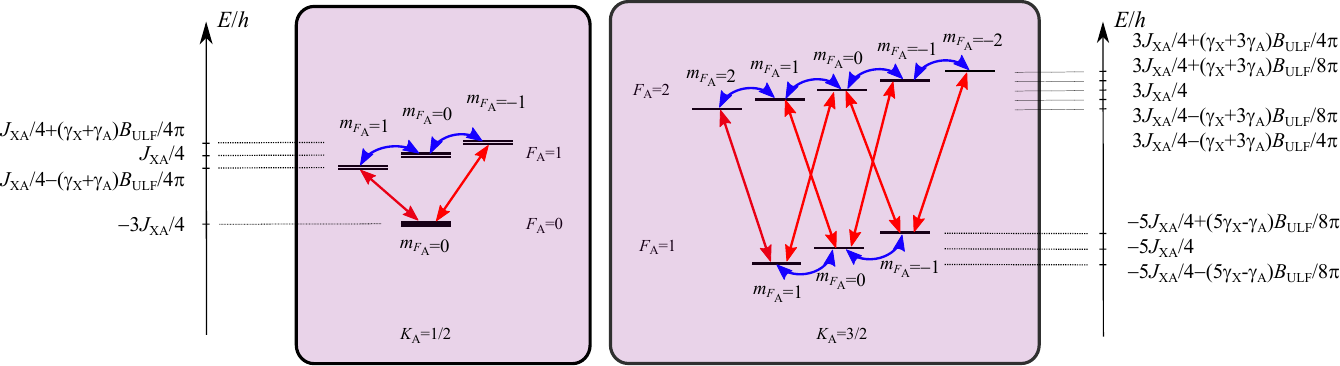}
    \caption{Energy-level diagram for a XA$_3$ spin system in the ultralow-field regime; a magnetic field is applied along the $z$-axis through which we define the projection quantum number $m_{\FA}$. Arrows correspond to transitions where magnetization is observed along a transverse axis; red arrows indicate transitions with $\Delta \FA=\pm 1$; blue arrows indicate transitions between magnetic sublevels of a given $\FA$ state.}
    \label{fig:XAnULF}
\end{figure*}

In \autoref{fig:XAnULF}, we show the energy-level structure of a representative XA$_3$ system at a magnetic field $B_{\rm ULF}$ applied along $z$. The field lifts the degeneracy of the $m_{\text{FA}}$ sublevels of the $F_\text{XA}$ state. As the value of the splitting depends also on $\FA$ [see \autoref{eq:EnergyZeeman}], the energy splittings between sublevels of the lower and higher states are modified. This manifests as splittings of the lines observed in ZULF NMR spectra at zero field.

\subsubsection{Selection rules}
\label{Subsubsec:ZeemanSelectionRules}
    
In principle, the presence of an ultralow magnetic field does not affect the selection rules for the observed transitions. Specifically, if the $z$-component of the magnetization $\hat{M}_{z}$ is detected, the selection rule $\Delta m_{\FA}=0$ remains unchanged.  Thus, it is often useful to align the quantization axis with the magnetic field direction, so that the field shifts the magnetic sublevels rather than couples them. This approach is utilized in \autoref{fig:XAnULF}, where the field splits the sublevels without generating additional coherences. If the magnetic field is not along the initial quantization axis, it may be useful to redefine the coordinate system. For example, if the field is initially oriented along the $x$-axis, rotation of the coordinate system transforms the sensitive direction from $z$ to $x$, i.e., instead of measuring longitudinal magnetization $\hat{M}_{z}$, one measures the magnetization $-\hat{M}_{x}$. As shown above, this observable has different selection rules than $\hat{M}_z$ and in this new rotated frame of reference, the selection rule is $\Delta m_F= \pm 1$.

The analysis of the selection rules for the total-angular-momentum change of the XA$_n$ spin system (\autoref{tab:example}) shows that $\Delta\FA=0$ transitions are generally allowed. However, at zero field, such transitions are at zero frequency due to the degeneracy of magnetic sublevels and manifest themselves as static magnetization and may be difficult to distinguish from the background magnetic fields. When the degeneracy is lifted by a magnetic field, these transitions manifest as slowly oscillating magnetic fields. As shown in \autoref{Subsubsec:Smallfieldspectra}, the transitions with $\Delta\FA=0$ give rise to so-called low-frequency resonances.

\subsubsection{Experimental spectra}
\label{Subsubsec:Smallfieldspectra}

The energy-level diagram presented in \autoref{fig:XAnULF} illustrates the XA$_3$ spin system subjected to a static magnetic field along the $z$-axis. The field shifts magnetic sublevels, lifting their degeneracy (for the sake of simplicity, only the linear terms, described by \autoref{eq:EnergyFirstOrderULFMT}, are displayed). As a consequence, distinct sets of transitions are observed, localized around $J$ or $2J$. These transitions are represented with red arrows, corresponding to the $M_x$ or $M_y$ magnetization. Additional low-frequency transitions, depicted with blue arrows, are also observed. The frequencies of these transitions are exclusively determined by the magnetic field and the effective gyromagnetic ratio [\autoref{eq:EffectiveGyromagneticRatio}].

\begin{figure}
    \centering
    \includegraphics[width=\columnwidth]{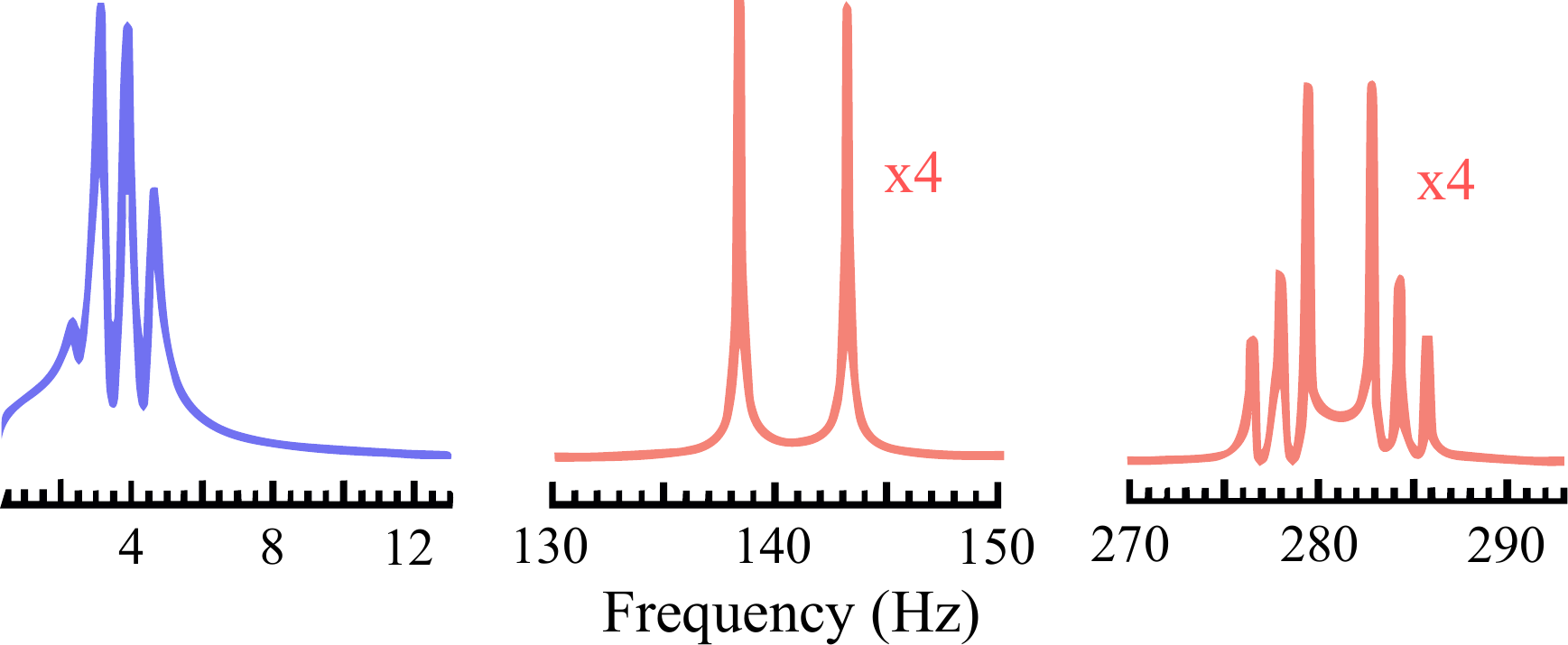}
    \caption{Experimentally measured ULF NMR magnitude spectrum of [$^{13}$C]-methanol, a representative XA$_3$ system, at a magnetic field of $\SI{129}{\nano\tesla}$. The spectrum is divided into three regions: low-frequency peaks, corresponding to the transitions between $m_{\FA}$ sublevels within the same $\FA$ state; two peaks around $J_{\rm XA}=\SI{141}{\hertz}$ corresponding to transitions in the $\KA=1/2$ manifold, and the six peaks around $2J_{\rm XA}=\SI{282}{\hertz}$ corresponding to transitions in the $\KA=3/2$ manifold. The largest low-frequency peak around $\SI{3.5}{\hertz}$ does not originate from the XA$_3$ system, but from the residual water in the sample. The schematic energy diagram of the system is shown in \autoref{fig:XAnULF}. Adapted from reference~\cite{PutThesis}.}
    \label{fig:XA3spectrum}
\end{figure}

Three regions of the ULF-NMR spectrum of the XA$_3$ spin system [$^{13}$C]-methanol, corresponding to the discussed transitions, are shown in \autoref{fig:XA3spectrum}. In the low-frequency part of the spectrum, three resonances, associated with the transitions between magnetic sublevels of the $\FA=1$ and $\FA=2$ states, and an additional peak at 4\,Hz, originating from the solvent (water), are visible. The amplitudes of these peaks are about a factor of 3 to 4 times larger than high-frequency peaks observed at around 141 (two peaks) and 282\,Hz (six peaks), corresponding to the $\Delta\FA\pm 1$ transitions, as shown in \autoref{fig:XAnULF}. 
%(a small peak at 135\,Hz is an artefact of imperfect field compensation and a small admixture of the $M_z$ magnetization) \MCDT{There is no longer a peak at 135 Hz.  Should this sentence be deleted?}.
The presented spectrum demonstrates the abundant analytical information provided by ULF NMR.

\subsection{Untruncated couplings}
\label{Subsec:Untruncated}

%\DB{The ability to detect spin couplings that are truncated in the case of high-field NMR is one of the hallmarks of ZULF NMR. Examples of such interactions include tenson $J$-couplings, .... Andreas has some of this below. The general parts should move here }

One of the most intriguing aspects of ZULF NMR is access to terms of interaction tensors that are suppressed in high-field NMR (see also the discussion in Sec.\,\ref{Subsec:Why}). In ZULF NMR, the dominant interactions are local spin--spin couplings, which involve coupling tensors with a different symmetry than the Zeeman Hamiltonian $\hat{\mathcal{H}}_Z$. At high field, interaction tensors are reduced due to the dominance of $\hat{\mathcal{H}}_Z$. That is, only parts that commute with $\hat{\mathcal{H}}_Z$ contribute to the observable signal. Such a truncation of interaction Hamiltonians can be understood as coherent averaging in the interaction representation of $\hat{\mathcal{H}}_Z$ \cite{Ernst1987NMR1D2D}. Note, however, that the truncated tensor coupling terms retain a role in relaxation.

In the absence of a dominant magnetic field, we have access to a wider range of terms. These include, for instance, the antisymmetric $J$-coupling terms \cite{King2017}, several terms of the direct dipole--dipole coupling \cite{Blanchard2015}, and a number of postulated but as-yet unobserved exotic interactions such as those mediated by pseudoscalar (axion-like) bosons \cite{Weinberg1978,Moody1984,Fadeev2019Pot}, which would lead to anomalous spin--spin tensor couplings, most of which do not commute with $\hat{\mathcal{H}}_Z$.

As a first example, consider the dipole--dipole interaction, described by the Hamiltonian $\hat{\mathcal{H}}_{DD}$, see \autoref{eq:HamDip}. For a spin-{\textonehalf} pair, $\hat{\mathcal{H}}_{DD}$ can be rewritten in the form of the so-called `dipolar alphabet' \cite{Abragam1961,SlichterBook}:

\begin{equation}
    \hat{\mathcal{H}}_{DD} = -\frac{\mu_0}{4\pi}\frac{\gamma_1 \gamma_2 \hbar}{r^3} \left(\hat{A}+\hat{B}+\hat{C}+\hat{D}+\hat{E}+\hat{F}\right)\,,
\end{equation}
where (with the internuclear vector $\bm{r}$ expressed in polar coordinates $\theta$ and $\phi$, and $\hat{I}_{j\pm} = \hat{I}_{jx} \pm \hat{I}_{jy}$)
%$\hat{\bm{\mathrm{I}}}^\pm$ = $\hat{\bm{\mathrm{I}_x}}\pm\hat{\bm{\mathrm{I}}_y}$

\begin{figure}
\centering
	\includegraphics[width=\columnwidth]{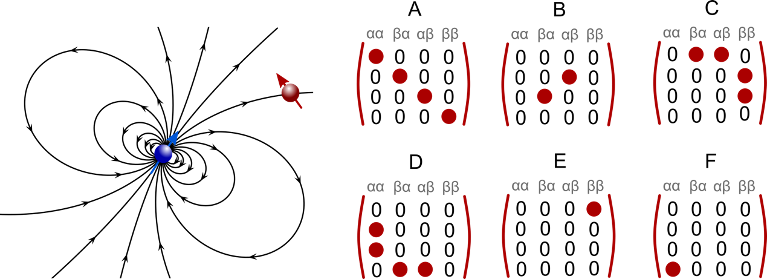}
	\vspace{-8pt}
	\caption{Dipole–dipole interactions of two nuclear spins and matrix plot of the dipolar alphabet (see main text). 
 %\DAB{I'm afraid column labels ab and ba are messed up  \SP{The labels above the matrices are also very small.}} %\DAB{For the submission, I would leave as is.}
	\label{fig:Dipolar_alphabet}}
		\vspace{-10pt}
\end{figure}

%\begin{eqnarray}\label{eq:DipAlphabet}
%    \nonumber\hat{A}&=&\hat{\bm{\mathrm{I}}}_{1z}\hat{\bm{\mathrm{I}}}_{2z}(3\cos^2(\theta)-1)\,,\\
%    \nonumber\hat{B}&=& \left( \hat{\bm{\mathrm{I}}}^+_1\hat{\bm{\mathrm{I}}}^-_2 + \hat{\bm{\mathrm{I}}}^-_1\hat{\bm{\mathrm{I}}}^+_2 \right) \frac{1-3\cos^2(\theta)}{4}\,,\\
%    \hat{C}&=& \left( \hat{\bm{\mathrm{I}}}^+_1\hat{\bm{\mathrm{I}}}_{2z} + \hat{\bm{\mathrm{I}}}_{1z}\hat{\bm{\mathrm{I}}}^+_2 \right) \frac{3\sin(2\theta)\mathrm{e}^{-\mathrm{i}\phi}}{4}\,,\\
%    \nonumber\hat{D}&=& \left(\hat{\bm{\mathrm{I}}}^-_1\hat{\bm{\mathrm{I}}}_{2z} + \hat{\bm{\mathrm{I}}}_{1z}\hat{\bm{\mathrm{I}}}^-_2 \right) \frac{3\sin(2\theta)\mathrm{e}^{\mathrm{i}\phi}}{4}\,,\\
%    \nonumber\hat{E}&=&\hat{\bm{\mathrm{I}}}^+_1\hat{\bm{\mathrm{I}}}^+_2 \frac{3\sin^2(\theta)\mathrm{e}^{-\mathrm{i}2\phi}}{4}\,,\\
 %   \nonumber\hat{F}&=&\hat{\bm{\mathrm{I}}}^-_1\hat{\bm{\mathrm{I}}}^-_2 \frac{3\sin^2(\theta)\mathrm{e}^{\mathrm{i}2\phi}}{4}\,.
%\end{eqnarray}
%\MCDT{I believe that in this case, the I quantities are simply operators and not vectors of operators, so there should be no bold/roman.}
%\AT{Agreed, thank you.}
\begin{eqnarray}\label{eq:DipAlphabet}
    \nonumber\hat{A}&=&\hat{I}_{1z}\hat{I}_{2z}(3\cos^2\theta-1)\,,\\
    \nonumber\hat{B}&=& \left( \hat{I}_{1+}\hat{I}_{2-} + \hat{I}_{1-}\hat{I}_{2+} \right) \frac{1-3\cos^2\theta}{4}\,,\\
    \hat{C}&=& \left( \hat{I}_{1+}\hat{I}_{2z} + \hat{I}_{1z}\hat{I}_{2+} \right) \frac{3\sin(2\theta)\mathrm{e}^{-\mathrm{i}\phi}}{4}\,,\\
    \nonumber\hat{D}&=& \left(\hat{I}_{1-}\hat{I}_{2z} + \hat{I}_{1z}\hat{I}_{2-} \right) \frac{3\sin(2\theta)\mathrm{e}^{\mathrm{i}\phi}}{4}\,,\\
    \nonumber\hat{E}&=&\hat{I}_{1+}\hat{I}_{2+} \frac{3\sin^2\theta\mathrm{e}^{-\mathrm{i}2\phi}}{4}\,,\\
    \nonumber\hat{F}&=& \hat{I}_{1-}\hat{I}_{2-} \frac{3\sin^2\theta\mathrm{e}^{\mathrm{i}2\phi}}{4}\,.
    \end{eqnarray}
A schematic matrix representations of these terms is given in \autoref{fig:Dipolar_alphabet}, with $\alpha$ and $\beta$ representing spin-up and spin-down states, respectively. Of the six terms, only $\hat{A}$ commutes with $\hat{\mathcal{H}}_Z$. Therefore, at high field and for a heteronuclear pair, $\hat{\mathcal{H}}_{DD}$ reduces to $\hat{A}$. (For a homonuclear pair, $\hat{B}$ also has to be taken into account, as the operator $\hat{I}_{1+}\hat{I}_{2-}+\hat{I}_{1-}\hat{I}_{2+}$ is time-independent in the rotating frame, provided that the spins are magnetically inequivalent.) By contrast, ZULF NMR provides ways to observing all six terms of the dipolar alphabet \cite{Blanchard2015}, thereby giving access to additional information on molecular structure and dynamics. This is discussed in \autoref{Subsec:Partial}.

An interaction that is entirely suppressed by truncation at high magnetic field is the rank-1 antisymmetric $J$-coupling, a nuclear-spin analog of the Dzyaloshinskii--Moriya interaction \cite{Dzyaloshinsky1958,Moriya1960,Robert2019_MagnetoEl}. This interaction is connected to molecular chirality, and it was proposed that it may be observed in ZULF-NMR experiments \cite{King2017}. Such experiments would enable differentiation between enantiomers without adding any additional chiral agent to the sample \cite{Wadhwa2024}. Possible routes to observing antisymmetric $J$-coupling terms with ZULF NMR are discussed in \autoref{Subsubsec:Chirality} (in the context of molecular chirality) and in \autoref{Subsubsec:Parity} (in the context of parity violation in nonchiral systems).

%\MCDT{Missing discussion of untruncated residual dipolar couplings.}
%\AT{I included some discussion of this in \autoref{Sec:Extensions}.}

\section{Spin polarization for ZULF NMR}
\label{Sec:Spin_pol}

%\MCDT{This section is progressing nicely, but we should take care not to be writing a review of hyperpolarization techniques.  I think that we are breaking the flow by discussing the basics of each technique in too much detail, at least without reference to ZULF.  I suggest where possible, readers should instead be referred elsewhere (e.g., other Progress in NMR Spectroscopy review articles) to read about PHIP, dissolution DNP basics etc.  Surely it is enough for our review to give a 1-2 sentence intro of the given technique, say what nuclear polarization levels are currently achievable, and then focus on the aspects most relevant to ZULF NMR.}
%\JE{Hmm, I disagree, I think we should provide sufficient detail so the reader has a sense of the experimental requirements to implement the HP methods in a lab, and the advantages/drawbacks of the methods. We can achieve that in 2-3 paragraphs for each. Otherwise the reader is reading about dDNP-ZULF or PHIP-ZULF with no understanding of why you would choose one method vs another. I agree all the historical-context parts can be dropped}

%\JE{Note: Put in an explanatory paragraph to make clear that we give a comprehensive review for polarization techniques and it's not just a description of hyperpolarization methods}

\subsection{Thermal polarization}
\label{Subsec:Thermal_pol}

%Spin systems become spontaneously, albeit typically weakly, ordered on reaching thermodynamic equilibrium, where the average population of each spin state is determined by Boltzmann statistics.  The equilibrium state moreover, in most cases, corresponds to an overall magnetization of the system with ZULF regime being a notable exception.  For example, a macroscopic number of $^1$H spins such as in a bulk solvent, a solution or a polymer at 300\,K in a magnetic field of strength $B_p$ yield a spontaneous magnetization per mole of $x B_p$, where $x$ is on the order of 2$\times10^{-5}$\, nuclear magnetons/mol/T.
%For a general case of spin $\bm{\mathrm{I}}$, one can derive that $ M = C \gamma_I^2 \hbar^2 I (I+1) B_0 /(3 k_B T)$ \cite{Abragam1961}
%$\mu_0$ is the vacuum permeability,
%Then one can define a nuclear susceptibility $\chi_N=\hbar^2 \gamma_I^2/(2 k_B T)$, such that $\mathbf{M}=N \chi_N \mathbf{B}.$
%For $^1$H spins at 300\,K, $\chi_N=$

%The equilibrium state moreover, in most cases, corresponds to an overall magnetization of the system with ZULF regime being a notable exception.  For example, a macroscopic number of $^1$H spins such as in a bulk solvent, a solution or a polymer at 300\,K in a magnetic field of strength $B_p$ yield a spontaneous magnetization per mole of $x B_p$, where $x$ is on the order of 2$\times10^{-5}$\, nuclear magnetons/mol/T. 
%[Feel free to replace x with another symbol.  I would also suggest converting the coefficient into a magnetic field at a specified distance, say 1 cm].

As already explored in \autoref{sec:Polarization1}, nuclear spin systems can become polarized by reaching thermodynamic equilibrium in a magnetic field $\bm{\mathrm{B}}_0$. In particular, for a sample consisting of particles with nuclear spin quantum number $I$, the magnetization $\bm{\mathrm{M}}$ (in SI units, \si{\joule\per\tesla\per\meter\cubed}) % \mathrm J \cdot \mathrm T^{-1} \mathrm m^{-3}$
is given by the product of individual nuclear magnetic moments $\gamma_I \hbar I$, the ensemble spin polarization $\bm{\mathrm{P}}$ (a dimensionless vector of magnitude between 0 and 1), and the spin number density $n$. Given Boltzmann statistics and in the high-temperature approximation (which implies that the thermal energy of the system, $k_B T$, is much larger than the energy gap between spin levels), the magnetization can be estimated as \cite{Abragam1961}\label{eq:ThermPol} 
\begin{equation} 
    \bm{\mathrm{M}} = n\cdot\gamma_I \hbar I\cdot \bm{\mathrm{P}}_{\rm thermal} = n\cdot \gamma_I \hbar I\cdot \frac{\gamma_I \hbar (I+1) \bm{\mathrm{B}}_0}{3 k_B T}\,,
\end{equation}
where $\gamma_I$ is the spin-$I$ gyromagnetic ratio. As a practical example, consider a sample of water (H$_2$O, $\gamma_{\rm H}/(2\pi)$ $\approx$ \SI{42.6}{MHz\per\tesla}, $I=1/2$) polarized at 3\,T and room temperature ($T=298$ K). Thermal nuclear polarization is on the order of $P_{{\rm thermal,}z} = 10^{-5}$, and the corresponding magnetization expressed in magnetic-field units ($\mu_0 M$) is on the order of 100\,pT, where $\mu_0$ is the vacuum permeability.  %Fields of such magnitude are large enough to be measured in conventional NMR experiments and further polarization steps are unnecessary. 
%\MCDT{I agree with this magnitude of magnetic field (somewhere from 100 pT to 1 nT).  Do you need to specify the sample geometry and size, e.g., 1 cm\textsuperscript{3}, sphere, cylinder ?}
%\MCDT{Another note: somewhere in the review, perhaps in the Section on detection, we should note that the magnetic field of the polarized nuclei is detected outside the sample, and depending on geometry/standoff may be much weaker than the value given above.} \DB{Let us discuss these points at our next meeting.} 
Geometry effects on the actual field measured will be discussed in \autoref{Subsec:Geometry}.

Under ZULF conditions, the thermal equilibrium polarization is so low that this essentially precludes direct acquisition of NMR signals. One solution is to temporarily apply a higher field to increase polarization prior to signal detection. This can be done, for example, by mechanically shuttling the sample between a permanent magnet array and the detection region \cite{Blanchard2013JACS,Tayler2017Halbach} (see Fig.\,\ref{fig:ZULF_NMR_Appa}a), or by applying millitesla fields using electromagnets \textit{in situ} \cite{Trabesinger2004,Tayler20192,Bodenstedt2021}. These methods are sufficient for observing highly concentrated species in solution ($\gtrsim 100$\,mM), but for samples at millimolar concentrations or below, dedicated hyperpolarization techniques are needed \cite{Eills2023spin_hyperpol)}.
Using such techniques, polarization levels above 50\% ($P>0.5 \gg P_{\rm thermal}$) for \textsuperscript{1}H \cite{JANNIN201299,Rayner2017,Korchak2018}, \textsuperscript{13}C \cite{JANNIN201299,Korchak2018} and above 20\% ($P>0.2 \gg P_{\rm thermal}$) for \textsuperscript{15}N \cite{milani2017,Barskiy2016jacs} have been demonstrated. 
% \SP{Shall we give here another section header entitled Hyperpolarization. This would give some intro to the discussion below? Now, this is in section entitled - Thermal polarization, which is misplaced.}

\subsection{From thermal polarization to hyperpolarization}

Generally, hyperpolarization implies nonthermal distributions of populations among spin energy levels, thus, all forms of spin order apart from thermal-equilibrium magnetization (which may include, for example, alignment, two-spin order, dipole order, etc.) are referred to below as hyperpolarization \cite{Kovtunov2018}.
We remark that the efficacy of a given hyperpolarization technique is often dependent on the magnetic field at which hyperpolarization is performed.  Several novel approaches specifically rely on the use of the ZULF regime. Therefore, not only can applications of ZULF NMR benefit from hyperpolarization, but a deeper understanding of hyperpolarization mechanisms can also be gained from NMR studies at zero and ultralow fields.

% In situations where polarization is proportional to the applied field, equilibrium magnetization expected particularly for the the case of ZULF NMR experiments may fall orders of magnitude short of the level necessary to be experimentally detectable above noise limits.  In the case of samples where spin concentration is highest---such as bulk solvents or biological tissue---temporarily applied fields of only a few millitesla may be sufficient to boost the magnetized state, where the fields are produced using electromagnetic field cycling \cite{Trabesinger2004,Tayler20192} or by mechanical shuttling of the sample into and out of persistent magnets \cite{Blanchard2013JACS,Tayler2017Halbach}.

Below we discuss the basic principles and applications in ZULF NMR of the most common hyperpolarization techniques: spin-exchange optical pumping (\autoref{Subsec:SEOP}), parahydrogen-induced polarization (\autoref{Subsec:PHIP}), dynamic nuclear polarization (\autoref{Subsec:DNP}), and chemically induced dynamic nuclear polarization (\autoref{Subsec:CIDNP}), along with two less commonly used approaches (\autoref{Subsec:Stat-pol} and \autoref{Subsec:Rotat-pol}).

\subsection{Spin-exchange optical pumping}
\label{Subsec:SEOP}

%\JE{This section still needs an intro to SEOP and text polishing, but the meat is there}\\
%\JE{Light is easy to polarize, lasers can be used to polarize electron spins in alkali metal atoms, the polarization of the light is transferred to the electrons, this generates an electron-polarized vapour of alkali metal atoms. This }

Light can be readily polarized. Circularly polarized light is composed of photons (spin-one particles) with the projection of the spin angular momentum of $+\hbar$ or $-\hbar$ in the direction of the light propagation. When polarized light is absorbed, the angular momentum is transferred to the absorbing medium, which can become spin-polarized.
%Circularly polarized light is one of the most readily available sources of spin polarization for nuclei and electrons. 
%Light with desired polarization can be prepared by using standard optical components: polarizers and waveplates. 
%It has been demonstrated in a variety of experiments that such ``hyper''-polarized light \MCDT{Some physicists may get upset with this term.  Better use ``circularly polarized light''} can relay its \MCDT{but be careful, the light is not losing its angular momentum} angular momentum to other spin-possessing particles \MCDT{including atoms and their nuclei.}such as electrons or nuclei. 
Typically, electrons are polarized with light and the polarization is transferred to nuclei via electron--nucleus interactions. Two examples are spin-exchange optical pumping (SEOP) and metastability-exchange optical pumping (MEOP), as reviewed in references \cite{Colegrove1963,nacher1985optical,Walker1997,GOODSON2002157,Barskiy2017gases}.
In SEOP, the nuclear spins of noble-gas atoms are polarized through 'spin-exchange' collisions with alkali atoms present in the same volume, where the alkali atoms are optically pumped. In MEOP, typically used to polarize $^3$He nuclei, atoms are excited in an electrical discharge to a metastable state with unpaired electron spins. In this state, optical pumping occurs, and the resultant nuclear polarization persists as the atoms relax back to the ground electronic state.

In the following section, we mention hyperpolarized noble gases
%, which are relevant to the development and infrastructure of passive (or active) magnetically-shielded environment 
with precessing magnetization detected by noninductive sensors below the Earth magnetic field. While by our own definition (\autoref{Subsec:Regimes}), this work qualifies as high-field NMR rather than ZULF NMR, these developments have occurred in parallel to those of ZULF NMR and with some of the similar (or same) hardware.

\begin{figure}
\centering
  \includegraphics[width=0.9\columnwidth]{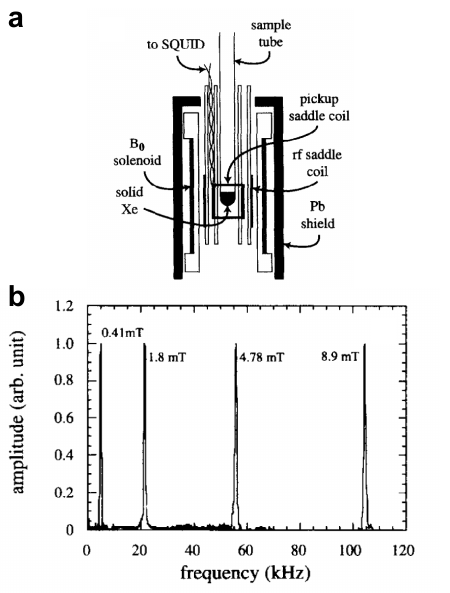} 
  \caption{a)~Schematic drawing of the SQUID-based NMR spectrometer. The sample is at the center of the orthogonal pickup and excitation saddle coils; a persistent-current solenoid provides a static magnetic field along the axis of the sample tube. Coils and sample are enclosed in a Pb tube. A two-layer $\mu$-metal shield (not shown) reduces the ambient magnetic field to below 0.1\,pT. b)~SQUID-detected $^{129}$Xe NMR spectra from isotopically enriched $^{129}$Xe (80\%) at four different magnetic fields. Reproduced from reference \cite{TonThat1997SQUID}, Copyright (1997), with permission from Elsevier.}
  \label{Fig:TonThat_129Xe}
\end{figure}

The earliest example of MEOP being used for sub-Earth-field NMR is the work of Cohen-Tannoudji \textit{et al.} in 1969 \cite{Cohen-Tannoudji1969}. The authors arranged two 6 cm diameter cells adjacent to one another, with one containing $^{87}$Rb (without a buffer gas) and the other 3\,torr of pure \textsuperscript{3}He. Inducing an rf discharge in the helium cell, one can populate the first excited metastable state of helium, in which these atoms can be directly pumped with light at a wavelength of 1.083\,$\mu$m, produced in the early experiments with a xenon discharge lamp (diode lasers for this wavelength are readily available today). The cells were held in a %\DB{I think it is necessary to mention that metastability was produced with a discharge (check) and that the wavelength of the light 1.083 nm (check) and how this light was produced.} \MCDT{They got the 1083 nm pumping light from a xenon discharge lamp, and yes, electrodes were needed to excite the metastable orthohelium state -- here is a related work from the same era: \url{https://ieeexplore.ieee.org/document/1067647}} 
magnetic shield with shim coils, which attenuated the external magnetic field to be on the order of 100\,pT. A beam of circularly polarized light was passed through the \textsuperscript{3}He cell to generate \textsuperscript{3}He nuclear-spin hyperpolarization via MEOP, and an orthogonal beam of circularly polarized light was passed through the \textsuperscript{87}Rb cell to optically pump the \textsuperscript{87}Rb. The Rb cell was used as a magnetometer to detect the magnetic signals from the \textsuperscript{3}He nuclei precessing about a 185 pT applied field, with a sensitivity on the order of \SI{100}{\femto\tesla\per\sqrt{\rm Hz}}. 
%\DB{Make Hz nonitalic \SI{100}{\femto\tesla\per\sqrt {\rm Hz} }}
Three decades later, a series of papers followed in which hyperpolarized \textsuperscript{3}He and \textsuperscript{129}Xe were detected in low magnetic fields using SQUID detectors. In the first of these \cite{TonThat1997SQUID}, TonThat \textit{et al.} collected hyperpolarized \textsuperscript{129}Xe in a sample tube at 4.2\,K, which solidified the \textsuperscript{129}Xe. The sample was held in a mu-metal shield and was surrounded by coils to provide a $B_0$ field and apply pulses. A pickup saddle coil was used to transfer the magnetic signals to a SQUID for detection. The experimental apparatus is shown in \autoref{Fig:TonThat_129Xe}a. The \textsuperscript{129}Xe signals were detected in $B_0$ fields ranging from 0.019 to 8.9\,mT. The authors also measured the solid-state \textsuperscript{129}Xe $T_1$ as a function of magnetic field, and saw it vary from approximately 8000\,s at 5\,mT to $\sim$2000\,s at fields below 0.05\,mT, where it eventually levelled off.

In a subsequent paper from the same group, the authors extended the scope of this work to perform 1D and 2D imaging and included experiments with hyperpolarized \textsuperscript{3}He \cite{Augustine1998}. A few years later, imaging techniques were applied to obtain low-field images of hyperpolarized \textsuperscript{129}Xe in porous media \cite{WongFoy2002}. Hyperpolarized \textsuperscript{129}Xe was flowed through a glass sample chamber containing a block of aerogel, a few millimeters in each dimension. The imaging experiments were conducted in a $B_0$ field strength of 2.26\,mT, with gradients on the order of 1\,mT/m applied for spatial encoding. The authors were able to image the penetration of \textsuperscript{129}Xe into the aerogel. In addition, they measured chemical shifts of xenon atoms in two different chemical environments at low field, thanks to the large chemical shift dispersion of \textsuperscript{129}Xe. Hyperpolarized \textsuperscript{129}Xe was flowed through powdered polypropylene, and the xenon adsorbed to the surface exhibited a 5 Hz (166 ppm at 2.55\,mT) shift compared to \textsuperscript{129}Xe in the gas phase.

A significant step forward came in 2003, when the scope of these experiments was expanded by using hyperpolarized \textsuperscript{129}Xe to polarize other molecules in solution for detection at low field \cite{Heckman2003enhancement} via an effect that became known as the spin-polarization-induced nuclear Overhauser effect (SPINOE). \textsuperscript{129}Xe was polarized to around 2\% and was bubbled into a protonated organic solvent (cyclopentane, acetone, or methanol). Cross-relaxation between the \textsuperscript{129}Xe atoms and the protons in the solvent molecules is the basis of SPINOE \cite{Navon1996}, leading to enhancements of the $^1$H NMR signals at the detection field of 0.95\,$\mu$T of around $10^6$. Although the corresponding polarization levels of the $^1$H spins are still low (on the order of $10^{-5}$), this method is a promising general approach to enhancing NMR signal, and would benefit greatly from the improved \textsuperscript{129}Xe polarization levels that can now be obtained.

Apart from the 1969 demonstration \cite{Cohen-Tannoudji1969}, all experiments measuring SEOP-enhanced low-field NMR until 2004 were conducted using SQUID magnetometers. In 2004, laser-polarized \textsuperscript{129}Xe in a 45\,$\mu$T field was detected using an optically pumped magnetometer with an experimental sensitivity of approximately 100\,fT/$\sqrt{\textrm{Hz}}$, with noise dominated by field fluctuations in the magnetic shielding \cite{yashchuk2004hyperpolarized}. This work was a significant step for ZULF NMR, as it marked the first use of a piercing solenoid, a solenoid that surrounds the sample and passes entirely through the magnetic shielding, with the ends of the solenoid outside. Passing a current through the solenoid generates a magnetic field inside it, while the field outside the solenoid but within the shielding is nominally zero. %\DB{It is slightly weird (but may be OK in the end---need to revisit this) that we talk about piercing solenoids in the SEOP Section... Also, while occasional naming names and years may be ok, in general, this should be minimized as this information is available in references}
%and the magnetic field outside the solenoid is carried back through the magnetic shielding (due to its high magnetic permeability compared hat inside the magnetic shielding, 
A magnetometer can be placed outside the solenoid next to the sample to detect magnetic fields from the sample without being affected by the much larger solenoid field. Experimentally, reduction factors in the field outside versus inside the solenoid of around 
%a factor of
$10^3$ are typically observed \cite{Bodenstedt2021,eills2023enzymatic}, with even larger factors obtainable by careful winding of the solenoid.

\begin{figure}
\centering
	\includegraphics[width=0.8\columnwidth]{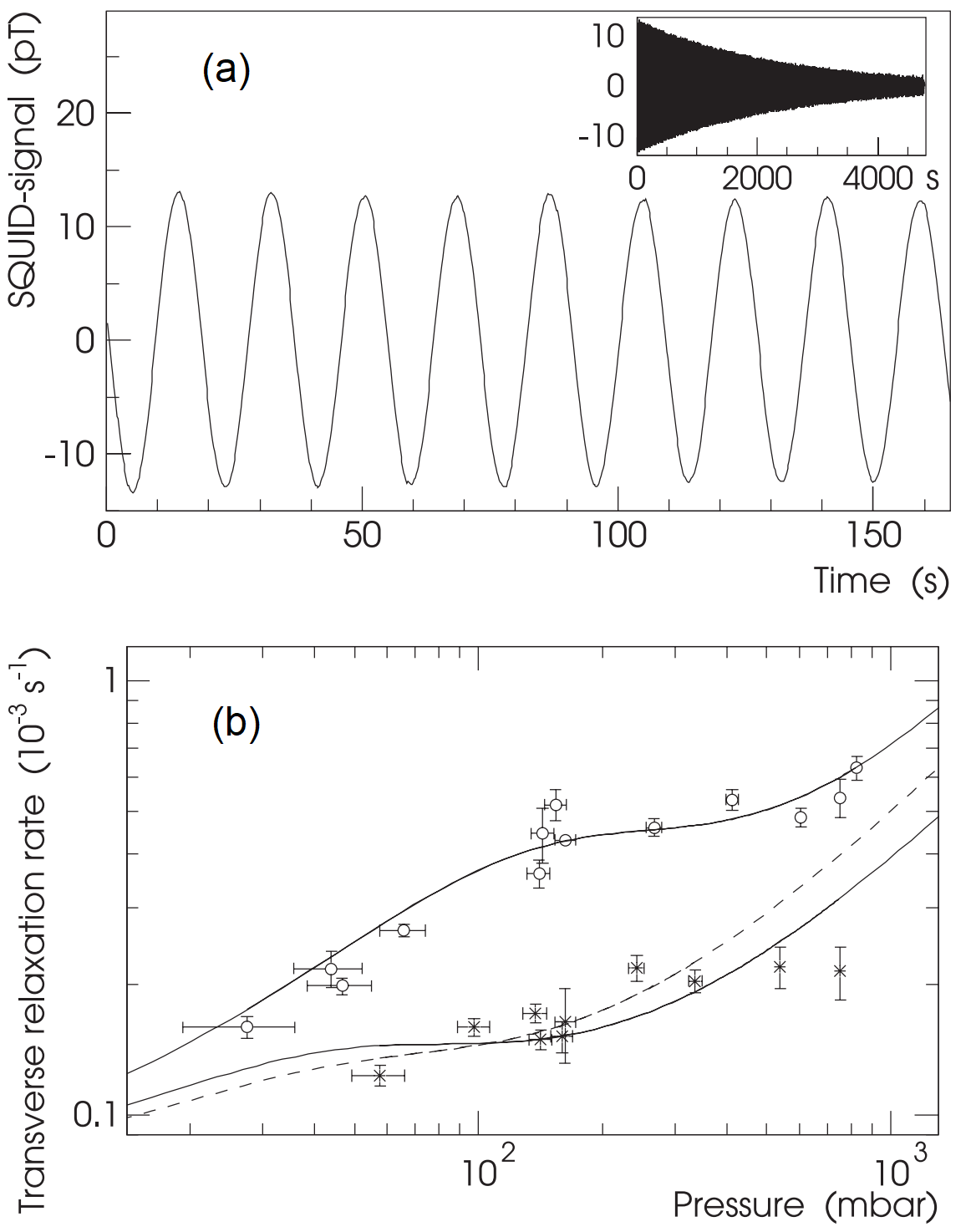}
	\caption{(a) Precession of hyperpolarized \textsuperscript{129}Xe nuclei in a 4.7 nT field, as measured by a SQUID magnetometer. (b) \textsuperscript{129}Xe $T_2$ relaxation rates as a function of pressure for samples in 4.5 nT (circles) and 15 nT (crosses) magnetic fields. The three lines represent fits to the data using the equations described in the original paper. Image adapted from reference \cite{kilian2007free}.}
	\label{fig:kilian-2007free}
\end{figure}

%\SP{Shouldn't we mentioned somewhere here Heil's work??} 
In 2007, a study of \textsuperscript{129}Xe relaxation times was conducted in the Berlin Magnetically Shielded Room (BMSR) \cite{kilian2007free}. The study involved measurements at various xenon pressures and an ambient magnetic field in the range of 4.5--15\,nT, with the hyperpolarized \textsuperscript{129}Xe held in spherical glass cells. In this work, SQUID magnetometers were employed to observe \textsuperscript{129}Xe $T_2$ relaxation times of up to 8000\,s in a 15 nT $B_0$ field. Figure\,\ref{fig:kilian-2007free} shows a plot of \textsuperscript{129}Xe $T_2$ times for various gas pressures. In a separate set of experiments from the same time at the BMSR \cite{burghoff2007squid}, the authors modified two SQUID systems (originally developed for biomagnetic applications) for ultralow-field NMR applications. They used hyperpolarized \textsuperscript{129}Xe as the target nucleus to optimize their system.

\begin{figure}
	\begin{center}
			\includegraphics[width=\columnwidth]{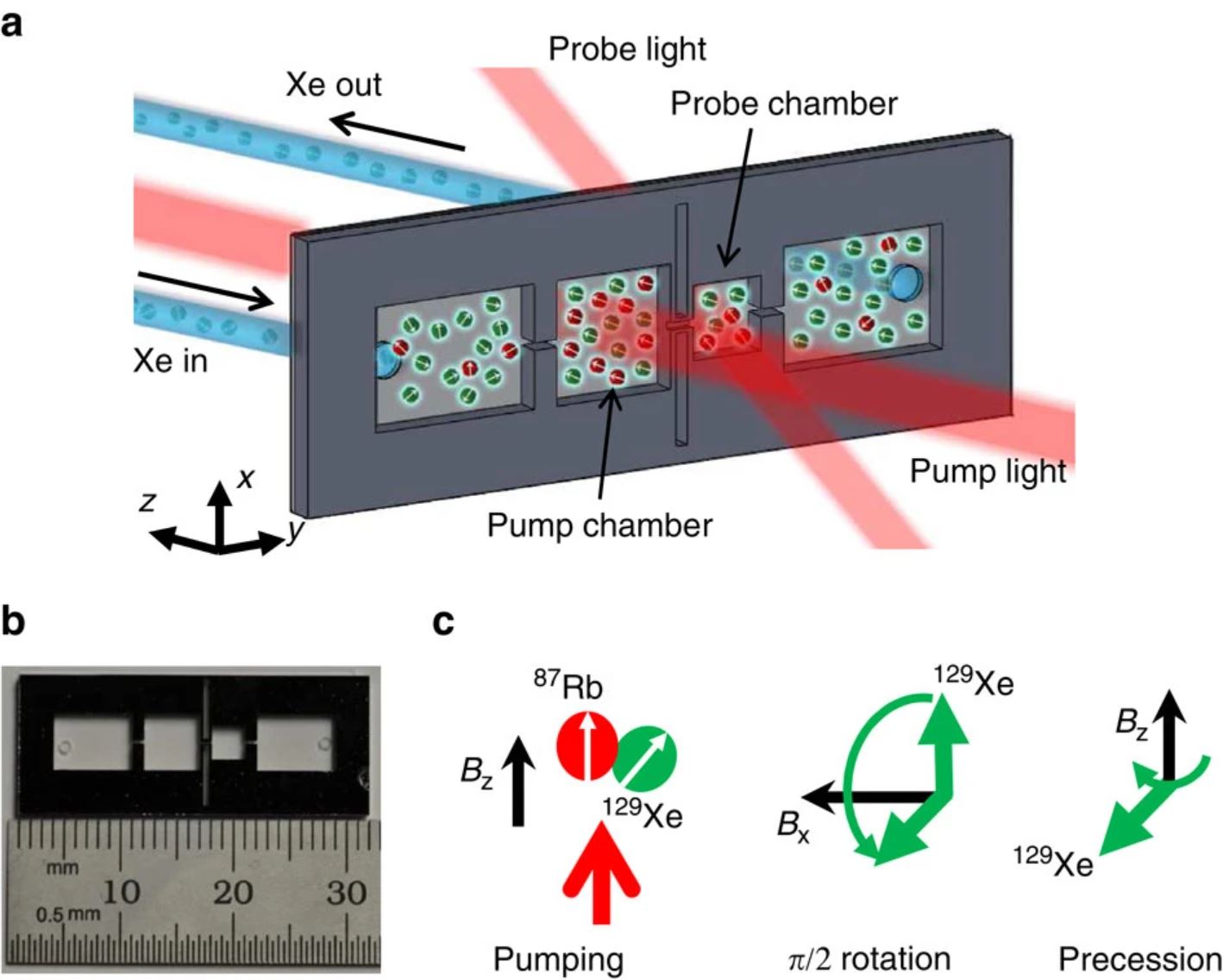}
	\end{center}
	\caption{The microfluidic device used by Jimenez-Martinez \textit{et al.}, combining optical pumping and detection of \textsuperscript{129}Xe on one platform. (a) The device, with lasers shown in red. (b) Size measurement of the chip. (c) The pumping and probing sequence. Adapted from reference \cite{jimenez2014optical} under terms of the Creative Commons CC-BY license 4.0.
	}
	\label{fig:jimenez-martinez2014}
\end{figure}

In the ensuing years, OPMs were further developed and miniaturized, which enabled the combination of SEOP with \textsuperscript{129}Xe and OPM readout on a single microfluidic device \cite{jimenez2014optical}.  The microfluidic chip was constructed of silicon sandwiched between layers of borosilicate glass, with chambers cut from the silicon for the SEOP of \textsuperscript{129}Xe and for optical readout. Figure\,\ref{fig:jimenez-martinez2014} shows the chip and the lasers used for optical excitation and readout. The total volume of this microfluidic chip was approximately 100\,$\mu$L. Xenon was polarized in the pump chamber, and in some experiments it was detected there, and in others in the probe chamber. Application of a $\pi/2$ pulse at a magnetic field ($B_0$) of 0.8\,$\mu$T resulted in the detection of NMR signals from \textsuperscript{129}Xe spins at 9\,Hz. Polarization levels of 0.7\% were measured in this device. %A $B_0$ field of 0.8 $\mu$T was applied %to provide a precession axis for the \textsuperscript{129}Xe, and rotating the \textsuperscript{129}Xe spins away from $\hat{z}$ with a magnetic pulse yielded an oscillating NMR signal at 9\,Hz. Polarization levels of 0.7\% were measured in this device.

A few years later, an improved version of this device was developed \cite{Kennedy2017}. By changing the method of rubidium production on the chip (using photolysis of rubidium azide instead of reacting barium azide with $^{87}$Rb chloride), the authors avoided the formation of barium chloride, drastically reducing contamination on the walls of the chip. This improvement enhanced \textsuperscript{129}Xe relaxation times by a factor of approximately 5, and the device achieved \textsuperscript{129}Xe polarizations of up to 7\%. Using this platform, the authors compared \textit{in situ} and \textit{ex situ} signal readout methods. In the \textit{in situ} experiments, the Xe and Rb atoms were in the same cell during signal detection, whereas in the \textit{ex situ} measurements, the Rb vapor cell was adjacent to a separate cell containing xenon. In the \textit{ex situ} measurement, the Rb electrons sense the dipolar field produced by the polarized xenon atoms. In the \textit{in situ} measurement, the Rb electrons sense the Xe nuclear spins predominantly via the Fermi contact interaction \cite{grover1978noble}. The overall signal enhancement compared to the \textit{ex situ} measurement was reported to be approximately 5300 \cite{Kennedy2017}.
%lectrons sense both the \textsuperscript{129}Xe dipole field, as well as experiencing a Fermi contact interaction with the \textsuperscript{129}Xe. The closer proximity of the Rb and Xe leads to the dipolar interaction being enhanced by a factor of $\sim$10 (which the authors verified with COMSOL magnetic field simulations), and the Xe--Rb Fermi contact interaction enhances the observed signal by a further factor of $\sim$500\,\cite{grover1978noble}. These factors together account for the authors' observation that the \emph{in situ} signal is $\sim$5300 times larger than the \emph{ex situ} signal.

%\begin{figure}[h]
	%\begin{center}
	%		\includegraphics[width=\columnwidth]{Progress_in_NMR_ZULF/kennedyOptimizedMicrofabricatedPlatform2017-fig3.jpg}
	%\end{center}
	%\caption{(a) A comparison between the signals in the \emph{in situ} and \emph{ex situ} modalities of the experiment. (b) The microfluidic chip used in these experiments, with lasers shown in red. AVAM = alkali vapour atomic magnetometer. Image adapted with permission from Ref.~\cite{Kennedy2017}
	%\label{fig:kennedy2017}
%\end{figure}

\subsection{Parahydrogen-induced polarization}
\label{Subsec:PHIP}

In 1986, Clifford Bowers and Daniel Weitekamp 
%of the California Institute of Technology (Caltech) 
proposed a thought experiment \cite{Bowers1986} in which they predicted the observation of enhanced $^1$H NMR signals in the spectra of products of a hydrogenation reaction with \textit{p}H$_2$, the spin-0 nuclear spin isomer of molecular hydrogen. Later, they experimentally demonstrated that the hydrogenation of acrylonitrile with \textit{p}H$_2$ using Wilkinson's catalyst at high magnetic fields indeed resulted in the predicted effect \cite{Bowers1987}. Termed PASADENA, the effect is characterized by enhanced antiphase multiplets in the $^1$H NMR spectra and can be observed both for the reaction products and for the hydride complexes of the metals used as hydrogenation catalysts \cite{Duckett2012}. PASADENA and the related ALTADENA effect discovered a year later \cite{Pravica1988} are now collectively known as parahydrogen-induced polarization (PHIP) \cite{Eisenschmid1987,Natterer1997}. %\MCDT{FYI the term PHIP was actually coined before ALTADENA in the same year as the PASADENA paper, by another group (including Bargon) who independently discovered parahydrogen-enhanced NMR signals. See Eisenschmid TC, Kirss RU, Deutsch PP, Hommeltoft SI, Eisenberg R, Bargon J, Lawler RG,
%Balch AL (1987) J Am Chem Soc 109(26):8089–8091. You may want to cite this original source too.} \DAB{I am aware of the story from a few different sources including Chekmenev and Bergman. Bargon was in the audience of Weitekemp's presentation at the Gordon conference. I will include the reference!}

At the core of PHIP and related techniques is the quantum-mechanical connection between the symmetry of the nuclear spin state and the allowed rotational states of molecules. This effectively gives the nuclear spin degrees of freedom access to the relatively large energy scales associated with rotations and thus enables much higher nuclear spin order than would be possible with spin interactions alone. The underlying physics is well described in the literature \cite{Farkas1935,GREEN20121} 
% \SP{Do we really want to quote this original 100 years old paper?} \MCDT{The author will probably not be upset about h-index if we do not cite, but I think we should keep this -- original works should be referenced.  Note very soon the first NMR papers will be 100 years old } 
and will not be addressed further here.  

Perhaps the greatest advantage of PHIP is the ease with which the source of polarization, \textit{p}H$_2$ gas, can be generated, stored, and used. Hydrogen gas at $>$99\% para-enrichment can be prepared by cooling the gas to below 25~K in the presence of a paramagnetic catalyst; if the catalyst is then removed, the hydrogen can remain in the para state for days at room temperature \cite{silvera1980solid}. 

In \emph{hydrogenative} PHIP, hyperpolarization can be engendered in a target molecule by chemically reacting \textit{p}H$_2$ with a suitable molecular precursor, with three important caveats: (1) the chemical reaction should proceed on a timescale not longer than that of nuclear spin relaxation; (2) the hydrogen addition should be pairwise, i.e., the hydrogen atoms from one molecule of \textit{p}H$_2$ should be incorporated into the same product molecule, so that their correlated spin state is preserved; and (3) the symmetry between the two \textit{p}H$_2$ protons must be broken (either in the final product or in a reaction intermediate) for hyperpolarized signals to be observable. In solution-state experiments, the first two considerations are addressed by the use of organometallic catalysts. If the magnetic equivalence of the two \textit{p}H$_2$ protons persists in the final product molecule, the spin order remains in the unobservable singlet state. This state can be long-lived \cite{carravetta2004long} and can support the spin order until further chemical reactions break the magnetic equivalence and hyperpolarized NMR signals are released \cite{zhang2014long,eills2021singlet}.

In \emph{non-hydrogenative} PHIP, also known as signal amplification by reversible exchange (SABRE), target molecules are polarized through temporary interactions with \textit{p}H$_2$ molecules during the reversible binding of these molecules to an organometallic catalyst \cite{Adams2009}. The $J$-couplings between the \textit{p}H$_2$-nascent hydrides and the bound-substrate nuclear spins in the complex allow for the redistribution of hyperpolarization onto the ligated substrates (molecules to be polarized) during the lifetime of the complex.

The low cost, small footprint, and ease of experimental implementation make PHIP an attractive hyperpolarization technique for combination with ZULF NMR. The majority of PHIP and SABRE experiments employing ZULF detection use a pressurizable NMR tube as the reaction vessel, in which the reaction solution is contained and \textit{p}H$_2$ gas is typically bubbled in at elevated pressure (5--10\,bar). After the chemical reaction has started, hyperpolarization-enhanced NMR signals can be detected either by applying a magnetic field pulse or by conducting the hydrogenation in the presence of a small bias magnetic field and generating hyperpolarized NMR signals by nonadiabatically switching off the field. In the first experiments combining ZULF NMR with PHIP, four different chemical systems were probed: the hydrogenation of [1-$^{13}$C]dimethyl acetylene dicarboxylate to [1-$^{13}$C]dimethyl maleate, styrene to ethylbenzene, 1-phenyl-1-propyne to 1-phenyl-1-propene, and 3-hexyne to 3-hexane \cite{Theis2011}. In each case, after the chemical reaction, $^1$H-$^{13}$C coherences were excited by applying a square magnetic field pulse with amplitude $B_P$ and duration $\tau_P$ such that  $B_P(\gamma_{\rm H}-\gamma_{\rm C})\tau_P =\pi/2$.  The pulse length $\tau_P$ is chosen to maximize the amplitude of the sine term in \autoref{eq:cyclic2a}, thereby optimizing the conversion between scalar spin order and observable spin order. The authors identified the spectral regions corresponding to particular coupled $^1$H-$^{13}$C groups, and simulated the zero-field PHIP spectrum of ethylbenzene, which was found to agree with the experiment.

%\JE{The low cost, small footprint, and ease of experimental implementation make PHIP an attractive hyperpolarization technique for combination with ZULF NMR. The majority of PHIP and SABRE experiments employing ZULF detection use pressurizable NMR tubes as the reaction vessels in which the reaction solution is held, and \textit{p}H$_2$ gas is bubbled in at elevated pressure (usually at 5-10 bar). After the chemical reaction has started, it is possible to detect the hyperpolarized NMR signals after applying a magnetic field pulse; alternatively, an observable signal can be generated after nonadiabatic switching off of the applied magnetic field. In the first experiments combining ZULF NMR with PHIP, four different chemical systems were probed: the hydrogenation of 1-${}^{13}$C-dimethyl acetylene dicarboxylate to 1-${}^{13}$C-dimethyl maleate, styrene to ethylbenzene, 1-phenyl-1-propyne to 1-phenyl-1-propene and 3-hexyne to 3-hexane. \cite{Theis2011} In each case, after the chemical reaction, ${}^{1}$H-${}^{13}$C coherences were excited by application of a magnetic field pulse with amplitude $B_P|\gamma_H-\gamma_C|=\pi/2$. The authors analyzed the spectra and were able to attribute the spectral regions to ${}^{1}$H-${}^{13}$C coupling partners, as well as simulating the zero-field PHIP spectrum of ethylbenzene.}

%%%%%%%%%%%%%%%%%%%%%%%%%%%%%%%%%%%%%%%
    \begin{figure}[t]
\centering
	\includegraphics[width=1\columnwidth]{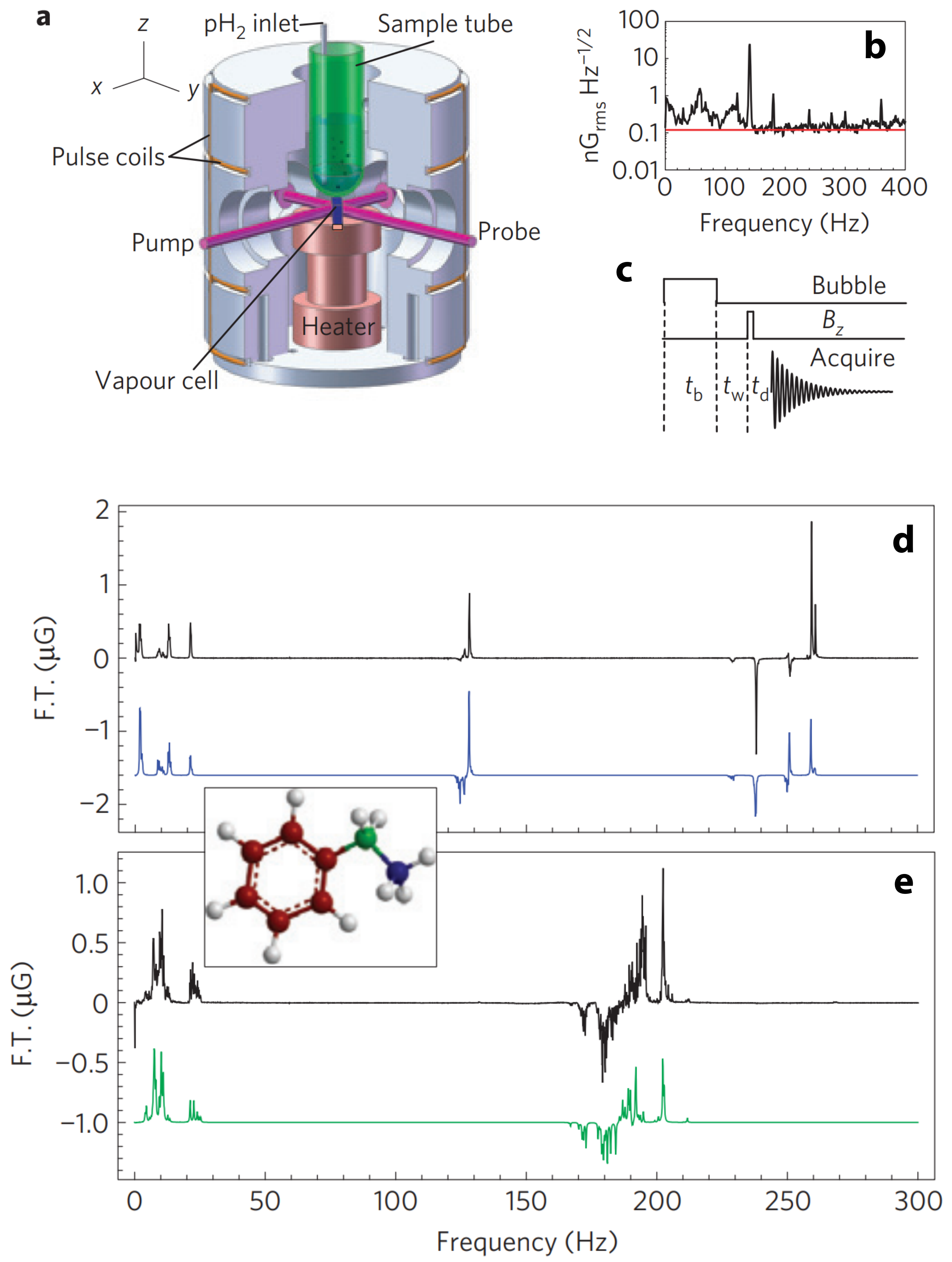}
	\vspace{-6pt}
	\caption{
(a) The zero-field NMR setup employed for the first parahydrogen-enhanced NMR experiments \cite{Theis2011}. (b) A background spectrum showing the noise floor of the experimental apparatus. (c) The pulse sequence employed for the experiments. (d,e) The PHIP-enhanced zero-field NMR spectra of ethylbenzene-$\beta$-$^{13}$C and ethylbenzene-$\alpha$-$^{13}$C polarized after the hydrogenation of  $^{13}$C-labelled styrene, with simulations shown in blue and green. Several features are seen in the spectra. For example, as expected (see Sec.\,\ref{Sec:Theory}), the $^{13}$CH$_2$ group produces a spectral feature around $3J/2$, while the $^{13}$CH$_3$ group produces features at $J$ and $2J$. Moreover, the spectral features for the group further from the aromatic ring are less perturbed by the ring protons compared to the features from the group closer to the ring. Reproduced from reference \cite{Theis2011}.}
	\label{fig:PHIP_Theis}
\end{figure}
%%%%%%%%%%%%%%%%%%%%%%%%%%%%%%%%%%%%%%%
Following this initial experimental demonstration, a theoretical description of parahydrogen-induced polarization at zero field was developed \cite{butler2013parahydrogen}. 
The incorporation of individual parahydrogen molecules into a target molecule is a nonadiabatic process. This means that the initial singlet order of \textit{p}H$_2$ is projected onto the populations and coherences of a new eigenbasis. As individual target molecules in the sample are hydrogenated at different time points, coherences among nondegenerate states are averaged to zero and only populations survive.  Mathematically, this corresponds to eliminating the nondiagonal elements of the spin density matrix in the new eigenbasis. The resulting magnetization remains zero. 
%Vector order, which can be detected by the magnetometer \DAB{[refs]}, is generated by applying a magnetic field pulse.}

%\MCDT{In this approach, spin order is described as arising from the initial singlet order of \textit{p}H$_2$ after it is incorporated into a molecule with a heteronuclear spin, e.g., \textsuperscript{13}C.  First, the addition of the extra spin results in a sudden change of eigenbasis (from the perspective of), so for a given molecule, the \textit{p}H$_2$ spin order is projected directly onto the populations and coherences of the new basis. Overall, since the reaction occurs with individual molecules in the sample being hydrogenated at different time points, the coherences are averaged to zero and only the populations survive.  Mathematically, this corresponds to eliminating the nondiagonal elements of the spin density matrix, in the new eigenbasis.}

In general, asymmetrical coupling between the hydrogen spins and the heteronucleus---i.e., broken magnetic equivalence of the \textit{p}H$_2$-sourced hydrogens in the hydrogenation product---is of interest, as then the process yields polarization of spin rank greater than zero.  Application of subsequent magnetic field pulses leads to the generation of observable vector order, which can then be measured \cite{goldman2005hyperpolarization,cavallari2015effects,eills2019polarization,joalland2019pulse}.

%An important advantage of \textit{p}H$_2$-based techniques compared to other hyperpolarization methodologies is its scalability. Indeed, it relatively straightforward to produce large boluses of hyperpolarized material (e.g., liters in seconds\MCDT{please give citation!  Sounds like a large bolus indeed}\DAB{Well examples of L/sec of HP material were never published but the meaning of the sentence is that it is easy to do by increasing the reactor size (while for dDNP one will always be limited by penetration depth of microwaves in the material).}, by flowing large volumes of \textit{p}H$_2$ gas through an optimized chemical reactor.  Delivery of parahydrogen to samples is relatively straightforward as compared to other techniques in which uniform supply of the source of spin order (visible light, microwave radiation etc.) to the whole to-be-polarized sample may represent an obstacle. For this reason, \textit{p}H$_2$-based hyperpolarization is currently considered a main candidate for several applications in fundamental physics (see Sec.\,\ref{subsec:phys} for details).

%\MCDT{I concur, but if it has not been done, we should not write `it is relatively straightforward'. Instead write something like `
The advantages of \textit{p}H$_2$-based polarization compared to other hyperpolarization methodologies are high polarization yield and simple, low-cost infrastructure. This should in principle allow straightforward scale-up to produce large boluses of hyperpolarized material (e.g., liters in seconds). For this reason, \textit{p}H$_2$-based polarization is currently considered a main candidate for applications in fundamental physics (see Sec.\,\ref{subsec:phys} for details).

\begin{figure*}[t]
\centering
	\includegraphics[width=0.9\textwidth]{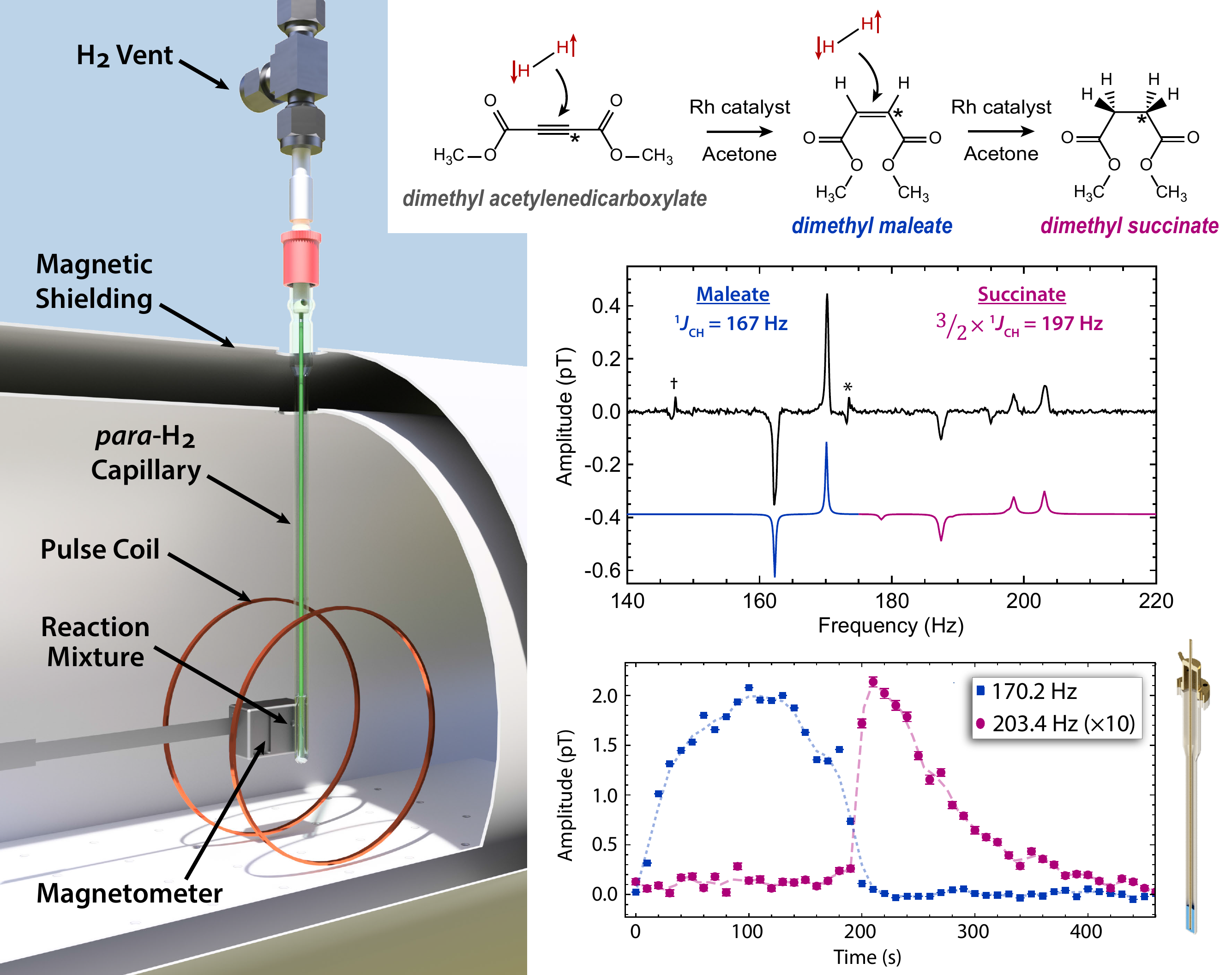}
	\vspace{-6pt}
	\caption{
Left: Parahydrogen can be bubbled into a vessel (in this case a 5-mm NMR tube) containing a reaction solution at zero field, and the NMR signals are excited and observed periodically using an OPM. Right, top: The two-step hydrogenation reaction carried out. Right, middle: A zero-field NMR spectrum showing resonances that correspond to [2-$^{13}$C]dimethyl maleate and [2-$^{13}$C]dimethyl succinate. Right, bottom: Integrals of the peaks at the specified frequencies over time as the reaction was carried out. The signal from [2-$^{13}$C]dimethyl succinate decays due to relaxation of the hyperpolarized signals. On the right the titanium tube is shown in which the reaction was carried out.
Reproduced from reference \cite{Burueva2020} under terms of the Creative Commons BY-NC license 4.0.
	}
	\label{fig:PHIP_RM}
\end{figure*}

%\DB{Here the transition from general theory to metal containers is way too sharp. Needs work!} \DAB{How about now?}
%\DAB{
Another important application of \textit{p}H$_2$-based hyperpolarization in the context of ZULF NMR is chemical analysis. One example is the demonstration that ZULF NMR can be used as a spectroscopic technique for chemical reaction monitoring, by observing the two-step hydrogenation of dimethyl acetylene dicarboxylate to dimethyl maleate to dimethyl succinate \cite{Burueva2020}. This is possible because ZULF NMR is a quantitative method, providing chemical specificity thanks to the unique $J$-spectra exhibited by different molecules. A schematic of the experimental apparatus used in this work and the key results are shown in Fig.\,\ref{fig:PHIP_RM}. It was additionally demonstrated that the reaction monitoring could be carried out under continuous bubbling and at high pressure ($>$10\,bar) in metal containers, something that is not possible using high-field NMR. It remains to be seen whether this technique has a role to play in industrial-scale process monitoring.
%%%%%%%%%%%%%%%%%%%%%%%%%%%%%%%%%%%%%%%
%%%%%%%%%%%%%%%%%%%%%%%%%%%%%%%%%%%%%%%

SABRE was also shown to be an effective way to hyperpolarize a chemical sample without the need for any magnets \cite{Theis2012_NH_PHIP}. 
%\DAB{
In this work, parahydrogen gas was bubbled at high pressure (5.8\,bar) into an NMR tube containing a solution of $^{15}$N-pyridine and Crabtree's catalyst in anhydrous methanol at zero field.
%Parahydrogen gas was bubbled at 5.8~bar into an NMR tube containing a solution of $^{15}$N-pyridine and Crabtree's catalyst in anhydrous methanol at zero field. 
After several seconds of bubbling, during which the singlet spin order was transferred from parahydrogen to pyridine, a magnetic-field pulse of duration $\tau_P$ and  amplitude $B_P$ was applied that satisfied the condition $B_P(\gamma_{\rm H}-\gamma_{\rm N})\tau_P=\pi/2$. 
%After some time in which spin order was transferred from parahydrogen to pyridine, a magnetic field pulse was applied such that $B_P(\gamma_{\rm H}-\gamma_{\rm N})\tau_P=\pi/2$.  
The resulting NMR spectrum was collected using an optically pumped magnetometer positioned beneath the sample. 
For comparison, the authors also collected a zero-field NMR spectrum of neat $^{15}$N-pyridine, thermally polarized prior to signal acquisition in a 1.6\,T permanent magnet. 
Peak shifts of $\approx$1\,Hz were observed between the two spectra, which was attributed to the $J$-couplings differing between the pyridine neat liquid and pyridine dissolved in methanol. 

Recently, another way of observing \textit{p}H$_2$-based hyperpolarization under ZULF conditions was demonstrated \cite{blanchard2021towards}. In this approach, which is conceptually similar to SABRE-SHEATH (SABRE in SHield Enables Alignment Transfer to Heteronuclei) \cite{Theis2015,Truong2015}, hyperpolarization of heteronuclei (for example, $^{15}$N) was detected at zero field followed by a sudden switching off of the sub-microtesla magnetic field. The authors also demonstrated that SABRE reactions could be carried out for longer periods of time by presaturating the parahydrogen gas with the solvent prior to bubbling into the sample, which helps to alleviate the problem of solvent evaporation.
This work paves the way for SABRE-enhanced ZULF NMR experiments that can be carried out for hours or even days with applications ranging from those in fundamental physics (Sec.\,\ref{subsec:phys}) to industrial ones (Sec.\,\ref{Sec:Chemistry}).
% \DB{Danila, please, please: let us state some numbers: volumes, concentrations, polarization levels, rates of polarization, at least crudely. Otherwise it is all a bit abstract...}\DAB{DONE}

Typical concentrations of substrates used in the  SABRE development work are on the order of 10--100\,mM. Specific demonstrations were performed with concentrations ranging from sub-$\mu$M (e.g., 0.5~$\mu$M in reference \cite{Eshuis2014}) to molar (e.g., polarization of a neat liquid used as a solvent, see reference \cite{Shchepin2015neat}). In general, due to the nature of SABRE mechanism, which is based on chemical exchange and formation of a specific polarization-transfer complex, polarization values drop with increasing substrate concentration \cite{BARSKIY2019sabre}. This limits applicability of SABRE to searches of new physics such as spin-gravity coupling (Sec.\,\ref{Subsec:GDM}) or dark matter (Sec.\,\ref{Subsec:Dark}), where often a near-unity polarization and a high spin density (on the order of 10$^{20}$\,cm$^{-3}$) are necessary. For such applications, `brute force' polarization of isotopically enriched compounds with as-high-as-possible magnetic field (for example, using superconducting magnets, ideally at low temperatures) could be a better choice.

One should note the difference in how the SABRE-based ZULF-NMR signals can be observed depending on whether the polarization is performed at zero or finite field. The zero-field case produces scalar spin order (spherical rank 0) by virtue of a lack of preferred field orientation, and requires a magnetic field pulse to generate observable coherences between ZULF eigenstates. The other case, where polarization occurs at finite field (e.g., 0.1--1\,$\mu$T, known as SABRE-SHEATH \cite{Truong2015}) generates \textit{z}-magnetization on the heteronucleus \cite{Theis2015,Barskiy2016jacs,ORTMEIER2024100149}, and requires only a sudden switching-off of the magnetic field to generate coherences.  Both of these approaches result in similar amplitudes of peaks in the zero-field \textit{J}-spectra and the choice of approach depends on the particular goal of the experiment.

Atomic magnetometers are not the only type of sensors used to detect molecules hyperpolarized by SABRE. Low-field NMR spectra of 18 different SABRE-polarized fluorinated $N$-heterocycles were measured using a SQUID magnetometer \cite{Buckenmaier2017squid}. Magnetic fields between 0.1 and 10\,mT were applied during the \textit{p}H$_2$ bubbling to induce spontaneous spin-order transfer from \textit{p}H$_2$ to the target molecules, and stepped down to 144\,$\mu$T for signal acquisition. Although this is not ultralow field per our definition, this work demonstrates the relative ease with which the SABRE experiment could be adapted for ZULF detection, as the SQUID magnetometer is also sensitive down to lower frequencies. Thanks to advances in homogeneous catalysis and progress in understanding of \textit{p}H$_2$-based spin dynamics, in recent years PHIP and SABRE have become routinely employed hyperpolarization methods. For this reason they were also used in the first demonstration of ZULF NMR carried out using commercial atomic magnetometers \cite{Blanchard2020}.

In principle, parahydrogen-enabled hyperpolarization allows extending all applications discussed above to imaging at magnetic fields below the Earth field. In reference \cite{lee2019squid}, the authors demonstrated $^1$H MRI at $\sim$30\,$\mu$T of a bubble-separated phantom by using a solution of SABRE-polarized pyridine. Strong signal enhancement of $^1$H signals provided by SABRE ($\sim$2600) shortened the MRI operation time, so that an image with resolution of 1.6$\times$1.6\,mm$^2$ %\DB{a 2D phantom? It seems some words are missing. What is the point of a small sample with a large FOV ?}  
of a 3D-printed phantom with a field of view of 60\,mm could be obtained in less than 50 minutes.

There is a considerable literature describing low- or Earth-field PHIP experiments \cite{Colell2013,hamans2011nmr}. Another topic involving PHIP is polarization-transfer methods that exploit level anticrossings in the ZULF regime \cite{BARSKIY2019sabre}. There have been a number of demonstrations involving both hydrogenative and non-hydrogenative PHIP: in some cases the reaction is carried out at ZULF, and in other cases the reaction is carried out at high/low field and a magnetic field cycle is applied to induce polarization transfer~\cite{reineri2015parahydrogen,kiryutin2016fast,shchepin2016efficient}. %Since these methods do not involve detection at ZULF, they are outside the scope of this review, although we point the interested reader to a recent review of this topic \cite{bengs2020manipulating}.  
These methods are typically applied with detection in high magnetic fields \cite{bengs2020manipulating} but are also compatible with detection under ZULF conditions \cite{EillsTayler2024}.

\subsection{Dynamic nuclear polarization}
\label{Subsec:DNP}

Dynamic nuclear polarization (DNP) is a term used to describe a set of hyperpolarization methods that rely on polarization transfer from unpaired electrons (typically present in the form of free radicals) to spin-active nuclei. This transfer typically takes place at elevated magnetic field, either at room temperature in solution (such as in Overhauser DNP) or within a low-temperature (a few K) amorphous solid, and is effected by microwave irradiation applied to electron spin transitions  \cite{Abraham1959prl,Griffin2010pccp,Larsen2003,GOERTZ200428}. DNP allows generating NMR signals increased by orders of magnitude compared to thermal equilibrium, resulting in dramatically decreased signal averaging times \cite{WOLBER2004173}. Below we discuss in more detail the DNP approaches that have been used in the context of ZULF NMR.

\subsubsection{Overhauser Dynamic Nuclear Polarization}
The Overhauser effect manifests itself as polarization transfer from electrons to nuclei, driven by dipolar relaxation. For the case of a coupled electron--nuclear system, dipolar relaxation in solution gives rise not only to individual $T_1$ relaxation of spins but also to cross-relaxation, a process in which both spins are flipped simultaneously \cite{WILLIAMSON2019264}. In such systems, upon saturation of electron transitions, cross-relaxation leads to a steady state with enhanced nuclear polarization, %\DB{Can order be populated? Maybe just: steady state with nuclear spin order?} nuclear spin order
the process known as Overhauser DNP \cite{hausser1968adv,WILLIAMSON2019264}. Saturation here refers to equalizing populations of spin energy levels upon applying a strong resonant microwave pulse. One can think of saturating EPR transitions as a way to block pure electronic relaxation. As this relaxation pathway is blocked, relaxation occurs via other channels in which electrons and nuclei relax together. The enhancement factor ($\epsilon$) for the nuclear spin signal can be calculated as
\begin{equation} \label{oDNP} 
    \epsilon = \frac{\braket{I_z}}{I_0} = 1 - \rho f s \frac{\abs{\gamma_{\rm S}}}{\abs{\gamma_{\rm I}}}\,,
\end{equation}
where $\braket{I_z}$ and $I_0$ are average spin polarizations for nuclei (for example, protons in a solvent such as H$_2$O) after saturating electron transitions and at thermal equilibrium, respectively; $\rho$ is a coupling constant, which can range from $-1$ assuming pure scalar interaction to $+0.5$ in the case of pure dipolar coupling between electron and nuclear spin; %\DB{Kindly write explicitly what coupling you mean; also need a reference to where one could look up the definitions of these ranges} 
$f$ is a leakage factor that
shows how effectively the proton spin is relaxed by the electron spin and ranges from 0 to 1; and $s$ is the saturation factor, which is equal to 1 for complete saturation \cite{GUIBERTEAU199647}. 
%When the conditions of %\DB{What is the eaning of ``it'' here? (A question in the style of Bill Clinton.) If you mean the equation, need to say when it holds...}
%Eq.\,\ref{oDNP} hold, it 
For these conditions, \autoref{oDNP} predicts the maximum polarization enhancement for protons to be $\sim$660 in the case of pure scalar coupling and $\sim$330 for pure dipolar coupling between the nucleus to be polarized and the `electron source', for example a nitroxide radical. Negative $\epsilon$ means that the nonequilibrium magnetization vector created with Overhauser DNP has the direction opposite to that of the thermal equilibrium magnetization in an external magnetic field \cite{zotev2010microtesla}.

\begin{figure}
\centering
	\includegraphics[width=\columnwidth]{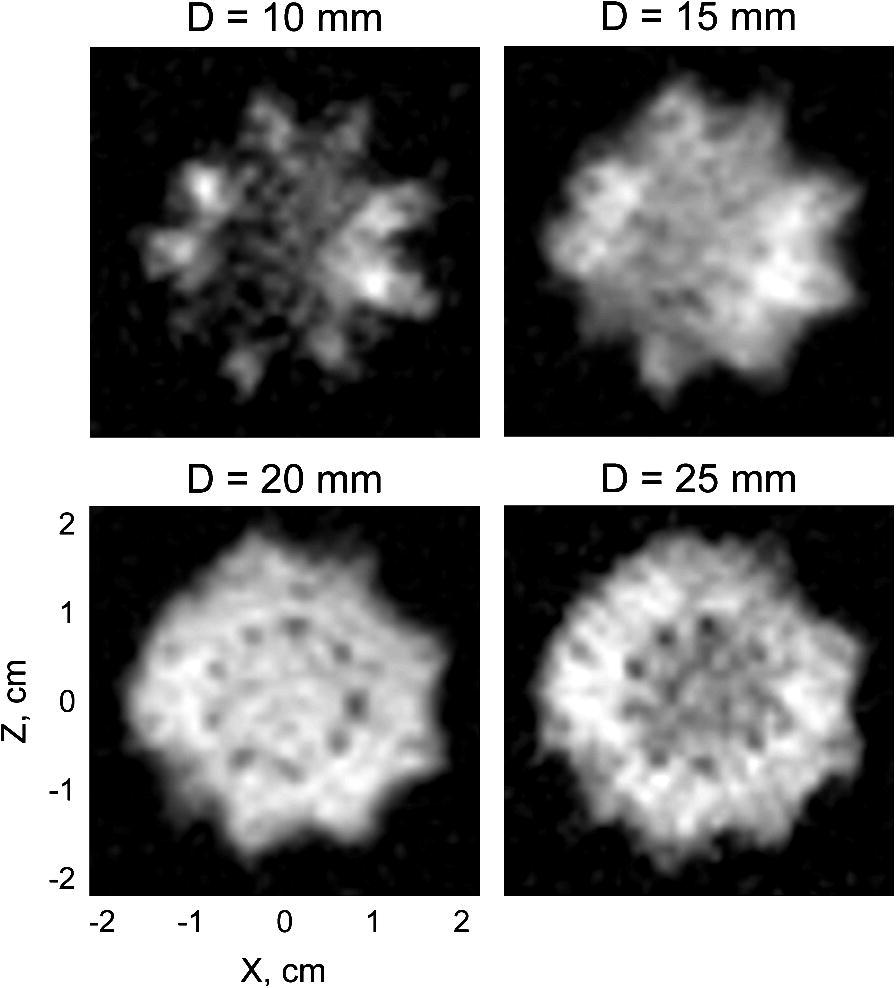}
	\caption{Overhauser-DNP-enhanced 3D image of a moon cactus, injected with TEMPO solution. The image was acquired at a field of 96\,$\mu$T in a single scan with 1-mm$\times$1-mm in-plane resolution. D is the depth of a given 5-mm-thick layer under the bottom of the cryostat. Reproduced from reference \cite{zotev2010microtesla}, Copyright (2010), with permission from Elsevier.}
	\label{fig:oDNP_ZULF_MRI}
\end{figure}

Overhauser DNP enhancement with direct detection below the Earth field was demonstrated in reference \cite{LEE2015}. As at fields below 1\,$\mu$T, electron and nuclear $^{14}$N/$^{15}$N-spins are strongly coupled in the nitroxide radical (the so called `hyperfine-field-dominant regime'), the saturated %\DB{Do we need to specify ``saturated'' here?} 
resonance frequencies are determined by both the electron--nucleus hyperfine coupling and the static magnetic field. By applying a 74.4 MHz radiofrequency field at the background fields where the hyperfine resonances occur (0.71, 2.04, 4.91, and 47.68\,$\mu$T) signal enhancements of protons in aqueous TEMPOL solutions on the order of a few hundreds were demonstrated. 
%\DB{It is unclear from what is written, why different values of static fields were used and whether the enhancement was observed at each of these values. Maybe explain what is the nature of the resonances at these fields... } \DAB{I think it is clear from what is written above: "the saturated resonance frequencies are determined by the hyperfine couplingand the static magnetic field".} 
The same authors also demonstrated \textit{in situ} generation and detection of a proton NMR signal (2 mM aqueous solution of TEMPOL) at 325\,nT \cite{HyunJoonLee2019}. An SNR of 32 at a linewidth of 0.7\,Hz was achieved after 16 averages, paving the way for novel applications of Overhauser DNP in ZULF NMR.

Not only spectroscopy is possible with Overhauser DNP at near-zero-field conditions. DNP-enhanced MRI at 96\,$\mu$T was demonstrated with a broadband SQUID sensors \cite{zotev2010microtesla}. Imaging of water phantoms and a cactus plant (Fig.\,\ref{fig:oDNP_ZULF_MRI}) was performed by using %\DB{irradiation Let us remove this word as there is no radiation :)} 
a hyperfine transition frequency of 120\,MHz. Enhancement factors as large as $-95$ for protons and $-200$ for $^{13}$C were demonstrated, which would require thermal polarization at 0.33\,T and 1.1\,T, respectively.

%\DAB{Mention chip-scale ODNP works! E.g., \url{https://assets.researchsquare.com/files/rs-124072/v1_stamped.pdf?c=1607708094}, \url{https://books.google.de/books?hl=en&lr=&id=HObGEAAAQBAJ&oi=fnd&pg=PA58&dq=chip+scale+overhauser+DNP&ots=-W3NU_sBE9&sig=-AnYc2DnieWAbMANzZBKQ3h61-I&redir_esc=y#v=onepage&q&f=false}. But this is not ZULF...}

\subsubsection{Dissolution Dynamic Nuclear Polarization}

%\DB{Danila agreed to to finalize 3.4.2. Dissolution Dynamic Nuclear Polarization, including the relevant material from the Roadmap}
An attractive hyperpolarization method is dissolution dynamic nuclear polarization ($d$DNP). In contrast to PHIP/SABRE or CIDNP (see \autoref{Subsec:CIDNP} below), this technique is general. This means that, in principle, any 
%small \DB{Why necessarily small? Does this need to be emphasized here?} 
molecule in solution can be hyperpolarized, as the polarization process does not rely on specific chemical reactions or interactions. To polarize a sample using $d$DNP, the molecule of interest is frozen in a glassy or polymeric matrix with a free radical dispersed homogeneously through the solid. The sample is cooled 
down to cryogenic temperatures and  
microwave irradiation is used to drive polarization transfer from the electrons of the free radicals to nuclear spins in the to-be-polarized molecules. There are several mechanisms by which this transfer occurs: solid effect, cross-effect, and thermal mixing \cite{thankamony2017dynamic}. All of these mechanisms rely on hyperfine interactions in combination with relaxation accompanying the microwave-driven electron spin flips.

In the \textit{solid effect}, transitions that are normally `forbidden' in the liquid-state (such as single- or double-quantum transitions) become `allowed' and, thus, can be pumped when anisotropic hyperfine interaction is present. By driving the forbidden transitions one can flip nuclear spins and hyperpolarize them \cite{HOVAV2010}. Importantly, electronic $T_1$ relaxation brings the system to a new equilibrium and the process can be repeated to polarize, via spin diffusion, other nuclei. As in the case of Overhauser DNP, the sign of the effect can be positive or negative, depending on whether a single- or double-quantum transition is saturated.

In the \textit{cross effect}, polarization transfer happens when a difference between electron Larmor frequencies matches the nuclear frequency: $\Delta \omega_e = \omega_n$. This is typically achieved when %concentration of free radicals in the matrix is increased \DB{Why does this lead to difference in electron Larmor frequencies?} \DAB{If, for example, they have different $g$-factors.} \DB{But what is the relation to concentration?} \DAB{Higher concentration result in statistically higher concentration of clusters containing pairs of radicals with the required properties on the difference between their g-factors - however this is my oversimplified explanation and there are more details, for example, quote from Ref. \cite{maly2008dynamic}: ``The CE is operative when 
the radicals have an inhomogeneously broadened EPR spectrum whose breadth 
is larger than the nuclear Larmor frequency \cite{hovav2012theoretical} 
%and, concurrently, the homogeneous linewidth  remains small⁠.'' We can remove that comment about concentration because it is hard to distinguish well between SE and CE in practice at levels of paramagnetic centers typically used (10-40 mM) yet I believe my uderstanding is correct \cite{hovav2012theoretical}...} 
or when special biradicals (molecules containing two chemical groups with unpaired electrons) are used in the process. Pairwise electron spin flips provide a field fluctuation at the resonance with nuclear spins. Once a nucleus is polarized, polarization can distribute to other nuclei via spin diffusion.

\textit{Thermal mixing} works when the concentration of free radicals is so high that an EPR line is homogeneously broadened (in other words, all spins `talk' to each other). It is hard to develop a mechanistic model for a situation like this, but thermodynamic arguments can be used to correctly predict and explain experimental observations in which nuclei are polarized upon 
%applying mw-saturation \DB{You mean ``microwave''? I would suggest: ... are polarized upon }
saturation of the microwave transitions at specific frequencies \cite{goldman197319f}. In practice, several mechanisms can contribute to the experimental observations and it is not always possible to identify a single mechanism responsible for polarization transfer from electrons to nuclei \cite{abragam1978principles,HOVAV2010,LILLYTHANKAMONY2017120}.

For the dissolution step, after sufficient nuclear spin polarization has been built up in the solid state (typically 20--60\% can be achieved on a timescale of tens of minutes), the sample is rapidly (within seconds) %\DB{(on the scale of nuclear spin relaxation)} 
dissolved in a hot solvent and flushed out of the polarizer as a room-temperature solution.

%\DB{\textbf{I would not go into great details here, but I think, some words along these lines above giving a flavor of what is happening are useful.}}

%%%%%%%%%%%%%%%%%%%%%%%%%%%%%%%%%%%%%%%
    \begin{figure*}[t]
% \begin{figure}
	\begin{center}	
\includegraphics[width=0.8\textwidth]
{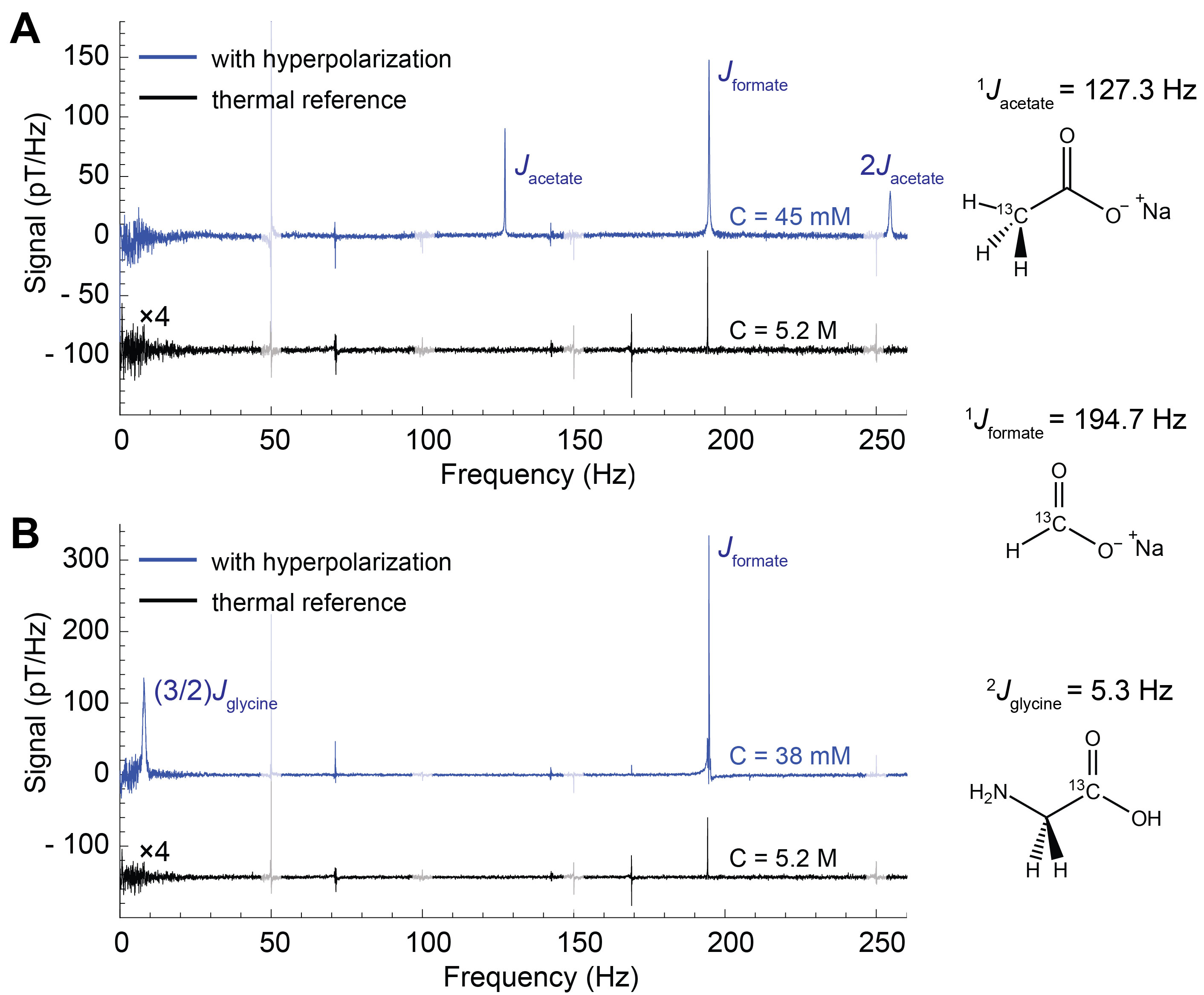}
	\end{center}
	\caption{ZULF-NMR spectra with $d$DNP hyperpolarization. (a) Single-scan ZULF-NMR spectrum of a hyperpolarized mixture of sodium [$^{13}$C]-formate and sodium [1-$^{13}$C]-acetate (blue trace) compared with a reference spectrum of 5.2\,M [$^{13}$C]-sodium formate thermally prepolarized at 2\,T after averaging 16 scans (black trace). (b) Single-scan ZULF-NMR spectrum of a hyperpolarized mixture of sodium [$^{13}$C]-formate and [1-$^{13}$C]-glycine (blue trace) compared with the same reference spectrum as in panel (a) (black trace). Note the scaling of the reference spectra. The noise peaks arising from the power line at 50\,Hz and overtones are grayed out, and probe-laser noise peaks are marked with asterisks; C refers to the concentration of [$^{13}$C]-formate. Reprinted (adapted) with permission from reference \cite{picazo2022dissolution}. Copyright 2022 American Chemical Society.%\MCDT{Isn't it more correct to say ``zero-field spectrum'', rather than ``ZULF spectra''? }
    }
	\label{fig:dDNP}
\end{figure*}

The $d$DNP method has been used to polarize substances such as the [1-${}^{13}$C]-pyruvic acid, [2-${}^{13}$C]-pyruvic acid, [${}^{13}$C]-formic acid, [2-${}^{13}$C]-acetic acid and [1-${}^{13}$C]-glycine molecules, or their anions, for detection at zero field using atomic magnetometers \cite{Barskiy2019,picazo2022dissolution,Mouloudakis2023JPCL}. The enhanced spin polarization made it possible to detect molecules in solution at micromolar concentration and natural \textsuperscript{13}C isotopic abundance.
As the $d$DNP method is general, it is possible to hyperpolarize multiple species at once and observe them simultaneously, as demonstrated in recent dedicated $d$DNP ZULF experiments \cite{picazo2022dissolution}.

Dissolution-DNP is a particularly promising method to pair with ZULF NMR since, as of 2024, several commercial polarizers exist. While $d$DNP usually involves a relatively
complex setup based on a high-field magnet and cryogenics, its combination with ZULF-NMR 
modalities offers opportunities in
a broad range of applications, 
including medical diagnostics, monitoring catalytic reactions within metal reactors, fragment-based drug screening, and others \cite{barskiy2023possible}, see Sec.\,\ref{Subsec:Biomed}. Another version of DNP, bullet-DNP, has been recently demonstrated, where a sample is quickly transferred outside of a DNP polarizer in its solid form prior to dissolution. The dissolution step takes place at room temperature  inside of the measurement apparatus, offering more experimental flexibility \cite{kouvril2019scalable}. %$\Delta J = \left( \gamma_{\rm {}^{14} N} / \gamma_{\rm {}^{15} N} \right) {}^1 J_{\rm {}^{15} N H} - {}^1 J_{\rm {}^{14} N H} = $ -58 mHz.

A drawback of the direct $d$DNP--ZULF combination is that the free radicals plus other additives (such as rare-earth ions) that are used in the polarization step remain in solution after the dissolution step, and these can induce paramagnetic relaxation of the nuclear spins at low fields \cite{Tayler20192}. However, a number of methods have been developed to circumvent this problem in the context of high-field NMR: (1) adding scavengers such as ascorbic acid to the solution, which quench the radicals \cite{mieville2010scavenging} or chelate paramagnetic ions; (2) phase extraction of the radicals into an organic solvent \cite{harris2011dissolution}; (3) the use of solid matrices that contain the free radicals, which are not dissolved into solution during the dissolution step \cite{vuichoud2016filterable}; and (4) the use of nonpersistent UV-generated radicals, which are rapidly quenched upon sample warming during the dissolution step \cite{eichhorn2013hyperpolarization}.

\subsection{Chemically induced nuclear polarization}
\label{Subsec:CIDNP}

%\MCDT{To me it seems this section is much longer than it needs to be. It starts off more like a review of CIDNP, rather than ZULF CIDNP.  It also has more than 40 references -- isn't that excessive?}
%\KS{Well, maybe the first three paragrpahs can be contracted, but the later text is about the zero-field CIDNP, it's mechanism and how it is different from the high-field regime. As for the number of references, I think it's Ok to have multiple references in a review. Zero-field and low-field CIDNP (detected indirectly) exist for several decades so there are naturally quite many papers came out, citing these works seems justified to me. Additionally, the third paragraph contains many references that are about some important developments made in the high-field CIDNP. Even if these are not about low-field, I think these are still good to have in terms of future applications: e.g. the idea that microfluidic is super well combined with photo-CIDNP and it should be implemented also in ZULF, powerful LED's - that's obvious, papers with the lists of biomolecules that can be hyperpolarized - I would not remove these either.}

In 1965, while investigating polymerization reactions by \textit{in situ} NMR, Joachim Bargon observed unusual high-field NMR spectra \cite{Bargon2010}. Whenever the polymerization of the maleic anhydride was initiated by using free radicals (i.e., molecules possessing unpaired electrons), intense absorption and emission lines appeared immediately upon the onset of reaction. If, instead, the polymerization was initiated ionically using pyridine, no such phenomena occurred. As the only seemingly related phenomenon known
at that time was DNP (see Sec.\,\ref{Subsec:DNP}) based on the Overhauser effect, the new phenomenon was named `chemically induced dynamic nuclear polarization' (CIDNP). Indeed, it seemed logical that DNP-like enhanced absorption and emission lines were `chemically induced' rather than induced by microwave irradiation of a system containing paramagnetic centers \cite{Bargon1967}.

However, it was quickly understood that the DNP-based theory could not explain all
experimentally observed features and only in 1969 did Kaptein and Oosterhoff \cite{Kaptein1969} and independently Closs find
the correct interpretation, namely, the `radical pair' (RP) theory of CIDNP. The RP-based explanation of CIDNP
took care of all previously unresolved problems; nonetheless, this currently accepted terminology of dynamic polarization remains in use today \cite{Closs1974}.

CIDNP is commonly detected at high fields in a wide class of reactions, for example, during thermal decomposition \cite{Kaptein1969} or in reversible reactions between photoexcited dyes and amino acids \cite{Salikhov1984}. This is one of the hyperpolarization methods that enables direct enhancement of NMR signals in macromolecules and their building units, including aromatic amino acids, proteins \cite{Hore1993}, nucleotides \cite{Kaptein1979}, and oligonucleotides \cite{Katahira1991}. Photo-CIDNP observation of NMR signals from nanomolar concentrations of proteins is one of the most sensitive experiments in bio-NMR \cite{Yang2022}. Moreover, photo-CIDNP can be used to screen chemical reactions on a microsecond timescale \cite{Closs1974,Morozova2019}, providing information on chemical rate constants \cite{Morozova2023} and  structures for transient radicals \cite{Morozova2018, Torres2021PCCP}. Recent developments in microfluidic probes \cite{Mompean2018, Gomez2023}, implementation of affordable systems with light-emitting diodes (LEDs) \cite{Yang2019}, discovery of new photosensitizers \cite{Okuno2016}, reductive radical quenchers for prolonging the stability of reversible photo-reactions \cite{Yang2021}, and the expanding application of the methodology for a growing list of molecules \cite{Torres2021MR, Torres2021PCCP, Torres2023JACS, Stadler2023} have further increased the interest in this technique. %However, it should be noted that photo-CIDNP has a complex dependence on the magnetic field in which the reaction is performed, and in many cases, an informative regime \DB{Hmm. I am not sure I understand the meaning of ``informative'' here} \DAB{probably informative means providing abundant information for extracting physicochemical parameters of the system such as electronic $g$-coupling tensors.} \DB{This seems plausible! Kirill, is this what you had in mind? Then we adopt this text.} \KS{Right, let's rewrite it, smth like 
However, it should be noted that photo-CIDNP
has a complex dependence on the magnetic field in which
the reaction is performed. Sign alternations and/or extremes in the magnetic field dependence that characterise the properties of the radicals such as exchange interaction, $g$-factors, and hyperfine couplings can be observed. For common organic radicals these features occur at fields of $<50$\,mT. In this section, we discuss the experiments developed for the observation of photo-CIDNP at (ultra)low magnetic fields. %\DAB{This paragraph sounds like an advertisement for CIDNP but probably could be significantly shrunk.} \DB{Yes, but why? The references seem useful...}

%The photo-CIDNP method is widely used to obtain information about the structure of macromolecules and elucidate dynamics of short-lived radical reaction intermediates \KS{\cite{Hore1993,Morozova2019}}. Some amino acids \KS{\cite{Torres2021MR,Torres2021PCCP}}, oligonucleotides \KS{\cite{Katahira1991}}, and macromolecules made up of those \KS{\cite{Yang2021}}, e.g. proteins and deoxyribonucleic acid (DNA) can be polarized in reactions with photoexcited flavins, aromatic ketones and compounds such as dipyridyl, all of which act as photosensitizers, i.e., molecules transferring energy from absorbed light to nearby molecules \AT{REF}.

%While it is necessary to use time-resolved techniques using special pulse sequences for the quantitative analysis of photoreactions [ref],

\begin{figure}
	\begin{center}\includegraphics[width=\columnwidth]{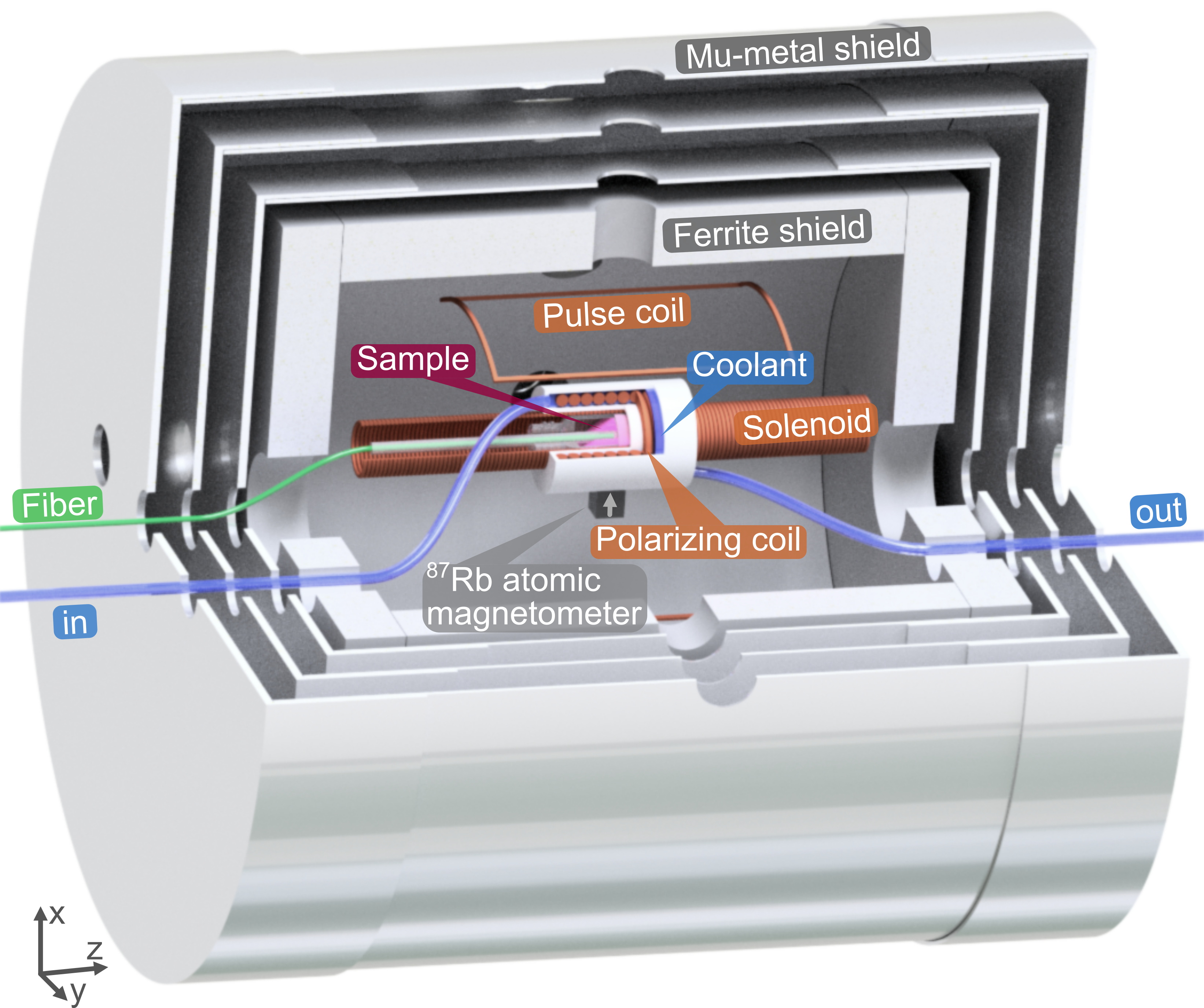}
	\end{center}
	\caption{Experimental setup for magnetometer-detected photo-CIDNP NMR measurements at nanotesla to microtesla fields. Reprinted (adapted) with permission from reference \cite{Chuchkova2023}. Copyright 2023 American Chemical Society.}
	\label{fig:CIDNPSetup}
\end{figure}

%\KS{Maybe find a new home for this paragraph?} It is possible to obtain a nuclear signal amplification of a few orders of magnitude in experiments with continuous irradiation, due to accumulation of nuclear polarization in  diamagnetic molecules \cite{Buchachenko1971}. The characteristic irradiation time required to achieve maximum amplification is of the order of the longitudinal nuclear spin relaxation.
Traditionally, low-field photo-CIDNP experiments are indirect, involving mechanical sample shuttling between two or more magnetic fields. Samples are illuminated in low magnetic field and then mechanically transported into a high-field NMR spectrometer for detection \cite{Hutchinson1975,Kaptein1972LowFields,Kaptein1979,Adrian1979,Stob1989,TARASOV1994,Bagryanskaya1997,Wegner2001,Ivanov2001APMR,Grosse2001JPCA,Miesel2004,PRAVDIVTSEV2015JMR,Zhukov2020JCP,Zhukov2021JCP,Zhukov2021MR}. Low-field photo-CIDNP profiles (i.e., dependences of the polarization on the magnetic field) obtained indirectly are useful for determining the exchange interaction in photoexcited electron donor--acceptor dyads, as these contain sharp features at magnetic fields corresponding to the level anticrossings due to exchange interaction \cite{Kanter1977,Stob1989,Wegner2001,Zhukov2020JCP,Zhukov2021JCP,Zhukov2021MR}. However, analysis of field dependence of photo-CIDNP measured indirectly is challenging for the cases of molecules containing multiple nuclear spins \cite{PRAVDIVTSEV2015JMR}. Recently, it was demonstrated that photo-CIDNP signals can be detected directly at nanotesla to microtesla fields using atomic magnetometers. The setup for this measurements is shown in \autoref{fig:CIDNPSetup}.   Fast field cycling was performed to observe photo-CIDNP occurring at fields up to 100\,mT \cite{Chuchkova2023}. It was shown that at ultralow fields, photo-CIDNP can be used to enhance the long-lived heteronuclear singlet order formed by directly bound $^{13}$C and $^1$H spins \cite{Sheberstov2021}, which can be readily detected with ZULF NMR. %Below we introduce the RP mechanism and how it is affected by the field for the case of small molecules in nonviscous liquid solutions.

%\DB{Kirill will revisit this paragraph; is this already changed?} \KS{I am on it writing a new paragraph instead of this one.} There are different mechanisms operating at different fields. At low field, where hyperfine couplings are much larger than Zeeman splittings, the photo-CIDNP mechanism is different from the one at intermediate fields, where these energy scales are comparable. In turn, these are different at high-field. %Also the dominant mechanism depends on the state of the matter. 
%They are also different for the solid-state (or large molecules with slow rotational diffusion) and liquid-state case, other mechanisms are responsible for the polarization in highly viscous media or in the case of linked radicals such as dyads/triads at low fields. All these different regimes are considered in great detail \DB{[refs]} and form a theoretical basis for spin chemistry. \DB{Kirill. this seems too general. What exactly do you want to say here?} \DAB{There is a field of spin chemistry providing a theoretical basis for understanding these phenomena.} %{Although CIDNP can also be observed in solids [ref],}

%\KS{Zero field regime in CIDNP is determined by the relative strength of the relevant interactions, e.g. the electron Zeeman interaction with the external field should be much weaker than the hyperfine interactions with magnetic nuclei present in the radicals. }

%As for detection of photo-CIDNP at ZULF, as of now, it has only been reported with nonviscous liquid solutions \DB{REFS ?}. Therefore, we focus on the theory for this case.

 \begin{figure*}[t]
\centering
	\includegraphics[width=0.9\textwidth]{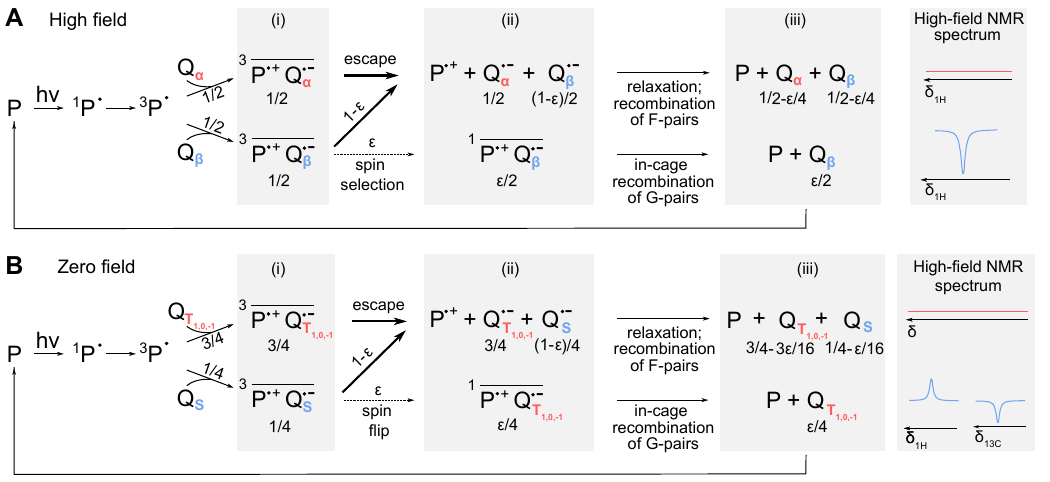}
	\vspace{-6pt}
	\caption{\textbf{A}: Photo-CIDNP in the high-field case: the geminal radical pair (G-pair) mechanism of photo-CIDNP. `Geminal' designates the primary P--Q pair produced upon photoexcitation of P followed by electron capture by Q. The photosensitizer (P) absorbs light and---after intersystem crossing occurring on a femtosecond timescale---becomes a relatively long-lived triplet radical (lifetime on the $\mu$s timescale or longer). (i) A quencher (Q) containing a nuclear spin (e.g., $^1$H), in either $\alpha$ or $\beta$ state with 50:50\% probability (neglecting weak thermal polarization), encounters P, producing an RP. (ii) In most of the cases, P and Q molecules separate, producing escape radicals, but sometimes the electron triplet state is converted into the singlet state, increasing the probability of in-cage recombination. Due to hyperfine interactions, the conversion preferentially  occurs for RPs with nuclei in one particular state (assumed $\beta$ here, as an example). (iii) Escape radicals exist for a relatively long time on the nuclear-spin relaxation time scale, leading to equilibration of nuclear-spin populations. By contrast, in-cage recombination occurs quickly, leading to accumulation of nuclear spins in the $\beta$ state. The recombination of escape radicals happens mostly in secondary radical pairs (F-pairs) that are formed upon encounter of two free P and Q radicals. The process is cyclic and, upon continuous irradiation, nuclear polarization builds up in the diamagnetic product, giving rise to the `emission-type' NMR spectrum. In steady state, F-pairs may further enhance photo-CIDNP. \textbf{B}: %\KS{Subplot in B is wrong. It should be any triplet state $T_{i}$ with 3/4 weight vs $S_{0}$ with 1/4. On top of it, the spin evolution in RP is more complicated but I have to check this in more details. Initial $S_{0}$ state in case of in-cage recombination must become $T_{0}$ or $T_{i}$ so that the total spin (electrons + nuclei) of the RP is conserved.} 
    Photo-CIDNP in the zero-field case. At least two different type of nuclei (e.g., $^1$H and $^{13}$C) having different hyperfine couplings to an electron should be present in Q to obtain hyperpolarization. Instead of $\alpha$ and $\beta$ states, heteronuclear singlet (S) and triplet (T$_i$, $i=0,\pm 1$) influence the RP dynamics. %\KS{Imbalance between singlet and averaged triplet populations is formed in this case.} 
    In both cases (\textbf{A} and \textbf{B}) high-field NMR detection is shown. In (\textbf{B}), an adiabatic transfer from zero to high field is assumed for the last step (iii).}
	\label{fig:CIDNP_RPM}
\end{figure*}

An important feature of the RP mechanism is that nuclear spin states of the radicals in a pair can affect electron-spin evolution and thus, the recombination rate in a given RP. Below we compare the simplest schemes for the RP mechanism operational at high field (\autoref{fig:CIDNP_RPM}A) and at zero field (\autoref{fig:CIDNP_RPM}B). Evolution of the RP in solution includes the following stages:
%The mechanism is illustrated in \autoref{fig:CIDNP_RPM}. 

a) \textit{Birth of the radical pair}. A radical pair in solution is a short-lived formation consisting of two radicals in proximity to one another, experiencing at least one re-encounter before they recombine or separate. This requirement is fulfilled in liquids due to the `cage effect': solvent molecules surrounding the two radicals trap them together for a certain time, limited by diffusion. During this time, the radicals experience multiple encounters, in the course of which nuclear-spin selection/flipping occurs. There are two types of radical pairs: those with a correlated spin state of the two electrons (geminal or G-pairs) and those with uncorrelated states (free pairs or F-pairs). The former are created in a single transfer of the electron from the excited photosensitizer to the substrate or (as in the case of thermal decomposition of the radical pair) obtained immediately after the breakage of chemical bond. After the formation of a G-pair, its spin state is the same as that of its precursor. In the case of thermal decomposition, the RP is in the singlet state, and in the case of a photochemical reaction, the RP can be in singlet (S) or triplet (T) state. F-pairs are obtained upon encounter of two `escape' radicals, for example those that were initially formed from G-pairs and then diffused apart in solution. F-pairs usually have more impact on steady-state CIDNP \cite{Sosnovsky2017}, which is achieved, for example, in the case of continuous light irradiation. This is the case for all ZULF-NMR experiments as of the time of writing. 

b) \textit{Spin dynamics in radical pairs.} Dynamics of singlet--triplet interconversion of the electron states in the RPs is of central importance for CIDNP. This dynamics depends strongly on the magnetic field, so that different scenarios may play out at different fields. Here we compare two cases: `spin-selection' occurs at high field and is responsible for the origin of the net polarization (\autoref{fig:CIDNP_RPM}A); `spin-flipping' occurs at low to ultralow and zero fields and is responsible for `multiplet' polarization of spin networks (\autoref{fig:CIDNP_RPM}B). These two mechanisms have different requirements on the spin system of the RP: spin-selection requires having at least one nuclear spin with nonzero hyperfine coupling in the RP; spin-flipping requires having at least two different nuclei with distinct hyperfine couplings in the RP. These requirements can be understood from symmetry considerations: whereas the high field sets the direction along which spin magnetization can build up, at zero field there is no preferred direction and the spin order produced by CIDNP must have spherical symmetry. This means that the terms in the density matrix describing hyperpolarization  must only be rank-zero spherical tensor operators, and those can be found only in systems of multiple spins, with the simplest example being the singlet order of a two-spin system. Note that we do not discuss multiplet CIDNP that also may occur under spin-selection at high fields,  neither do we consider net CIDNP occurring at low fields due to spin-flipping. There are dedicated reviews on these and other regimes of CIDNP \cite{Adrian1979,Kaptein1971,Kaptein1972,Yusuke2017,Morozova2019}.
 
 Let us now proceed with the analysis of the spin-selection mechanism. We assume here that RPs are initially excited in the triplet states (stage i in \autoref{fig:CIDNP_RPM}A) and that only one nuclear spin is present in the spin system of the RP. Once the two radicals separate in space, the exchange interaction between electrons essentially vanishes due to negligible overlap of the electron wavefunctions between the two radicals. They still constitute a radical pair `trapped' in the `cage', but now they are a bit further apart: a separation of 5 to 10\,$\AA$ is large enough to almost fully suppress short-range exchange interaction in common radicals \cite{Kaptein1972}. Therefore, upon abrupt separation, the correlated triplet state is no longer an eigenstate of the high-field Hamiltonian in the radical pair, $\hat{\mathcal{H}}_{\rm RP}^{\rm HF}$, and starts to evolve. The RP Hamiltonian can be written in the form \cite{Kaptein1972}
 \begin{eqnarray}\label{eq:RPHamHighField}
 \hat{\mathcal{H}}_{\rm RP}^{\rm HF} &=& \hat{\mathcal{H}}_{\rm RP}^{\rm HF,0} + \hat{\mathcal{H}}_{\rm RP}^{\rm HF,1}\,,\\
 \nonumber\hat{\mathcal{H}}_{\rm RP}^{\rm HF,0} &=& \frac{g_1 + g_2}{2} \mu_{\rm B} B_0 \left(\hat{S}_{1z}+\hat{S}_{2z}\right)\\
 \nonumber &&-2\pi\hbar J_{\rm ex}\left(\frac{1}{2} + 2\oper{S}_{1}\cdot\oper{S}_{2} \right)\\
 \nonumber &&+ \pi\hbar A_{1} \left(\hat{S}_{1z}\hat{I}_{z} + \hat{S}_{2z}\hat{I}_{z} \right)\,,\\
 \nonumber\hat{\mathcal{H}}_{\rm RP}^{\rm HF,1} &=& \frac{g_1 - g_2}{2} \mu_{\rm B} B_0 \left(\hat{S}_{1z}-\hat{S}_{2z}\right)\\
 \nonumber &&+ \pi\hbar A_{1}\left(\hat{S}_{1z}\hat{I}_{z} - \hat{S}_{2z}\hat{I}_{z} \right) \,.
 \end{eqnarray}
% \DB{What are the units we want the Hamiltonian to be in? Have a look at the mess with Eqs. (1) and (4). I am not sure whether $\hbar^{-1}$ should be in the formulas} \KS{Well spotted, it should be fine now.}
 Here $\mu_{\rm B}$ denotes the Bohr magneton, $g_{i}$ the $g$-factor of the electron $i$, $\hat{\boldsymbol{S}}_{i}$  the spin operator of the $i$-th electron, $\hat{I}_{z}$  the $z$-projection of the spin operator of the nucleus, $A_{1}$  the hyperfine coupling between the nuclei and the first electron, 
%\DAB{I would just use $A_1$ to denote HF coupling, no need to keep these large letters, $S$ operator for electrons, $I$ for nuclei, no need for e.} \DB{Kirill: could you kindly accept this suggestion or explain why we should not} \KS{Yes, I've implemented this "lighter" notation}, 
and $J_{\rm ex}$ residual exchange interaction (we keep it here to remain general, although often it can be neglected for nonviscous solutions \cite{IvanovPCCP2003}). Both the exchange interaction and the hyperfine coupling are expressed here in hertz. The $\hat{\mathcal{H}}_{\rm RP}^{\rm HF,0}$ term is diagonal in the basis of singlet--triplet electron functions, but the $\hat{\mathcal{H}}_{\rm RP}^{\rm HF,1}$ term mixes the triplet $T_{0}$ with the singlet $S_{0}$ electrons states. Therefore, the overpopulated  $T_{0}$ state starts evolving into $S_{0}$. The frequency of this evolution depends on the nuclear state ($\alpha$ or $\beta$) and is proportional to the corresponding matrix elements $\langle S_{0}^{el}\alpha^{nuc} |\hat{\mathcal{H}}_{\rm RP}^{\rm HF,1}|T_{0}^{el}\alpha^{nuc} \rangle$, and $\langle S_{0}^{el}\beta^{nuc} |\hat{\mathcal{H}}_{\rm RP}^{\rm HF,1}|T_{0}^{el}\beta^{nuc} \rangle$:
 \begin{eqnarray}\label{eq:RPHighFieldMatrixElements}
 \omega_{TS}^\alpha &=& \frac{\Delta g \mu_{\rm B} B_0}{2 \hbar} + \frac{ \pi A_{1}}{2}  \,,\\
\nonumber\omega_{TS}^\beta &=&\frac{\Delta g \mu_{\rm B} B_0}{2 \hbar} -  \frac{\pi A_{1}}{2} \,.
 \end{eqnarray}
% \begin{eqnarray}\label{eq:RPHighFieldMatrixElements}
% \omega_{TS}^\alpha &\propto & \left(\Delta g \mu_B B_0 + (1/2)\hbar A_{HF1} \right)/2  \,,\\
%\omega_{TS}^\beta &\propto&\left(\Delta g \mu_B \hbar^{-1}B_0 - (1/2) A_{HF1} \right)/2  \,.
% \end{eqnarray}
 % \begin{eqnarray}\label{eq:RPHighFieldMatrixElements}
% \nonumber \omega_{ST}^\alpha \propto \langle S_{0}\alpha |%\hat{\mathcal{H}}_{RP}^{HF,1}|T_{0}\alpha \rangle &= &\\
% \nonumber&& \left(\Delta g \mu_B \hbar^{-1}B_0 + (1/2)A_{HF1} \right)/2  \,,\\
 %\omega_{ST}^\beta \propto\langle S_{0}\beta |\hat{\mathcal{H}}_{RP}^{HF,1}|T_{0}\beta \rangle &=& \\\nonumber&&   \nonumber\left(\Delta g \mu_B \hbar^{-1}B_0 - (1/2) A_{HF1} \right)/2  \,.\\
 %\,
 %\end{eqnarray}
 %}
% \DB{I rather dislike how Eqs. (61-63) are formatted: nested parentheses, inline fractions, ... Could you kindly format these nicely?} \KS{Done.} 
Assuming $\Delta g = \left(g_1 - g_2\right)>0$ and $A_{1} > 0$, it can be seen that $T_{0}$ converts into $S_{0}$ faster with nuclear spin up than with the nuclear spin down by a value dependent on the hyperfine splitting. %\DB{I have an impression that this is written for the high-field case. Is this correct?} \KS{Yes, we compare high-field case to zero-field case that goes right after.}
%\KS{Changing the sign of either the difference of the g-factors or of the hyperfine coupling alternates increase/decrease of the frequency of the triplet-to-singlet evolution for the spin-up/spin-down nuclei states, which ultimately determines the sign of CIDNP polarization. In the general case, the sign of CIDNP is determined by the set of Kaptein's rules \cite{Kaptein1971}.} 
 Triplet-to-singlet conversion is never complete because typical lifetimes of the RPs are too short; however, the lifetimes can be long enough that the probability of an RP being in the singlet (triplet) state upon the re-encounter of the radicals is correlated with the nuclear spin state being up (down) \cite{Kaptein1972,Adrian1979}.
 % average amount of RP having higher probability to be in the singlet state upon the reencounter of the radicals is correlated with the nuclear spin state. Spin-up states on the other hand are somewhat have higher probability to be found in the triplet RP's as the time passes on. 
 This is called spin sorting.

Now let us consider the spin-flipping mechanism responsible for multiplet CIDNP occurring at zero field. Zero-field condition for the RP evolution means that the energies of the hyperfine couplings greatly exceed the energy associated with the difference between $g$-factors in the RP;
%DB{Kirill: we are stepping on the same rake the n-th time here. It is not the difference in the g-factors but rather the difference in the Zeeman energies!} \KS{Agree, so it will be "the differences in the Zeeman energies".} 
for typical organic radicals, this occurs at fields below 100\,$\mu$T. Again, we consider the case when the RPs are excited in the triplet state of the electrons (stage (i) in \autoref{fig:CIDNP_RPM}B). The quencher molecule contains two distinct $J$-coupled nuclei (denoted as 1 and 2). The Hamiltonian in this case can be written as \cite{Kaptein1972LowFields}: 
%\MCDT{Why are the vector operators italicized, not roman like in the theory section? See for example \autoref{eq:KAKB}}
 \begin{eqnarray}\label{eq:RPHamZeroField}
 \hat{\mathcal{H}}_{\rm RP}^{\rm ZF} &=& \hat{\mathcal{H}}_{\rm RP}^{\rm ZF,0} + \hat{\mathcal{H}}_{\rm RP}^{\rm ZF,1}\,,\\
 \nonumber\hat{\mathcal{H}}_{\rm RP}^{\rm ZF,0} &=& -2\pi\hbar J_{\rm ex}\left(\frac{1}{2}+ 2\oper{S}_{1}\cdot\oper{S}_{2} \right)\\
 \nonumber &&+ \pi\hbar\left(\oper{S}_{1} + \oper{S}_{2} \right) \left( A_{1}\oper{I}_{1} + A_{2}\oper{I}_{2} \right)\,,\\
 \nonumber\hat{\mathcal{H}}_{\rm RP}^{\rm ZF,1} &=& \pi\hbar\left(\oper{S}_{1} - \oper{S}_{2} \right) \left( A_{1}\oper{I}_{1} + A_{2}\oper{I}_{2} \right) \,.
 \end{eqnarray}
The nonzero hyperfine couplings $A_{1}$ and $A_{2}$ are between electron 1 and the two nuclei. The $\hat{\mathcal{H}}_{\rm RP}^{\rm ZF,1}$ term mixes the singlet with the three triplet states of the electrons, whereas $\hat{\mathcal{H}}_{RP}^{ZF,0}$ does not. Therefore, the presence of the $\hat{\mathcal{H}}_{\rm RP}^{\rm ZF,1}$ term drives the conversion of the overpopulated electron-triplet states into the singlet state, and the speed of this conversion depends on the nuclear spin state. Kaptein and Hollander showed that mixing the nuclear spin states does not affect spin dynamics in the RP \cite{Kaptein1972LowFields}, but the total spin of the nonequivalent nuclei is what matters. The frequency of the conversion is different for the nuclei in the singlet state and triplet states:
 \begin{eqnarray}\label{eq:RPZeroFieldMatrixElements}
 \omega_{TS}^{S} &\propto& \frac{\pi|A_{1}-A_{2}|}{2}\,,\\
 \nonumber\omega_{TS}^{T} &\propto& \frac{\pi|A_{1}+A_{2}|}{2} \,.
 \end{eqnarray} 
Thus in the case of hyperfine couplings of opposite sign, the electron conversion into the singlet state is faster for the RPs having nuclei in the singlet state, as shown in \autoref{fig:CIDNP_RPM}B. By considering the mixing of different states induced by $\hat{\mathcal{H}}_{\rm RP}^{\rm ZF,1}$, it can also be shown that the states of the nuclei may change depending on the change of state of the electron. For example, the RP in the $T_{+1}^{el}S_{0}^{nuc}$ state evolves into $S_{0}^{el}T_{+1}^{nuc}$. Generally speaking, this occurs because the Hamiltonian of the RP [\autoref{eq:RPHamZeroField}] commutes with the operator of the total spin of the nuclei and electrons but not with the operators representing the total spin of either the electron or nuclei. Therefore the total spin of the RP is conserved, but not the total spin of the nuclei or electron. This peculiarity is the reason behind the name of the mechanism (spin-flipping).

c) \textit{Recombination.} Nuclear hyperpolarization obtained upon recombination of the geminal RPs can be understood as follows. According to the model, an RP recombines only in the singlet state forming `in-cage' diamagnetic products, P and Q, that regenerate to the initial form but may acquire nuclear polarization \cite{Kaptein1972}. Only a small fraction, $\varepsilon$, of geminal RPs---not even all singlet RPs---recombine in the cage as shown in  \autoref{fig:CIDNP_RPM} (extent of reaction between stages i and ii). The remaining RPs separate, forming escape radicals. In high field, spin selection applies, see \autoref{eq:RPHighFieldMatrixElements}, and those nuclear states that convert the electron triplet state into the singlet state faster gradually become overpopulated in the in-cage products. In the example considered here $\omega_{TS}^{\beta} > \omega_{TS}^{\alpha}$, so that on average after evolving under $\hat{\mathcal{H}}_{\rm RP}^{\rm HF}$, the RP has a higher probability to be found in the singlet state for the RPs with the nuclei being in $\beta$ state and therefore the $\beta$ nuclear state is said to be selected in the in-cage products. This is illustrated in \autoref{fig:CIDNP_RPM}A, stage iii. Escape radicals, on the other hand, are enriched immediately after separation with the nuclear $\alpha$ states. However, the lifetime of the escape radicals is much longer than the lifetime of RPs, and over this time efficient paramagnetic relaxation occurs so that on average nuclear spins in the escape products quickly relax to the initial 50:50\% distribution of $\alpha$ and $\beta$ nuclear states. Here we ignore the relaxation dynamics \cite{Morozova2019} and for clarity consider that these states do not contribute to the NMR signal. Diamagnetic products in the considered example are therefore enriched with the $\beta$ nuclear states, giving rise to the emission type of NMR spectrum (\autoref{fig:CIDNP_RPM}A right-hand side).

Geminal hyperpolarization at zero-field occurs similarly, but in this case, nuclear spins are flipped. The example shown in \autoref{fig:CIDNP_RPM}B assumes that having singlet nuclear states in the RP stimulates the conversion of the electronic triplet into the singlet state ($\omega_{TS}^{S} > \omega_{TS}^{T}$), and a fraction $\varepsilon$ of these RPs recombine. Because of the nuclear spin flip, in-cage recombination products become enriched with nuclei in the triplet states. Considering ZULF studies, the obvious example of a spin pair would be a directly bound $^{13}$C and $^1$H spin pair; in this case, conventional NMR after an adiabatic increase of the field should reveal enhanced $^1$H and $^{13}$C NMR signals, one being of the absorption type, the other of the emission type (or vice versa, depending on the sign of the $J$-coupling) as shown on the right-hand side of \autoref{fig:CIDNP_RPM}B. Indeed, such spectra of hyperpolarized $^1$H and $^{13}$C were observed experimentally for photo-CIDNP of para-quinone \cite{Sheberstov2021}. Alternatively, one can perform direct detection of photo-CIDNP ZULF $J$-spectra. At the time of writing, direct observation of photo-CIDNP spectra of $^1$H magnetization precessing in nT fields has been demonstrated \cite{Chuchkova2023}, but not yet photo-CIDNP ZULF $J$-spectra.

\subsection{Statistical polarization}
\label{Subsec:Stat-pol}

If an ensemble of $N$ independent spin-{\textonehalf} nuclei is prepared in a random fashion with each nucleus in either spin-up or spin-down state with respect to an arbitrarily chosen quantization axis, then an excess of either spin-up or spin-down nuclei on the order of $\sqrt{N}$ is expected in each realization of the ensemble. This kind of random polarization enables magnetic resonance noise spectroscopy and even imaging \cite{Mueller2013SpinNoise,Mueller2006}, which becomes particularly advantageous for small-$N$ samples. This polarization approach that does not require any action on the part of experimentalists is especially effective in the case of single-spin NMR (see, for instance, \cite{Budker2019JMR} and references therein), where the signal size in the case of stochastic polarization is no smaller that that of a fully polarized sample. ZULF NMR with stochastic polarization is being pursued using detection with single-spin sensors based on color centers in diamond \cite{Lenz2021_Single}, see Sec.\,\ref{Subsec:NVdiamond}.

\subsection{Spin polarization induced by rotating magnetic fields}
\label{Subsec:Rotat-pol}
An intriguing and somewhat counter-intuitive way to generate static thermal spin polarization without a static magnetic field was first demonstrated in 1957 by Whitfield and Redfield \cite{whitfield1957rotating}, who applied a circularly polarized rf field to a paramagnetic solid and showed that a stationary magnetization component perpendicular to the plane of the rf field emerges. The effect can be understood as relaxation toward the instantaneous direction of the rf field and subsequent precession of the magnetization about the rotating rf field. The result is a nonzero magnetization component perpendicular to the plane in which the rf field rotates.

Half a century later, an analogous experiment was performed for nuclear spins in a liquid \cite{lee2006rotating} using audio-frequency rotating fields (9.6\,kHz). By adiabatically reducing the strength of the rotating field, the steady-state polarization was transformed into laboratory-frame magnetization and detected with a SQUID. In contrast to prepolarization with static magnetic fields (see \autoref{Subsec:Thermal_pol}), the use of rotating fields---which have zero time average---avoids magnetising materials in the vicinity of the sample. This might be of practical relevance in a number of applications \cite{lee2006rotating}.

\section{Spin evolution in ZULF NMR}
\label{Sec:SpinEvolution}
Polarization of nuclear spins is generally followed by further application of magnetic fields to redistribute the spin order from one nucleus to another or to create detectable coherences.  Below we discuss the techniques most widely used in ZULF NMR.

\subsection{Initiating free evolution}

Radiofrequency (rf) pulses are closely associated with the modern NMR performed in persistent high-field magnets, where small (typically on the order of ppm of the leading field) alternating magnetic fields are applied to the sample at an eigenfrequency of the observable magnetization. Countless textbooks and reviews explaining aspects of these pulses---from the most basic to most elaborate---often do not explicitly state that this form of external perturbation is the only option, since as the name `persistent’ suggests, the strength and orientation of the magnet are hard or impossible to change. 
In contrast, there is generally the opportunity to switch the magnetic field in low-field and ZULF NMR.  One example is the experiment of Varian and Packard \cite{Packard1954}, described in \autoref{Subsec:History}, in which the magnetic fields for spin polarization and precession were applied along separate (ideally, orthogonal) axes.  In that case, the latter field is provided by the Earth and therefore the `pulse sequence' involves merely switching off the polarizing field.  The procedure may be emulated in ultralow field by placing two orthogonal coils around the sample inside a magnetic shield, so as to polarize the spins along the first axis and then observe Larmor precession at much lower frequencies about the second \cite{Tayler20192}.  McDermott \textit{et al.}\ implemented a spin echo this way by nonadiabatically inverting the bias field, rather than the magnetization, during free precession \cite{McDermott2002}.  Field reversal of this type is in analogy to the gradient echoes used in MRI, where the polarity of the encoding field is changed.  In place of the polarizing coil, one can also use a piercing solenoid passing through the shield walls to introduce the prepolarized sample from the laboratory outside \cite{yashchuk2004hyperpolarized,Savukov2005PRL}.

There is a further minimalist approach to omit the second coil, so that the total field is switched to zero after prepolarization.  Fast quenching of the magnetic field amplitude with respect to the Larmor frequency, while keeping the same axis, can lead to coherent oscillations of the sample magnetization in heteronuclear compounds under dipolar and/or \textit{J}-coupling interactions (see \autoref{Subsec:VectorModel}). Experimental examples that fit within the modern definition of `zero to ultralow field' appear in the literature as early as 1961, where sub-Earth-field NMR studies of HPO(OH)$_2$ and other heteronuclear compounds are reported \cite{Brown1961,Thompson1962}.  More recent examples use much the same approaches, combined with coil-based signal detection at high field \cite{ZAX1984550,Zhukov2020JACS} or magnetometer-based detection in the ZULF regime \cite{McDermott2004,ledbetter2009optical}.  % Danila suggested to cite \cite{Anderson1956}, but I don't see what this has to do with ZULF, anyway I will re-read the paper and reconsider.
In order to produce a measurable signal, the spins must have unequal magnetic moments and initial polarization, as for instance occurs starting from Boltzmann equilibrium of a heteronuclear spin pair I--S.  Here polarization is proportional to $\gamma_I I_z + \gamma_z S_z $, so that when switching to zero field, the component $(I_z – S_z)$ does not commute with the $J$-coupling or dipolar Hamiltonians and leads to a time-dependent overall magnetization (see \autoref{Subsec:GenericExperiment}).  Pulses or more advanced prepolarization techniques can increase the initial amplitude of the component $(I_z – S_z)$ and thus improve overall SNR.

\subsection{Coherent control}
\label{CoherentControl}

\subsubsection{Pulsed-field excitation}

Most of the ZULF NMR spectroscopy performed to date involves spin excitation via direct current (dc) pulses.  As in the experiments described in the previous section, excitation relies upon nonadiabatic switching of the magnetic field axis to induce Larmor precession---and/or $J$-coupling evolution---of the spins, so the flip angle of the pulse is given by $\theta = \gamma_i B_{\rm pulse} \tau$ for pulse length $\tau$, dc field amplitude $B_{\rm pulse}$ and spin species $i$. 

A typical ZULF NMR instrument employs three orthogonal coils such as a triple set of Helmholtz coils \cite{Tayler2017}. These allow one to change the direction of the pulsed field, for coherence-order-based signal-selection approaches akin to the `phase cycling' techniques used in conventional NMR. These and other error-correction techniques can be useful in two-dimensional and multidimensional implementations of ZULF NMR spectroscopy \cite{THAYER1986,Sjolander2020}.  Because more than one coil can be energized at a time, such that the pulsed field is applied along any arbitrary direction in space, original pulses can be developed for which there is no direct analog at high field.  The option of arbitrary pulse axis offers other advantages, for example direct point-to-point rotations in the Bloch sphere, which may be important where speed matters in a ZULF-NMR pulse sequence.

\begin{figure}
    \centering
    \includegraphics[width = \columnwidth]{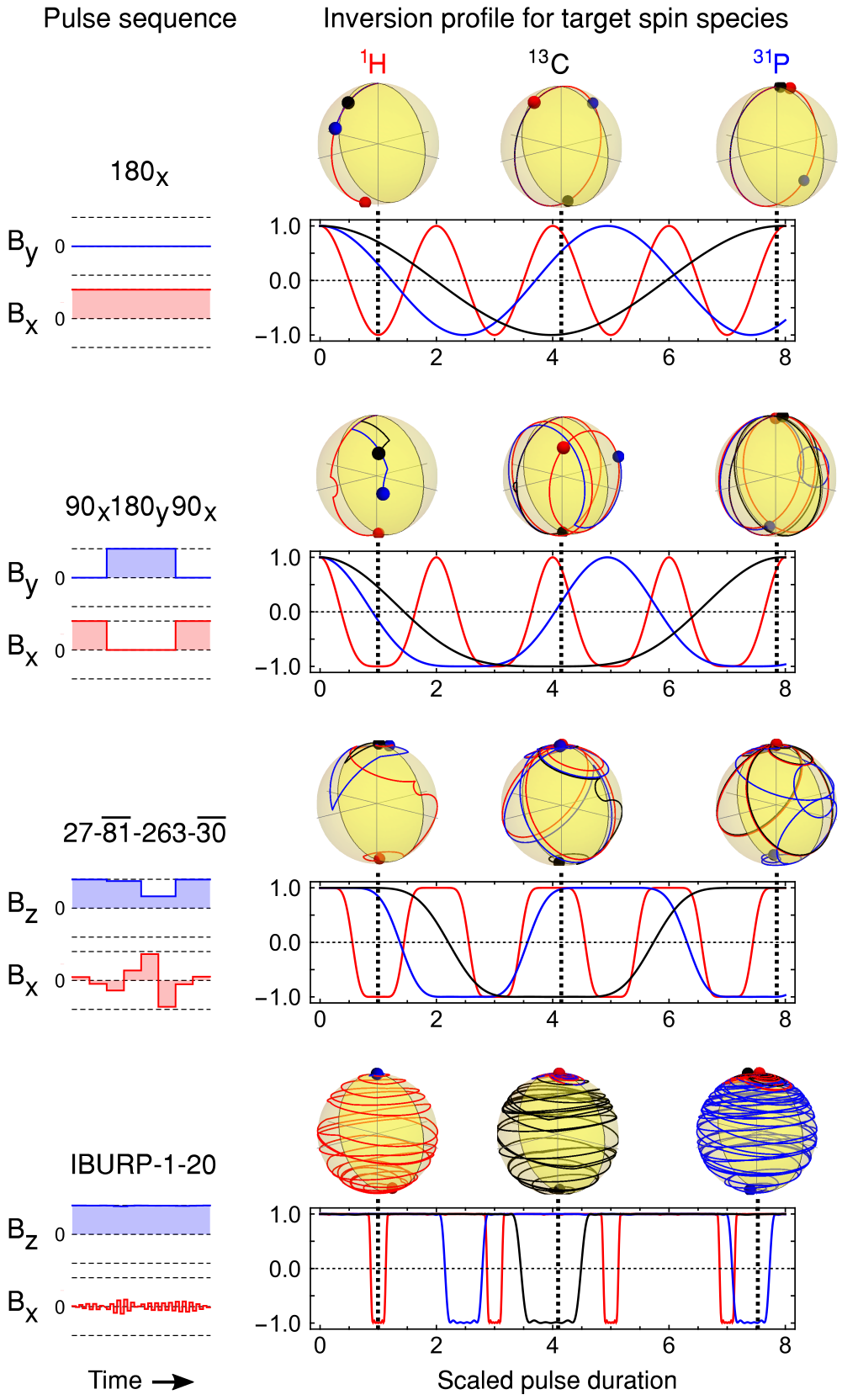}
    \caption{Sequences of constant-field pulses and their spin-species selectivity for control of magnetization along the $z$-axis (red: \textsuperscript{1}H, black: \textsuperscript{13}C, blue: \textsuperscript{31}P).  Large yellow spheres represent the 3D Bloch-vector trajectory during the pulse, starting from the North pole where the magnetization vector lies along the $z$-axis, $(0,0,1)$.  Colored spheres indicate the final position of the magnetization vectors at the end of each pulse. The plots below the large spheres indicate the $z$-projections of the polarization of the three nuclear species. The pulse durations are scaled to $\pi$ rotation on $^1$H.}%\KS{COnsider changing colors for color-blind readers.} %\MCDT{I tested the figure using \url{https://www.color-blindness.com/coblis-color-blindness-simulator/} and didn't see any color-blindness-related issues with the color scheme.  So either the test is wrong, or the red/blue/black combination is fine. }}
    \label{fig:dccompositepulses}
\end{figure}

A disadvantage of single dc pulses is their limited spin selectivity.  Under a field of constant amplitude and direction, spins rotate through an angle proportional to their gyromagnetic ratio, therefore the ratio of flip angles $\theta_I/\theta_S$ between any two spin species $I$ and $S$ is a constant ratio, $\gamma_I/\gamma_S$, and a noninteger. For instance, in a system of $^{1}$H and $^{13}$C, the ratio is $\gamma_{1\rm H} / \gamma_{13\rm C} = 4.02$.  Even rounding to the nearest integer, $4$, this is an inconvenient ratio because while it is possible to rotate either spin by a net-odd multiple of $\pi$ radians, it is impossible to rotate both by a net-odd multiple of $\pi$ at the same time---for a graphical representation, see the top row of \autoref{fig:dccompositepulses}. For similar reasons it can also be difficult to apply selective pulses to spin species with close gyromagnetic ratios, such as $^{19}$F and $^{1}$H ($\gamma_{19\rm F} / \gamma_{1\rm H} \approx 0.94$), and the problem is worsened by having three or more spin species because conditions for spin-selective rotation occur much less frequently.  As an example, \autoref{fig:dccompositepulses} shows that there is no common intersection of inversion profiles under a single dc pulse for nuclei \textsuperscript{1}H, \textsuperscript{13}C and \textsuperscript{31}P, where $\gamma_{31\rm P} / \gamma_{1\rm H} = 2.47$. 

An approach towards improving the selectivity of dc pulses is to chain several rotations together, thus creating a composite dc pulse \cite{Bodenstedt2022jpcl,Thayer1986CompositePI}.  A successful composite may be regarded as one that is sufficiently tolerant of flip-angle variations to absorb an `inconvenient’ residual fraction of the gyromagnetic ratio into the error-compensation bandwidth of the pulse \cite{Bodenstedt2022jpcl}.  Candidate pulses are dc analogs of high-field NMR composite pulses, where flip angles in the rf rotating frame map to the lengths of the dc pulses, and the rf phases correspond to the laboratory-frame axes of the pulsed fields.  One example is $\pi/2_X - \pi_Y - \pi/2_X$, which in high field is implemented with a single coil and three pulses with relative rf phases of $0$, $\pi/2$ and $0$.  In ZULF NMR (see second row of \autoref{fig:dccompositepulses}), the composite pulse could be applied as square pulses through separate coils oriented along $x$-, $y$- and $x$-axes, respectively \cite{Thayer1986CompositePI}. An alternative, flexible approach is provided by dc-composite-pulse analogs of high-field finite-band pulses, such as Broadband Uniform-Rotation Pure-phase pulses (BURP pulses \cite{Geen1991jmr,Freeman1998pnmrs,Bodenstedt2022Meridional}).  These pulses aim to produce finite-band excitation over a relatively narrow but tunable range of gyromagnetic ratios $\gamma$, whose width depends on the number of simple dc pulses in the composite pulses \cite{Bodenstedt2022Meridional}.  The bottom two rows of \autoref{fig:dccompositepulses} illustrate how such pulses can produce spin-species-selective inversion in a system containing \textsuperscript{1}H, \textsuperscript{13}C and \textsuperscript{31}P, which have relative gyromagnetic ratios 1\,:\,0.252\,:\,0.405.  In general, these dc composite pulses can produce spin-selective rotations of arbitrary flip angle, including the special angles of $\pi/2$ and $\pi$.  If for some reason the composite-pulse approach is inconvenient, one can alternatively temporarily apply a higher magnetic field and use conventional pulses to achieve spin selectivity, for example, audiofrequency ac pulses \cite{Tayler2016}.

\subsubsection{Selective excitation} % using low-amplitude ac fields
\label{Subsubsec:Selective}
 
Oscillating fields of nanotesla amplitude can be used to excite a small subset of transitions or even single transitions in the ZULF spectrum of a compound \cite{Kreis1988}.  Such pulses that achieve selective population inversion across a transition $|A\rangle\leftrightarrow|B\rangle$ can be used to assign the characteristic $F$ and $K$ quantum numbers (or other conserved quantum numbers, see \autoref{Sec:XAn}) for any connected transition in the ZULF-NMR $J$-coupling spectrum \cite{Sjolander2016}.  For example, in an XA$_n$ system at zero field, selection rules are $\Delta K_A = 0$ and $\Delta F_XA=0,\,\pm 1$.  Selective population inversion across eigenstates $|A\rangle$ and $|B\rangle$ prior to measuring the zero-field spectrum leads to perturbed amplitudes of all allowed transitions to connected states $|C\rangle$, namely $|A\rangle\leftrightarrow|C\rangle$ and $|B\rangle\leftrightarrow|C\rangle$.  From these changes in amplitude, one can deduce that $|A\rangle$, $|B\rangle$ and $|C\rangle$ belong to a common manifold of $K_A$. This approach also applies where energy states are split by dipolar couplings \cite{AgrawalThesis}. 
%\AT{Should be include also a citation to \cite{AgrawalThesis} (see file AgrawalThesis2004.pdf in this folder)?} 
%\MCDT{In response to Andreas, one could write: ``This approach also applies where energy states are split by dipolar couplings \cite{AgrawalThesis}.''  The thesis is cited because there was no journal publication, right?} \AT{Perfect, thank you. And yes, the work was never formally published, but the author has recently made some noise about her work not being acknowledged. Including a citation here might be a good way to assure her that this is not the case.}

Another type of selective pulse can be used when the ZULF spectrum contains multiple transitions: a pulse of a rotating (i.e. circularly polarized) magnetic field.  Fields rotating in the positive or negative sense in the $x$-$y$ plane respectively excite only positive or negative $\Delta m_F$ transitions \cite{Sjolander2016}.  Such fields can be introduced by applying a sine current waveform to two orthogonal coils, with relative phase $\pi/2$.  Rotating fields can also be used to saturate the hyperfine transitions of coupled electron--nucleus systems, leading---in analogy to optical pumping---to a mechanism of dynamic nuclear polarization via the Overhauser effect \cite{HILSCHENZ2019138}.

\subsubsection{Scaling and suppression of spin interactions}
\label{Subsubsec:Decoupling}

The majority of the early ZULF-NMR experiments in the 1980s and 1990s were focused on obtaining high-resolution NMR spectra of powdered solid samples, by measuring the `untruncated' and rotation-invariant dipole--dipole spin couplings (see \autoref{Subsec:History}).  To observe only a specific subset of spin couplings while ensuring an overall invariance to sample rotation, coherent-averaging techniques were introduced and developed.  In most of these, delta-function pulses were applied between periods ($\tau$) of free evolution in ZULF.  One of these sequences, XY4 ($\tau/2-\pi_x-\tau-\pi_y-\tau-\pi_x-\tau-\pi_y-\tau/2$) was particularly effective at suppressing the first-order average-Hamiltonian contributions to linear operator terms of the spin Hamiltonian, such as the magnetic field, while preserving bilinear operator terms such as those involved in homonuclear dipole--dipole couplings \cite{Lee1987TheoryOM}.  The XY4 sequence therefore provided a zero-field analog of a spin echo and Carr--Purcell--Meiboom--Gill (CPMG) decoupling, where it allowed nominally zero-field spectra to be acquired even in the presence of a small dc field $|B|\ll 1/|\gamma\tau|$.  One application of coherent averaging in ZULF NMR was therefore to lower the experimental demands on field shimming, by coherently decoupling the spins from residual background magnetic fields.  Related sequences for averaging higher-order terms of the Hamiltonian were found via recursive expansion of the XY4, for example the WHH  or WAHUHA sequence \cite{whh1968}.

Around the same time, a few groups showed that isotropic averaging could be performed entirely in high magnetic field by combining rapid magic-angle spinning of the sample with synchronous rf pulses, to select terms of the Hamiltonian of a given spherical tensor rank \cite{Tycko1990JCP9210,TYCKO1990205,CHANDRAKUMAR1992372}.  These `zero-field in high-field' experiments have the advantage that conventional spin-selective (between isotopes, e.g.,\ $^2$H and $^1$H) rf pulses can be employed, and they do not require field cycling.  They are limited, however, by a need to use spinning speeds on the order of several tens of kHz.  At the time these were considered somewhat high; today, sample speeds above 100\,kHz are considered to be routine for commercial magic-angle-spinning NMR probes.

\begin{figure}
    \centering
    \includegraphics[width=0.7\columnwidth]{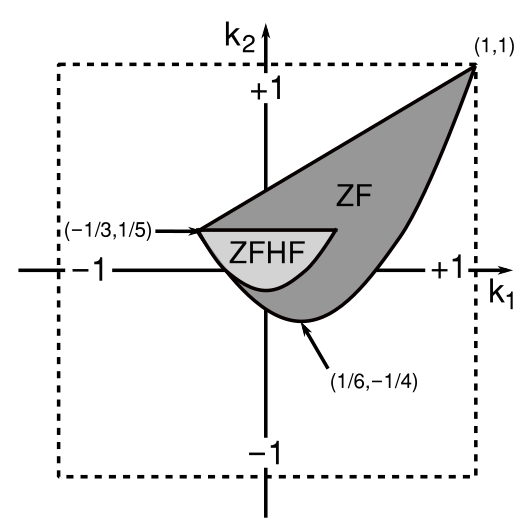}
    \caption{Allowed isotropic scaling factors $k_1$ and $k_2$ in the first-order average Hamiltonian for spin spherical rank-1 and rank-2 interactions, respectively, under true-zero-field (ZF, dark gray) and zero-field-in-high-field (ZFHF, light gray) conditions. The curved boundary corresponds to the most general of the regular-polyhedron sequences. XY4 and bilinear rotation decoupling methods developed in \cite{Lee1987TheoryOM} also fall within the ZF area. After reference \cite{Llor1995a}.}
    \label{fig:Llor2dplot}
    \vspace{0.5cm}
    \includegraphics[width=\columnwidth]{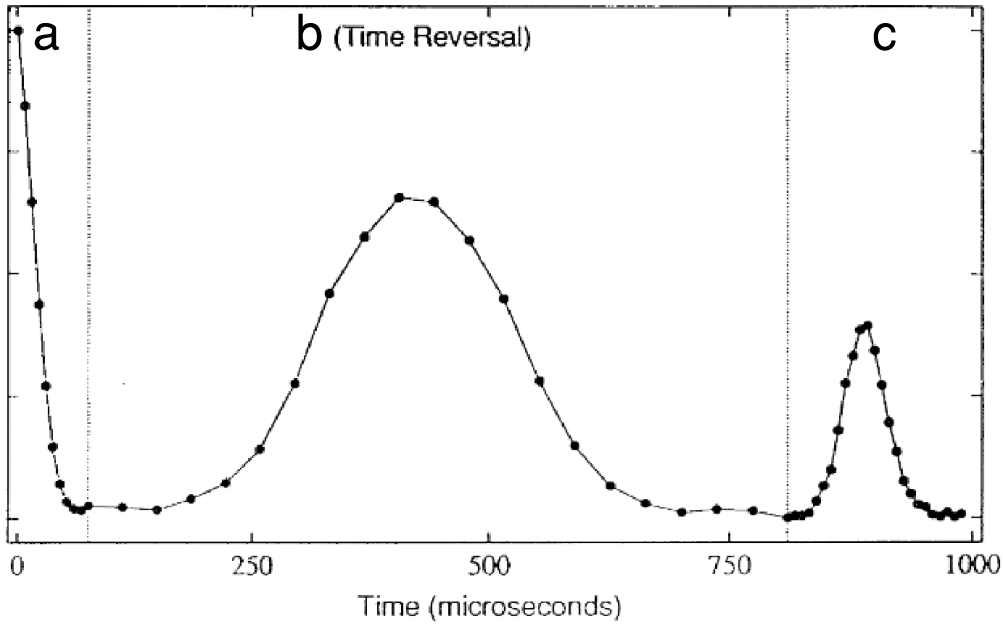}
    \caption{An example of an isotropic spin echo produced by a polyhedral symmetry-based pulse sequence at zero field.  In (a), transverse magnetization (signal on the vertical axis) of \textsuperscript{1}H in a sample of polycrystalline adamantane decays due to \textsuperscript{1}H-\textsuperscript{1}H dipole--dipole couplings.  In (b), starting at around 70\,$\mu$s and lasting until 810\,$\mu$s, a sixteen-element decoupling sequence imposes scaling factors $k_1=+1/5$ and $k_2=-1/5$.  Time reversal of the strong rank-2 interaction produces an echo at $\sim$350 $\mu$s or around $|k_2|^{-1} = 5$ times later.  In (c), when free evolution resumes, a second echo appears at 880 $\mu$s.  Reprinted (figure) with permission from \cite{Llor1991}. Copyright 1991 by the American Physical Society.}
    \label{fig:Llorecho}
\end{figure}

Families of pulse sequences for true-zero-field and `zero-field-in-high-field' coherent averaging were analyzed extensively by Llor, who developed an analytical formalism for Hamiltonian engineering using the theory of rotation quadrature in polyhedral symmetry groups \cite{Llor1991,Llor1995a,Llor1995b}.  The objective of Llor's approach was to find a series of pulses that scale a given spin interaction in the first-order average Hamiltonian by a fixed numerical factor that depends only on its (spin) spherical tensor rank, in a way that is independent of principal axis system.  It turned out that the Euler rotation angles of the pulses, when analyzed in a convenient frame of reference, i.e., the `toggling' interaction frame in which the frame is rotated cumulatively by all of the previous pulses, % [extra ref necessary? I think it should be covered already by the papers above \AT{I also don't think that additional references are needed here.}], 
could be equated to the polar coordinates of regular polyhedron vertices, for instance vertices of an octahedron---hence the name polyhedral isotropic averaging.  %These pulses may or may not be modelled as delta function pulses.  
It was then shown that, depending on the symmetry group of the polyhedron, spin interactions of tensor rank 1, 2 or higher could be independently scaled to zero, therefore finding conditions under which a given spherical component of the spin Hamiltonian is averaged completely.  For example, a rank-1 decoupling pattern (or in other words, a sequence that scales the rank-1 average Hamiltonian by a factor $k_1=0$) should always scale the rank-2 interactions between a factor $-1/5 \leq k_2\leq +2/5$.  The flip angle of the corresponding pulses is $2\pi/3$.  The allowed set of scaling factors is represented by the dark-shaded region in \autoref{fig:Llor2dplot}.  A negative scaling factor corresponds to time reversal of the corresponding spin interaction.  Experimentally, this can be observed in the form of a signal echo, see for example rank-2 dipole--dipole coupling reversal shown in  \autoref{fig:Llorecho}. 

The application of isotropic averaging under true ZULF conditions as a means to obtain ultrahigh-resolution NMR spectra of solid samples is nowadays almost obsolete, following the major increase in magic-angle-spinning speeds achievable and magnetic field strengths that make `zero-field-NMR in high field' the preferred alternative.  Recently, however, ZULF decoupling techniques have become of interest and have been applied successfully to liquid-state samples, where it can be desirable to average out heteronuclear scalar couplings \cite{Sjolander2017,Bodenstedt2022jpcl}.  It can be convenient, for example, to decouple `unwanted' spins during the course of a hyperpolarization procedure, in order to steer the polarization towards a particular spin species of interest.  The decoupling of residual magnetic fields can also prolong spin lifetimes by suppressing unwanted spin-decoherence pathways, or provide a margin of error to tolerate poor shimming of the background field \cite{Bodenstedt2022jpcl}.  As another application, decoupling can also be used to implement quantum gates in the ZULF regime \cite{Jiang2018}. 

\subsubsection{Magnetic field sweeps in the ZULF regime}
\label{Sububsec:polarization_transfer}

\begin{figure}
    \centering
     \includegraphics[width=\columnwidth]{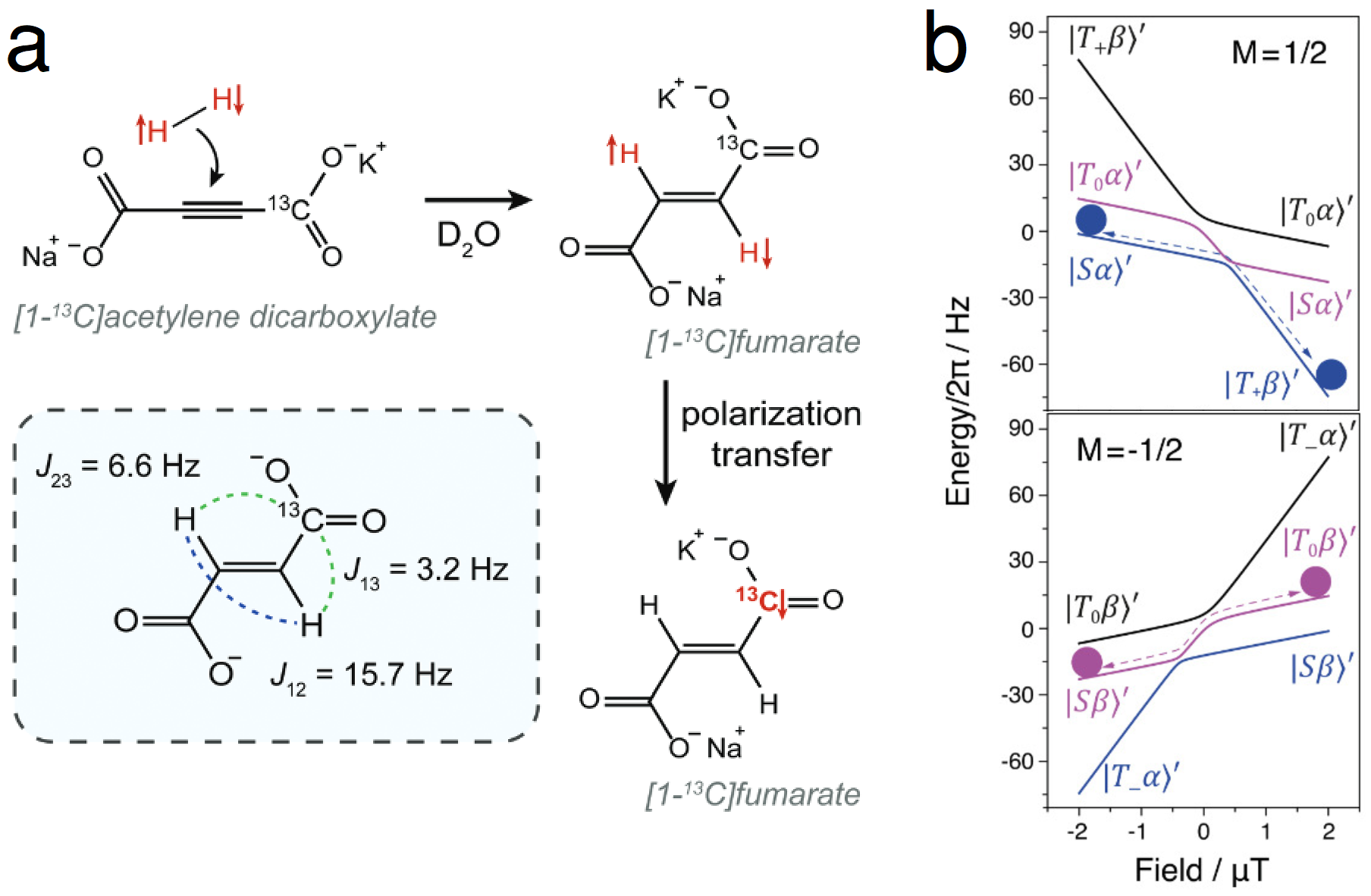}
    \caption{A procedure for microtesla-field hyperpolarization of [1-\textsuperscript{13}C]-fumarate, an XAA' spin system (X=\textsuperscript{13}C, A=\textsuperscript{1}H). (a) Reaction scheme for production of spin-singlet polarized AA' spins in fumarate by chemical addition of parahydrogen to [1-\textsuperscript{13}C]-acetylene dicarboxylate.  (b) Representation of population transfer within the $m_F=\pm1/2$ state manifolds. Colored balls represent the initial and final state populations during an adiabatic magnetic field sweep from \SI{-2}{\micro\tesla} to \SI{2}{\micro\tesla} along the z axis.  The result of the sweep is to generate opposite Zeeman polarization on the \textsuperscript{13}C and \textsuperscript{1}H spin species.  Adapted from references \cite{Rodin2021PCCP} and \cite{Rodin2020MR} under terms of the Creative Commons CC-BY license 4.0.}
    \label{fig:fieldsweepfumarate}
\end{figure}

The ultralow-field regime is used in PHIP experiments to transfer the \textsuperscript{1}H spin-singlet order into \textsuperscript{1}H or \textsuperscript{13}C magnetization. The attraction of \textsuperscript{13}C as a target nucleus includes a wide chemical-shift range for spectroscopic dispersion (at high field), combined with long relaxation times on the order of minutes to facilitate perfusion of the hyperpolarized compound through a living organism. There are a few mechanisms to convert the singlet order into \textsuperscript{13}C magnetization which involve avoided crossings that occur at magnetic field values on the order of $B \approx 2 \pi J_{IS}/|\gamma_I-\gamma_S|$, where ZULF eigenstates transition between zero-field (total angular momentum) and high-field (Zeeman) eigenstates. In the simplest case we can consider a three-spin system such as the [1-\textsuperscript{13}C]-fumarate molecule as shown in \autoref{fig:fieldsweepfumarate}a (see also \autoref{fig:Magnetogenesis}), containing two protons weakly coupled to a \textsuperscript{13}C spin. In \autoref{fig:fieldsweepfumarate}b, the relevant ultralow-field avoided crossings are shown. In one singlet-to-magnetization transfer approach the magnetic field is increased adiabatically from negative to positive field, and the spin system passes through four avoided crossings (of which three are relevant for the evolution of state populations) \cite{eills2019polarization}.  In another approach, the field is increased adiabatically from zero field upwards and the spin system passes through two avoided crossings (of which one is relevant for the evolution of state populations) \cite{goldman2005hyperpolarization}. A third approach is to carry out the hydrogenation at the avoided-crossing field itself, which can yield up to 50\% \textsuperscript{13}C polarization in a three-spin system \cite{rodin2021constant}.  The situation is more complex for molecules with additional spins in the $J$-coupling network, but these methods are still employed albeit with some need for optimization of the precise magnetic field used. These procedures are closely related to singlet-to-magnetization pulse sequences used to convert homonuclear spin-singlet order into observable coherence in high-field NMR \cite{pileio2017singlet}.

An alternative type of system for which such ultralow-field transfers of singlet polarization to magnetization are employed is SABRE-SHEATH \cite{Theis2015,Truong2015}. The major difference in this case is that breaking of the magnetic-equivalence symmetry occurs due to chemical exchange and therefore involves modulation of $J$-couplings during the transfer process. In typical cases,
%\MCDT{the acronym is already defined in the PHIP section}) 
the sample is held at an avoided-crossing field during parahydrogen bubbling, so that the proton singlet order is %spontaneously 
converted into heteronuclear (\textsuperscript{15}N or \textsuperscript{13}C) magnetization. This approach has been modified in recent years to include modulating magnetic fields for obtaining signal enhancements under conditions beyond those described by level anticrossings \cite{lindale2019unveiling,pravdivtsev2021coherent,Li2022,Eriksson2022}.

\section{Detection in ZULF NMR}
\label{Sec:Detection}

% Traditional NMR incorporates detection of nuclear polarization based on Faraday induction. This was also how NMR was detected in the early ZULF experiments. Since then, a variety of other (especially, noninductive) sensors have been used with most ZULF NMR experiments today using atomic magnetometers. However, other technologies (e.g., diamond or anisotropic magnetoresistive sensors) may become the detectors of choice for future ZULF-NMR microfluidic applications.
% \JE{Possible rewrite of the text above: 
In high-field NMR, Faraday induction is used for signal pick-up, providing high sensitivity at signal frequencies of megahertz and above. Inductive detection was used in early ZULF experiments, but is less common now since alternative detectors that are sensitive at low signal frequencies (kilohertz and below) are available, such as optically pumped magnetometers (OPMs), superconducting quantum interference devices (SQUIDs), color centers in diamond, and others. In particular, the emergence of SQUIDs in the 1980s and advances in OPM technology in the early 2000s were critical points for the development of ZULF NMR. 

\begin{figure}
    \centering   \includegraphics[width=\columnwidth]{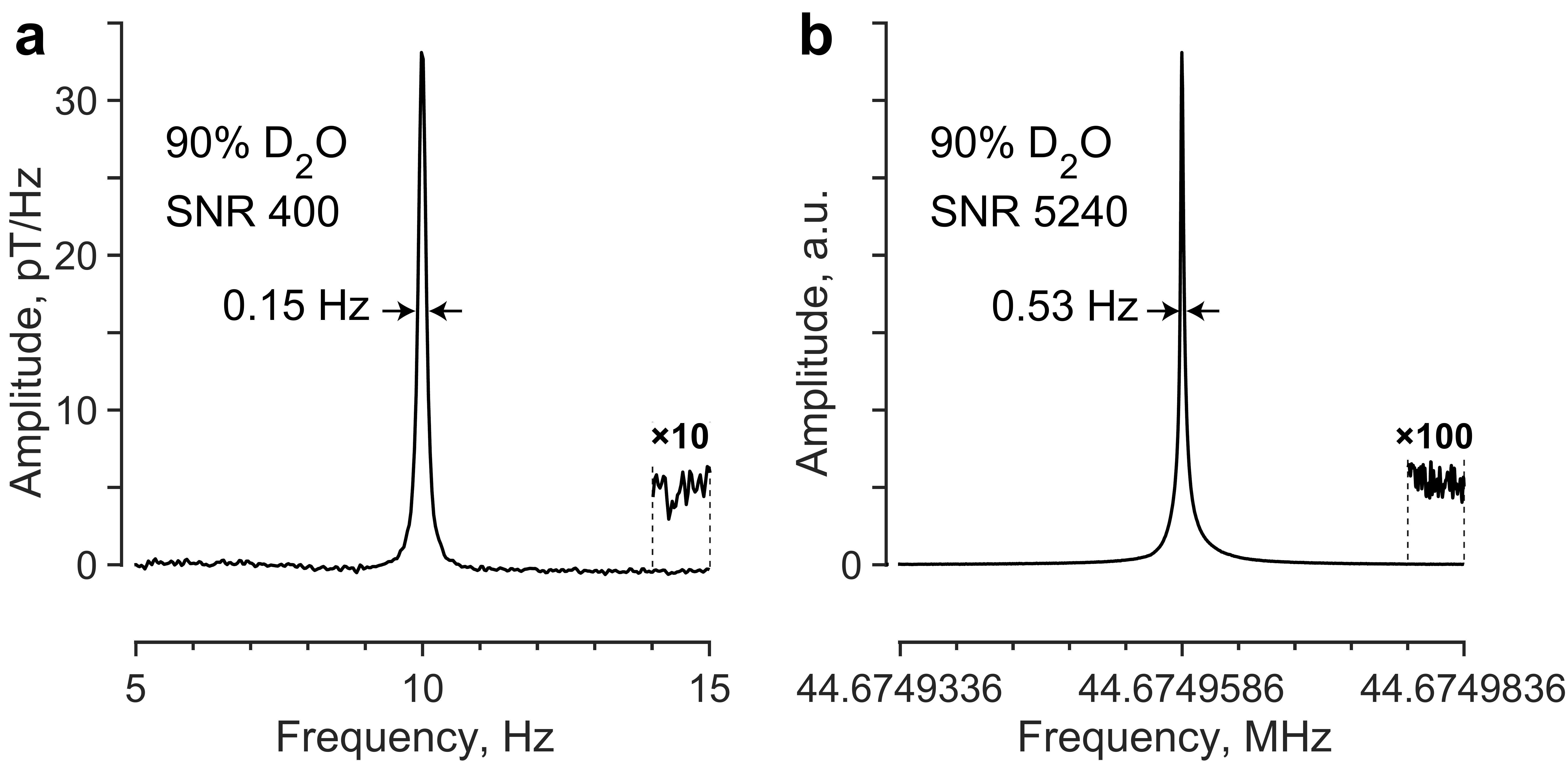}
    \caption{  
    (a) Atomic magnetometer-detected single-scan $^1$H NMR spectrum of a 1.5 mL quantity of 10~\% H$_2$O in D$_2$O, recorded $\sim$100 ms after remote prepolarization at 1\,T. A four-channel ``gradiometric quadrature'' configuration of two commercial magnetometers was used, as described in reference \cite{Fabricant2024}. The spectrum was obtained from a 10 s trace sampled at 2 kHz; no filtering or other lineshape correction was applied. SNR was calculated as the ratio of peak amplitude to standard deviation of noise in the frequency region 14-15 Hz (inset). (b) Single-scan \textsuperscript{1}H NMR spectrum of a 0.5\,mL %\MCDT{Likely that the sensitive volume of the rf coil is much smaller, around 0.1\,mL} 
    quantity of 10~\% H$_2$O in D$_2$O, recorded using a 1 T benchtop spectrometer (Magritek ULTRA, 40 MHz). The spectrum, obtained from a 6 s acquisition, shifted to 5 Hz oscillation frequency and sampled at 2 kHz; no filtering or other lineshape correction was applied. SNR was calculated as the ratio of peak amplitude to standard deviation of noise in the rightmost 5 Hz frequency window (inset). %\MCDT
    %{Sorry, there must be something wrong with my understanding, or the data. Why is the SNR in (b) less than 10x greater than in \autoref{fig:ZULF-HF-3}(b), where the sample contains around 100x fewer protons?} \DAB{My assumption would be this is because there is less magnetic noise in the environment at this frequency compared to the low-frequency region of the spectrum.}
    }
    \label{fig:ZULF_HF_fin}
\end{figure}

These gains in sensitivity mean that, at the time of writing, detection in the ZULF regime is only one to two orders of magnitude less sensitive than detection using a high-field benchtop NMR spectrometer. We illustrate this in \autoref{fig:ZULF_HF_fin}, comparing the signal-to-noise obtained for single-scan detection of 10\,\% H\textsubscript{2}O in D\textsubscript{2}O samples in the two regimes. The benchtop measurement was performed on a 0.5\,mL static sample in a 1-T field; the ZULF measurement on a 1.5-mL sample pre-polarized in a 1-T field, shuttled down ($\sim$100\,ms) into the ZULF region for quadrature detection using two commercial OPMs.
Although these results show an approximate comparison between ZULF versus benchtop NMR, they depend on experimental factors and details of the detector used.

In this section we describe the available detection modalities in greater detail.

\subsection{Detection modalities}
\label{Subsec:Det_modal}

\subsubsection{Inductive detection}
\label{Subsec:Inductive_det}

% \paragraph{Indirect detection}
% \label{Subsec:Indirect_det}

%\MCDT{According to our definition of ZULF, this \cite{DeVience2021SLIC} does not count? The point is that if we include this type of rotating frame NMR, then there will be a lot of other examples to discuss.  I don't think this counts as ZULF, personally.} \DB{Danila and Dima say: let us mention these sort of experiments, there are not too many of them dealing with J spectroscopy, but not dwell on these too much...} 

The first NMR experiments at zero external field that we are aware of were performed by detecting the unusually large line splittings due to dipolar interactions in solid hydrogen, which are on the order of hundreds of kilohertz \cite{Reif1953earlyzero}. These experiments used the `traditional' method of absorption of energy from the radiofrequency field in a cavity \cite{Pound1953} (see also \autoref{Subsec:History}), but could also be performed with modern inductive detection. However, these experiments rely on the specific properties of solid hydrogen.

The first ZULF NMR experiments based on a generally applicable method were carried out using indirect inductive detection: after a polarization period at high field the samples were allowed to evolve under ZULF conditions, and then were rapidly shuttled back to high field for coil-based signal detection \cite{Weitekamp1983,Zax85,thayer1987zero}. This approach carries the advantage of the high sensitivity of high-field inductive detection, which is usually below 1\,fT/Hz$^{1/2}$ at frequencies above 1\,MHz \cite{Savukov2007sensitivity}, although the challenge comes from the need for shuttling the sample a relatively long distance (order of 1\,m) to the center of a high-field NMR magnet.

An alternative approach is to carry out inductive detection \textit{in situ}. This helps to simplify the experimental apparatus and avoid the need for a SQUID or OPM \cite{matlashov2010squids,espy2011progress}. Using a 1400 turn 90 mm o.d., 20 mm i.d. coil wound with AWG24 (0.51\,mm) wire, a detection sensitivity of 20\,fT/Hz$^{1/2}$ at 3.3\,kHz was achieved, and this sensor was used to carry out imaging and relaxation measurements on liquid samples. The footprint was similar to that of a SQUID, although the drawback is the drop in sensitivity at lower frequencies relevant for the ZULF regime, meaning detection must be carried out at moderate $B_0$ field strength.
\
% \AT{However, we are aware of only one case where magnetic-resonance signals of nonquadrupolar nuclei have been inductively detected in the ZULF regime: in experiments with solid hydrogen, as first performed by Frederick Reif and Edward Purcell in 1953 \cite{Reif1953earlyzero}. The unusually large line splittings due to dipolar interactions, combined with high polarization at cryogenic temperatures, make it possible to obtain resolved spectra at zero magnetic field.
% 1953 work of Reif and Purcell \cite{Reif1953earlyzero} on inductively detected NMR in solid hydrogen (see also \autoref{Subsec:History})}

% \AT{As argued in Sec.\,\ref{Subsec:Regimes}, for spins without measurable couplings to each other, the only field regimes are high- and zero-field. The majority of Earth-field experiments, including those described above, therefore belong in the former category. However, there are several works in which heteronuclear $J$-coupling interactions between ${}^{1}$H--${}^{29}$Si \cite{Appelt2006}, ${}^{1}$H--${}^{19}$F \cite{Qiu2008,Qiu2009} and ${}^{1}$H--${}^{31}$P \cite{Liao2010} species were resolved. These fall into the low-field regime.}

% \DB{This means that inductive detection occurs in the same field as the evolution. Andreas will work on this}

%     \begin{itemize}

%     \item Tycko's high-field emulation of ZULF conditions
    
%     \item SLIC

%     \item Alex Pines work of 80-s
    
%     \item Appelt's work

%     \item 1953 work of Reif and Purcell \cite{Reif1953earlyzero} on inductively detected NMR in solid hydrogen (see also \autoref{Subsec:History})
    
%     \end{itemize}

\subsubsection{Superconducting interferometers}
\label{Subsec:SQUID}

The pioneering experiments in zero-field NMR \cite{Weitekamp1983,Zax85} were based on indirect inductive detection (see \autoref{Subsec:Inductive_det}), but the necessity of repeatedly returning the sample to a high field meant that these experiments were both technologically challenging and time-consuming. An alternative emerged in the form of detection with SQUIDs, magnetometers sensitive enough to detect nuclear magnetization directly and therefore independently of precession frequency.

The operation of a SQUID is based on the effect of flux quantization in superconducting loops and the Josephson effect. The magnetic flux through a loop is quantized; that is, it is an integer multiple of the elementary flux quantum
% Through a superconducting ring only magnetic fluxes with a magnitude that is an integer multiple of the elementary magnetic flux quantum 
$\Phi_0 = 2.07\times10^{-15}$\,V$\cdot$s (1\,V$\cdot$s=1\,Wb is the SI unit of magnetic flux, the weber). When an external magnetic field penetrates the ring, an electric current flows to increase or decrease the magnetic flux in the ring to the nearest multiple of $\Phi_0$. The corresponding supercurrent is measured by exploiting the Josephson effect, which describes the behavior of the supercurrent when Cooper pairs tunnel through a layer of insulating material from one superconductor to another. The insertion of one or two such Josephson junctions gives rise to a measurable voltage change that depends on the strength of the magnetic field \cite{Clarke2006SQUIDs}.
%\DB{, which is..., Andreas is constructing this...}.

Magnetometers based on the Josephson effect have been used for the detection of NMR signals since the late 1960s (see reference \cite{Greenberg1998} and \autoref{Subsec:History} for an overview). In the context of liquid-state NMR, work in the early 2000s that demonstrated SQUID detection of thermally prepolarized samples in microtesla fields \cite{McDermott2002,Trabesinger2004} paved the way for $J$-spectroscopy, as discussed in this review. More recently, SQUID-detected NMR was extended to 2D-NMR studies and combined with hyperpolarization techniques \cite{Shim2014_2D,Lee2023_2D,Buckenmaier2019MQ, Myers2024PRB}.

% \DB{What about the present work in Korea and PTB Berlin: Wolfgang Killian} 
% \MCDT{This is a recent one: \cite{Myers2024PRB}}

\subsubsection{Atomic magnetometers}
\label{Subsec:OPM}

% \MCDT{Highly relevant paper not cited yet: 
% \url{https://doi.org/10.1002/cmr.a.20134}. Should also be mentioned in the `Effects of geometry' section.
% }

SQUIDs are among the most sensitive magnetometers; however, even these are not without competition. Already in 1969 Claude Cohen-Tannoudji and co-workers in Paris used an atomic magnetometer to detect magnetization from a gas of polarized $^3$He nuclei \cite{Cohen-Tannoudji1969}. In an atomic magnetometer, atomic spins are polarized by optical pumping. The spins then evolve under the influence of the magnetic field and the resulting evolution is detected (nowadays, most frequently, optically). The practical use of atomic magnetometers for NMR detection had to wait for some three decades. The potential of atomic magnetometry in zero-field spectroscopy and imaging was recognized by Alexander Pines and co-workers at Berkeley, who initiated a collaboration with the local atomic-magnetometry group led by one of us (D.\,Budker). The developments proceeded from the initial demonstration of detection of hyperpolarized gaseous xenon with an optical atomic magnetometer \cite{yashchuk2004hyperpolarized} and low-field remote-detection imaging \cite{xu2006magnetic_PNAS,Xu2008b} to detection of $J$-couplings in the ZULF regime \cite{ledbetter2009optical} and combining ZULF with parahydrogen hyperpolarization techniques \cite{Theis2011,Theis2012_NH_PHIP}.
Atomic magnetometers have also been successfully used by Michael Romalis and collaborators for detection of nuclear quadrupole resonance (NQR) \cite{lee2006subfemtotesla}, sometimes referred to as `the original zero-field NMR' for its lack of necessity of a bias field. 
%Incidentally, SQUIDs also left their mark on the field of NQR \cite{Augustine1998b}.

Since around the early 2000s, the seasoned field of atomic magnetometry has experience a significant boost of activities, in part associated with developments in diode-laser technology, and in part due to the widening use of techniques to obtain narrow (at times, sub-Hz) electron spin resonance lines, with a corresponding boost in the sensitivity to magnetic fields \cite{Budker2007,Yi2009_Portable}. These techniques include magneto-optical rotation in alkali-metal vapors contained in antirelaxation-coated cells as well as the use of high-density alkali vapors in the so-called spin-exchange-relaxation-free (SERF) regime, see a detailed discussion of the techniques in the book \cite{budker2013optical}. 

\begin{figure}
\centering
	\includegraphics[width=0.9\columnwidth]{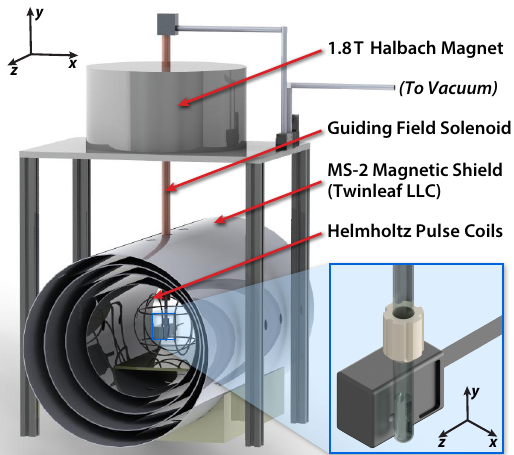}
	\caption{Experimental apparatus for ZULF NMR comprising a commercial alkali-vapor-cell magnetometer (QZFM, QuSpin Inc.) for detection, and a permanent magnet for remote prepolarization of the sample.
	The inset shows a magnified view of the sample and sensor.
	For experiments utilizing parahydrogen, the permanent magnet and guiding-field solenoid are removed and the sample is placed in a custom NMR tube assembly that permits bubbling of parahydrogen through the solution.
	The end caps on each of the cylindrical magnetic shielding layers are omitted for clarity.
	%The coordinate system is chosen to be consistent with the labels on the MS-2 magnetic shield -- it should be noted that these are different from the axes defined by the QZFM manufacturer.
	Reproduced from reference\,\cite{Blanchard2020}, Copyright (2020), with permission from Elsevier. %\MCDT{I would move this to the detection section. OK?}
	}
	\label{fig:QuSpin-schematic}
\end{figure}

In recent years, atomic magnetometers with sufficiently high sensitivity (10--20\,fT/$\sqrt{\textrm{Hz}}$) and small size (a few milliliters) have become available commercially from a number of manufacturers (for example, TwinLeaf and FieldLine), with the vast majority of the `market share' for ZULF NMR applications held by QuSpin. In fact, QuSpin sensors, in combination with high-quality magnetic shielding (e.g., from TwinLeaf) have greatly `democratized' ZULF NMR, making it accessible to scientists without expertise in atomic physics or optics \cite{Blanchard2020,TaylerNMRduino,Andrews2024arxiv}.  
The development of these sensors was driven mostly not by NMR but rather applications in magnetoencephalography (MEG), magnetocardiography (MCG) and other ultralow-frequency-sensing applications. 
Examples of compact NMR apparatus built using these off-the-shelf components are illustrated in Figs.\,\ref{fig:QuSpin-schematic},~\ref{fig:PortaZULF} and~\ref{fig:PortaZULF2}. The development of SERF magnetometers continues, sometimes specifically directed towards ZULF NMR applications \cite{hong2024femto}. Reaching the reported $\sim$1\,fT/$\sqrt{\textrm{Hz}}$ sensitivity level in the $~1\,$kHz bandwidth encompassing the ZULF NMR frequencies of commonly studied molecules will greatly boost the range of applications, eliminating the need for isotopically labeled compounds.

While SERF magnetometers typically operate at fields up to several microtesla, other commercial atomic magnetometers are becoming available with centimeter-scale dimensions and sensitivities on the order of a few 100\,fT/$\sqrt{\textrm{Hz}}$. These can already be used for NMR detection and have the advantage of enabling operation in unshielded environments (see also \autoref{Subsec:fieldable}). 

Despite the availability of commercial sensors, some researchers continue to prefer building their own atomic magnetometers, driven either by the desire for particular sensor features (e.g., full control of feedback loops for better stability, for example against temperature drifts, larger bandwidth, fast recovery following an NMR control-pulse sequences, or producing synthetic gradiometers of sensor arrays), or lower cost for a home-made sensor with sufficient sensitivity for less demanding applications. 
%\DB{Michael, anything to add here? For example a reference to the sensor you showed at HYP23?} 
%\MCDT{I would say that we developed our own sensor because we wanted full control of the feedback loops, for better stability -- for example against temperature drifts.  That doesn't fit with the last sentence about lower cost although in principle, if the sensors were made on a massive scale, price could be another motivation.  The NMRduino paper (now submitted\cite{TaylerNMRduino}) also demonstrates the use of QuSpin sensors in a magnetic shield, so we could somehow mention it together with John's paper in the previous paragraph.}

We also mention here a particular modality for the detection of nuclear spins (typically, of noble gases) with co-located alkali atoms (see, for example, \cite{Terrano2021comagnetometer} and references therein). An advantage of this approach is that the interaction of the alkali spins and the nuclear spins under study is enhanced due to the Fermi-contact interactions. This has two benefits. First, the nuclear spins can be efficiently hyperpolarized in collisions with the alkali atoms by optical pumping. Second, the strong coupling of the nuclear and alkali spins enables efficient probing of the nuclear spin state by interrogating the alkali atoms.
One possible application of such `comagnetometers' is the detection of nonmagnetic perturbations such as apparatus rotations or effective fields due to exotic interactions that may arise due to background dark-matter fields  \cite{huang2023axionlike} or due to fifth-force interactions \cite{SBDJ18,Terrano2021comagnetometer}.
Alkali-noble gas comagnetometers can be configured to work in the self-oscillating regime---a spin maser \cite{Jiang2021Floquet}.

\subsubsection{Color centers in diamond}
\label{Subsec:NVdiamond}

%\paragraph{Single-NV ZULF}

Conventional NMR relies on the creation of spin polarization and the detection of the associated magnetization. 
However, it has also been shown that systems with no net time-averaged polarization produce an NMR signal through statistical fluctuations of the nuclear magnetization. 
This effect is known as `spin noise' \cite{Bloch1946,Sleator1985} (see also \autoref{Subsec:Stat-pol}). 
The spin noise signal scales as $\sqrt{N}$ (with $N$ being the number of nuclear spins in the detection volume), whereas Boltzmann polarization scales linearly with $N$, meaning that in strong fields and for typical sample sizes, the spin-noise signal is significantly weaker than that arising from the magnetization. 
Whereas spin-noise detection can offer distinct advantages even in relatively conventional settings---for example, the spins do not have to be excited using RF irradiation \cite{Mueller2006}---the greatest advantages emerge for microscopic samples \cite{Crooker2004}. 
In the extreme case of $N=1$, there is no difference in signal strength for polarized and unpolarized samples, and in fact the sample can be thought of as being fully polarized \cite{Muller2014}.

In proof-of-principle experiments, NV centers just a few nanometers from the diamond surface have been used for NMR spectroscopy of nanoscale samples \cite{Mamin2013,devience2015nanoscale}. 
In recent years, this approach has been harnessed to detect signals from single molecules \cite{Lovchinsky2016}. 

As spin noise is independent of magnetic field strength, statistical polarization can be detected at arbitrary fields (including ZULF) without the need to `hyperpolarize' the sample. 
This leads to the ongoing experimental effort to detect nanoscale ZULF NMR with a stochastically polarized sample producing a signal detected with a single NV center \cite{Budker2019JMR}.
Measurement of $J$-spectroscopy in liquids would be a natural goal, enabling measurement of metabolic products from a single cell, or other mass-limited samples.
However, for liquid samples measured with single NV centers, a new problem arises, which is that spectra are broadened by diffusion \cite{Kong2015}.
The spin noise signal arises from correlated fluctuations, and high-resolution spectra require long correlation times (analogous to spin coherence times in conventional NMR).
While spin fluctuations in a given molecule may remain correlated for many seconds, this is only beneficial if the molecule stays within the sensing volume of the NV---if one molecule diffuses away and is replaced by another, the spins fluctuations of the new molecule are generally uncorrelated with those of the previous one.
A number of approaches for spatial confinement of analytes are being explored, including encapsulation in host--guest complexes, intercalation in layered materials, and reversible binding to a functionalized diamond surface. Molecules encapsulated in endofullerinces \cite{Levitt2013_Endof} could be an interesting option; single (electronic) spins within endofullerine have been detected with NV centers \cite{Pinto2020}.
A more general, albeit technologically demanding, strategy may be to control the movement of analytes by nanofluidic (di)electrophoresis. 
%\AT{Are there any papers or preprints that we can cite here (whether for electrophoresis or other possible routes to spatial confinement)?}

% \begin{figure}
% 	\includegraphics[width=\columnwidth]{NV-ZULF_Concept.pdf}
% 	\caption{Conceptual representation of `extreme NMR:' $J$-spectroscopy of a single molecule performed with a single-color-center sensor (an NV$^-$ center located several nanometers below the diamond surface), conducted at ZULF.}
% 	\label{fig:NV-ZULF}
% \end{figure}

Another interesting area is the application to solid-state systems, where molecular diffusion is considerably slower. Particularly appealing candidates include those that are largely inaccessible to conventional NMR, including thin films, two-dimensional materials, multiferroics and topological insulators. 
One notable demonstration of the potential applications of NMR detected with single NV centers is the measurement of ${}^{14}$N and ${}^{11}$B in atomically thin hexagonal boron nitride \cite{Lovchinsky2017}, where the authors observed changes in the ${}^{11}$B nuclear quadrupolar coupling when studying monolayer, bilayer and bulk samples, establishing the capability of NMR measurements to observe nanoscale effects.

\subsubsection{Anisotropic magnetoresistive sensors}
\label{Subsec:AMR}

%\DB{DB will write this.}
%\DB{The Abstract of the one paper \cite{Verpillat2008remote}:}
Among the magnetic sensors that do not rely on Faraday induction are solid-state sensors that are based on anisotropic magnetoresistance (AMR). These devices, with sizes down to micrometers, feature sufficiently high bandwidth to cover the range of frequencies (up to tens of kilohertz) of interest to ZULF NMR. However, their drawback is a sensitivity that is typically several orders of magnitude lower compared to atomic magnetometers (\autoref{Subsec:OPM}).

Reference \cite{Verpillat2008remote} reported NMR detection using an AMR sensor. A `remote-detection’
arrangement (see \autoref{Subsubsec:remote}) was used in which protons in flowing
water were prepolarized in the field of a superconducting high-field NMR
magnet, adiabatically inverted, and subsequently detected in a magnetically shielded region downstream from the magnet.  A fast-adiabatic-passage
inverter was used to periodically modulate the magnetization of the protons in the flowing water.

A potential future use of AMR sensors is for NMR detection in
microfluidic `lab-on-a-chip’ applications. An estimate of the optimized sensitivity \cite{Verpillat2008remote}
indicates that $\approx 6\times10^{13}$ protons in a
volume of 1,000\,$\mu$m$^3$, prepolarized in a 1 T magnetic field, can
be detected with a signal-to-noise ratio of 3 in a 1 Hz bandwidth.
This level of sensitivity is competitive with that of
microcoils in superconducting magnets and with the
sensitivity of microfabricated atomic magnetometers. 

Further progress in AMR-sensor technology may result in higher-sensitivity devices; however, an important issue to be dealt with, especially in the ZULF-NMR context, is the stray magnetic field produced by the sensor that can affect the sample.

\subsubsection{Radioactively detected ZULF NMR}
\label{Subsec:Radioactive}

An interesting noninductive NMR-detection modality involves detection of the products of radioactive decay of NMR-active unstable nuclei \cite{Jancso2017_Gamma_Beta}. $\beta$-NMR relies on preferential emission of the $\beta$-decay electron or positron in a direction along or opposite to the spin orientation. This effect has its origin in nonconservation of parity in weak interactions. In $\gamma$-NMR, another type of decay asymmetry is exploited---for quadrupolar nuclei ($I>1/2$), the probability of $\gamma$-emission is different along the alignment axis and perpendicular to it. Note that in this case, in contrast to $\beta$-decay, there is no preferred spatial direction, only a preferred (bidirectional) axis.  

As the detection efficiency for radioactive decay products is high (in principle, down to a single particle), radioactively detected NMR has the potential to increase the sensitivity compared to both inductive and noninductive NMR detection techniques \cite{Jancso2017_Gamma_Beta}. Furthermore, with radioactive nuclei introduced into a sample,  intrinsically background-free detection is possible. The `flip side' (that comes together with the high selectivity in NMR spectroscopy and imaging) is that the method is restricted to available radioactive nuclei that need to be produced and then utilized before the radioactivity has decayed.  

%We note that 
$\beta$-NMR has been performed in the ZULF regime for the study of quadrupolar nuclei  \cite{salman2006beta, voss2011development}. %\DB{Yes, please!}.
A `marriage' of ZULF with $\gamma$- or $\beta$-NMR \cite{Blanchard2021_LtL} may open up interesting possibilities for practical materials studies, including batteries, thin films, and biological samples. Proof-of-concept studies are ongoing at CERN in collaboration with researchers at Mainz.

\subsubsection{Nuclear spin optical rotation}
\label{Subsec:NSOR}
\medskip

When linearly polarized light propagates co-linearly with the magnetization axis of a magnetized medium, the polarization of the light rotates as a consequence of the Zeeman effect. 
%\AT{Is it possible to give here an intuitive one-sentence explanation of why this is so?} \MCDT{
A classical example is the Faraday effect, discovered in heavy-atom (e.g., lead) glasses as early as the 1840s. The Faraday effect is usually discussed in conjunction with electrons. Nuclear spin optical rotation (NSOR) is also observable in the case of polarized nuclei \cite{Savukov2006NSOR}. The authors of  \cite{Savukov2006NSOR} studied optical rotation in polarized water and liquid xenon and, in the case of the latter, found the magnitude of optical rotation of $(5.8 \pm 0.6)\cdot10^{-6}$\,rad cm$^{-1} $M$^{-1}$, given for fully polarized $^{129}$Xe with molar concentration M (in units mol/l). This magnitude of the effect is a factor of five or so smaller than naive estimates \cite{Savukov2006NSOR}, however, is adequately reproduced by more sophisticated theory taking account of the cancellation of contributions from various excited states \cite{Savukov2015_Xe_NSOR}.

While an aesthetically beautiful and conceptually simple method to detect nuclear magnetization and, correspondingly, NMR, the small magnitude of the NSOR effect limits its utility to highly concentrated liquid samples. The technique becomes practically useful in combination with hyperpolarization techniques. For example, NSOR of dilute analytes (90\,mmol/l solutions of pyridine and pyrazine) hyperpolarized using SABRE (see Sec.\,\ref{Subsec:PHIP}) was reported in reference \cite{Stepanek2019NSOR}. The work of reference \cite{Zhu2021_DNP_NSOR}
combined hyperpolarization using $d$DNP (see Sec.\,\ref{Subsec:DNP}) with NSOR detection, where NSOR signals from protons and $^{19}$F nuclei were observed in dilute compounds.  %\MCDT{A question: does NSOR require any particular wavelength?  At least given the choice of UV-vis-IR.}
The magnitude of NSOR typically increases towards shorter wavelengths as long as the material remains transparent.

The ideas for future application of NSOR in NMR include correlating optical and, for example, inductive detection as well as MRI without the use of magnetic gradients, the latter based on the ability to localize a laser beam (in the transverse plane). 

While we are not aware of application of NSOR for detection of ZULF NMR, exploring this combination in future work appears well motivated.

\subsubsection{Remote detection}
\label{Subsubsec:remote}
% \DB{Do we want to have such a (brief) section here?}

In conventional NMR experiments, the spin-polarization, information-encoding and signal-detection stages typically happen at the same physical location, within a magnet and with the sample residing in a radio-frequency coil. Simultaneous optimization of these three stages may lead to trade-offs. In ZULF NMR, the three steps are often naturally separated spatially and temporally, which provides an additional degree of flexibility in designing and optimizing experiments. In particular, NMR information can be stored in longitudinal magnetization following the encoding stage and then be detected at another physical location. This  is known as remote detection \cite{moule2003remote, granwehr2005remote, seeley2004remote} and has been employed, for instance, in the context of low-field imaging \cite{xu2006magnetic_PNAS,Xu2008b} and detection with AMR sensors \cite{Verpillat2008remote}. 

Intriguing directions for remotely detected ZULF NMR include microfluidics-based NMR-on-a-chip \cite{Ledbetter2008} and the concentration of, for example, gaseous samples to optimize detection. A recent development of this idea is to use long-lived spin states \cite{Levitt2012reviewLLSS,Pileio2020bookLLSS} 
%\DB{(NEED REFERENCES TO LEVITT)} 
% \MCDT{\url{https://doi.org/10.1146/annurev-physchem-032511-143724}\cite{Levitt2012reviewLLSS} and \url{https://doi.org/10.1039/9781788019972} \cite{Pileio2020bookLLSS} } 
to deliver the polarization to, say, the brain, and then convert the long-lived spin states that can be used for detection of metabolism and imaging. In connection with ZULF NMR, this has been recently discussed in \cite{Eills2021Singlet-Contrast_MRI}.

\subsection{Practical considerations}
%\JE{Paragraph 1:\\
%- The first consideration for detection modality is sensitivity in the desired frequency bandwidth.\\
%- The majority of detection modalities are sensitive at frequencies below.. 1 MHz..\\
%- [give real sensitivity vs frequency estimates for each sensor?]\\
%- Point to comparison between ZULF and benchtop NMR sensitivity in \autoref{fig:ZULF-HF-3}\\
%- Inductive detection becomes highly insensitive at frequencies below.. 100kHz.. and hence it is not useful for directly detecting ZULF signals in situ. However, indirect detection is possible.}\\

%The sensitivity of inductive coil-based sensing that is used for high-field NMR scales as the power between 3/4 and 1 of the signal frequency \cite{Hoult1976signal} \JB{[This is true at higher frequencies (from reference, $>5$\,MHz) where you're limited by skin depth of the coil -- it's often closer to linear.]}, 
The sensitivity of inductive-coil-based sensing that is used for higher-field NMR scales linearly with signal frequency (decreases to the power of 3/4 above ca. 5\,MHz) \cite{Hoult1976signal, Savukov2007sensitivity}, which explains why inductive detection is less commonly used for low-frequency signals in the ZULF regime. The alternative magnetometry methods described in \autoref{Subsec:Det_modal} scale better for low-frequency signal detection, and in particular SQUIDs and OPMs make up the majority of sensors used for ZULF NMR.

An important factor is the background magnetic field operation range of the sensor.
Sensors that are based on the interaction of spin-polarized electrons with a magnetic field (e.g., OPMs and color centers in diamond) are sensitive to the size of background fields, as these may perturb or destroy the electron-spin polarization.
Other types of sensors based on properties such as magnetoresistance or flux quantization in a superconducting loop may also saturate under background fields.
This is notably not the case for sensing based on magnetic induction, or for the scintillators used for RD-NMR (see \autoref{Subsec:Radioactive}), which operate independent of the background-field amplitude.

Most detection modalities rely on recording the dipolar field outside the spin-polarized sample, although this is notably not the case for nuclear spin-induced optical rotation (NSOR, \autoref{Subsec:NSOR}) detection in which a light beam probes inside the sample itself.
But for other detection approaches it is advantageous to encompass the sample with a detector or an array of detectors to maximise sensitivity.
This is essentially the case in high-field NMR and MRI experiments, in which a coil can be wrapped around a sample to maximise the `filling factor', and also in $\beta$-NMR, where the scintillators usually surround the sample.
Detectors such as OPMs, AMR sensors, and diamond supporting color centers are small in size (typically millimeter scale), and many can be placed around a milliliter-size sample.

Another consideration for detector choice is the logistics of operation.
SQUIDs are exquisitely sensitive \cite{MYERS2007182} but they require cryogenic cooling and usually operate in a magnetically shielded room, meaning they may not be easily transportable.
OPMs are most sensitive in the SERF regime, at temperatures of the vapor cell exceeding 100\,ºC, so they must be thermally isolated from most samples.
Thermal isolation might also be necessary for diamond-based sensors, as they can heat under laser irradiation.
Light-based sensors such as OPMs, color centers in diamond, and NSOR require optical setups and penetration of light into the experimental apparatus, although some detectors operate with a millimeter-scale laser next to the sensing medium, obviating the need for external optical components \cite{schwindt2007chip}.
Some sensors employ flux concentrators to enhance sensitivity (this is common for AMR-based sensors, but concentrators are can also be used with OPMs \cite{Griffith2009FluxConc} and diamond sensors \cite{Fescenko2020_Flux_Conc_NV}), but these produce a dc magnetic field, and the gradient of this field can cause line broadening in ZULF spectra.

\subsection{Effects of geometry in detection}
\label{Subsec:Geometry}

\begin{figure*}[t]
\centering
	\includegraphics[width=0.8\textwidth]{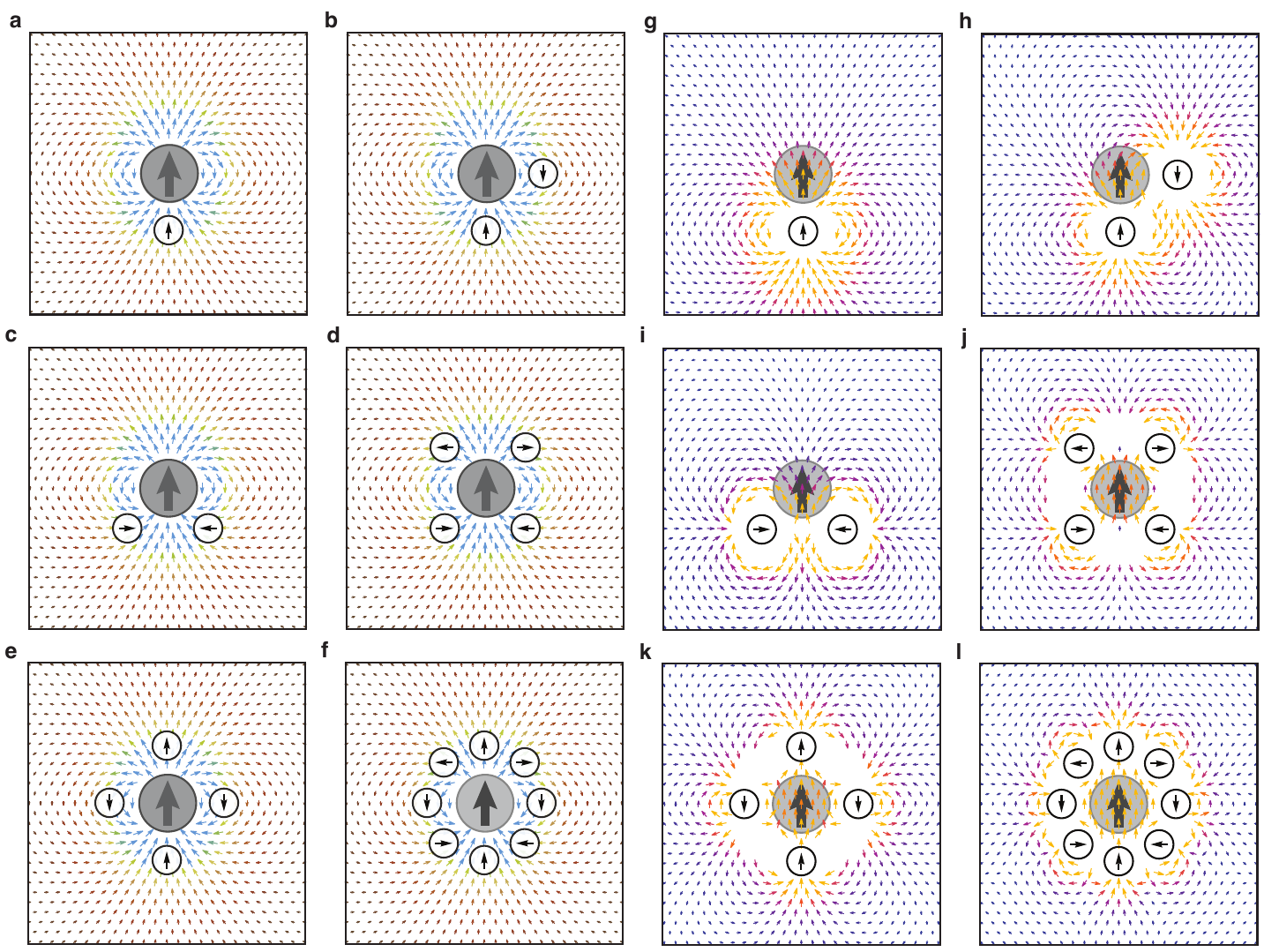}
	\caption{(a--f) Magnetometers (white circles) in gradiometric detection arrays to measure the magnetic field (vector fields) produced by a magnetized sample (grey circle). (g--l) Sensing field (see text for explanation) $\bm{\mathrm{B_{sense}}}$ (vector fields) for these magnetometer arrays.
    (a,g) A simple, commonly used detector arrangement (see Figs.\,\ref{fig:ZULF_NMR_Appa},\,\ref{fig:PHIP_Theis}) where the magnetometer is offset from the sample in the direction of the sensitive axis.
    (b,h) A simple gradiometer arrangement---note that common-mode magnetic noise is cancelled out while magnetic signals from the sample are added.
    (c,i) Another two-magnetometer gradiometer arrangement. Note that this arrangement is sensitive to magnetization in the vertical direction (with respect to the page) even though neither magnetometer is sensitive to fields in the vertical direction.
    (d,j) Higher-order gradiometer produced by combining two gradiometers of the kind shown in panel (c).
    (e,k) Higher-order gradiometer produced by combining two gradiometers of the kind shown in panel (b).
    (f,l) `Halbach-type' gradiometer produced by combining the gradiometers in panels (d) and (e). 
%\DB{John, please add an explanation of color coding and the meaning of the size of the arrows in the caption. I am not sure I understand, why this is called a ``doubled figure but maybe I will once the explanations are given...}
 % \MCDT{John will double the figure!.}
	}
	\label{fig:arrays}
\end{figure*}

%\JB{Is this just to comment on the fact that an ensemble of spins produces a field that is the sum of a bunch of dipole fields? So you can't measure anything inside of a uniformly magnetized sphere.  And if you've got an infinitely long cylinder magnetized along the cylindrical axis, you can't measure anything outside of that cylinder.} \MCDT{Surely this second comment is wrong.  If you have a uniformly magnetized sphere, the external field behaves as a point dipole located at the center of the sphere.  What about the transverse magnetization of a cylindrical sample?  I think that is also nonzero outside of the sample.} \DB{Yes, correct. But if you read what John wrote, it is exactly right!}

The efficiency of detection of nuclear magnetization is generally dependent on sample and sensor geometry.
%Inductive detection: filling factor, or projection of sample magnetization onto the field produced by a unit current in the pickup coil
%SQUID would be the same thing?
%Atomic magnetometer detection: projection of sample magnetization onto the field produced by the atomic spins in the sample
%Radiation detection: You just have to catch the emitted particles. Still wouldnt work with a beta detector in the center of a spherical sample (this would be pretty hard to do, anyway)?

For conventional inductive sensing of NMR, the signal is proportional to the emf ($\xi$) induced in the pickup coil,
\begin{equation}
\xi= - \iiint \frac{\partial}{\partial t}{\left[\bm{\mathrm{B_1}} \left( x,y,z \right)\cdot\bm{\mathrm{M}}\left( x,y,z,t \right)\right]} dx dy dz\,,
\end{equation}
% \DB{I was trying to make the outer parentheses larger but failed...}\MCDT{This is normal. The parentheses automatically adjust to the enclosed content. You could use [] instead of ().}
where $\bm{\mathrm{M}}\left( x,y,z,t \right)$ is the nuclear magnetization and $\bm{\mathrm{B_1}} \left( x,y,z \right) $ is the field produced by a unit current applied to the pickup coil \cite{Hoult1976signal}. The integral over the three spatial dimensions is taken over the entire space, but is inherently zero in places without sample magnetization, i.e., where $\bm{\mathrm{M}}\left( x,y,z,t \right)=0$.

One can follow the same reciprocity-based argument for noninductive sensors by replacing $\bm{\mathrm{B_1}}$ with a sensing field $\bm{\mathrm{B_{sense}}}$.
For SQUIDs, for example\footnote{SQUIDs are sometimes also considered to be inductive detectors but that does not alter the argument.}, the sensing field is equal to the field produced by a unit supercurrent in the pickup coil.
Then the SQUID signal is proportional to the flux coupled into it,
\begin{equation}
\phi= \iiint \left[\bm{\mathrm{B_{sense}}}\cdot\bm{\mathrm{M}}(x,y,z,t)\right] dx dy dz\,.
\end{equation}
For atomic magnetometers, $\bm{\mathrm{B_{sense}}}$ is the magnetic field that would be produced by the sensing volume if it were replaced with a medium magnetized in the direction of its sensitive axis (for spherical samples, this is a dipole field). 
%\DB{I wonder if this is the case for all cell geometries...}
%\JB{I think this should be true for any infinitesimal volume within the cell, so you should be able to make any geometry you want.} \DB{I think I can give a counter-example, but we need to chat. It may boil down to the definition of ``sensitive direction''}
In many cases, the magnetometer sensing volume is approximately spherical, so that $\bm{\mathrm{B_{sense}}}$ can be treated as a point dipole located at the center of the cell.
A similar analysis was recently performed for micron-scale NMR with ensembles of NV centers in diamond \cite{Bruckmaier2021}.

In all of these cases, the important quantity is the projection of the magnetization onto the sensing field.

% \DAB{I would not focus on SQUIDs (since our review is mostly focused on using atomic magnetometers) and instead would write down the equation for the signal measured by a vector-OPM. In the first approximation, this would be a field along a specific direction at the position of the sensor integrated over the sample volume. For uniform magnetization, this could be simplified further by assuming a single dipole at a distance R from the sample and a geometry-based coefficient.} \DB{It is fine to add this; I find the discussion above mentioning different sensors balanced and useful and would not cut from it...} \MCDT{Agree.}

\subsubsection{Detector arrangement}
\label{Subsubsec:Gradiometers}

For a spherical sample and a single point-like magnetic field sensor, the optimal arrangement is to have the sensitive axis of the magnetometer aligned with the sample magnetization and to have the sensor as close as possible to the sample along the magnetization axis.
If one has access to multiple sensors, a linear combination of different sensor signals allows for tailoring of the overall $\bm{\mathrm{B_{sense}}}$.
This is particularly useful for applications that benefit from spatial selectivity, either for imaging or rejection of background magnetic noise \cite{Yi2009_Portable}.
For example, a pair of magnetometers is often arranged as a gradiometer: the first sensor is close to a sample of interest, the second is further away (such that the contribution from the sample is negligible), and one measures the difference in signals between the two.
This helps to reject common-mode noise such as background magnetic fields, while retaining the sensitivity to fields from the sample.
One downside to this particular arrangement is that by adding an additional sensor, one is measuring twice as much uncorrelated noise. If the common-mode noise is not the limiting factor, this arrangement is actually less sensitive than a single sensor. 
%\DB{I disagree with this.  You have two times the signal and two times the noise. But the noise adds under square root. So you get a root two gain in SNR :) } \MCDT{If the noise source is a magnetic field, isn't it correlated to some degree?}
%%\JB{If the second magnetometer measures a signal less than $\sqrt{2} -1$ times the signal of the nearby magnetometer, you end up with less SNR.}

Rather than arranging magnetometers to explicitly measure magnetic field gradients, one can arrange them to maximize the signal from a sample, while still minimizing unwanted signals (such as background fields).
Examples of different arrangements are shown in \autoref{fig:arrays}, for a spherical sample with uniform magnetization.
One approach to designing such `gradiometer' arrangements is to start with the expected field produced by the sample, as in \autoref{fig:arrays}a--f.
When placing sensors, simply choose the sensitive axis to match that of the field from the sample, and ensure that there are equal numbers of sensors pointing along any given direction so that they can cancel out background fields.

Another strategy is to design a magnetometer array by considering the overall $\bm{\mathrm{B_{sense}}}$ that results from adding up the sensing fields from all of the sensors.
For the same detector arrangements as in \autoref{fig:arrays}a--f, \autoref{fig:arrays}g--l shows the resulting $\bm{\mathrm{B_{sense}}}$.
For example, \autoref{fig:arrays}a,g show the simplest case, where we consider either the dipole field produced by the sample, or the dipole field produced by the sensor---keeping them as close as possible and oriented the same way gives the best overlap.
A simple gradiometric arrangement where both sensors are near the sample, but measuring in opposite directions, is shown in \autoref{fig:arrays}b,h.
In the three-dimensional case where the sample is a sphere, the signal is only 3/4 that of the most sensitive case (one sensor above and one sensor below), but has the advantage that it is insensitive to uniform background magnetic fields.
For the `two-dimensional' case where the sample is a long cylinder with transverse magnetization (a common sample geometry), this is an ideal arrangement for two magnetometers.
Such an arrangement has been demonstrated in \cite{eills2023enzymatic}.

Another two-sensor arrangement is shown in \autoref{fig:arrays}c,i.
This is an illustration of the fact that, in general, having magnetometers sensitive along a given axis does \emph{not} mean that one is measuring magnetization in that direction.
It is also worth noting that $\bm{\mathrm{B_{sense}}}$ falls off much faster in the arrangements of \autoref{fig:arrays}h and \autoref{fig:arrays}i than in the single-magnetometer case (\autoref{fig:arrays}g).

The arrangements in \autoref{fig:arrays}d,j and \autoref{fig:arrays}e,k are higher-order gradiometers formed by doubling up the arrangements in \autoref{fig:arrays}b,h and \autoref{fig:arrays}c,l, respectively.
In addition to the improved sensitivity coming from increasing the number of sensors, these arrangements offer better rejection of background fields, and also rejection of lower-order field gradients.

The arrangement in \autoref{fig:arrays}f,l combines the previous two gradiometers and is reminiscent of a Halbach cylinder. 
Just as increasing the number of permanent magnet elements in a Halbach dipole magnet increases the strength and homogeneity of the internal field, while minimizing the external field, such magnetometer arrays are more uniformly sensitive to internal samples, and insensitive to external field sources.

%\DB{Points to make around John's beautiful figure}
%\begin{itemize}
%    \item a) Common and simple detector arrangement. he magnetometer is offset from the source in the direction of the sensitive axis
%    \item A simple gradiometer arrangement; however, the signal is 3/2 of max for the 3D cas. Perfectly ideal for 2 D case. Make a point of the difference between 2D and 3D case
%    \item c) Another gradiometer arrangement. This is a very good illustration that, in general, a magnetometer sensitive along a cerain direction is NOT measuring the magnetization in that direction
%    \item d,e,f) are higher-order gradiometers, meaning that they have better suppression of ambient fields and even gradients
%    \item f) illustrates how Halbach array works; Also, we note that if we start rotating the source dipole in some direction, the magnetic field at a *fixed point* rotates in the *opposite* direction with twice the frequency 
%    
%\end{itemize}

In principle, such detector arrays can be used for image reconstruction as in beamforming analysis for magnetoencephalography \cite{HillebrandMEG}.
For image reconstruction, having magnetometers sensitive to multiple axes is particularly valuable.

\subsubsection{Sample geometry}
\label{Subsec:SampleGeometry}
The magnetic field produced by nuclear magnetization also depends on the shape of the sample.
For a uniformly magnetized sphere, the resulting magnetic field at position $\bm{\mathrm{r}}$ outside of the sample is equal to that of a point dipole (having the same total magnetic moment, $\bm{\mathrm{m}}$) located at the center of the sphere:
\begin{equation}
\bm{\mathrm{B}}=\frac{\mu_0}{4\pi}\left(\frac{3\bm{\mathrm{m}}
\cdot\bm{\mathrm{r}}}{r^5}\bm{\mathrm{r}} - \frac{\bm{\mathrm{m}}}{r^3}\right)\,.
\label{eq:dipolefield}
\end{equation}
%where $r=|\bm{\mathrm{r}}|$.
Noting the $r^{-3}$ dependence in \autoref{eq:dipolefield}, one might come up with the idea of placing a sensor \emph{inside} the sample, given that the field inside a uniformly magnetized sphere is $\bm{\mathrm{B}}=2\mu_0 \bm{\mathrm{M}}/3$, uniform over the volume of the sphere \cite{Griffiths2023}.
%\DAB{[references, e.g. Purcell]} \AT{Add reference to Griffiths book, 5th ed. Ch. 6.2}
The reason why this does not work is related to the so-called sphere of Lorentz.
The idea is that actually measuring the field inside the sample necessarily requires removing some of the magnetized material, and removing this material has an effect on the magnetic field.
Making use of the superposition principle, 
%\DAB{(is it necessary to use this ambiguous term?)}
the trick is to subtract the magnetic field of a `removed sphere' from that of the larger magnetized volume.
For the case of a uniformly magnetized sphere:
\begin{equation}
\bm{\mathrm{B_M}} = \bm{\mathrm{B_{sample}}} -\bm{\mathrm{B_{sphere}}} = 2\mu_0 \bm{\mathrm{M}}/3 -2\mu_0 \bm{\mathrm{M}}/3=0\,,
\label{eq:Lorentzsphere}
\end{equation}
so anywhere you measure inside of a magnetized sphere, the magnetic field (from the sphere) is zero. 
%\DB{One may say here: why cut out a sphere and not some other shape?}
%A useful consequence of this is that, for uniform spherical samples, the sample magnetization does not couple to the spins inside the sample.

For an infinitely long cylinder magnetized transverse to its symmetry axis, the magnetic field within a small spherical cavity 
%(the ``sphere of Lorentz'') 
inside the cylinder is $\bm{\mathrm{B_M}}=\mu_0 \bm{\mathrm{M}}/2 - 2\mu_0 \bm{\mathrm{M}}/3= -\mu_0 \bm{\mathrm{M}}/6$, with nuclear magnetization $\bm{\mathrm{M}}=n P \hbar \gamma_I \bm{\mathrm{I}}$, where $n$ is the number density of nuclear spins, 
$P$ is the spin polarization, and $\gamma_I$ is the nuclear gyromagnetic moment of spins $\bm{\mathrm{I}}$.
For highly magnetized samples, peak shifts arising from this bulk magnetization effect have been used to estimate sample polarization \cite{Eichhorn2022}.

The magnetic field outside of a uniformly transversely magnetized cylinder is
\begin{equation}
% \bm{\mathrm{B_M}}=\frac{\mu_0 \bm{\mathrm{M}}a^2}{2r^2}\left(\cos \theta \, \hat{\bm{\mathrm{r}}}+\sin \theta \, \hat{\bm{\mathrm{\theta}}}\right)\,,
\bm{\mathrm{B}}=\frac{\mu_0 a^2}{2}\left(\frac{2\bm{\mathrm{M}}\cdot\bm{\mathrm{r}}}{r^4}\bm{\mathrm{r}} - \frac{\bm{\mathrm{M}}}{r^2}\right)\,,
\end{equation}
% \DB{Please define $\theta$ or, better, rewrite in the same spirit as Eq.\,(69)}
where 
%$\theta$ is the angle relative to the magnetization direction and 
$a$ is the radius of the cylinder.
If it is possible to place a sensor at a distance $r<\sqrt{3}a$, then it is still preferable to measure the signal outside rather than inside the sample.

\begin{figure}
\centering
	\includegraphics[width=\columnwidth]{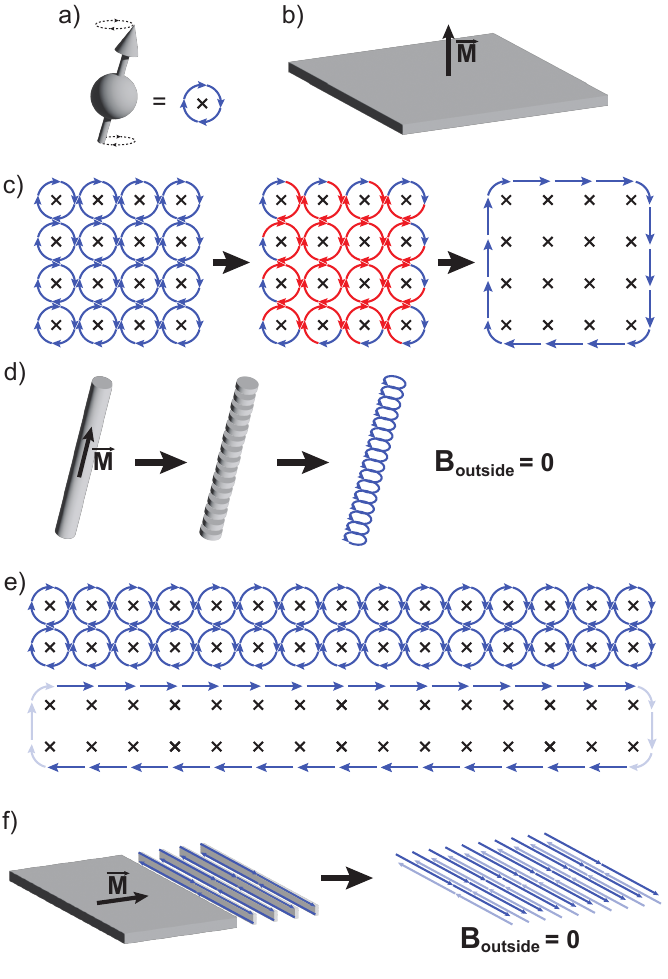}
	\caption{(a) A point dipole---such as a nuclear spin or an infinitesimal volume of magnetized material---can be considered equivalent to a small loop of current.
    (b) A thin slab of magnetized material, magnetized normal to the plane.
    (c) Representing a magnetized slab by the sum of many small current loops, we see that most of the currents cancel out, leaving only a finite current density along the edge of the slab. As the geometry approaches that of an infinite plane, the resulting magnetic field at any point outside of the material approaches zero.
    (d) A cylinder magnetized along its axis
    %like that in (c) 
    can be broken down into a sum of thin discs of material magnetized normal to the plane of the disc, exhibiting a field equivalent to that produced by a circumferential current loop. The sum of these current loops is equivalent to a long solenoid. As the length of such a solenoid approaches infinity, the external magnetic field goes to zero.
    (e) For an infinitely long strip of material, magnetized normal to the plane of the strip, the field produced is equivalent to that of a pair of currents in opposite directions on either side of the strip.
    (f) A slab of magnetized material can be treated as the sum of infinitesimal strips of material, magnetized normal to the plane of the strip. The field produced is equivalent to that produced by a pair of opposing current sheets. In the limit of an infinitely broad slab of magnetized material, the external magnetic field goes to zero.}
	\label{fig:MagnetizedPlane}
\end{figure}

For an infinitely long cylinder magnetized along its symmetry axis, the magnetic field within a small spherical cavity is 
$\bm{\mathrm{B_M}}=\mu_0\bm{\mathrm{M}} - 2\mu_0 \bm{\mathrm{M}}/3= \mu_0 \bm{\mathrm{M}}/3$.
Note that the magnitude of this field is twice that of the cylinder with transverse magnetization.

It is important to note that for an infinitely long cylinder magnetized along the cylindrical axis, the field outside of the cylinder is zero.
A graphical explanation for this is provided in \autoref{fig:MagnetizedPlane}.
We can represent a point dipole such as an infinitesimal volume of magnetized material as a small loop of current, where the current is proportional to the magnetization.
Similarly, the field of any two-dimensional object magnetized normal to the plane is equivalent to the field produced by a sum of such current loops.
As shown in \autoref{fig:MagnetizedPlane}c, that sum is equivalent to a larger current loop, running along the perimeter of the object.
Because a magnetized cylinder is equivalent to a sum of magnetized discs, the field produced is equivalent to field produced by the sum of the corresponding current loops.
For our infinite longitudinally magnetized cylinder, the field is therefore equivalent to that of an infinite solenoid, which is zero outside.
Many ZULF-NMR researchers made the mistake of trying to measure signals with such a sample geometry.

Another interesting geometry is a slab of material approximating an infinite plane, magnetized normal to the plane.
Again, following the approach of \autoref{fig:MagnetizedPlane}, we see that the field of such a slab of material is equivalent to that of a current loop running around its perimeter.
Recall that the field produced by a loop of current with radius $R$ is
$B = \mu_0 I/2R$ at the center of the loop.
As the radius tends to infinity, the field measured in the center tends to zero.

For a slab of material with in-plane magnetization, we can follow a similar thought process to that for a magnetized cylinder.
In this case, the slices normal to the magnetization consist of thin strips of material, and the field of each slice is equivalent to that of two opposing lines of current above and below (neglecting the turns at the edges, which are assumed to be far away), as in \autoref{fig:MagnetizedPlane}e.
Adding these up, we see that the field of an infinitely broad slab of material with in-plane magnetization is equivalent to that of two opposing sheets of current.
For an infinite sheet of current with linear current density $J$, the field is $B=\mu_0 J/2$, parallel to the sheet and perpendicular to the direction of the current.
The sum of the fields from the two opposing current sheets is then zero.

Note that the result for an infinite slab is independent of magnetization direction, as any magnetization can be expressed as a sum of the in-plane and out-of-plane cases, which both result in zero external field.
This is particularly relevant, among other cases, for microfluidic NMR measurements using NV$^-$ centers in diamond, where one might be tempted to produce a wide fluid channel in order to integrate over a larger area---in reality, signals will only come from the edges of the channel, so multiple thin channels would produce more signal than one broad channel.
Note also that magnetization along the axis of a long channel looks like an infinite magnetized cylinder; it is therefore best to arrange the experiment such that the magnetization to be detected is transverse to the axis of the channel.

\subsubsection{Applied magnetic field}

%\JB{Add something about how the magnetization projection affects selection rules?} \DB{Yes,  if there is something intelligent to say...}

In section \autoref{Sec:Theory}, we assumed that the sensor was sensitive in the direction of sample magnetization, and was displaced along the same axis.
More generally, we can replace the magnetization vector of \autoref{eq:MagnetizationMz1} with the field produced by the sample, $\bm{\mathrm{B_M}}$.
Then the field direction relative to $\bm{\mathrm{B_{sense}}}$ defines selection rules, as discussed in Sec.\,\ref{Subsubsec:ZeemanSelectionRules}.
Depending on the geometry of the sample and arrangement of the detector(s), one may measure spectra with $\Delta m_F = 0$ or $\pm 1$, or a combination of the two.

\section{Extensions of ZULF NMR}
\label{Sec:Extensions}

\subsection{Two-dimensional spectroscopy}
\label{Subsec:2D}

\begin{figure}
\centering
	\includegraphics[width=0.8\columnwidth]{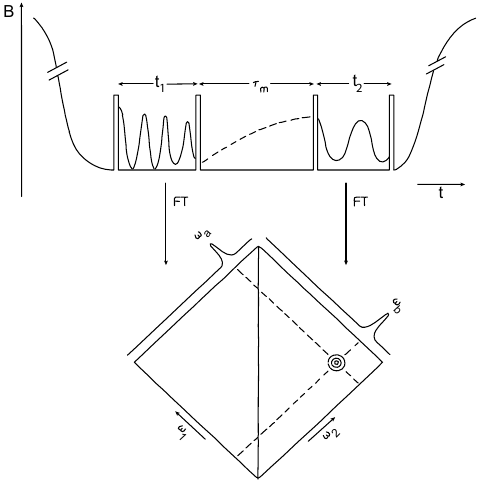}
	\caption{Schematic diagram of the time-dependent field profile used for early two-dimensional zero-field exchange experiments to monitor spin diffusion. Reprinted from reference  \cite{JarvieThesis} with the author's permission.} 
% reported in reference \cite{THAYER1986}.  \MCDT{This is not the right reference! The figure does not come from that paper.  Does anyone know which is the correct one?} \AT{While I do not know where the version used here comes from, the same figure appears as Fig. 4.5 in Thomas Jarvie's thesis (\url{https://escholarship.org/content/qt8qf4t5dg/qt8qf4t5dg.pdf}), with no journal reference being given there, as far as I could see.}}
	\label{fig:2DSchematic}
\end{figure}

Just as for high-field NMR, ZULF NMR spectroscopy can be extended to two (or more) dimensions by introducing an additional evolution time into the experiment.
%If necessary:
%1. Preparation into a chosen coherence
%2. t1 evolution
%3. mixing
%4. t2 evolution
Figure \ref{fig:2DSchematic} shows an example of an early two-dimensional\footnote{Technically, as the detection phase of the experiment was carried out in a point-by-point fashion by shuttling the sample into a high-field spectrometer, one could argue that this was actually a three-dimensional experiment.} zero-field NMR experiment \cite{JarvieThesis}, designed to measure spin diffusion.
This experiment showed that high `solution-like' resolution is achievable as line broadening due to anisotropic powder averaging is absent, given that all crystallite orientations are equivalent because of the absence of a preferred direction at zero magnetic field \cite{Suter1987}.

\begin{figure*}[t]
\centering
	\includegraphics[width=16cm]{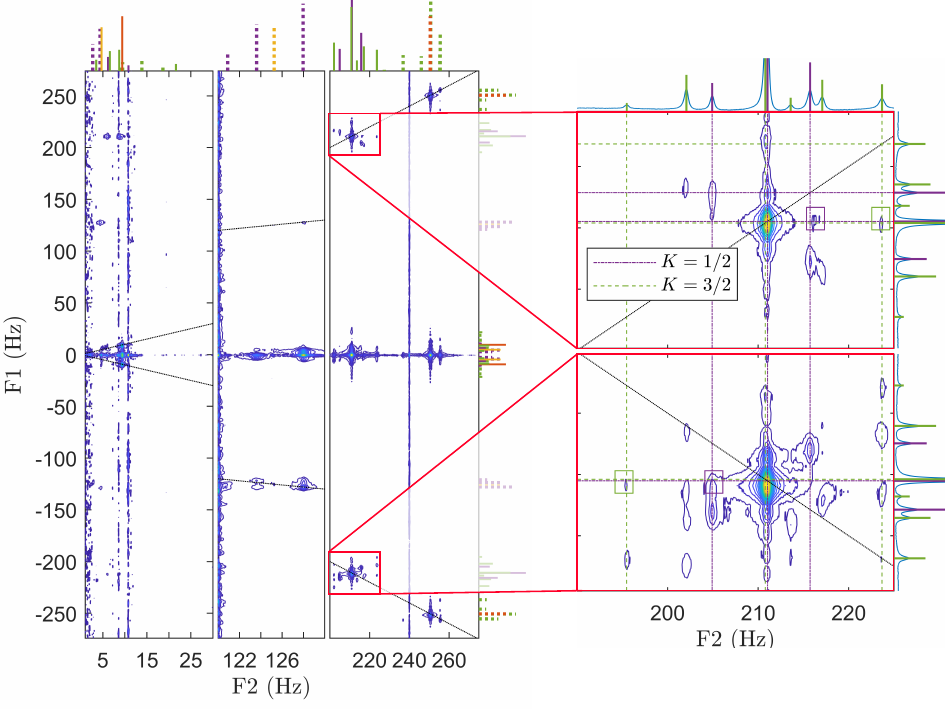}
	\caption{A two-dimensional ZULF $J$-spectrum of a mixture of [1-${}^{13}$C] and [2-${}^{13}$C]ethanol, with cross-peaks appearing only between peaks corresponding to the same isotopomer. The zoomed-in portion of the [1-${}^{13}$C]ethanol spectrum shows how peaks overlapping in a 1D spectrum can be resolved in the 2D spectrum if the peaks correspond to different angular-momentum manifolds. Reprinted from reference \cite{Sjolander2020}, Copyright (2020), with permission from Elsevier.
	}
	\label{fig:2DEthanol}
\end{figure*}

Zero-field multiple-quantum $J$-spectroscopy was demonstrated in \cite{Sjolander2020}, producing two-dimensional spectra like the one shown in Fig.\,\ref{fig:2DEthanol}. 
More specifically, Fig.\,\ref{fig:2DEthanol} is an example of correlation spectroscopy for a mixture of the two ${}^{13}$C isotopomers of ethanol.
As expected for correlation spectroscopy, cross-peaks are present only between peaks belonging to the same isotopomer.
Furthermore, as expected in ZULF NMR, cross-peaks are only present for resonances belonging to the same spin manifold, differentiated in this case by the total angular momentum, $K$, of the methylene and methyl protons.
For example, the right-hand side of Fig.\,\ref{fig:2DEthanol} zooms in on one multiplet of the [1-${}^{13}$C]ethanol spectrum, dominated by a large peak at 211\,Hz.
This peak in fact consists of two overlapping resonances belonging to the $K=1/2$ and $K=3/2$ manifolds, which are not resolved in the one-dimensional spectrum, but can be identified in the two-dimensional spectrum by the presence of distinct cross-peaks.

%\JB{Add one paragraph briefly summarizing other nice 2D stuff from Tobias?
%For more details, see \cite{2DZULFBook}.
%}

In addition to these examples of 2D NMR in the ZULF regime, experiments at the boundary between the low- and ultralow-field regimes have been performed in SQUID-detected NMR at microtesla fields \cite{Shim2014_2D,Lee2023_2D,Buckenmaier2019MQ}. An in-depth review of two-dimensional methods for ZULF NMR, including field-cycling experiments between ZULF and high field, is given in \cite{2DZULFBook}.

\subsection{Imaging}
% \DB{Let us all write this collectively by contributing paragraphs; for now please consider the whole section colored}

% \DB{(do not forget long-lived spin states) Singlet‐contrast magnetic resonance imaging: unlocking hyperpolarization with metabolism. Angew. Chem \cite{eills2021singlet}---unfortunately, this is not ZULF...}
Imaging is one of the two major subdivisions of the NMR field, along with spectroscopy primarily discussed in the previous sections. Indeed, ZULF imaging, while being well-motivated and holding significant promise, is still in its infancy, with most of the imaging work so far done not in the ZULF conditions, albeit using ZULF-compatible equipment.

A significant body of work was devoted to imaging at fields on the order of 100\,$\mu$T utilizing SQUID detection \cite{McDermott2004,Lee2005,Myers2005Concomitant,SQUID-Pepper-Can,Zotev2008MRIMEG,Zotev2008Parallel,zotev2010microtesla,Inglis2013MRI,lee2019squid,HILSCHENZ2019138,de1999nmr,clarke2007squid} with a systematic account presented in a monography \cite{kraus2014ultra}. Of particular practical importance is MRI at low fields (1--80\,mT) \cite{sarracanie2015low,deoni2022development,kimberly2023brain}. MRI with atomic magnetometer detection has been already demonstrated at such fields \cite{hori2022magnetic}.

%Detection of nuclear signals for MRI purposes at such fields with atomic magnetometers was demonstrated \cite{hori2022magnetic}.

Recently, $^{13}$C imaging of 50\,mM [1-$^{13}$C]pyruvate solution was conducted at 120\,$\mu$T \cite{kempf202413c}, about twice the Earth’s magnetic field, utilizing two different SABRE variants: SABRE-SHEATH and LIGHT (Low-Irradiation Generates High Tesla)-SABRE. The three-dimensional images of a phantom were obtained using a SQUID detector with submillimeter resolution \cite{kempf202413c}. This work demonstrates what kind of experiments can be done \textit{in vivo} using similar methodology.

The important problem of the so-called concomitant gradients needs to be dealt with at low fields. Maxwell's equations tell us that the divergence of magnetic field is zero. This means that if we apply a gradient, say, $\partial B_z/\partial z$, there necessarily appear the $\partial B_x/\partial x$ and $\partial B_y/\partial y$ gradients so that the sum of all the three terms is zero. At high field, the transverse gradients can usually be ignored as the Larmor frequency depends on the total magnetic field and, to first order, the latter is the sum of the leading field and the $z$ component of the gradient field. However, the transverse gradients become important when the product of the applied gradient and the size of the system becomes comparable to the (small) leading field and the image suffers significant distortions \cite{Myers2005Concomitant}. Various 
%ingenious 
techniques have been developed to mitigate the concomitant-gradient problem \cite{Kelso2009Low-field_MRI}. 

In the case of detection using optical atomic magnetometers, one early imaging demonstration \cite{xu2006magnetic_PNAS} used remote detection and phase (rather than frequency) encoding with point-by-point detection. While painfully slow, with typical image-acquisition times of hours, the technique has nevertheless resulted in imaging of fluid-flow patterns inside porous metals \cite{Xu2008b} and reached millimeter-scale spatial resolution \cite{xu2008submillimeter}. An imaging modality combining prepolarization with a permanent magnet and encoding was implemented \cite{Crawford2008Fluid}, which was, in fact, magnetic imaging without resonance.

Current efforts to develop ZULF MRI are focused on several directions \cite{barskiy2023possible}, including imaging with sensor arrays (similar to how one records magnetoencephalograms, see Sec.\,\ref{Subsubsec:Gradiometers}) but with an important additional capability of providing spectroscopic information in addition to spatial information. 

Another approach was dubbed 
`sweet-spot-scanning', where one takes advantage of the fact that a characteristic ZULF spectrum vanishes at nonzero field. The zero-field spot can be produced with coils in the anti-Helmholtz configuration. The `sweet spot' can be moved with respect to the sample to produce its molecular image \cite{hegyi2014nanodiamond}.
To speed up imaging, one can consider using arrays of zeros of the field for parallel imaging.

Finally, as concomitant gradients are a problem at ultralow fields but not at higher fields, to achieve spatial encoding, one can apply a gradient of the `dc' magnetic-field pulse used to initiate the ZULF-NMR signal. 

An alternative approach for high-field hyperpolarization-enhanced imaging is to use a singlet state formed by coupled spins prepared in the process of hyperpolarization. In this approach, the singlet state itself is a nonmagnetic singlet state. In this state, the molecules do not produce an observable NMR signal, but if the singlet state is broken, the hyperpolarization can be observed. The advantage is that the singlet state can be long-lived, which means that the signal can be imaged over a longer time \cite{tayler2012direct,marco2013hyperpolarized,pileio2013recycling}. 

This approach was demonstrated for a DNP-polarized $^{13}$C-$^{13}$C singlet pair in maleic acid derivatives, using rf pulse sequences to `release' the singlet order gradually over time \cite{pileio2013recycling}.
Another variant called singlet-contrast imaging relies on a chemical reaction of the contrast agent to break the singlet state and release observable hyperpolarization-enhanced NMR signals \cite{eills2021singlet}. This was demonstrated with a parahydrogen-polarized $^1$H-$^1$H singlet state in fumarate, which was metabolised to malate by an enzyme to break the $^1$H chemical equivalence and allow subsequent $^1$H imaging.
At high field, these singlet-state approaches only work for homonuclear spin pairs, where the two nuclei do not have a Larmor frequency difference. Under ZULF conditions it should be possible to extend this approach to heteronuclear spin pairs \cite{Emondts2014}, which may significantly broaden the range of potential contrast agents for singlet-contrast imaging.

\subsection{Relaxometry}
\label{Subsec:Relaxometry}

Throughout the history of NMR, the magnetic field dependence of nuclear spin relaxation rates has been extensively studied and widely applied to elucidating topics ranging from protein and polymer dynamics, to contrast agents for \textit{in vivo} MRI (including SQUID-detected ULF MRI \cite{Clarke2007ARBE}), to food spoilage \cite{AugustineTomatoes2019} and petrology \cite{Kowalewski2019book,Kowalewski2020}.  The interest in relaxometry is largely motivated by the utility of relaxation rates in the determination, ideally quantitative, of physical and chemical properties of a system. These may include, for instance, translational or rotational diffusion rates of molecules in solution, or concentrations and relaxivities of dilute paramagnetic species \cite{Michalak2011MRM,Bodenstedt2021,Put2021}.  
%The common explanation of the field dependence of spin-relaxation rates involves the parameterization of each responsible relaxation mechanism with a `correlation time' $\tau_{\rm c}$.  These correlation times allow each mechanism to be weighed relative to the nuclear Larmor period to determine which ones most efficiently relax the nuclear spins at a given field $B$. \JE{Or: 
%The common explanation of the field dependence of spin relaxation rate involves the correlation time ($\tau_{\rm c}$) associated with a physical mechanism that can induce relaxation.

\subsubsection{$T_1$ relaxometry of uncoupled nuclear spins}

Nuclear spin-relaxation rates in the case of uncoupled spins can be quantitatively described by 
parameterizing each of the relaxation mechanisms with a corresponding correlation time $\tau_{\rm c}$. 
%Correlation time refers here to a characteristic timescale of molecular motion. 
For gas-phase molecules, the relevant process is rotational-state relaxation with a correlation time related to the interval between collisions
(on a nanosecond timescale at room temperature and atmospheric pressure). Correlation times of molecules in liquids are associated with molecular reorientation and are on the picosecond to nanosecond time scale. %\SP{This reads as there is only one. But the next sentence talks about probing other mechanisms. We do talk abuot only the coorelataion times that are associated with rotation, right? So we should state it more clearly.}.
%The correlation times associated with different relaxation mechanisms allow one to
%allow each mechanism to be weighed relative to the characteristic nuclear evolution times to 
%determine which of the possible mechanisms dominate the relaxation of nuclear spins at a given field $B$.
%, and its size relative to the nuclear Larmor period. 
Processes with correlation times up to the inverse Larmor frequency, $\tau_{\rm c} \approx 1/|\gamma B|$, affect relaxation rates most. 
When magnetic fields are large enough so that $\tau_{\rm c} \gg 1/|\gamma B|$, the collisional relaxation mechanism becomes ineffective and other mechanisms dominate relaxation. 
%In this regime, 
%changes in the correlation time affect the relaxation rates only weakly. \MCDT{What was written here before?  I do not quite agree with the `weakly' part of this statement.  In the motional narrowing regime the relaxation rates are usually proportional to $\tau_c$, so the dependence is not necessarily weak.  For example, simply changing the viscosity of the solvent can substantially change the relaxation rates.}
Therefore, low-field relaxation studies are valuable for gathering information about slow processes, for instance the intermolecular dynamics of intermediate- to large-size molecules.  We note that a not-too-common exception exists where chemistry allows \textsuperscript{1}H/\textsuperscript{2}H isotopic replacement to be performed; in this case, the frequency dependence of relaxation can be probed as a result of changes in gyromagnetic ratios between the two isotopes rather than changes in magnetic field \cite{Lankhorst1982}.
%Processes with short correlation times, $\tau_{\rm c} \ll 2\pi/|\gamma B|$, are the most important ones to consider.  In the case where $\tau_{\rm c}$ is long compared to the nuclear evolution period ($\tau_{\rm c} \gg 2\pi/|\gamma B|$, for example, in the high-field regime), the induced relaxation  is relatively weak because when seen from the frame of nuclear spin evolution the interaction is efficiently averaged out. \JE{I don't follow this.. I thought short $\tau_c$ meant the mechanism is averaged out (but it still affects spin relaxation), and long $\tau_c$ means in the frame of spin precession the mechanism is static.} Low-field relaxation studies are therefore a necessity to gather information about slow correlation processes, for instance, intermolecular dynamics.
 
It is most common to determine $\tau_{\rm c}$ via a series of relaxation measurements across several decades of magnetic field, say, from millitesla to tesla, and fit the resulting `nuclear magnetic relaxation dispersion' (NMRD) curve to dynamics models \cite{KimmichFieldCycling1979,Kimmich2004,KimmichFieldCyclingNMR-RSCbook}. 
%\SP{It seems we are only citing FFC papers here, while the statement is more general. I suggest citing our papers in low-field/ULF relaxoemtry.}.\MCDT{The "It is most common..." phrase refers to what is conventionally done.  Most people are not using the kind of ULF relaxometry you refer to.} \DB{Michael: Kindly add the references and resolve this discussion!}  \MCDT{I looked into this, and the book already cited as \cite{KimmichFieldCyclingNMR-RSCbook} gives plenty of examples using nonFFC techniques.  So I think this covers the concern already.  So far, there is no review paper on ULF relaxometry, right?  The specific papers you mention should be cited below, in this paragraph.}
The range of magnetic fields used translates into the range of correlation times that can be probed.  Various techniques have been developed to improve the convenience, speed and consistency of NMRD experiments achievable on a single instrument.  One of the techniques in use is electromagnetic fast-field-cycling (FFC) pioneered in the 1990s to show adsorption correlation times of several microseconds in microporous glasses \cite{Stapf1996,Kimmich1994}. More recently, FFC was extended into the 1--10 \si{\micro\tesla} regime to investigate MRI contrast at such fields \cite{Anoardo2003,Kresse2011}.  Lately, utilization of atomic magnetometers to detect sub-kilohertz NMR signals enabled FFC measurements to be made inside magnetic shielding \cite{Ganssle2014}, so that NMRD data can extend the accessible range of correlation times into the millisecond range \cite{Bodenstedt2021}. Here the NMRD measurements can also be combined with spectrally resolved ZULF-NMR detection of several chemical species at the same time, to facilitate analyses of dynamics in mixtures \cite{Tayler2018}.  
A remaining challenge, however, is to understand the applicability of current relaxation models at the lowest fields.  The authors of reference \cite{Bodenstedt2021} measured $T_1$-NMRD for n-octane (C$_8$H$_{18}$) confined in the $\sim$5\,nm diameter pore spaces of a commercial $\gamma$-alumina sample, showing relaxation rates of $1/T_1 \sim 30$\,s$^{-1}$ at $^1$H Larmor frequencies below 10\,Hz. Wangsness--Bloch--Redfield (WBR) and other conventional NMR relaxation theories cannot strictly describe these scenarios since several key assumptions are not fulfilled, including the assumptions of $T_1 \gg \tau_c$ %\DAB{this condition is always fulfilled, $\tau_c$ is never on the order of sec} \MCDT{I disagree.  It depends on the relaxation mechanism, which is the whole point.  
(for surface jumps in porous materials %(the example given in the preceding sentence), 
$\tau_c$ can be the order of 10 ms and so can $T_1$) 
%, so the condition is not met} 
and $\tau_c \ll 1/|\gamma B|$ in the Redfield model \cite{Bodenstedt2021}. %\DAB{I don't fully understand this; I think instead of $B$ here should be RMS of the fluctuating field amplitude. Tbh I think Redfield theory is overall fine for ZULF, one just has to use appropriate bases states. So we may want to rewrite this text a bit...} \MCDT{I would not rewrite}.  
Nevertheless, there are several known examples where Redfield theory does seem to work in cases where it should not, for instance reproducing the predictions of stochastic Liouville models of relaxation at low field \cite{Kruk2022RSCbook}.  It will be interesting to see the progression of these theories and their predictions of whether correlation times accessible by measurements in the ZULF regime are limited by spin-diffusion dynamics or chemical dynamics.  For now, there remains much to be learned by the application of ZULF-NMRD techniques beyond proof of principle, for instance, to microporous materials of industrial importance \cite{Tayler2018} or relevant biomolecules \cite{Put2021}, biological fluids (including, for example, blood \cite{Alcicek2023relaxometry}) and plants \cite{Fabricant2024FrontiersPlantSci}. %\DB{The positioning of the references looks strange. Is this intended?}

\subsubsection{Relaxometry of coupled nuclear spins} \label{sec:relaxometry}

\begin{figure}
\centering
	\includegraphics[width=\columnwidth]{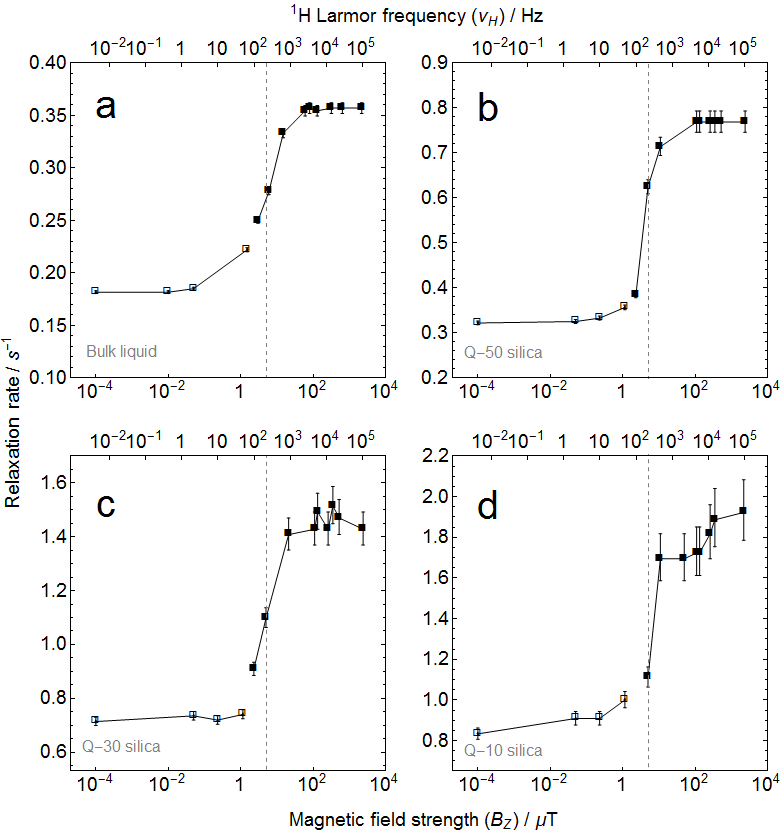}
	\caption{Relaxation rates of spin order originating from \textsuperscript{1}H magnetization in [\textsuperscript{13}C]-formic acid (\textsuperscript{13}CHOOH) samples: (a) pure liquid, (b)--(d) pure liquid inside silica pores of mean diameter (b) 50\,nm, (c) 30\,nm, and (d) 10\,nm.  The dotted vertical line in each plot indicates the magnetic field where the \textsuperscript{1}H Larmor frequency equals the \textsuperscript{1}H-\textsuperscript{13}C scalar coupling constant, around 220\,Hz. Above and below this field are the high-field and ZULF relaxation regimes, respectively.  The slower relaxation within the ZULF regime results from the long-lived singlet state of formic acid, which outlives longitudinal magnetization by a factor of between 2 and 3 even in the porous confinement.  Dependence of the vertical error bars on the magnetic field is due to a systematic uncertainty in the estimation of the relaxation times, as described in the original paper.  Reprinted from reference \cite{Tayler2018}, Copyright (2018), with permission from Elsevier. 
	}
	\label{fig:FAporous}
\end{figure}

So far, our discussion of ZULF relaxometry has focused on uncoupled $^{1}$H spin systems.  In heteronuclear spin systems, there are some other field dependencies to review, because in much the same way %that we have shown 
that the coherent behavior of spins is different between ZULF and high-field NMR regimes, incoherent spin evolution or relaxation in ZULF is also strongly influenced by the spin eigenbasis.  One case is heteronuclear spin singlet ($F=0$) eigenstates in the ZULF regime (homonuclear singlet-state relaxation has been extensively studied over the past decade by many groups \cite{LongLivedSpinOrderBOOK2020}).  Examples include the singlet-pair state of \textsuperscript{1}H\textsuperscript{13}C in [\textsuperscript{13}C]-formic acid (H\textsuperscript{13}COOH) \cite{Emondts2014,Tayler2018}, the E-internal rotational symmetry state of \textsuperscript{13}CH$_3$ groups \cite{Ledbetter2011near,Dumez2015}, and the pseudo-singlet eigenstates of \textsuperscript{13}C-substituted aromatic compounds, such as [\textsuperscript{13}C]-benzene \cite{Blanchard2013JACS} and [2--\textsuperscript{13}C]-benzoquinone \cite{Sheberstov2021}.  Such singlet states are immune to relaxation by spin interactions that are perfectly correlated across the spin system, for example, the intramolecular \textsuperscript{1}H-\textsuperscript{13}C or intermolecular dipole--dipole couplings \cite{Kiryutin2020}.  Pseudo-singlet eigenstates may also persist with long lifetimes in nonzero field, in the ultralow-field regime, where the Zeeman interaction is small compared to $J$-couplings \cite{Tayler2018,Zhukov2019} (see \autoref{fig:FAporous}).  These states may offer a potential `safe haven' for hyperpolarized spin order, as discussed for [1-\textsuperscript{13}C]-fumarate in reference \cite{Eills2021Singlet-Contrast_MRI}.

As total angular momentum states are eigenstates of nuclear spin Hamiltonians for molecules in ZULF (see \autoref{Sec:XAn}), detectable coherences may involve transitions to and from singlet ($F=0$) eigenstates. These can exhibit decoherence-free properties that are partly responsible for the ultra-narrow spectral linewidths observed in ZULF NMR, on the order of millihertz \cite{Zhukov2018}.

\begin{figure}
\centering
	\includegraphics[width=\columnwidth]{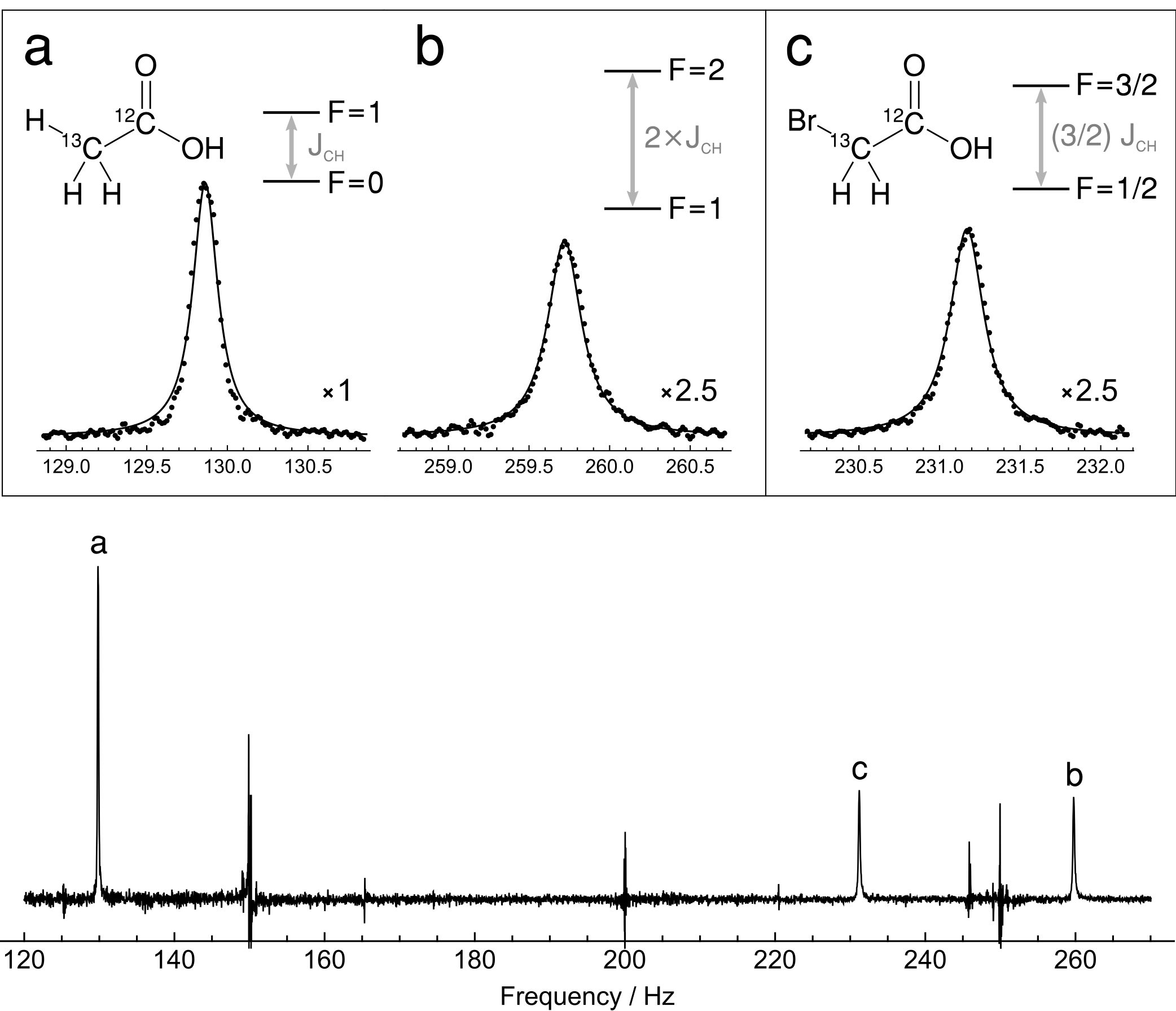}
	\caption{The zero-field $J$-coupling NMR spectrum of a mixture of [2-\textsuperscript{13}C\textsubscript{1}]-acetic and bromo-[2-\textsuperscript{13}C\textsubscript{1}]-acetic acids.  This shows an example of halogen self-decoupling in the ZULF regime, where scalar couplings between \textsuperscript{13}C and Br are slow compared to the Br relaxation rate, with the result that the presence of Br does not broaden the resonance line.  Reprinted from reference \cite{Tayler2019}, Copyright (2019), with permission from Elsevier.}
	\label{fig:bromoaceticacid}
\end{figure}

A separate important consideration for ZULF NMR is the joint relaxation of spins that are strongly coupled, where one nucleus can act as a relaxation sink for other spins in the system \cite{ivanov2008high,korchak2010high,chiavazza2013earth,ALTENHOF2023107540}.
This is commonly seen for directly-bonded \textsuperscript{1}H-\textsuperscript{13}C spin pairs where the \textsuperscript{1}H spin dominates the \textsuperscript{13}C $T_1$ relaxation, 
%\DAB{(We need references here or rewriting to remove focus from 1H-13C systems)}
%\MCDT{I don't see a major emphasis being placed here on 1H-13C}
but is particularly detrimental when quadrupolar nuclei with sub-millisecond relaxation times are part of the spin system.
Nuclear spins with a high quadrupole moment, including the halogens \textsuperscript{35}Cl, \textsuperscript{37}Cl, \textsuperscript{79}Br and \textsuperscript{81}Br, however, do not usually exhibit this effect because they relax so rapidly that they effectively self-decouple from the spin system \cite{Tayler2019}, as indicated by the narrow-line spectra shown in \autoref{fig:bromoaceticacid}.  Low quadrupole-moment nuclei such as \textsuperscript{2}H, \textsuperscript{14}N and \textsuperscript{17}O \cite{Hartwig2011} on the other hand, depending on their position in the molecule, may introduce fast relaxation of coupled \textsuperscript{1}H spins, for example in [2,2,2-d\textsubscript{3}]-ethanol \cite{Tayler2019,Bodenstedt2022jpcl}.  This is unfortunate as \textsuperscript{1}H/\textsuperscript{2}H isotopic substitution is commonplace in high-field solution-state NMR to simplify \textsuperscript{1}H scalar coupling patterns and mitigate intramolecular dipole--dipole relaxation.  To benefit from the same tactic in ZULF NMR it would be necessary to suppress the deleterious effect of the quadrupolar relaxation sink.  A possible solution in some molecules may be to perform the ZULF-NMR experiment at a magnetic field where \textsuperscript{1}H and quadrupolar spins (e.g., \textsuperscript{14}N) are weakly coupled, but other spins (e.g., \textsuperscript{1}H and \textsuperscript{19}F) remain strongly coupled. 
An alternative may be to dynamically decouple \textsuperscript{2}H spins using suitable pulse sequences \cite{Bodenstedt2022jpcl}.

The joint-relaxation effect is not only problematic for detecting ZULF-NMR spectra, but also for driving coherent spin-system evolution in the ultralow-field regime with pulse sequences or field manipulations. This is particularly relevant for PHIP/SABRE experiments, where it is common to use ultralow fields to transform \textsuperscript{1}H singlet order into magnetization of a %heteronuclear
nonproton spin. Partial or full deuteration of the target molecule can be highly beneficial for enhancing polarization levels outside of the strong-coupling regime as this reduces dipolar relaxation during the PHIP/SABRE process, but can cause problems for the polarization transfer step in the strong coupling regime. Current experimental practice shows that it is often better to avoid the strong coupling regime altogether by carrying out the polarization transfer step at low field (i.e., not ultralow field, so that the \textsuperscript{2}H spins are not strongly coupled to spin-1/2 nuclei), using a resonant AC-pulse sequence to drive the transfer. This was demonstrated for polarizing the $^{13}$C spin in [1-$^{13}$C]-succinic acid-$d_2$ by applying a $B_0$ field on the order of 10\,µT and using pulse sequences such as WOLF (Weak Oscillating Low Field) \cite{Dagys2021wolf}, STORM (Singlet-Triplet Oscillations through Rotating Magnetic Fields) \cite{Dagys2022pccp}, or a combination of the two \cite{dagys2022deuteron}. Similar approaches from the technical standpoint employing alternating magnetic fields have been demonstrated for SABRE-polarized acetonitrile \cite{lindale2024multi}.

\begin{figure}
    \centering
	\includegraphics[width=\columnwidth]{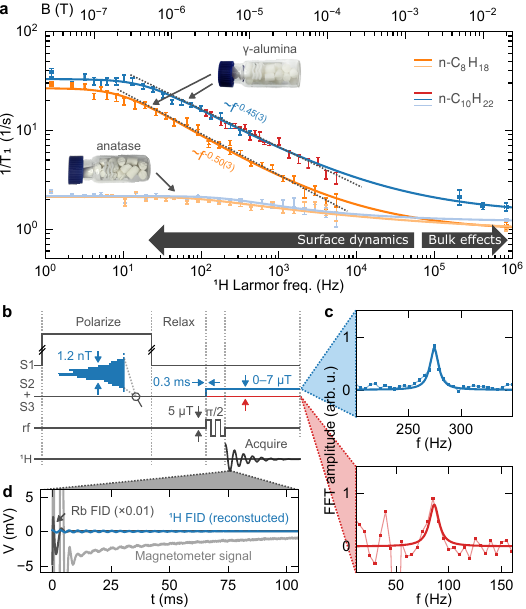}
	\caption{
 %\DB{Please add a proper caption!} \DAB{Done!} 
    (a) Field dependence of \textsuperscript{1}H magnetization decay rates $1/T_1$ for n-octane and n-decane in porous $\gamma$-alumina and anatase titania. For n-decane in $\gamma$-alumina, colors indicate the NMR detection field, using fast-field switching (blue) or fixed field (red). These correspond to the colors also used in (b) and (c);
    (b) Pulse sequence details. The signal is detected with an atomic magnetometer; 
    (c) Frequency- and (d) time-domain \textsuperscript{1}H-NMR signal for n-decane in $\gamma$-alumina after relaxation at \SI{2}{\micro\tesla} (85 Hz Larmor frequency). The reconstructed time-domain signal (blue curve) equates to the Lorentzian line shape fitted in (c). Reproduced from reference \cite{Bodenstedt2021} under terms of the Creative Commons CC-BY license. 
	}
%	\label{fig:GDM}
\end{figure}

\subsection{Partial molecular alignment and orientation} 
\label{Subsec:Partial}
%\AT{This text was originally in %\autoref{Subsec:Untruncated}.}
% \KS{I will review this section.} \DAB{Per Kirill's request, DAB has done this.}

Independently of truncation by a dominant magnetic field (see \autoref{Subsec:Untruncated}), terms of the nuclear spin Hamiltonian can be suppressed by isotropic molecular tumbling. Because of this, in typical liquid-state NMR experiments, the magnetic dipole--dipole coupling term, Eq.\,\eqref{eq:HamDip}---a valuable source of molecular information---is averaged to zero. However, under partial alignment, this information is recovered in the form of the so-called residual dipolar couplings (RDCs).

ZULF NMR measurements of RDCs were performed in [2-$^{13}$C]-acetonitrile aligned in stretched polyvinyl acetate gels \cite{Blanchard2015}. These experiments constituted the first investigation of dipolar couplings as a perturbation on the spin--spin $J$-coupling in the absence of applied magnetic field. Future work will address the temperature- and electric-field-dependent phase transitions in liquid crystals.

To observe RDCs or other anisotropic  interactions, motion of molecules should be restricted. This can be done in several ways. One method is based on the use of alignment media, such as stretched polymer gels \cite{Luy2005_StretchedGels} or lyotropic liquid crystals \cite{Thiele2008_RDC}. These approaches have been carefully studied as the RDCs extracted from the high-field NMR spectra of such partially oriented samples are directly related to the distances between atoms and their mobility in molecules \cite{Tjandra1997,Prestegard2004}. Another method is based on the use of diastereomorphous interactions between chiral analyte and chiral alignment media, which in turn can be observed through the RDCs \cite{Kobzar2005}.
%This is the most-straightforward way to obtain structural information from NMR data, making it possible to study even highly mobile proteins in solutions \DB{17}. Furthermore, chiral liquid crystals or gels can be used to perform enantiodifferentiation \DB{[refs]}. This method is based on different diastereomorphous interactions between chiral analyte and chiral alignment media, which in turn are observed through the RDCs.}

Apart from alignment media, external electric fields can be used to partially order molecules in solutions \cite{bastiaan1987high,Riley2000}. Among the benefits of the latter approach is that allows gradual and reversible introduction of anisotropy, facilitating sensitive experiments.
In addition, with electric fields molecules can be oriented, not only aligned.

Distinguishing between alignment and orientation is important. \textit{Alignment} of molecules implies that they are preferentially ordered along a director axis, without distinction between `up’ and `down’ directions along that axis. \textit{Orientation}, on the other hand, implies a preferential direction along the director axis, and molecules are ordered either parallel or antiparallel relative to this axis.

%Apart from alignment media, external electric fields \DB{[refs]} \KS{Augustine} can be used to partially order molecules in solutions. Among the benefits of the latter approaches is that they lead to less-truncated nuclear spin Hamiltonians. In addition, with electric fields molecules can be oriented, not only aligned. 
%Distinguishing between alignment and orientation is of central importance for our work. Alignment of molecules implies that they are preferentially ordered along a director axis, without distinction between `up’ and `down’ directions along that axis. Orientation, on the other hand, implies a preferential direction along the director axis, and molecules are ordered either parallel or antiparallel relative to this axis. This differentiation is illustrated in \DB{Figure X1}, where different distributions of ordered vectors are shown. In order to characterize the degree of ordering along the ordering axis, Legendre polynomials are used. The Legendre polynomial of first order is a scalar product of a vector \DB{which vector?} and the unit vector along the ordering axis. It is nonzero for oriented vectors. It is zero for the aligned ones, unless orientation and alignment coexist in the sample. The degree of alignment is characterized by a Legendre polynomial of second order. 
\subsection{ZULF NMR of solids} 
\label{subsec:solids}

Zero-field NMR is ideally suited for studies of oriented samples including the use of electric field to achieve partial alignment and orientation (see Sec.\,\ref{Subsubsec:Chirality} for application in the direct determination of molecular chirality). In the ZULF NMR of solid-state samples, evolution under dipole couplings does not depend on the orientation of crystallites as there is no preferred direction in zero field, and the resolution of ZULF NMR spectra of powders can be comparable to that of high-field NMR spectra of single-crystal samples. Furthermore, ZULF NMR provides ways of observing all six terms of the ‘dipolar alphabet’ (see reference \cite{abragam1978principles} and Sec.\,\ref{Subsec:Untruncated}), whereas even in principle only one term for nuclei of different species is visible in high-field NMR (Fig.\,\ref{fig:Dipolar_alphabet}). ZULF-NMR measurements of the dipolar alphabet are thus expected to give additional (compared to high-field NMR data) dynamical information on spin evolution in the solid state.

One way to reintroduce dipolar couplings normally averaged in a liquid is to use stretched polymer gels. Using different degrees of polymerization, one can effectively smoothly transition between a liquid and a solid. In the context of ZULF NMR, such experiments were performed by the authors of \cite{Blanchard2015} who reported measurements of residual dipolar couplings in [2-$^{13}$C]-acetonitrile aligned in stretched polyvinyl acetate gels. They investigated dipolar couplings as a perturbation on the indirect spin-spin $J$-coupling in the absence of an applied magnetic field and observed terms of the dipole-dipole coupling Hamiltonian that are invisible in conventional high-field NMR. 

A conceptual approach to moving from ZULF NMR of fluids to that of solids, which complements the cross-gel technique, is the study of ZULF NMR of viscous liquids as a function of viscosity \cite{Shimizu2015}. In that work, ethylene glycol was used, a liquid with a sharp dependence of viscosity on temperature in the convenient range of 30--100$^{\circ}$C. 
Prospects for ZULF NMR of solids were recently revisited \cite{kurian2022solid}. The conclusions of the authors are optimistic despite the significant challenges associated with relatively broad spectral lines and the presence of dipole--dipole interactions.

%OLD text: Among all the methods mentioned above only the one based on electric field can orient molecules. Polar molecules have the tendency to be slightly oriented by the field, so that the direction of the dipole moment coincides with the applied electric field. Small organic molecules with a dipole moment between 1 and 5 D have orientation factors $ \langle cos \theta \rangle $ on the order of 10$^{-3}$–10$^{-2}$ in electric fields of 50 kV/cm (references ...). This is the optimal regime where the RDCs between neighboring spins become comparable to the \textit{J}-couplings (10–100 Hz) [ref]. Zero-field NMR is ideally suited for studies of oriented samples. Originally it was developed to observe high-resolution spectra of powders23. In zero field there is no preferred axis, so evolution under dipole couplings does not depend on the orientation of crystallites, and the resolution of ZULF NMR spectra of powders is comparable in resolution to that of high-field NMR spectra of single-crystal samples. Furthermore, ZULF NMR provides ways to observing all six terms of the ‘dipolar alphabet’, [ref] whereas even in principle only one secular term \AT{We decided to avoid the terms `secular' and `nonsecular' and use `commuting' and`noncommuting instead'.} (for nuclei of different species) is visible in high-field NMR (Figure X3). This gives additional information on molecular structure and dynamics [ref]. 

\section{Applications of ZULF NMR}
\label{Sec:Applications}

\subsection{Physics} 
\label{subsec:phys}

There are several directions where large-scale hyperpolarization is particularly important in fundamental science. Examples include spin-polarized nuclear targets and NMR-based searches for dark matter and the fifth force (e.g., spin gravity, Sec.\,\ref{Subsec:GDM}), and parity violation (Sec.\,\ref{Subsubsec:Parity}). There are several important parameters for this work, including, first and foremost, the total number of polarized spins. The density of polarized spins is characterized by molar polarization, which is a product of concentration and polarization. Other relevant parameters include the degree of polarization and the relaxation rates.

% \DB{Production of large quantities [Danila, John]} \DAB{I would simply delete this. Work \cite{blanchard2021towards} is discussed in depth in the HP section.}

\subsubsection{Polarized nuclear targets}
\label{Subsec:PolTargets}

Polarized targets for accelerator-based nuclear experiments prepared using PHIP (\autoref{Subsec:PHIP}) and specifically taking advantage of the ZULF regime were discussed in \cite{Budker2012PolTar}. There are multiple potential advantages for using hyperpolarized liquids in this regime. First of all, liquids undergo thermal convection and liquid targets are therefore particularly attractive when the target needs to take significant energy deposition from high-energy beams. For some experiments, near-zero magnetic fields are a regime of choice for extracting the relevant information from the scattering process. Finally, if one prepares a coherent superposition of ZULF states, the target polarization would undergo oscillation at the frequencies of $J$-coupling resonances. This latter feature can be thought of as a fast reversal, extremely useful for precision measurement and isolation of polarization-dependent effects. The use of nonhydrogenative parahydrogen-based hyperpolarization (SABRE) was also discussed in \cite{Budker2012PolTar} and is currently being pursued in conjunction with experiments at the MAMI accelerator facility at Mainz (B.\,Collins and S.\,Duckett, private communication, 2023). 

\subsubsection{Dark matter searches}
\label{Subsec:Dark}

While ZULF-NMR spectroscopy was originally developed with no apparent connection to fundamental physics, exciting applications in this area were introduced later
\cite{Ledbetter2012liquid,Wu2018}. 

Among these are searches for `ultralight bosonic dark matter' that produced significant exclusions of the possible dark-matter parameter space \cite{Garcon2019,Wu2019} (now further improved by other experiments). Dark-matter fields oscillating with periods as long as months or even years produce signals that are a priori hard to distinguish from slowly drifting spurious magnetic fields. 
% With ZULF NMR, this problem is solved using liquid-state co-magnetometry \cite{Ledbetter2012liquid,Wu2018}. In this technique, one that takes advantage of different ZULF NMR transitions of the same molecule having different relative sensitivities to dark matter (or other nonmagnetic entities) and magnetic field, enabling searching for exotic fields oscillating with periods of years \cite{Wu2019}.
% \JE{Last sentence may be changed to:
With ZULF NMR, this problem is solved using co-magnetometry \cite{Ledbetter2012liquid,Wu2018}: different NMR transitions of the same molecule can have different relative sensitivities to dark matter (or other nonmagnetic entities) and magnetic fields, allowing them to be distinguished. This has enabled the search for dark-matter fields oscillating with periods as long as years \cite{Wu2019}.

\subsubsection{Parity violation}
\label{Subsubsec:Parity}

One of the attractive features of ZULF NMR is the sensitivity to untruncated couplings (see \autoref{Subsec:Untruncated}). An intriguing application is the detection of `apparent chirality' or, more precisely, parity violation in intrinsically nonchiral systems such as atoms or diatomic molecules. This phenomenon originates in fundamental weak interactions responsible for nuclear beta decay. Parity violation manifests in a small mixing of eigenstates of opposite parity in atoms and molecules, an effect observed and well studied in atoms since 1970s, but so far eluding observation in molecules
\cite{Berger2019PNC,SBDJ18}. 

Of particular interest are the nuclear-spin-dependent parity nonconserving contributions to the molecular Hamiltonian.  These result in indirect nuclear spin--spin coupling and can be distinguished from the parity-conserving couplings in molecules of appropriate symmetry, including diatomic molecules \cite{Blanchard2020PNC}. The estimated magnitude of the coupling is in the millihertz range for molecules such as TlF \cite{Blanchard2023_TlF_PNC}. Couplings of this magnitude can be observed with ZULF NMR in liquid- or gas-phase samples (for molecules existing in gas/liquid form). However, it appears that the most promising approach is molecular-beam experiments with electrically oriented molecules \cite{Blanchard2023_TlF_PNC}.

A related application, detection of molecular chirality, is discussed in \autoref{Subsubsec:Chirality}.

\subsubsection{Search for spin--gravity coupling}
\label{Subsec:GDM}
% \JE{Does this experiment specifically involve ZULF NMR?} \DB{YES}

\begin{figure}
\centering
	\includegraphics{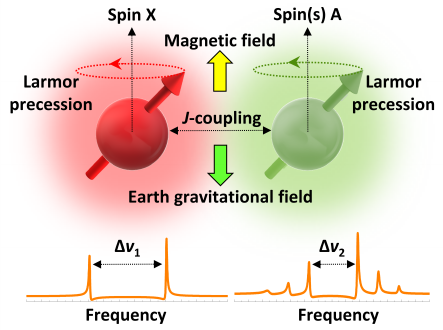}
	\caption{Fundamental-physics experiments such as  searches for the exotic spin--gravity couplings benefit from molecular nuclear spin co-magnetometry: the splittings of different Zeeman $J$-multiplets of the same molecule generally have different dependencies on the applied (ultralow) magnetic field and the new-physics coupling. This allows the effects to be disentangled via analysis of the overall spectrum \cite{Wu2018}. Reprinted from reference \cite{Blanchard2021_LtL}, Copyright (2021), with permission from Elsevier.
	}
	\label{fig:GDM}
\end{figure}

An ongoing experiment at Krak\'ow and Mainz is using ZULF NMR to search for the hypothetical nuclear gravitational dipole moment (GDM) that would violate a number of fundamental symmetries 
%, including parity (P), time-reversal invariance (T), and Einstein equivalence 
\cite{kimball2017constraints}. GDM may arise from exotic beyond-standard-model fields sourced by the Earth, in which case the `GDM' terminology would be a misnomer and one should rather talk about an exotic `monopole--dipole' coupling (having nothing to do with gravity), where monopole refers to the Earth and dipole refers to the spin. Another source is a modification of gravity beyond general relativity, in which case the term GDM would be accurate.

Figure\,\ref{fig:GDM} depicts the idea of the experiment.  The interaction of the GDM with the gravitational field is analogous to the interaction of a nuclear magnetic dipole moment with a magnetic field. The key to isolating the subtle GDM effect from systematics (e.g., magnetic field drifts) is the single-component co-magnetometry technique \cite{Wu2018} that has already proven itself in a search for dark matter \cite{Wu2019}. The `trick' is to simultaneously record multiple groups of spectral lines that depend differently on magnetic field and exotic physics.

Experiments of this type operate at the boundaries of sensitivity of ZULF-NMR experiments and require samples with the highest possible numbers (`a bucket') of polarized spins. The other requirement is the presence of an XA$_n$ group (e.g., XA$_3$).   Recent work at Krak\'ow employed on the order of a milliliter of neat $^{13}$C-methanol prepolarized in the field of a 1.5\,T Halbach magnet. Modern hyperpolarization techniques such as those based on parahydrogen (see Sec.\,\ref{Subsec:PHIP}) may not yet offer a significant improvement, due to dilution; however, brute-force polarization in a stronger field may yield several orders of magnitude in the number of polarized spins, enabling a new level of sensitivity in GDM searches.

\subsubsection{Quantum control}
While NMR of spin ensembles was once one of the central paradigms for the development of quantum computing, there is less effort in this direction today because it is not easy to scale an NMR platform and   
%\AT{To me, this sounds too much like guessing --- I think it's well established what the weaknesses are and I reckon that the most important one (violating DiVincenzo's first criterion) is the lack of scalability of the NMR platform (at least when working with molecules). The discussion seems to ignore that point at the moment.} because in NMR one is usually forced to work with so-called `pseudo-pure states' rather than well-defined quantum states of individual qubits. 
because working with pseudo-pure states necessitates the use of phase-cycling protocols with exponential time scaling \cite{Warren1997Usefulness}.
Nevertheless, NMR remains an active playground for research in quantum control, if technically as a `classical simulation of a quantum computer'.

As ZULF NMR represents a radically different modality compared to traditional high-field NMR, it is interesting to examine what opportunities may be afforded for quantum control in the ZULF regime. Let us say we wish to look at the simple two-spin qubit comprised by $J$-coupled $^1$H and $^{13}$C spins. 
Either of the spins can be manipulated by application of dc pulses; however, such pulses necessarily affect both spins at the same time. 
There are various ways to overcome this limitation and effectively address only one of the spins (see \autoref{Subsubsec:Selective}). 
One technique takes advantage of the ratio of gyromagnetic ratios of $^1$H and $^{13}$C being close to 4. Thus, for example, applying a $4\pi$-pulse to protons is an approximation of a $\pi$-pulse on the carbons. 
The residual differences can be compensated by employing composite pulse sequences \cite{Bian2017Universal}.

Experimental demonstration of universal quantum control in ZULF NMR was presented in reference \cite{Jiang2018}. 
The work utilized composite pulses for both
arbitrary one-spin rotations and a two-spin controlled-not (CNOT) gate in the heteronuclear ${}^{1}\rm{H}$-${}^{13}\rm{C}$ two-spin system. %, demonstrating universal quantum control (a complete set of two-qubit gates).
%\JB{I'm not quite sure what the following sentence is supposed to mean... Just saying that random benchmarking and process tomography were used to quantify gate fidelity?} \DB{Yes, we are just stating that high fidelities are possible.}
Using randomized benchmarking inspired by quantum information processing protocols and partial quantum process tomography,  single-spin control for $^{13}$C was achieved with an average fidelity of 0.9960(2) and
two-spin control via a controlled-not (CNOT) gate with a fidelity of 0.9877(2). Applications to materials science, chemical analysis, and fundamental physics were envisioned; however these have not been developed as yet. We are confident that quantum-information processing and quantum optimal control \cite{Ansel2024_QOC} can inspire methods to improve ZULF NMR spectroscopy.

\subsubsection{Quantum simulation}
\label{Subsubsec:quantumsim}

An intriguing development currently emerging in the field of quantum science and technology is the quantum simulation of NMR experiments \cite{demler2023quantum, babbush2022quantum, elenewski2024quantumsimulation}. In quantum simulation, quantum systems with constituents that can be, ideally, initialized, controlled and read out individually are used to emulate the quantum dynamics of other, less controllable systems. In an early example of the approach \cite{demler2023quantum}, a four-qubit trapped-ion quantum computer was used to simulate the zero-field NMR spectrum of the methyl group in acetonitrile \cite{Ledbetter2011near}.

The motivation for choosing ZULF NMR as an early test case was the complex dynamics even for relatively small biomolecules \cite{elenewski2024quantumsimulation}. Given the rapid progress in several quantum technologies and on quantum algorithms for Hamiltonian learning \cite{babbush2022quantum}, we expect that in the foreseeable future system sizes can be reached where classical numerical simulation of ZULF-NMR spectra is not practical. Experiments can then provide a direct means of verification. In the longer term, quantum simulations might guide the choice of molecules to be studied experimentally.

\subsection{Chemistry}
\label{Sec:Chemistry}

%\DAB{For applications in chemistry, we want a spectroscopic technique that (1) can distinguish various chemicals based on their resonance characteristics, (2) does not suffer terribly from  (such }

Generally speaking, for a spectroscopic method to deliver insight into the properties and behaviour of molecules, it should, first, yield detectable signatures that distinguish different molecules of interest, and, second, be applicable in `typical' environments in which chemical reactions and transformations take place. In high-field NMR, the spectroscopic signatures are typically chemical shifts. As these can be obscured by bulk susceptibility effects related to sample inhomogeneity, `good' samples for high-resolution NMR have to be homogeneous, or at least symmetric \cite{vanderhart2007}. Variations in the magnetic susceptibility---for example, discontinuities due to bubbles or precipitates in liquid samples---lead to distortions of the applied magnetic field and thereby degrade spectral resolution.

\begin{figure}
    \centering
    \includegraphics[width=\columnwidth]{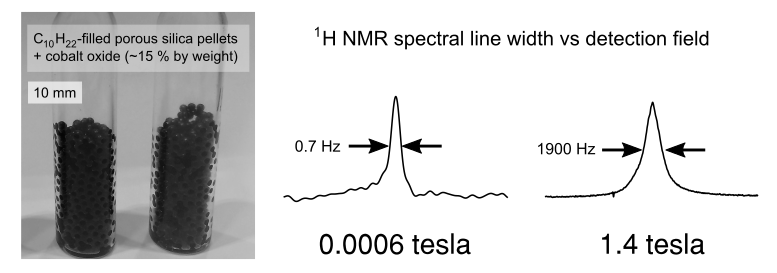}
    \caption{A comparison of magnetic susceptibility broadening in low and high-field NMR for a highly heterogeneous pellet sample, shown in the photograph.  The spectra on the right show the \textsuperscript{1}H-NMR resonance in \textit{n}-decane, which is  imbibed in the porous pellets; these comprise silica of mean diameter \SI{50}{\nano\meter} with a 15 wt.\% surface coating of cobalt (II/III). Spectra are detected in a field of \SI{600}{\nano\tesla} (sum of 64 scans, prepolarization field \SI{0.02}{\tesla}) using an atomic magnetometer and at \SI{1.4}{\tesla} (single scan) using a commercial NMR instrument. Reprinted (adapted) from reference \cite{Tayler2018}, with the permission of AIP publishing.}
    \label{fig:CobaltULFvsHF}
\end{figure}

By contrast, ZULF NMR spectra are largely unaffected by susceptibility, due to the absence of a strong applied magnetic field. This is true even for highly heterogeneous samples \cite{Blanchard2015, Tayler2018}, such as porous materials supporting paramagnetic catalyst species on the surface, see \autoref{fig:CobaltULFvsHF}. Such robustness against susceptibility broadening greatly helps relating information derived from ZULF NMR spectra to changes in molecular electronic structure.

In addition, as at typical frequencies used in ZULF NMR the skin effect is negligible even for metals---for example, the skin depth of aluminum at 1\,kHz is about 2.5\,mm, and for nonmagnetic 316 stainless steel it is more than a centimeter---applications become possible that are out of the reach for high-field NMR, including the monitoring of reactions inside metallic chemical reactors (see \autoref{Subsec:ReactionMonitoring}) and the study of electrochemical cells (discussed in \autoref{Subsec:Batteries}). 

Chemical reactions occurring in the ZULF regime exhibit different nuclear spin dynamics compared to working at high field. At high magnetic field, the Zeeman interaction dominates. When observing, for example, molecules undergoing a chemical transformation, the chemical shifts and $J$-couplings change, but these changes are on the scale of hertz to kilohertz and are small compared to the Larmor precession frequencies of hundreds of megahertz. Nonetheless, these relatively small changes are sufficient for NMR to be a powerful tool for chemical fingerprinting and identification. In zero to ultralow field the situation is different: the $J$-coupling interaction dominates and the Zeeman interaction is a perturbation to this, making chemical shifts negligible. In this case, the relative effect of changing $J$-couplings, by a chemical reaction, on the spectra is much greater, as the dominant interaction is modified. This situation is helpful for chemical identification as the spectra now depend  strongly on the $J$-couplings, but can have a negative impact on the polarization of molecules undergoing transformations at zero or ultralow field.

In order to understand this negative impact, consider nonequilibrium nuclear spin order in a molecule whose \textit{J}-coupling network is subsequently modified, for example by changing the couplings and/or adding or removing spins during a chemical reaction. In a ZULF-NMR experiment, such modifications can amount to a relatively large change in the spin eigenbasis. Molecular transformations typically occur much faster than the nuclear spin system can evolve, which means that they are nonadiabatic processes with respect to the nuclear spin dynamics. Therefore, if initial spin order is distributed across a spin network and it is then subjected to a change in the \textit{J}-couplings and spin topology of the network, spin order can be lost during subsequent evolution in the part of the molecule that experiences the \textit{J}-coupling change \cite{Barskiy2019}.  This is typically not the case at high field, where changes in \textit{J}-coupling values have almost no effect on the eigenbasis and, therefore, molecular changes do not lead to polarization loss. This potentially undesirable feature of ZULF NMR experiments can be addressed by maintaining a magnetic field around the sample during chemical transformations such that the polarization survives until zero-field read-out \cite{eills2023enzymatic}. Alternatively, one can exploit the chemical reaction itself to repolarize the spin system, for example, by using parahydrogen \cite{Burueva2020,VanDyke2022}.

\subsubsection{Study of chemical exchange}

ZULF-NMR spectra may display a strong dependence on chemical exchange---an important kind of chemical dynamics. Zero-field NMR spectra of deuterated solids may display effects of conformational exchange, for example two-site flips, allowing one to extract information about the dynamics \cite{Jonsen1986JCP85}. %For instance, in studies of 
In other studies using [$^{15}$N]-ammonium ions ($^{15}$NH$_4^+$) in aqueous solution as a model system, it was shown that pH-dependent proton exchange substantially alters ZULF-NMR $J$-spectra \cite{Barskiy2019}.  %In some cases, exchange can lead 
Well-resolved $J$-spectra can be observed for slow proton exchange rates in solution at pH 0 and below, allowing for measurements with unprecedented precision and in particular the analysis of $^1$H-$^{14}$N and $^1$H-$^{15}$N $J$-coupling ratios \cite{picazo2024zero} (see Sec.\,\ref{Subsec:Premet}).

\subsubsection{Chemical reaction monitoring}
\label{Subsec:ReactionMonitoring}
%\DAB{The promise of ZULF NMR is becoming reality with several recent developments. In \cite{Burueva2020}, heterogeneous/biphasic chemical reactions were monitored by ZULF NMR,
%exploiting the fact that magnetic susceptibility broadening is insignificant in this regime.
%In this experiment, hydrogenation of dimethyl acetylenedicarboxylate with para-enriched hydrogen gas followed by hydrogenation of the product (dimethyl maleate) was studied while parahydrogen was continuously bubbled through the sample \ref{fig:PHIP_RM}.
%Measurements were carried out both in conventional glass NMR tubes and in a titanium tube, where, due to the low NMR frequencies, there was no significant signal attenuation due to shielding by the electrically conductive sample container. This opens up the possibility of monitoring catalytic reactions in practical operational conditions.}

Low-field NMR can be used for reaction monitoring, carrying many of the advantages of high-field NMR but with greater instrument portability and magnetic field control, for example, in relaxometry measurements \cite{dalitz2012process}. The fact that low-frequency NMR signals pass through electrically conductive materials (including metals) and the absence of line broadening at low fields from magnetic susceptibility heterogeneity make the ZULF regime particularly promising for NMR-based reaction monitoring in industrial reactors. The authors of reference \cite{Burueva2020} demonstrated monitoring of biphasic chemical reactions with ZULF NMR.
In this experiment, the two-step hydrogenation of dimethyl acetylenedicarboxylate to dimethyl maleate to dimethyl succinate with para-enriched hydrogen gas was studied while continuously bubbling parahydrogen through the sample (see Fig.\,\ref{fig:PHIP_RM}).
Measurements were carried out both in conventional glass NMR tubes and in a titanium tube, to more closely replicate practical operational conditions. A consideration for process monitoring with ZULF NMR is that the sample needs to be polarized for detection.

In another example, low-field NMR relaxometry measurements were carried out on tomato samples in 1000 L, metal-lined containers, to observe the spoiling of the tomatoes over time \cite{pinter2014towards}. The containers were lined with 75\,µm of aluminium, which would act as an rf shield for typical high-field NMR experiments. By carrying out experiments in a $B_0$ field of $\sim$130\,mT, the proton 
%NMR signal 
Larmor frequency was $\sim$5\,MHz; at this frequency the skin depth of aluminium is $\sim$40\,µm, which was sufficient for signal penetration.

Finally, many common metals including some alloys of stainless steel or even aluminum \cite{Tayler20192} are weakly ferromagnetic, imposing an additional magnetic field when used in proximity to NMR samples.  For ZULF NMR, this magnetic field may be large enough to perturb the signals, and also, if spatially inhomogeneous, degrade spectral resolution through gradient broadening.  To avoid these problems, metal sample chambers or reactors should be kept as magnetically uniform as possible.  Preferably they should be made from metals without ferromagnetic alloying elements, such as high-purity titanium.
Ferromagnetic materials may be degaussed prior to experiments to remove residual magnetization, although the dc pulses used to manipulate nuclear spins in the sample may cause some degree of remagnetization. %This topic has not been investigated.

\subsubsection{Battery diagnostics}
\label{Subsec:Batteries}

Despite significant progress in rechargeable-battery technologies over the past decades, nondestructive measurement modalities compatible with realistic commercial batteries are limited \cite{ilott2018r}. While mapping battery susceptibility \cite{Hu2020} and
detection of small ($\mu$A) currents during charging/discharging cycles have been demonstrated \cite{romanenko2020accurate,Hatano2022high}, it was not possible to detect changes in electrolyte composition directly.

Recent work demonstrated measurements of a battery electrolyte
%in realistic volumes
performed with aluminium battery enclosures containing tens of $\mu$L of electrolyte \cite{Fabricant2024batteries}. Electrolyte concentrations as well as changes in composition and degradation were quantified with ZULF NMR. 
%The detection and assignment of electrolyte signals such that concentrations, as well as changes in composition and degradation, were quantified with ZULF NMR. 
NMR precession measurements conducted at 2.7\,$\mu$T revealed clear signatures of both protonated solvent EC/DMC and electrolyte LiPF$_6$. Further steps are expected in the direction of \textit{operando} monitoring of electrolyte degradation during the  charging--recharging cycles. While data acquisition times in \cite{Fabricant2024batteries} were on the order of a day (due in part to the small sample volumes used), with realistic improvements outlined by the authors (e.g., using a superconducting pre-polarization magnet), a practical diagnostic system for commercial batteries is within reach.  

A further discussion of applications of ZULF NMR in battery research and diagnostics can be found in reference \cite{Blanchard2021_LtL}.

\subsubsection{Detection of molecular chirality}
\label{Subsubsec:Chirality}

% \DB{Dima is taking a stab at finishing this Section; Some material may overlap/be redundant with the PNC section; we need to streamline this}

% \AT{Andreas will read and edit.}

One interaction that is suppressed by truncation at high magnetic field is the rank-1 antisymmetric $J$-coupling, a nuclear-spin analog of the Dzyaloshinskii--Moriya interaction \cite{Dzyaloshinsky1958,Moriya1960,Robert2019_MagnetoEl}. This interaction is connected to molecular chirality, and it was proposed that it might be observed in ZULF-NMR experiments \cite{King2017}. Such experiments would enable differentiation between enantiomers without adding any additional chiral agent to the sample. 

%The antisymmetric $J$-coupling is only observable in liquid or gas samples if the interaction tensor $\bm{J}^{(1)}$ has a nonzero projection along the electric dipole moment of a molecule, $\bm{d}$. One way to achieve this is to orient molecules in an external electric field. 
%\DB{This sentence is offending} If parity is conserved, then the observable $J^{(1)} =\bm{J}^{(1)}\cdot\bm{d}$ \DB{The notation is not so good here. Do we actually need a letter for the observable? If we do, let us not use $J$!} has to transform as a pseudoscalar, \DB{Let us discuss how to write this properly. 
%If parity is conserved, observables should be scalar, not PS} which means that it is odd under improper symmetry transformations such as inversion. This means that $J^{(1)}$ can only be measured in chiral molecules, as all achiral point groups contain improper symmetry elements. At the same time, as chiral molecules are related by reflection, the sign of $J^{(1)}$ carries direct information about the chirality of the molecule under investigation. 

A chiral object has two forms that are mirror images of each other that cannot be transformed into one another by 3D rotation. For chiral molecules, the mirror images are called R- and S-enantiomers. Many chemicals, notably amino acids and sugars, exhibit chirality, which is often of key importance for bioactivity. It is well known that almost all amino acids in organisms are left-handed, while sugars are predominantly right-handed; their left-handed counterparts are not metabolized. Chirality also defines chemical reactivity in many organic reactions, making this property important for organic synthesis.

Chirality is, however, difficult to probe. For instance, chirality itself does not affect conventional NMR spectra. This is because all the internal interactions commonly observed in NMR spectroscopy, isotropic and nonisotropic, are identical for both enantiomers. Sometimes chiral solvents or chiral derivatizing agents are used to induce measurable enantiomer-specific chemical shifts \cite{Yang2016_Chiral} or RDCs \cite{Berger2012_Chiral} yielding the chirality of a compound relative to a chiral addition. There are several proposals suggesting that molecular chirality can be observed directly by NMR via terms in the spin Hamiltonian arising from an electric-field-induced pseudovector. These proposals involve either detection of chiral nuclear magnetic shielding through an induced oscillating molecular electric dipole moment \cite{Pizzirusso2012_Alignment,Yang2016_Chiral,Buckingham2004_NMR_Chirality,Buckingham2015_Chiral,Garbacz2015_Chiral,Soncini2016_Chiral,Garbacz2016_loop_gap} or a magnetic dipole signal induced by an applied electric field \cite{Walls2008_Chirality,Walls2014_Chiral,Harris2006_Note_on_Chirality}. Chiral effects might also manifest in the electric polarizability of molecules with $J$-coupled nuclear spins \cite{Garbacz2016_with_Buckingham,King2017}.

% \DAB{All of these methods rely on detecting a nuclear-spin interaction that transforms as a pseudovector under 3D transformations. We will focus here on the antisymmetric part of the \textit{J}-coupling tensor. 
The general form of the \textit{J}-coupling Hamiltonian for two spins can be written as
\begin{equation}
\hat{\mathcal{H}}_J = 2 \pi \hbar\,  \hat{\bm{\mathrm{I}}} \cdot \bm{\mathrm{J}} \cdot \hat{\bm{\mathrm{S}}}\,.
\end{equation}
Here $\hat{\bm{\mathrm{I}}}$ and $\hat{\bm{\mathrm{S}}}$ are the vector spin operators of the interacting spins, and $\bm{\mathrm{J}}$ is the \textit{J}-coupling tensor. It is convenient to group the terms of this tensor according to their transformation properties under 3D rotations:
\begin{equation}
\bm{\mathrm{J}}=J^{(0)}+J^{(1)}+J^{(2)}\,.   
\end{equation}
Here, the first term is the isotropic part of the \textit{J}-coupling, which does not depend on the molecular orientation; it is a scalar quantity (a zero-rank tensor). $J^{(0)}$ is the part that is commonly observed in liquid-state NMR experiments. The $J^{(1)}$ and $J^{(2)}$ components, on the other hand, do depend on molecular orientation, and they always average out due to fast molecular tumbling in the liquid state. They can be measured, however, when the reorientation of the molecules is restricted. The $J^{(1)}$ component transforms as a vector upon rotations, corresponding to a first-rank tensor, whereas the $J^{(2)}$ component transforms as a second-rank tensor. The $J^{(2)}$ component has been studied and observed experimentally in NMR of molecules dissolved in liquid crystals \cite{Vaara2002} as well as, indirectly, inferred from hyperfine structure observed in molecular-beam or high-resolution microwave spectroscopy \cite{Bryce2000}.

Importantly, dipole--dipole interactions cannot produce a first-rank (vector) component of the interaction tensor, even in the presence of an applied electric field. Therefore, observation of the rank-one interaction is a signature of molecular chirality or parity violation. This can be seen in several ways \cite{Blanchard2020PNC,Blanchard2023_TlF_PNC}. Let us consider a general interaction between spins $\hat{\bm{\mathrm{I}}}$ and $\hat{\bm{\mathrm{S}}}$. The rank-zero (scalar) interaction should be proportional to the only scalar that can be built out of these vectors, $\hat{\bm{\mathrm{I}}}\cdot\hat{\bm{\mathrm{S}}}$, while the first-rank (vector) interaction should be proportional to $\hat{\bm{\mathrm{I}}}\times\hat{\bm{\mathrm{S}}}$. The latter term is a (pseudo)vector, and it can only enter the Hamiltonian as part of the scalar triple product with another vector available in the problem. In the presence of an external electric field, $\bm{\mathrm{E}}$, this vector is the average direction of the molecular axis, which is along $\bm{\mathrm{E}}$. With this, the triple product becomes
$a\bm{\mathrm{E}} \cdot \hat{\bm{\mathrm{I}}}\times\hat{\bm{\mathrm{S}}}$,
where $a$ is a proportionality constant specific for the system under study. The quantity $\bm{\mathrm{E}} \cdot \hat{\bm{\mathrm{I}}}\times\hat{\bm{\mathrm{S}}}$ is pseudoscalar, as it changes sign under spatial inversion (which transforms as $\bm{\mathrm{E}}\rightarrow -\bm{\mathrm{E}}$, $\hat{\bm{\mathrm{I}}}\rightarrow \hat{\bm{\mathrm{I}}}$, and $\hat{\bm{\mathrm{S}}}\rightarrow \hat{\bm{\mathrm{S}}}$).  In the absence of parity violation, the only way for the Hamiltonian term to be parity-even is for $a$ to be a pseudoscalar, which is possible in the case of chiral media. We note that an external magnetic field cannot substitute for an electric field because the quantity $\bm{\mathrm{B}} \cdot \hat{\bm{\mathrm{I}}}\times\hat{\bm{\mathrm{S}}}$ is odd under time-reversal.

A practical and important consequence of this symmetry argument is that there is a distinct way, at least in principle, to distinguish the chirality and parity-violation interactions from larger terms in the Hamiltonian. The sought-after effect should behave in a distinct manner with respect to reversals, for example, it should change sign under the reversal of $\bm{\mathrm{E}}$.

The existence of antisymmetric \textit{J}-coupling, which transforms as a first-rank spherical tensor under symmetry operations, was predicted more than four decades ago \cite{Buckingham1982_Symmetry}. However, so far this interaction has never been observed, despite experimental efforts to do so \cite{Harris2009nmr}. 
% Experimental observation of the similar antisymmetric spin–spin couplings is well-known for electrons [ref] — the Dzyaloshinskii−-Moriya interaction.
Calculations show that in some cases the antisymmetric component $J^{(1)}$ can be substantial, of the same order, or even larger than the corresponding isotropic part of the \textit{J}-coupling. Recently, it was found in quantum calculations for a series of molecules \cite{Garbacz2018_Computations} that the antisymmetric component of the spin--spin coupling tensor has a magnitude on the order of a few hertz (or possibly even orders of magnitude larger \cite{Garbacz2021_Chiral_Discr})  for commercially available chemical compounds. The largest values of the antisymmetric \textit{J}-coupling are expected between the heteronuclear pairs containing $^{19}$F or $^{17}$O. In ZULF NMR, such \textit{J}-couplings between heteronuclei can be detected, but they are truncated
% secularized \AT{We decided to avoid the terms `secular' and `nonsecular' and use `commuting' and`noncommutiong instead'.} 
at high field.

The antisymmetric \textit{J}-coupling tensor can be represented as a 3$\times$3 matrix in Cartesian coordinates:
\begin{equation}
J^{(1)}=\left (
    \begin{matrix}
     0 & J_{xy} & J_{xz}\\
    -J_{xy} & 0 & J_{yz} \\
    -J_{xz} & -J_{yz} & 0
\end{matrix} 
\right ) \,.
\end{equation}
A 3D transformation given by a matrix $\bm{\mathrm{T}}$ transforms any tensor as $\bm{\mathrm{T}}^{-1} J^{(1)} \bm{\mathrm{T}}$; the three independent components ($J_{xy}$,$J_{xz}$,$J_{yz}$) transform as coordinates of a vector upon rotation. These components behave as components of a pseudovector under reflections.
%, as illustrated in Fig. X4. 
The component perpendicular to the mirror plane is conserved, whereas the other components are inverted. Therefore, different enantiomers of a molecule possess differently oriented antisymmetric components, whereby they can be distinguished. 
As discussed in Sec.\,\ref{Subsec:Partial}, an electric field can orient polar molecules. In this situation, the $J^{(1)}$ component that is parallel to the molecular dipole moment is not averaged to zero and can be detected. For a pair of coupled spins, 
%(represented in green and blue in Figure X4), 
the component parallel to the orienting axis 
%(here $J_{xy}$) 
has different signs for R- and S-enantiomers. This causes opposite phases of the ZULF-NMR signal for the enantiomers \cite{King2017}.
Experiments to check these predictions are currently under way in several laboratories.

\subsection{Precision metrology}
\label{Subsec:Premet}
%\DB{Please read the whole section}

An advantage of ZULF NMR is in the intrinsically narrow resonances, as the spectral lines are not broadened by field inhomogeneity. Typically, the statistical uncertainty in determining the spectral line position is given by the width of the line divided by the SNR of the signal. In the case of ZULF NMR it has been possible to push the precision of the line-position determination and the corresponding extraction of $J$-coupling parameters down to  the sub-millihertz level \cite{Wilzewski2017}. Such levels of uncertainty opens up possibilities to study the subtle effects on spin--spin couplings of environmental effects such as temperature, solvent composition, and viscosity \cite{Shimizu2015}. 

Another interesting direction is the study of isotopic effects in spin--spin coupling. The strength of spin--spin coupling between nuclei of light elements is mainly determined by the Fermi contact interaction (i.e., a second-order hyperfine interaction effect based on finite probability of finding an electron inside a magnetic nucleus \cite{Ramsey1953_J-coup}). For example, for amino groups and ammonia, the main difference in the experimentally measured ${}^{14}$N-H and ${}^{15}$N-H $J$-coupling values is expected to be due to the Fermi contact interaction. 
In particular, the coupling strength between $^1$H and $^{14}$N/$^{15}$N spins is expected to be directly proportional to the product of the corresponding gyromagnetic ratios (assuming identical electronic structure).
However, a careful analysis of the ZULF-NMR spectra of [$^{14}$N]- and [$^{15}$N]-ammonium revealed that the $| J_{\rm 15NH} / J_{\rm 14NH} |$ ratio differs from that of gyromagnetic ratios, and this difference is statistically significant (Fig.\,\ref{Fig:NH4}). 
From 36,000 measurements of a NH$_4$Cl sample containing 50:50 mixture of $^{15}$N and $^{14}$N isotopes, the ratio of the $J$-couplings, ($| J_{\rm ^{15}NH} / J_{\rm ^{14}NH} |$), was determined to be within the range of 1.4009--1.4013, as compared to the literature data for gyromagnetic ratios $|\gamma_{\rm ^{15}NH} / \gamma_{\rm ^{14}NH} | =  1.4027548(5)$ \cite{stone2005table}. A specific value of $| J_{\rm 15NH} / J_{\rm 14NH} |$ depends on the specific peaks used for such estimation and ranges from 1.4009(7) to 1.40129(9) \cite{picazo2024zero}.

The amplitude of the primary isotope effect may be defined as
\begin{equation}
\Delta J = \left( \frac{\gamma_{{}^{14}\rm NH}}{\gamma_{{}^{15}\rm NH}} \right)J_{{}^{15}\rm NH}-J_{{}^{14}\rm NH}\,,
\end{equation}
quantifying the gyromagnetic-ratio-independent effect of replacing ${}^{14}$N with ${}^{15}$N in a molecule. Experimentally,
%extracted $J$-coupling values, $\Delta J$ was estimated to be $ -58$\,mHz \cite{picazo2024zero}.
%\DAB{Using $J_{{}^{14}\rm NH}=52.4019(17)$ and $J_{{}^{15}\rm NH}= -73.4093(9)$, $\Delta J \approx -58$\,mHz.}
%\DAB{Honestly, because of many reasons mentioned in the paper, I would simply replace with: "Using extracted $J$-coupling values, $\Delta J$ was estimated to be $ -58$\,mHz \cite{picazo2024zero}".}
a primary isotope effect $\Delta J \approx -58$\,mHz was deduced by analysis of the proton--nitrogen $J$-coupling ratios \cite{picazo2024zero}.
The negative sign for $\Delta J$ is consistent with computational investigations for tetrahedral molecules \cite{Sergeyev1975,Jameson2007isotope}.

While $J$-couplings are well studied for bound molecules, the second-order hyperfine interactions are also manifest in unbound van der Waals complexes. This effect was first observed in a solution of hyperpolarized $^{129}$Xe and pentane \cite{Ledbetter2012observation}. Measurements were performed with SQUID detection in a field of 1\,µT. The shifts of xenon and proton NMR lines were analyzed and the average value of the $J$-coupling was found to be statistically significant, $-2.7(6)$\,Hz.

\subsection{ZULF NMR to prepare specific hyperpolarized spin states}
\label{Subsec:ZULFpreparation}
A common use of the ZULF regime is to manipulate hyperpolarized spin states within molecules, which can then be used in applications such as biomedical imaging and protein--ligand binding studies. This is particularly relevant for PHIP-polarized molecules, as the \textsuperscript{1}H singlet order of parahydrogen is nonmagnetic and would not produce enhanced NMR signals after being incorporated into molecules. It has to be first converted into a magnetic state, and the apparatus for ultralow-field NMR is convenient for this task. Due to the large coils that encompass entire samples and provide high field homogeneity, the fidelity of state transformations is typically higher than is achieved in a high-field NMR system. The ZULF regime has also been used to manipulate polarization in dissolution DNP-polarized samples \cite{stern2023rapid}, but this is less common as the DNP process itself yields a magnetic state in the substrate (in contrast to PHIP).

As discussed in \autoref{Subsec:PHIP}, level anticrossings can exist between spin states at ultralow fields where Larmor frequency differences between different atoms approximately match molecular $J$-couplings. These level anticrossings can be used to transform and manipulate the spin order of nuclei in molecules, which enables the creation of magnetization (magnetogenesis) from the nonmagnetic starting state of parahydrogen. In the case of PHIP, it is typical to use adiabatic passage through level anticrossings to transform \textsuperscript{1}H singlet order into \textsuperscript{1}H and \textsuperscript{13}C/\textsuperscript{15}N magnetization \cite{goldman2005hyperpolarization,eills2019polarization,marshall2023radio}.  
In SABRE experiments, it is common to carry out the SABRE process at ultralow field to match the field to a level-anticrossing condition, so that polarization is spontaneously generated on the heteronuclei \cite{Truong2015}.
Reference \cite{nantogma2024MATRESHCA} provides a useful guide for constructing ULF NMR apparatus for the task of polarization transfer in liquids.

In NMR experiments it is usually not possible to observe the state of the nuclear spin system while driving evolution (as the detector saturates under strong driving fields), and therefore these methods to generate hyperpolarization have been optimized by experimental repetition. The experiment is repeated on a freshly polarized sample, and each time a parameter of the transfer sequence is incremented. The state that is generated by the sequence is then measured (usually at high field), and in this way the parameter space can be sampled, albeit slowly.

\begin{figure*}
\centering
	\includegraphics[width=0.95\textwidth]{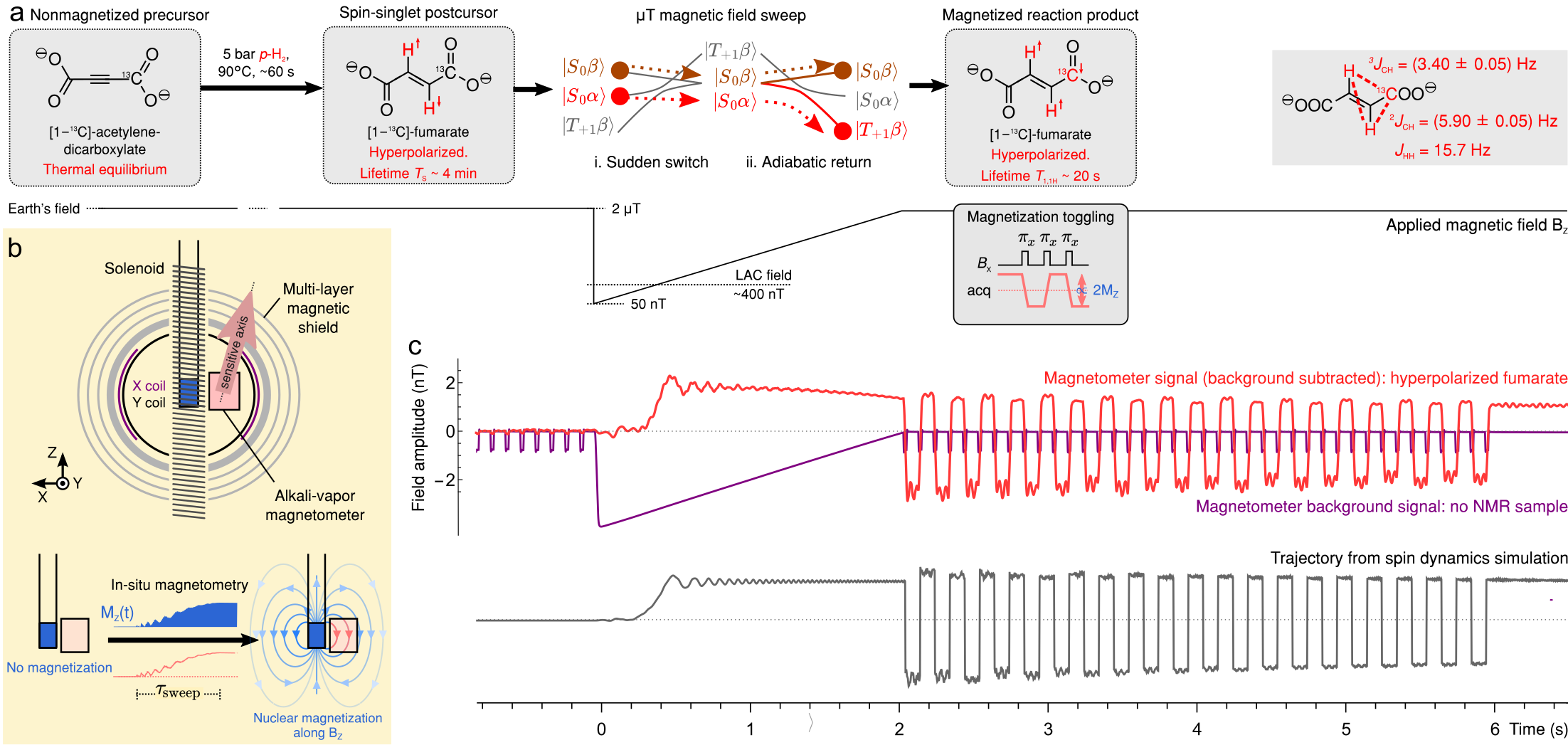}
	\caption{Observation of magnetogenesis in nuclear-spin-hyperpolarized molecules. 
    (a) Hyperpolarized fumarate is formed by chemical reaction of [1--\textsuperscript{13}C]-acetylenedicarboxylic acid with para-enriched H\textsubscript{2}, and a 0 to 2 µT magnetic field sweep is applied to convert the \textsuperscript{1}H singlet order ($\ket{S_0\alpha}+\ket{S_0\beta}$) into a state of positive \textsuperscript{1}H and negative \textsuperscript{13}C magnetization ($\ket{T_{+1}\beta}+\ket{S_0\beta}$).
    (b) An alkali-vapor OPM is situated next to the hyperpolarized [1--\textsuperscript{13}C]-fumarate (1\,mL, 60\,mM) sample, inside a multi-layer magnetic shield. A `piercing solenoid' coil produces the field sweep at the sample location, but this is attenuated by a factor of $\sim$500 at the position of the OPM.
    (c) The signal from the OPM during the magnetic field sweep, and the time before and after the sweep while toggling pulses are applied.
    Reprinted from reference \cite{EillsTayler2024} under terms of the Creative Commons CC BY-NC-ND license 4.0.
	}
	\label{fig:Magnetogenesis}
\end{figure*}

It was recently demonstrated that it is possible to directly measure the conversion of \textsuperscript{1}H singlet order into \textsuperscript{1}H and \textsuperscript{13}C magnetization in [1-\textsuperscript{13}C]fumarate during adiabatic passage using ZULF NMR.\cite{EillsTayler2024} This is illustrated in \autoref{fig:Magnetogenesis}. Using a piercing-solenoid arrangement to provide a magnetic field at the sample location but not at the position of the magnetometer (so the magnetometer remains sensitive throughout), it is possible to detect the magnetization of the NMR sample while applying a field sweep from 0 to \SI{2}{\micro\tesla} to induce singlet-to-magnetization conversion. The geometry of the ZULF apparatus is shown in \autoref{fig:Magnetogenesis}b. In this example, toggling pulses are applied before and after the field sweep using composite $\pi$-pulses (90\textsubscript{{X}}180\textsubscript{{Y}}90\textsubscript{{X}}) selective to \textsuperscript{1}H, to measure the \textsuperscript{1}H magnetization without background-field contributions. Before the field sweep, essentially no magnetization is observed during toggling. During the linear field sweep, the sample magnetization is observed to increase rapidly as the applied field passes 400\,nT, corresponding to the level anticrossing. Following the field sweep, the toggling pulses reveal a sample magnetization of a few nanotesla as measured at the position of the magnetometer. These results are shown in \autoref{fig:Magnetogenesis}c. Related efforts are currently underway to characterize the spin dynamics in the SABRE-SHEATH hyperpolarization process by SQUID-detected ZULF NMR \cite{Myers2024PRB,myers2024}.

ZULF methods are not limited to solution-state samples: although less common, in the solid state, low-field thermal mixing (LFTM) is used to obtain polarization transfer between heteronuclei. For LFTM, the external magnetic field is temporarily reduced (typically for millisecond timescales) so that different spin species within a solid become strongly coupled (i.e., the dipolar couplings dominate over Larmor frequency differences), and polarization can diffuse from one nuclear species to another \cite{cherubini2003hyperpolarizing,hirsch2015brute,peat2016low}. 

\subsection{Biomedicine}
\label{Subsec:Biomed}

Magnetic resonance has played an important role in medicine since the development of MRI in the 1970s and 80s, for both diagnostics and providing image contrast to see inside the body. This form of radiology is attractive because it is noninvasive, does not use ionizing radiation, and can be both quantitative and chemically specific. There is a wide range of applications such as structural imaging, measuring diffusion, real-time imaging of organs, and more recently metabolic MRI where metabolism is observed in real time using hyperpolarized tracers (e.g., [1--\textsuperscript{13}C]-pyruvate). 
There has been a recent drive towards lower-field, portable MRI systems \cite{o2019three,he2020use}, to make this tool more widely accessible for clinicians, particularly in less economically developed parts of the world. Zero and ultralow-field NMR (as defined in this review) has not yet been employed for animal or human imaging, but given the inherent portability and low cost of ZULF spectrometers and recent progress in ZULF-imaging, it may in the future be used for biomedical imaging.

\subsubsection{Studies with biomolecules}
\label{Subsec:biomolecules}

Given the portability and noninvasive nature of ZULF NMR, it is well-placed for studies of biological systems, although with the caveat that at present the sensitivity needs to be boosted using hyperpolarization. Though to date no observations have been made of molecules reporting on changes or flux in biological systems, a number of papers have explored ZULF NMR for the detection of different biomolecules.

One example was the detection of different isotopologues of urea, to understand how the zero-field NMR spectra change depending on proton exchange and isotope labelling \cite{Alcicek2021}.
[$^{15}$N$_2$,$^{13}$C]-urea and [$^{15}$N$_2$]-urea were measured in water, at varied pH, to understand the effect of chemical exchange on the spectra, and simulations supported the results. Additional experiments were carried out with varied deuteration level of the solvent, which produced different spectral patterns due to the changing spin topology.

Hundreds of biomolecules used in fragment-based drug discovery can be hyperpolarized by photo-CIDNP to perform drug-screening experiments \cite{Torres2023JACS, Stadler2023}. 
A first step in this direction was made when benzoquinone polarized via photo-CIDNP was detected using an OPM at low field \cite{Chuchkova2023}. For these experiments the benzoquinone target was at 5\,mM concentration, and the polarization obtained by photo-CIDNP was approximately 0.01\%.
With further advances in instrumentation, such as parallel detection of multiple samples, and with high-field prepolarization in excess of 10\,T, such screening methods have the potential for high throughput even without hyperpolarization \cite{Andrews2024arxiv}.

An important class of biomolecules is that of phosphorus-containing compounds, since these play a vital role in biological functions such as pH homeostasis, cell metabolism and bone mineralization. A first step towards ZULF-NMR characterization of organophosphorus compounds was made when a set of such molecules were measured at zero field, after thermal prepolarization in a 1.8\,T magnet \cite{Alcicek2021Ogranophosphorus}. It was shown that subtle $J$-coupling differences of less than a hertz between, for example, trimethyl phosphate and trimethyl phosphine, can be detected to differentiate between the species.

The library of ZULF spectra of biomolecules was expanded further by the use of dissolution DNP to polarize metabolites for ZULF detection, including [2-\textsuperscript{13}C]pyruvate, [2-\textsuperscript{13}C]acetate and [1-\textsuperscript{13}C]glycine \cite{Barskiy2019,picazo2022dissolution}. Spectra have also been obtained for samples of thermally-polarized [1-\textsuperscript{13}C]glycine, [1,2-\textsuperscript{13}C\textsubscript{2}]fumarate and [1-\textsuperscript{13}C]-D-glucose \cite{Put2021}.

Furthermore, metronidazole and fampridine molecules were polarized with SABRE and detected with ZULF NMR \cite{Burueva2024}. This is significant because these molecules are regulatory-approved drugs, fampridine being a specific blocker of potassium ion channels used to treat patients with multiple sclerosis, while metronidazole is used to combat amoebal infections.

A goal of ZULF NMR has been to detect \textsuperscript{13}C-hyperpolarized metabolites used for \textit{in vivo} metabolic imaging in animals and humans. There are only a small number of compounds used in this way, and two key examples are the injection of hyperpolarized [1-\textsuperscript{13}C]pyruvate, which is converted into [1-\textsuperscript{13}C]lactate (and sometimes detectable quantities of [1-\textsuperscript{13}C]alanine and [\textsuperscript{13}C]bicarbonate), and the injection of hyperpolarized [1-\textsuperscript{13}C]fumarate, which is converted into [1-\textsuperscript{13}C]malate. At the time of writing, no \textit{in vivo} imaging or spectroscopy in the ZULF regime has been performed, but the metabolic reactions described have been observed in NMR tubes at both zero and low field \cite{eills2023enzymatic}. The metabolites [\textsuperscript{13}C]pyruvate and [\textsuperscript{13}C]fumarate were formed via PHIP, and enzymes were used for the metabolic reactions to obtain sufficient concentrations for detection of the downstream metabolites [1-\textsuperscript{13}C]lactate and [1-\textsuperscript{13}C]malate.

\subsubsection{Dynamic nuclear polarization for \textit{in vivo} studies}
\label{Subsubsec:DNPvivo}
%\JE{This section belongs in Applications}

The work employing DNP for \textit{in vivo} as well as proof-of-principle studies can be generally divided into two categories, one employing Overhauser DNP for enhancing solvent (typically, water-proton) signals \cite{Buckenmaier2018}, and another employing $d$DNP in combination with detection employing OPMs \cite{picazo2022dissolution,Mouloudakis2023JPCL} or SQUIDs (to be demonstrated). 
%\DB{Colleagues, have you maybe heard of any work in this direction?}

Applying ZULF NMR for \textit{in vivo} molecular imaging and/or spectroscopy in combination with DNP methods seems reasonable \cite{barskiy2023possible}. Since $d$DNP has been extensively used for both clinical and pre-clinical metabolic imaging for more than two decades \cite{Larsen2003,nelson2013metabolic}, 
sufficiently enhanced NMR signals can in principle be measured \textit{in vivo} with ZULF-NMR. Importantly, in ZULF NMR, not only can magnetization of selected spin groups  be detected (for example, not only $^{13}$C of carboxylic groups or $^1$H NMR signals of methyl groups) as in high-field NMR, but also spin orders between various nuclei within molecules, for example in AX$_2$ or AX$_3$ groups (see Sec.\,\ref{Subsec:XAn}). These spin orders can be prepared by varying $d$DNP preparation conditions \cite{tayler2012direct,bornet2014long}. Their lifetimes are expected to be different from state to state and to have pronounced magnetic field dependence at hypogeomagnetic conditions: this can be used as an additional source of  information about dynamics and function of molecules.

It is important to emphasize that currently the main driver of atomic magnetometer technology is magnetoencephalography (MEG). ZULF-NMR spectra have been collected using the same devices (see, for example, \cite{boto2018moving}). Detection of metabolic changes associated
with cognitive processes simultaneously with MEG is an interesting direction of research related to brain activity. Combination of ZULF NMR with MEG for simultaneous mapping of brain activity and brain chemistry could be used for detecting abnormalities and
improving our general understanding of cognition \cite{barskiy2023possible}.

We foresee applications of affordable and scalable ZULF NMR coupled with hyperpolarization to study chemical-exchange phenomena \textit{in vivo} and in situations where high-field NMR detection is impractical or impossible.

\subsection{Portability and fieldable devices}
\label{Subsec:fieldable}

%\DB{Talk about oil/gas exploration: mention the Chinese review and Paul Ganssle work; NMRduino\cite{TaylerNMRduino}}
A ZULF-NMR setup is fundamentally simple. 
One needs a magnetic shield to achieve the ZULF conditions and screen out environmental noise, but such a shield is more straightforward to construct and is more lightweight than a high-field NMR magnet (which requires a homogeneous magnetic field; note, however, the developments in portable magnets, for example, \cite{Danieli2010Small}). 
% \MCDT{This is debatable. There are some very compact, high-homogeneity permanent magnets out there (for example DOI 10.1002/anie.201000221), and these are similar in weight/size to current commercial magnetic shields.  The term `easier' is quite subjective.  Most likely the `easier' part is because of fewer parts, and the existence of companies who will fabricate these shields at a relatively low cost, because mumetal shields are used everywhere in industrial electronics, e.g. electric car chargers}.
The magnetic field readout can now be done with commercial magnetometers \cite{Blanchard2020,savukov2020detection,Put2021}, which have a size of typically 1\,cm$^3$ and sensor standoff distances of only a few millimeters. These sensors exist thanks to other application areas such as magnetoencephalography that have motivated their path to commercialization.
The greatest barrier to portability is in achieving nuclear spin polarization, though this is already possible through the use of permanent magnets (which in contrast to high-field NMR magnets do not need to be homogeneous) and sample shuttling.
Significantly larger signals are achievable using other hyperpolarization techniques, which are rapidly approaching the degree of maturity that would enable their use in such devices. 
\textit{In-situ} hyperpolarization of the sample, for example using PHIP, as demonstrated in \cite{Theis2011,Theis2012_NH_PHIP}, eliminates the need for sample shuttling, which further simplifies the setup. The particular advantage of PHIP for portability is that the polarization source (parahydrogen) can be transported in gas cylinders or in liquid form.

\begin{figure}
\centering
	\includegraphics[width=0.85\columnwidth]{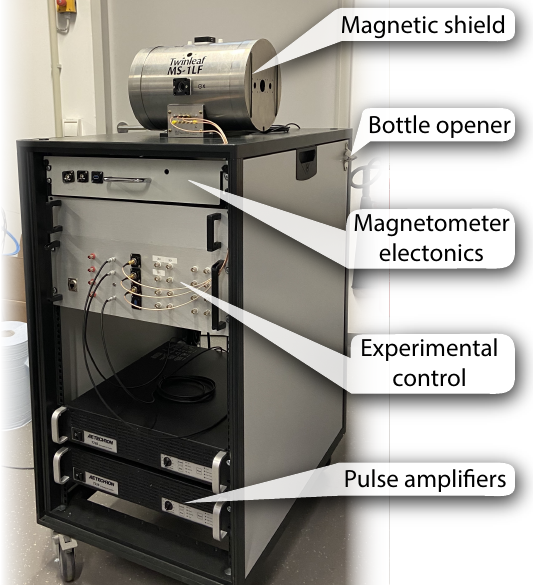}
	\caption{A mobile ZULF NMR spectrometer developed in Mainz.
%	The rack size was chosen to provide a convenient working height -- smaller implementations are straightforward.  \MCDT{This caption is exactly the same as in the original article (LtL). We should change it so it is not copied verbatim.} \AT{Done.}
   Reprinted from reference \cite{Blanchard2021_LtL}, Copyright (2021), with permission from Elsevier.
	}
	\label{fig:PortaZULF}
 	\includegraphics[width=0.9\columnwidth]{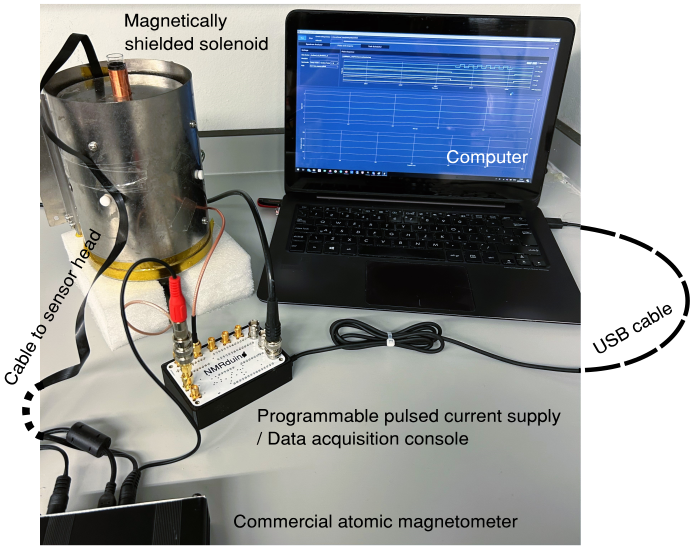}
	\caption{An even more portable ZULF NMR spectrometer currently operating in Barcelona \cite{TaylerNMRduino}.
	}
	\label{fig:PortaZULF2}
\end{figure}

By eliminating the requirement of high magnetic fields for NMR spectroscopy, one can imagine the possibility of developing a lightweight, portable instrument based on ZULF NMR, potentially much smaller even than those illustrated in \autoref{fig:PortaZULF} and \autoref{fig:PortaZULF2}.
The ultimate goal for portability and miniaturization may be `ZULF on a chip' along the lines of `NIST on a chip' \cite{Kitching2018}, towards which the first steps have already been made \cite{Ledbetter2008,Kennedy2017,eills2023enzymatic}.
Recent developments with chip-scale PHIP \cite{Eills2019Chip} should interface particularly well with ZULF-NMR detection.

The portability of ZULF devices makes them appealing for fieldable NMR, meaning carrying out spectroscopy outside the lab, in an applications setting. In many cases, the high resolution of a standard NMR system is not needed. For portable chemical characterization it is often sufficient to carry out relaxometry or diffusion measurements to measure the presence or quantity of a specific chemical. Examples would include measuring the oil content of grains, or the water content in a hydrocarbon mixture.
A fieldable NMR system was shown in 2019 when a helicopter-borne NMR coil weighing 1000\,kg and 6\,m in diameter was used to measure oil in jerry cans to imitate measurements under sea ice \cite{altobelli2019helicopter}.
A simple ZULF spectrometer operating with a 50 µT background field (to imitate that of the Earth) was also shown to have sufficient sensitivity and resolution to differentiate between water and different hydrocarbons \cite{Ganssle2014}. Relaxation measurements showed $T_1$ and $T_2$ differences between water, alcohols, and different hydrocarbons, and diffusion measurements could readily distinguish decane from water. The results serve as a proof-of-principle demonstration that low-field NMR may be a suitable fieldable spectroscopy for characterizing hydrocarbon mixtures in oil wells.

\section{Conclusions and outlook}
\label{Sec:Conclusions}

\begin{figure}
\centering
	\includegraphics{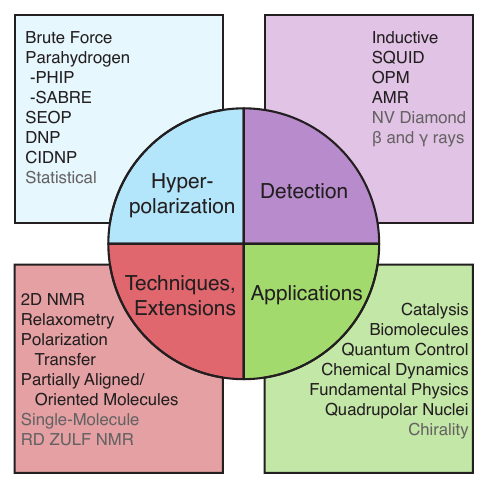}
	\caption{ZULF NMR at a glance. The techniques and extensions that have already been implemented are shown with dark font, while the ones that are in progress are shown in grey. }
	\label{fig:Summary}
\end{figure}

% \begin{itemize}

% \item Crystal ball

% \item Conclusions [Eills,Barskiy\ldots]

% \end{itemize}

We conclude our tour of ZULF NMR with a diagram (Fig.\,\ref{fig:Summary}) summarizing the status and some of the ongoing development directions in this relatively young field, which, however, has already established itself as a vibrant and rapidly developing subfield of NMR. As for all other NMR branches, progress in ZULF NMR depends on the development of hyperpolarization---a must for ZULF!---, encoding, and detection techniques. Naturally, we cannot know for sure which of the many directions discussed in this review article will turn out most fruitful, and what other interesting developments and application will eventually emerge. However, we can now say with certainty that ZULF NMR will be used and developed for years to come.
%However, we can now say with certainty that ZULF NMR will be used and developed for years to come.}

%\DB{In this article, we have discussed the roots, history, and motivation of ZULF NMR. We have also tried to foresee some of the areas where ZULF NMR is likely to be used in the future. 
%One cannot know for sure which of these directions will turn out most fruitful, and what other interesting developments and application will eventually emerge.
%However, we can now say with certainty that ZULF NMR will be used and developed for years to come. }

What makes us particularly confident in this is the fact that recent years have seen several dozen new doctoral students specifically trained in ZULF NMR. Here a quote from Felix Bloch, one of the founding fathers of NMR, is perfectly fitting. In his Nobel Banquet speech \cite{bloch1952banquet} he said:\\
\textit{``Our indebtedness to youth has for me two different aspects. One originates from the daily contact with my students: Their interest and enthusiasm have been a constant stimulus and a great source of inspiration and the spirit of my young collaborators has been an important factor in the success of our work. The other aspect is of more personal nature. I am sure my fellow-scientists will agree with me if I say that whatever we were able to achieve in our later years had its origin in the experiences of our youth and in the hopes and wishes which were formed before and during our time as students.''}

The authors of this article agree wholeheartedly.

% \subsubsection{Dynamic nuclear polarization (DNP) and in-vivo studies}
% \label{Subsec:DNPvivo}
% \DB{It looks like we will not actially have much af a crystal ball in this section; I propose moving this part to the end of the DNP section, what do you think?}

% \DAB{DAB will finish this section.}
% \DB{One interesting direction for ZULF NMR is combining it with a variety of hyperpolarization techniques (in addition to prepolarization in a strong magnetic field and parahydrogen-based methods), for example, with dynamic nuclear polarization. The first step in this direction was described in \cite{Barskiy2019}, which reported a ZULF-NMR spectrum of [2-$^{13}$C]pyruvic acid hyperpolarized via dissolution DNP. We foresee applications of affordable and scalable ZULF NMR coupled with hyperpolarization to study chemical exchange phenomena in vivo and in situations where high-field NMR detection is impossible or impractical.}

% \DB{Mention: \cite{de1999nmr,clarke2007squid}. These are references about SQUID detection.}

\section*{Acknowledgements}
We are grateful 
% to Prof.\,Alexander Pines for introducing us to unusual NMR, 
to Prof.\,Malcolm Levitt, Prof.\,Igor Koptyug, Dr.\,Seyma Alcicek, and Dr.\,Piotr Put for helpful discussions, and to Anne Fabricant and Jingyan Xu for performing the comparison of sensitivity between ZULF- and benchtop-NMR modalities.
This work was supported by the DFG (Project ID 465084791), by the DFG/ANR grant BU 3035/24-1, and by the Cluster of Excellence Precision Physics, Fundamental Interactions, and Structure of Matter (PRISMA+ EXC 2118/1) funded by the DFG within the German Excellence Strategy (Project ID 39083149). 
DAB acknowledges support from the Alexander von Humboldt Foundation in the framework of the Sofja Kovalevskaja Award. 
MCD Tayler acknowledges financial support from
%through the Junior Leader Postdoctoral Fellowship Programme from ``La Caixa'' Banking Foundation (project LCF/BQ/PI19/11690021).  
the Spanish Ministry of Science MCIN with funding from European Union NextGenerationEU (PRTR-C17.I1) and by Generalitat de Catalunya ``Severo Ochoa'' Center of Excellence CEX2019-000910-S; 
the Spanish Ministry of Science projects, MARICHAS (PID2021-126059OA-I00), SEE-13-MRI (CPP2022-009771) plus RYC2022-035450-I, funded by MCIN/AEI /10.13039/501100011033.
This project has received funding from the European Union’s Horizon 2020 Research and Innovation Programme under the Marie Sk\l{}odowska-Curie Grant Agreement 101063517.

\nomenclature{ZULF}{Zero to Ultralow Field}
\nomenclature{ULF}{Ultralow Field}
\nomenclature{NMR}{Nuclear Magnetic Resonance}
\nomenclature{NQR}{Nuclear Quadrupole Resonance}
\nomenclature{MRI}{Magnetic Resonance Imaging}
\nomenclature{EPR}{Electron Paramagnetic Resonance}
\nomenclature{SQUID}{Superconducting Quantum Interference Device}
\nomenclature{PHIP}{Parahydrogen-Induced Polarization}
\nomenclature{SEOP}{Spin-Exchange Optical Pumping}
\nomenclature{MEOP}{Metastability-Exchange Optical Pumping}
\nomenclature{SPINOE}{Spin-Polarization-Induced Nuclear Overhauser Effect}
\nomenclature{BSMR}{Berlin Magnetically Shielded Room}
\nomenclature{OPM}{Optically Pumped Magnetometer}
\nomenclature{PASADENA}{Parahydrogen and Synthesis Allow Dramatically Enhanced Nuclear Alignment}
\nomenclature{ALTADENA}{Adiabatic Longitudinal Transport After Dissociation Engenders Nuclear Alignment}
\nomenclature{SABRE}{Signal Amplification By Reversible Exchange}
\nomenclature{SABRE-SHEATH}{Signal Amplification By Reversible Exchange in SHield Enables Alignment Transfer to Heteronuclei}
\nomenclature{DNP}{Dynamic Nuclear Polarization}
\nomenclature{TEMPOL}{4-Hydroxy-2,2,6,6-tetramethylpiperidine-1-oxyl}
\nomenclature{TEMPO}{2,2,6,6-tetramethylpiperidine-1-oxyl}
\nomenclature{SNR}{Signal-to-Noise Ratio}
\nomenclature{CIDNP}{Chemically Induced Dynamic Nuclear Polarization}
\nomenclature{RP}{Radical Pair}
\nomenclature{BURP}{Broadband Uniform-Rotation Pure-Phase}
\nomenclature{CPMG}{Carr--Purcell--Meiboom-Gill}
\nomenclature{WAHUHA}{Waugh--Huber--Haeberlen}
\nomenclature{AWG}{American Wire Gauge}
\nomenclature{SERF}{Spin-Exchange Relaxation-Free}
\nomenclature{MEG}{Magnetoencephalography}
\nomenclature{MCG}{Magnetocardiography}
\nomenclature{NV}{Nitrogen Vacancy}
\nomenclature{AMR}{Anisotropic Magnetoresistance}
\nomenclature{NSOR}{Nuclear Spin Optical Rotation}
\nomenclature{RD-NMR}{Remote Detection Nuclear Magnetic Resonance}
\nomenclature{NMRD}{Nuclear Magnetic Relaxation Dispersion}
\nomenclature{FFC}{Fast Field Cycling}
\nomenclature{WBR}{Wangsness–Bloch–Redfield}
\nomenclature{WOLF}{Weak Oscillating Low Field}
\nomenclature{STORM}{Singlet-Triplet Oscillations Through Rotating Magnetic Fields}
\nomenclature{RDC}{Residual Dipolar Coupling}
\nomenclature{GDM}{Graviational Dipole Coupling}
\nomenclature{CNOT}{Controlled NOT}
\nomenclature{EC}{Ethylene Carbonate}
\nomenclature{DMC}{Dimethyl Carbonate}
\nomenclature{LFTM}{Low-Field Thermal Mixing}

%\AT{Below is a AI-generated list, with some dublicate entries}

\nomenclature{DC}{Direct Current}
\nomenclature{RF}{Radio Frequency}
\nomenclature{ZULF}{Zero to Ultralow Field}
\nomenclature{AC}{Alternating Current}
\nomenclature{FID}{Free Induction Decay}
\nomenclature{SNR}{Signal-to-Noise Ratio}
\nomenclature{SQUID}{Superconducting Quantum Interference Device}
\nomenclature{ODMR}{Optically Detected Magnetic Resonance}
\nomenclature{DNP}{Dynamic Nuclear Polarization}
\nomenclature{EPR}{Electron Paramagnetic Resonance}
\nomenclature{MRI}{Magnetic Resonance Imaging}
\nomenclature{QFT}{Quantum Fourier Transform}
\nomenclature{QIP}{Quantum Information Processing}
\nomenclature{QND}{Quantum Non-Demolition}
\nomenclature{MAS}{Magic Angle Spinning}
\nomenclature{AC}{Alternating Current}

\printnomenclature

%\section*{LtL Introduction: what is ZULF NMR?}

%It has been by now demonstrated in a formidable body of work that ZULF NMR experiments are, indeed, possible. But why would one want to do them? As we discuss below, there are various attractive features that make ZULF NMR largely complementary to its high-field counterpart, opening up new possibilities for NMR in fields ranging from chemistry and chemical engineering, to medical diagnostics, to fundamental physics and cosmology.

%We focus on addressing all parts of the canonical polarization--encoding--detection sequence, thus completely eliminating the need for strong magnetic fields. The work and perspectives discussed here therefore stand apart from the beautiful pioneering works concerned with individual periods of the triad, which includes stunning advances from Bernhard Bl{\"u}mich and his group over the years. 
%To name a few, these contributions cover work on NMR in the Earth-field range \cite{Bene1980,Mohoric2009}, most notably perhaps the demonstration of chemical analysis by ultrahigh-resolution NMR in the Earth's magnetic field \cite{Appelt2006} (whereas his contributions to MRI velocimetry \cite{Amar2010,Perlo2015} have, perhaps unexpectedly, roots in Earth-field NMR as well \cite{Hahn1960}), work on parahydrogen-enhanced low-field NMR \cite{Colell2013} as well as helping us understand the power of $J$-spectroscopy \cite{Ledbetter2011near}.

\bibliography{ZULF_Progress}

\appendix

\end{document}